\documentclass[english,3p,floatfix]{elsarticle}
\newcommand{\keywords}[1]{\textbf{Keywords:}\quad #1}
\usepackage{lineno,hyperref}
\usepackage{verbatim}
\usepackage{graphicx}
\usepackage[english]{babel}
\usepackage{epsfig}
\usepackage{color}
\usepackage{amsmath}
\usepackage{amssymb}
\usepackage{subfigure}
\usepackage{bm}% bold math
\usepackage{float}

\journal{Physics Reports}

\begin{document}

%\linenumbers
%\modulolinenumbers[2]
%\begin{frontmatter}

\title{Electronic phase separation: recent progress in the old problem}

\author[HSE,IFP]{M.Yu. Kagan}

\ead{kagan@kapitza.ras.ru}

\author[ITAE,HSE]{K.I. Kugel}

\author[ITAE,MIPT]{A.L. Rakhmanov}

\address[HSE]{National Research University Higher School of Economics, Moscow 101000, Russia}

\address[IFP]{Kapitza Institute for Physical Problems, Russian Academy of Sciences, Moscow 119334, Russia}

\address[ITAE]{Institute for Theoretical and Applied Electrodynamics, Russian Academy of Sciences, Moscow 125412, Russia}

\address[MIPT]{Moscow Institute of Physics and Technology (National Research University), Dolgoprudny, Moscow region 141700, Russia}

%\address[VNIIA]{All-Russia Research Institute of Automatics, Moscow 127055, Russia}

\date{today}

\begin{abstract}
We consider the nanoscale electronic phase separation in a wide class of different materials, mostly in strongly correlated electron systems. The phase separation turns out to be quite ubiquitous manifesting itself in different situations, where the itineracy of charge carriers competes with their tendency toward localization. The latter is often related to some specific type of magnetic ordering, e.g. antiferromagnetic in manganites and  low-spin states in cobaltites. The interplay between the localization-induced lowering of potential energy and metallicity (which provides the gain in the kinetic energy) favors an inhomogeneous ground state such as nanoscale ferromagnetic droplets in an antiferromagnetic insulating background.
The present review article deals with the advances in the subject of electronic phase separation and formation of different types of nanoscale ferromagnetic (FM) metallic droplets (FM polarons or ferrons) in antiferromagnetically ordered (AFM), charge-ordered (CO), or orbitally-ordered (OO) insulating matrices, as well as the colossal magnetoresistance (CMR) effect and tunneling electron transport in the nonmetallic phase-separated state of complex magnetic oxides. It also touches upon the compounds with spin-state transitions, inhomogeneous phase-separated state in strongly correlated multiband systems, and electron polaron effect. A special, attention is paid to the systems with the imperfect Fermi surface nesting such as chromium alloys, iron-based pnictides, and AA stacked graphene bilayers.
%We briefly discuss also the recent results on the formation of stripes and zig-zag structures in the Jahn--Teller systems and other systems with strong electron--lattice coupling, charge and orbital ordering.
\end{abstract}

\maketitle


\tableofcontents{}

%\end{frontmatter}

%\section*{Notation used in the text}
%
%\input{TEX_FILES/notation}

\section{Introduction. Spontaneously formed nanoscale inhomogenieties in different materials}
\label{Intro}

The phase separation is a universal and ubiquitous phenomenon, which is present not only in electronic liquid in metals, semiconductors~\cite{KeldyshPSSA1997}, heavy-fermion~\cite{CastroNetoPRB2000,ChandraNature2002} and high-$T_c$ superconductors, but also in different types of mixtures (Fermi--Bose, Fermi--Fermi, and Bose--Bose mixtures), in bismuthates (BaKBiO), liquid and solid solutions of $^3$He in $^4$He, mixtures of $^6$Li and $^7$Li, $^{40}$K and $^{87}$Rb isotopes, in spin-polarized (imbalanced) Fermi gases, and in two kinds of Bose gases and other atomic condensates in magnetic traps and on optical lattices, as well as in low-dimensional organic compounds and in mixed states (different pasta-like structure) in nuclear physics~\cite{CastellaniPRL1995,CastellaniJPhChSol1998,PeraliPRB1996,
Zaanen_stripes1989,MachidaPhysC1989,MachidaJPSJ1990,MachidaPRL2005,ZwierleinScience2006,MenushenkovJETP2001,
MenushenkovJScNMag2016,BashkinAdPh1981,KaganUFN1994,
MaruyamaPRC2005,MaruyamaPRC2006}. In condensed matter physics, it usually manifests itself either in the form of electronic phase separation (with the formation of paramagnetic spin bags or strings, ferrons, or stripes~\cite{KampfPRB1990,BulNagKhomJETP1968, BrinkmanRicePRB1970}, etc.) or the phase separation in ionic subsystems (e.g. charge density wave formation in superconducting cuprates such as LaBaCuO and LaSrCuO for some specific values of hole doping~\cite{CastellaniPRL1995,CastellaniJPhChSol1998,PeraliPRB1996,
TranquadaNature1995,BianconiPRL1996}). While the latter phenomena usually require a reconstruction of the crystal lattice or formation of the superstructures (e.g. doubling of the lattice period), the former one does not necessarily correspond to the instability of the crystal lattice. In our review, we are dealing with the electronic phase separation. Moreover, we will mostly consider the nanoscale phase separation in the electronic subsystem and only briefly discuss other types of electronic phase separation (such as a total phase separation giving rise to two large clusters, formation of stripes, etc.). We will also focus on strongly correlated electron systems with a strong interplay between kinetic and potential energy of electrons. The gain in kinetic energy favors delocalization and metallicity in the electron subsystem. At the same time, the potential energy promotes localization and insulating behavior (as e.g. in the Hubbard model at strong onsite Coulomb repulsion $U$ or in the $t-J$ model in the limit of large $J/t$ ratios~\cite{HubbardPrRoySocA1963,EmeryPRL1990,KaganRiceJPCM1994}).

We stress again that the instability of the ground state with respect to the nanoscale phase separation is an important feature characteristic of numerous strongly correlated electron systems. It corresponds, in particular, to the creation of small ferromagnetic (FM) droplets inside different insulating host materials. The host materials can be antiferromagnetic \cite{nagaev2001colossal,DagottoPhysRep2001,KaganUFN2001,KaganKhomPhB2000,
KaganUFN2003,KaganJPhA2003,KaganJPCM2006,KaganBookSpringer}, charge-ordered (CO)  \cite{KaganJETP2001,KaganFNT2001,KaganPhC2001,LorenzanaPRB_II_2001}, or orbitally ordered (OO) ones \cite{KugKhomUFN1982,KugSboKhomPRB2008orbitals} The FM droplets are randomly distributed within these hosts and are usually metallic ones. The shape of the droplets is usually a spherical or an ellipsoidal one~\cite{nagaev2001colossal,DagottoPhysRep2001,KaganUFN2003}. We can say that the droplets are indeed real nanoscale objects looking like raisins in a cake. Their radius is of the order of $R$ = 10--15 \AA, which corresponds to 2--4 interatomic distances). Inside droplets, there are typically 30--100 local spins, which are FM oriented and owing in fact to one additional conduction electron or hole. Hence, a FM droplet exhibits a large magnetic moment of the order of 15--50 Bohr magnetons. Usually, the phase-separated state in complex magnetic oxides (such as manganites, cobaltites, and nickelates) is a superparamagnetic one. This fact is due to the residual magneto-dipole interaction between the FM droplets. Meanwhile, the electron transport here is governed by the tunneling of conduction electrons between the neighboring metallic droplets through the tunneling barriers, which are formed in the insulating host. The tunneling usually corresponds to the regime of Coulomb blockade   \cite{RakhmanovPRB2001resistivity,RakhmanovFMM2001,SboychakovJETP2002tunMR,
SboychakovJMMM2003,kugel2004characteristics,SboychakovJPCM2003,
YmryBook2008,LikharevChapter1992,BienakkerChapter1996}, and it is spin-dependent, which could be important for the applications in spintronics.

In this review article, we will consider different types of free and bound magnetic polarons arising in AFM, CO, and OO matrices both in the isotropic and anisotropic cases on regular (cubic, quadratic) and frustrated (triangular) lattices. We ananalyze the thermodynamic and  transport (kinetic) characteristics of the phase-separated state including magnetization and magnetic susceptibility, resistivity, magnetoresistance, and the spectrum of $1/f$-noise in it \cite{KoganBook1996,DuttaRMP1981,ShklEfrBook1984,PodzorovPRB2000}.

As we have already mentioned, in cobaltites, for example, the phase separation is usually accompanied by the transitions between high- and low-spin states~\cite{SboychakovPRB2009}. For cobaltites, the electronic phase separation was considered not only in 3D and 2D systems but in quasi-one-dimensional magnets such as BaCoO$_3$ as well (see   \cite{GonzalezPRB2004,SboychakovGonzPRB2005,PardoAPL2005,CastroEPJB2004}).
These papers deal with the creation of magnetic polarons in spin chains with Heisenberg AFM interaction upon doping. For such 1D systems, it was possible to demonstrate the formation of rather long-range spin distortions around the polaronic core. The long-range tail of slowly decaying spin distortions (usually they decay in a power-law fashion) is also formed around the ferrimagnetic core of the magnetic polaron bound by donor impurity. This type of spin distortions arise in manganites at very low doping levels in 2D and 3D cases in the model, which combines the Hund's rule and Heisenberg type magnetic interactions (as well as the one-site magnetic anisotropy) with electron--ion Coulomb attraction potential between Sr (or Ba) impurity ion and conduction electron~\cite{OgarkovPRB2006,KugelOgarPhB2008}.

Another type of interesting mesoscopic phenomena in strongly correlated electron systems corresponds to the electron polaron effect (EPE). EPE usually takes place in two-band systems with strongly differing band masses. This situation is typical of many mixed valence electron systems  \cite{NewnsAdvPh1987,ColemanPRB1987,TsvelikBook2007,
VarmaPhB2006,FuldeBook2002,KeiterGrewe1981}, including uranium-based heavy fermion compounds. The most general model, which should be used here, is a two-band Hubbard model with one narrow band \cite{KaganCzJPh1996, KaganJETP2011,KaganValFNT2011,KaganSchoenBook2012,KaganValJScNM2012}. The EPE usually describes the strong enhancement of a heavy mass due to many body effects. They are predominantly related to the dressing of a heavy particle in a coat of low energy (infrared) electron--hole pairs of light particles and lead to a strong additional narrowing of the heavy particle bandwidth. The EPE has a lot of similarities with the physics of infrared divergences \cite{KondoPrThPh1964,IcheNozPhA1978,NozieresPR1969} for the Brownian motion of a heavy particle in the Fermi liquid of light particles \cite{KaganProkJETP1986,KaganProkJETP1987}). Another close analogy here is that with the Andeson physics of the orthogonality catastrophe  \cite{AndersonPRL1967,AndersonPR1967}, which describes a 1D chain of $N$  electrons in the presence of one impurity in the system.

The electron polaron effect not necessarily leads to the nanoscale phase separation (though this scenario is also possible, at least, at the level of the strong density redistribution for the large initial mismatch between the densities of heavy and light components in agreement with the predictions of  \cite{sboychakov2007phase}).

At the same time, in all the cases (with and without phase separation), the electron polaron effect  leads to the anomalous temperature dependence of the resistivity in 3D and especially in quasi 2D (layered) case providing possible alternative explanation to  the mechanism of spin dependent tunneling conductivity when we try to explain resistivity characteristics in manganese silicides \cite{DemishevJETPL2016} or layered manganites. In manganese silicides  MnSi and in the stoichiometric alloys Mn$_{1-x}$Fe$_x$Si, we have a very peculiar power-law (almost of the square root type) dependence  of the resistivity $R(T)$ (for temperatures in the range from 30 to 250 K) and magnetization $M(H)$ (in moderate and high magnetic fields from 2--4 T up to 50 Tesla) in the paramagnetic phase.

In our review, we also consider the role of a more standard electron--phonon polaron effect as well as a long-range part of Coulomb interaction  in the formation of different inhomogeneities and nanoscale phase-separated structures. Usually the electron polaron effect and polaron effect of the phonon origin play in concert leading to the multiplicative effect for the bandwidth narrowing. In this connection, it is important to mention the giant oxygen isotope effects in manganites~\cite{ZhaoNature1996,IsaacPRB1998,BabushkinaPRB2000} (see also \cite{TaldenkovJETP2018} and references therein) and in cobaltites~\cite{KalinovPRB2010}, which are closely related both to the polaron effect and phase separation. Moreover, the oxygen isotope exchange can even induce the metal--insulator transition in manganites~\cite{BabushkinaNature1998,ZhaoSSC1997,BabushkinaJAP1998}. Electron--phonon (electron--ion) interactions are responsible for the creation of different types of superstructures in manganites and other Jahn--Teller systems ~\cite{KhomKugEPL2001,KhomKugPRB2003,SboychakovPRB2011zigzag}. In manganites, two conductive bands (delocalized orbitals) of $e_g$ electrons are additionally split by sufficiently large Jahn--Teller gap, while three localized $t_{2g}$ orbitals form the local spin $S=3/2$ at them in accordance with the generalized Hund's rule.

For the physics of phase separation and the formation of different anisotropic  structures (stripes, zig-zags, spin ladders, etc.), a very important role is played by the orbital degrees of freedom (often related to the Jahn--Teller type effects), as well as by the relativistic (magneto-dipole or spin-orbit) interactions. The anisotropic magneto-dipole interaction could be also a source for the formation of anisotropic elongated structures (such as stripes) in many strongly correlated systems~\cite{KhomKugEPL2001}. Another possibility for a strongly anisotropic situation and formation of stripes arises when we take into account the interplay between nanoscale phase separation and charge ordering. At the same time, the formation of zig-zags and spin ladders  \cite{DagottoRiceSience1996,KaganRicePhC1999} takes place even when we consider stable crystallographic distortions.

An important role of orbital degrees of freedom also manifests itself in multiorbital systems such as transition metals compounds with several $e_g$ and $t_{2g}$ orbitals for $d$ electrons~\cite{KugKhomUFN1982}. In such a case, it is possible to formulate a spin--orbital model, which adequately describes entangled spin--orbital excitations in the system~\cite{KaganMikhJETPL2014}. The phase separation in Jahn--Teller systems with localized and itinerant electrons was analyzed in detail in  \cite{KugelPRL2005}.

Finally, in the last sections of this review, we consider the systems with the imperfect nesting  such as chromium-based alloys, iron-based superconductors, stacked AA graphene bilayers, hydrogenized graphene, and  graphene-based heterostructures  \cite{KatsnelsonNatPhys2006}. These systems are inherently unstable toward the electronic phase separation  \cite{RozhkovPhRep2016,WeImperf,Sboychakov_PRB2013_PS_pnict,
Sboychakov_PRB2013_MIT_AAgraph} already in the framework of the simplest Rice model~\cite{RicePRB1970}, which was originally proposed for the description of spin density waves in chromium-based alloys. This model captures the band structure effects in itinerant antiferromagnetism and contains repulsive Coulomb interaction between almost spherical electron and hole pockets of different radii (imperfect nesting).

An idealized graphene monolayer as well as an idealized  AB graphene bilayer  are also the subjects of anomalous Kohn--Luttinger type of superconductivity \cite{KohnLattPRL1965,KaganKorovUFN2015,KaganJETPL2016}. The same type of anomalous superconductivity and nanoscale phase separation with the formation of small metallic droplets inside the CO matrix (similar to phase separation in the Verwey model \cite{VerweyNature1939,VerweyPhys1941} for manganites) occurs in the strong-coupling limit of the Shubin--Vonsovsky model   \cite{ShubVonsPrRoySoc1934} for strong Coulomb interaction at nearest-neighbor sites, $V \gg t$,  and  $n=1/2$  (near the quarter filling) \cite{KaganEfrJETPL2011}. The Shubin--Vonsovsky model can be important for cuprates, iron-based superconductors, and idealized monolayer and bilayer graphene. The electron polaron effects are clearly pronounced in monolayer and bilayer graphene as well as in the other quasi-2D and 3D (Weyl) semimetals with the Dirac spectrum such as bismuth, for example  \cite{ValenzuelaNJP2008,MishchenkoPRL2007,RodionovPRB2015}. It is interesting here to study more thoroughly a combined effect of strong Coulomb correlations and disorder for the conductivity in this class of materials.

One of the main goals of our review is to describe the recent progress in the subject of nanoscale phase separation. It is quite surprising that a lot of intriguing results on this subject obtained during about fifteen last years have not been reflected yet in comprehensive review articles. Indeed, on the wave of the so called ``manganite boom'' at the end of nineties, there appeared several reviews and books~\cite{DagottoPhysRep2001,
nagaev2001colossal,KaganUFN2001,NagaevBook2002,DagottoBook2003,DagottoSci2005} related mostly to the phase separation in manganites, and nearly nothing after that. In this paper, we are trying to fill somehow such evident ``blank spot".

The structure of the review is as follows: it consists of 10 Sections including Introduction and Conclusions. In the next Section~\ref{MagPolaronTrans}, we consider ferromagnetic polarons (ferrons) in magnetic semiconductors and manganites as well as the colossal magnetoresistance, the tunneling conductivity, and $1/f$-noise spectrum in the phase-separated nonmetallic state of these systems. In Section~\ref{ModelSys}, we analyze the shape and internal structure of free and bound magnetic polarons, which arise in different model systems in 3D, 2D, and 1D cases on regular and frustrated lattices. In Section~\ref{ChargeOrder}, we discuss the interplay between charge ordering and nanoscale (as well as large-scale) phase separation in the Verwey and Shubin--Vonsovsky models. In Section~\ref{twobands}, we consider charge density redistribution in strongly correlated multiband systems, especially having in mind the possible application to cuprates and heavy-fermion systems as well as to the systems near the topological Lifshitz phase transitions. In this section, we also consider electron--lattice interactions, nonuniform distortions, and the formation of the superstructures in manganites and other Jahn--Teller systems. At the end of this section, we consider EPE and anomalous resistivity in the two-band model with one narrow band. In Section~\ref{SpinState}, we analyze magnetic instabilities and spin-state transitions in cobaltites and other complex magnetic oxides. In Section~\ref{Orbitals}, we discuss the important effects of orbital degrees of freedom in transition metals and oxides as well the formation of the orbital polarons, stripes, and zig-zag structures in the framework of degenerate two-band Hubbard model and orbital $t-J$ model.  In Section~\ref{imperfect}, we describe the formation of the electronic phase separation in chromium alloys, iron-based superconductors, and other systems with the imperfect nesting. In Section~\ref{graphene}, we are dealing with the phase separation in graphene-based systems with a special emphasis on AA graphene bilayers and hydrogenated graphene. %In Section 10, we discuss the electron polaron effect and anomalous resistivity characteristics in mixed-valence systems described by the two-band Hubbard model with one narrow band.
In Section~\ref{Concl}, we present our conclusions and discuss some interesting unsolved problems. %The extensive Reference list covers not only the important papers of our groups but also the essential papers of  other groups on the fascinating subject of the electronic phase separation.

\section{Magnetic polarons and related transport phenomena}
\label{MagPolaronTrans}

\subsection{Magnetic polarons in magnetic semiconductors and manganites}
 \label{MagPolarons}

%\subsection{Introduction}\label{Intr1}

\begin{figure}[H] \begin{center}
\includegraphics*[width=0.5\columnwidth]{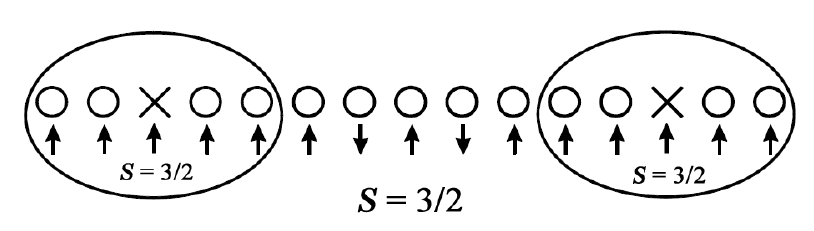}
\end{center} \caption{\label{Fig_FMpolaron} FM polarons with one conduction electron inside the insulating AFM matrix~\cite{KaganUFN2001}.
}
\end{figure}

Historically, nanoscale phase separation with the formation of small FM droplets inside AFM insulating matrices was first predicted at the end of  nineteen sixties in seminal papers of Mott, Nagaev, and Kasuya on magnetic semiconductors~\cite{MottBook1971,NagaevJLett1967,NagaevBook1983,KasuyaSSC1970,
KasuyaSSC1970a}. In this context, it is important to mention also the papers of Krivoglaz~\cite{KrivoglazUFN1974} on heterophase inhomogeneities (so called fluctuons) arising, e.g., in paramagnetic insulating matrices at finite temperatures, as well as a very important paper by de Gennes \cite{deGennesPR1960} on the double exchange model, canted spin states, and other type of (bound) magnetic polarons (or magnetic impurities), which can arise in such model. Later on, analogous predictions for bound magnetic polarons on regular and frustrated lattices were made in \cite{OgarkovPRB2006,KugelOgarPhB2008,KaganOgarJPCM2008}, in the framework of the specific model, which effectively combines exchange interaction, magnetic anisotropy, and attractive electron--ion Coulomb interaction for manganites in the limit of very small doping (see Section~\ref{ModelSys} for more details). The nanoscale electronic phase separation in manganites  and other systems exhibiting the phenomena of the colossal magnetoresistance (CMR) \cite{JonkerPhys1950,JonkerPhys1956,WollanPR1955,JinScience1994,
DagottoBook2003}, was experimentally confirmed by several groups \cite{HennionPRL1998,AllodiPRB1997,BabushkinaPRB1999,
VoloshinJETPL2000,YakubovsPRB2000,FathScience1999,MoritomoPRB1999}, which utilized different experimental techniques including elastic and inelastic neutron scattering, NMR, STM, electron diffraction, dark image spectroscopy, and so on. Let us recall that the CMR phenomenon represents a colossal decrease of resistivity (often by a factor of 100--10000) in moderately strong magnetic fields $H=2-4$ T.  Physically it is often related to the increase in the size of magnetic inhomogeneities (magnetic polarons or ferrons) in the applied magnetic field, which, in its turn, increases the metallic fraction in the sample. As a result, we have a strong decrease in the resistivity. The magnetoresistance is negative in this case
\begin{equation}\label{MR}
MR = \frac{\rho(H) - \rho(0)}{\rho(H)}\, .
\end{equation}
Let us remind that  according to the Mott--Nagaev--Kasuya approach, the radius of a 3D spherical droplet in the absence of magnetic field is (see the details below)
\begin{equation}\label{Rpol}
R_{pol} = a\left(\frac{\pi t}{2zJ_{ff}S^2}\right)^{1/5} \, ,
\end{equation}
where $a$  is the period of the lattice, $z$ is the number of nearest spins  in the given lattice,  $S$ is the value of a local  spin, $J_{ff}$ is the Heisenberg (AFM) exchange interaction between local spins at the neighboring sites, and $t$ is the hopping matrix element between the neighboring sites for conduction electrons. In the 2D case for FM droplets of the circular shape, the power 1/5 in Eq.~(\ref{Rpol}) is replaced by the power 1/4. The droplet radius in Eq.~(\ref{Rpol}) can be found from the minimization of the energy of the nanoscale phase-separated state, which reads~\cite{KaganUFN2001}
\begin{equation}\label{Epol}
E_{pol} = -tna\left(z -\frac{\pi^2a^2}{R^2}\right)+
\frac{4}{3}\pi\frac{zJ_{ff}S^2}{2}\left(\frac{R}{a}\right)^3n
-\frac{1}{2}zJ_{ff}S^2\left[1- \frac{4}{3}\pi\left(\frac{R}{a}\right)^3\right]n \, .
\end{equation}
From the minimization condition $dE_{pol}/dR = 0$, we find the optimum droplet radius given by Eq.~(\ref{Rpol}). The structure of a spherical FM droplet within the AFM matrix is illustrated in Fig.~\ref{Fig_FMpolaron}

Let us now show that the homogeneous canted state
%both in the case of classical de Gennes canting and quantum Nagaev canting
for the typical values of  parameter $\alpha =t/J_{ff}S^2$  of the order of 80--100 (which is typical of manganites) is less favorable in energy than the (inhomogeneous) nanoscale phase-separated state. Moreover, in a wide range of parameters, the homogeneous canted state has a negative compressibility $\kappa = d^2E/dn^2$, which signals the instability of the ground state toward the electronic phase separation. This instability can be easily demonstrated for the classical canted state proposed by de Gennes. In the classical picture, an electron cannot move being surrounded by of AFM-ordered  local spins, and that is why, it is favorable for the electron to cant one sublattice relative to another for having a possibility of hopping to a neighboring site. As a result, the homogeneous canted state has the energy
\begin{equation}\label{Ecant}
E = -tn\cos\frac{\theta}{2}+\frac{1}{2}zJ_{ff}S^2\cos\theta \, ,
\end{equation}
where $\theta$ is the canting angle between two AFM sublattices and the arising effective hopping integral  $t_{eff} = t\cos(\theta/2)$ is related to the spinor nature of the electron wave function~\cite{AndersonHasePR1955}. Minimizing the energy in Eq.~(\ref{Ecant}) with respect to  $\cos(\theta/2)$, we get
\begin{equation}\label{Ecant1}
\cos\frac{\theta}{2} = \frac{tn}{2J_{ff}S^2} \, ,
\end{equation}
and, consequently,
\begin{equation}\label{Emin}
E = -\frac{zt^2n^2}{4J_{ff}S^2} - \frac{zJ_{ff}S^2}{2} \, .
\end{equation}
As a result, the compressibility of the classical canted state reads
\begin{equation}\label{kappa}
\kappa = -\frac{zt^2}{2J_{ff}S^2} < 0 \, .
\end{equation}
It is obviously negative, thus suggesting a tendency toward the phase separation.

It is important to stress that the energies of both the nanoscale phase-separated state (\ref{Epol}) and the homogeneous canted state (\ref{Ecant1}) can be considered as different approximations to the exact energy of the ground state in the double-exchange model,
%of de Gennes,
which is specified by the Hamiltonian
\begin{equation}\label{Hexch}
\hat{H} = -J_H\sum_{i}{\bf S}_i{\bf \sigma}_i
-t\sum_{\langle i,j\rangle}Pc_{i \sigma}^{\dag} c_{i \sigma}P
+ J_{ff}\sum_{\langle i,j\rangle}{\bf S}_i{\bf S}_ j\, ,
\end{equation}
where $J_H$ is a strong onsite Hund's rule coupling, ${\bf \sigma} = c^{\dag}{\bf \tau}c/2$  is the spin operator of conduction electrons, ${\bf \tau}$ are Pauli matrices, $J_{ff}$ is the AFM Heisenberg exchange between local spins at the neighboring sites, ${\dag c}$ and $c$  are creation and annihilation operators, $W=2zt$ is the electron bandwidth. %which is substantially reduced by the electron--phonon polaron effect%,
In manganites, projection operator $P$ is responsible for the single occupation of  $e_g$ orbitals, for which a strong onsite Coulomb repulsion $U$ prevents two conduction electrons from occupying the same site. In actual manganites, we have $J_H \simeq  1$ eV, $t  \simeq 0.3$ eV, $J_{ff}\simeq 0.001$ eV, $z=6$ for the simple cubic lattice, $S=3/2$  and thus we are in the strong coupling limit of the FM Kondo lattice model (FM KLM) $J_H \gg t\gg J_{ff}S^2$, which in the literature is just called the double-exchange model of de Gennes. In this limit, a conduction electron and local spin form together a coupled state with a total spin $S_{tot} = S + 1/2$  at one site. Both in the energy of the phase-separated state (\ref{Epol}) and  the energy of the classical canted state (\ref{Ecant1}), we omit the large constant term $J_HS/2$ (related to the parallel alignment of the local and conduction electron spins) since this term does not change the difference between the energies of these states. The result for $t_{eff}$ in the classical canted state was derived in \cite{AndersonHasePR1955} just in the strong coupling limit of the FM Kondo lattice model.

\begin{figure} [H]
\begin{center}
\includegraphics*[width=0.5\columnwidth]{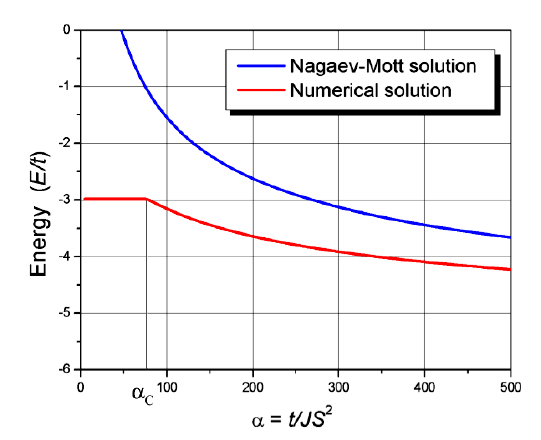} \end{center}
\caption{\label{Nagaev-Mott} (Color online) Comparison of the Nagaev--Mott solution  (taking into account the quantum effects) and the exact numerical solution for the energy of a FM polaron~\cite{KaganUFN2003}.}
\end{figure}

In the quantum canted state introduced by Nagaev~\cite{NagaevJLett1967}, there are two bands $t_+$  and $t_-$ for the motion of conduction electrons corresponding to two different projections of the total spin $S_{tot}^z = S + 1/2$ and $S_{tot}^z = S - 1/2$. The de Gennes classical description of the canted state corresponds to the limit $t_- \rightarrow 0$, whereas, in a quantum treatment for the canting angle $\theta = \pi$ (which corresponds to the collinear AFM structure), both hopping integrals are equal to each other $t_- = t_+ = t/\sqrt{2S+1}$. In this point lies the main difference between quantum and classical canted state. In the quantum picture, the hopping integral for large values of local spin $S \ll 1$ is small only as $/\sqrt{S}$, while in the de Gennes classical limit, it is small as $1/S$. In quantum description, the physical mechanism allowing the electron hopping in the AFM environment (even for the angle $\theta = \pi$,  where it is classically forbidden) is related to the possibility of coherent string-type motion (see~\cite{ZaanenPRB1992}). This type of motion occurs when an electron hops from the site with a total spin projection $S_{tot}^z = S - 1/2$ to the neighboring site (where it forms the state with the total spin projection $S_{tot}^z = S - 1/2$), then hops back to the site where it forms the state with $S_{tot}^z = S + 1/2$, and so on. As a result of this motion, the coherent band of the width $W =2zt/\sqrt{2S+1}$ is formed. However, even in the quantum case, it is possible to show that, at least for substantially large values of the parameter $\alpha =1/J_{ff}S^2$,  the nanoscale phase-separated state is more favorable in energy~\cite{KaganEPJB1999}. The homogeneous string-like motion in collinear AFM environment in exact quantum treatment (which includes the spatial variations of the canting angle $\theta(x)$ and its gradients) can be stabilized only when $\alpha < \alpha_c$, where $\alpha_c$  is the threshold value of the order of 75 (see \cite{KaganUFN2003} and Fig.~\ref{Nagaev-Mott}). For cuprates, we have $\alpha =1/J_{ff}S^2 \approx 2-3$  (for $S=1/2$) and instead of FM polarons, the AFM strings discussed in   \cite{BulNagKhomJETP1968} and \cite{BrinkmanRicePRB1970} can be stabilized in the system at low hole doping.

In the applied magnetic field, an effective Heisenberg exchange decreases due to the Zeeman splitting and reads
\begin{equation}\label{Zeeman}
J_{ff}^{eff}S^2 = J_{ff}S^2 - g\mu_BHS \,
\end{equation}
where $g$ is the gyromagnetic ratio and  $\mu_B$ is the Bohr magneton. Correspondingly, the ferron radius in Eq.~(\ref{Rpol}) and its volume
\begin{equation}\label{vol-pol}
\Omega = \frac{4\pi}{3} \left(\frac{R_{pol}}{a} \right)^3 \, ,
\end{equation}
also increase due to the decrease in the effective exchange. %In the absence of magnetic field, we have $R_{pol}\simeq (2-4)a$ and typically can get the values of 10--15 {\AA}. Correspondingly, the number of FM oriented local spins inside the ferron can be of the order of 30--100 thus forming a large magnetic moment of the droplet.

In the case of electron-doped system, the central spin $\sigma =1/2$  of the conduction electron in a FM droplet is surrounded by local spins $S = 3/2$, whereas in  case of hole doping (typical for manganites), the central unpaired spin $S = 3/2$  is surrounded by paired spins $S_{tot} = 2$.

The similar effect of the FM polaron growth in magnetic field takes place  for the so called temperature ferrons, which arise in the paramagnetic (PM) phase. In 3D systems, their radius in the absence of magnetic field is given by
\begin{equation}\label{temp-pol}
R_T \approx a\left(\frac{\pi t}{2T\ln(2S+1)}\right)^{1/5} \, ,
\end{equation}
and can be obtained from the minimization of the corresponding change in the free energy $\Delta F$. This change takes place due to the formation of the FM droplets above the Curie temperature $T_C$ in the PM matrices and can be written as~\cite{KaganUFN2001}
\begin{equation}\label{free-en}
\Delta F  = -tn\left(z -\frac{\pi^2a^2}{R^2}\right)+
\frac{4}{3}\pi T\left(\frac{R}{a}\right)^3n \ln{(2S+1)} \, .
\end{equation}
In the denominator of Eq.~(\ref{temp-pol}), instead of the Heisenberg exchange term $J_{eff}S^2$, we have the term $T\ln{(2S+1)}$, which determines the contribution of the spin entropy to $\Delta F$. This term also decreases under effect of the applied field due to the Zeeman effect, which reduces the spin degeneracy, thus increasing again the radius and the volume of the 3D spherical droplet.

It should be emphasized again that in the case of the nanoscale phase separation, electron transport
is described by the spin-dependent tunneling of conduction electron between the neighboring metallic FM droplets separated by AFM insulating barriers. When we increase the concentration of droplets, we get closer to the percolation threshold. In this case,  the average distance between the droplets becomes strongly reduced and we can effectively consider the system as a network of the tunneling contacts (or bridges).  This limiting case is described by the physics of the tunneling magnetoresistance (TMR)~\cite{JullierePLA1975}. If we apply an external  magnetic field,  we can control the contact resistance. In this way, we can control the electric current flowing across contact and hence, can read and write the information. This is a main goal and the main challenge for the applications in spintronics~\cite{KaganPrague2014}.

\begin{figure} [H]
\centering
\includegraphics*[width=0.5\columnwidth]{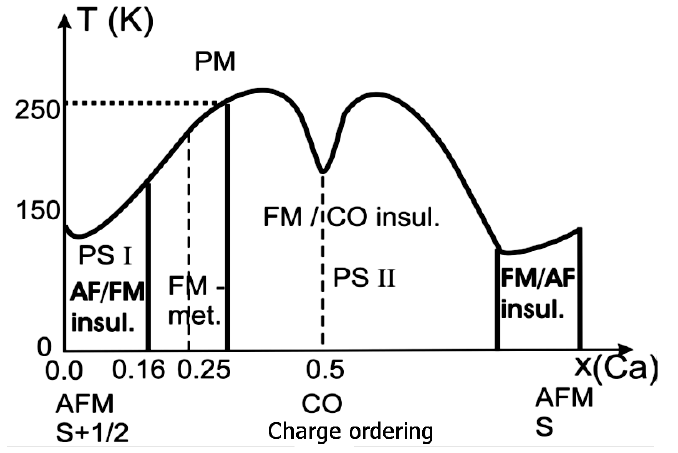}
\caption{ \label{glob-diag} Schematic illustration of the global phase diagram of manganites [18]. PSI and PSII are different kinds of the phase-separated state: FM metallic droplets in the AFM and charge ordered (CO) insulating matrices, respectively.
}
\end{figure}

In the system of 3D spherical droplets, the percolation threshold is achieved at the critical density of charge carriers
\begin{equation}\label{crit-dens}
n_c = 1/\Omega = \frac{3}{4\pi}\left(\frac{a}{R_{pol}}\right)^3 \, .
\end{equation}
In real 3D manganites, the percolation threshold corresponds to the critical density of about 0.16. For larger densities, the ferrons start to overlap forming the infinite FM metallic cluster. The CMR phenomenon is especially pronounced for the charge carrier density within the $n=0.2-0.4$ range (see Fig.~\ref{glob-diag} illustrating the global phase diagram of manganites). Here, we have two phase transitions, which occur simultaneously at the Curie temperature $T_C$, namely the metal--insulator and FM--PM transitions. That is why, even within this density range, for temperatures above $T_C$, the electrical resistivity is mainly determined by the presence of temperature ferrons inside the PM matrix.  At the same time, the temperature ferrons overlap at the critical density
\begin{equation}\label{crit-dens-T}
\delta_c = \frac{3}{4\pi}\left(\frac{a}{R_T}\right)^3 \, .
\end{equation}

Substitution of Eqs.~(\ref{Rpol}) and (\ref{temp-pol}) to Eqs.~(\ref{crit-dens}) and (\ref{crit-dens-T}) yields for the ratio
\begin{equation}\label{crit-dens-ratio}
\frac{n_c}{\delta_c} \varpropto \left[\frac{T\ln{(2S+1)}}{zJ_{ff}S^2}\right]^{3/5}
\varpropto \left[\frac{T_C\ln{(2S+1)}}{T_N}\right]^{3/5}\,
\end{equation}
near the triple point between FM, AFM, and PM phases. Here, $T_C$ and $T_N$ are the Curie and  N\'eel temperatures, respectively. For the hole-doped manganites, $T_C  \backsim T_N \approx 120-150$ K and $\ln{(2S+1)}\approx 1.6$ for $S=2$. Thus, $\delta_c < n_c$ in agreement with the experimental data.

\subsection{Electron transport in inhomogeneous phase-separated manganites} \label{el-transp}

\subsubsection{Electrical conductivity in the phase-separated systems}
\label{conduct-PS}

Let us consider in more detail the transport properties of the phase-separated manganites. As we will show below, the electrical resistivity in the nanoscale phase-separated state reads
\begin{equation}\label{rho}
  \rho(T) = BT\exp\frac{A}{2T} \, ,
\end{equation}
where $A$ is the energy corresponding to the Coulomb repulsion of two conduction electrons occupying the same FM metallic droplet. In the model of spherical droplets, $A$ is given by
\begin{equation}\label{Coulomb-A}
  A \approx \frac{e^2}{\varepsilon R_{pol}} \, ,
\end{equation}
where $\varepsilon$ is the static dielectric constant. In manganites, it can be of the order of 10--20, so the typical values of $A$ are of the order of (3--4)$\times 10^3$ K. In Eq.~(\ref{Coulomb-A}), the droplet radius $R_{pol}$  is given by Eq.~(\ref{Rpol}).

The temperature-independent factor $B$ in Eq.~(\ref{rho}) in the spherical droplet model can be written as \cite{RakhmanovPRB2001resistivity}
\begin{equation}\label{factorB}
  B = \frac{1}{128\pi e^2 \omega_0l^5n^2k} \, ,
\end{equation}
where $e$ is the electron charge, $n$ is the concentration of the FM droplets, $l$ is the characteristic tunneling length, $\omega_0$  is a frequency corresponding to the characteristic energy of an electron in a droplet, and  $k$ is the number of itinerant charge carriers in the droplet. Let us emphasize that in small droplets with the radius given by Eq.~(\ref{Rpol}), we have $k = 1$,  whereas in large droplets usually observed in transport measurements \cite{BabushkinaFTT2003,BabushkinaPRB2003,WagnerEPL2002,ZhaoJPCM2001,
ZhaoPRB2002}, we are dealing with $k \gg 1$. A typical size of a large droplet can be (7--8)$a$ and hence, the number of charge carriers can reach such values as $k = 50-100$. Thus, effectively we can visualize a large droplet as an object which can be divided into many  Wigner--Seitz spheres of small radius, each sphere containing one charge carrier to minimize the Coulomb energy.

Let us try to understand qualitatively the results of Eqs.~(\ref{rho})--(\ref{factorB}). For this purpose, we should following  \cite{RakhmanovPRB2001resistivity,RakhmanovFMM2001} consider a sample of volume $V_S$ in external electric field $\mathbf{E}$. We assume that the sample is in the insulating state and there are $N$ metallic droplets (polarons) embedded in the sample. Then, for the density of the droplets, we find $n = N/V_S$. In the simplest case, upon doping, the number of droplets is equal to the number of electrons (or holes). We neglect the conductivity of the insulating phase and thus consider the case, when we have charge carriers only inside the droplets. In this case, a charge transfer is governed by the motion of the (heavy) droplets as a whole or (if the droplets are considered to be immobile) by the electron tunneling between the droplets. The first mechanism is usually ignored because the motion of a droplet as a whole is related to a strong reconstruction of the local spins surrounding the droplet. Hence a droplet acquires a large effective mass. Moreover, quite often the droplets are localized at dislocations and at other lattice defects.  As a result, we can consider only the tunneling of electrons between almost immobile droplets.

If the size of the droplet is small enough, the polaron typically posses only one electron (or hole) in the ground state of the droplet. However, if we take the tunneling processes into account, then we should consider also the droplets with more than one electron (e.g., with two electrons) as well as empty droplets (assuming that their lifetimes are quite large). Let the energy of the empty droplet be zero, $E(0) = 0$. In this case, the energy of a one-electron droplet is $E(1) \approx t(a/R_{pol})^2$. This estimate corresponds to the kinetic energy of electron localization inside a spherical droplet of radius $R_{pol}$. Accordingly, for the energy of a two-electron droplets, we can give an estimate $E(2) \approx 2E(1) + U$, where $U$ is the energy of interaction between two electrons. Here, we neglect the surface energy. By doing this, we assume that the surface energy for the FM droplets (or ferrons) is small according to the estimates given in  \cite{KaganEPJB1999,KaganKhomPhB2000,KaganUFN2003}. Therefore, we obtain $E(2) + E(0) > 2E(1)$. This means that to form the two-electron droplets, we need to overcome a potential barrier with the energy $A = E(2)-2E(1) \approx U$. It is easy to understand that $U$ is governed by the Coulomb repulsion between two electrons localized at one droplet. In this way, we come to the estimate for $A$ given by Eq.~(\ref{Coulomb-A}). We assume that we are working in the low-doping limit, where the mean distance between the droplets  is $n^{-1/3} \gg R_{pol}$ (the polarons do not overlap). Thus, in accordance with our estimates and  transport measurements (see below), the typical values of $A \simeq$ 3500--3700 K are larger than the average Coulomb energy $e^2n^{1/3}/\varepsilon_0$. Moreover, $A \gg T$ for all realistic temperatures. That is why, we can neglect the creation of the droplets with the number of electrons larger than two. It is possible to demonstrate that even if the many-electron droplets are stable, their contribution to conductivity is strongly reduced by the Coulomb interaction and hence, it is very small for low droplet densities (when the system is far from the percolation threshold).

Let us evaluate factor $B$ in Eq.~(\ref{factorB}), which is essentially temperature independent \cite{kugel2004characteristics}). It is convenient to introduce parameters $N_1$, $N_2$,  and  $N_3$ for the numbers of droplets with one electron, two electrons, and of empty droplets, respectively.  According to our model, we assume that the total number of the droplets in the system is constant. Thus, $ N_2 = N_3$ and  $N_1 + 2N_2 = N$, and the partition function of the system is given by
\begin{equation}\label{part-func}
  Z = \sum_{m=0}^{N}{C_N^m C_{N-m}^m \exp{(-m\beta)}}, \,\,\,\,\, \beta = A/T \, ,
\end{equation}
where $C_N^m$ are the binomial coefficients. Using the Stirling formula, replacing the summation by the integration, and calculating the corresponding integral by the steepest descent method, we obtain the following expressions for the average values of $N_1$, $N_2$,  and  $N_3$
\begin{eqnarray}\label{N-aver}
\overline{N_2}=\overline{N_3}= N\exp{\left(-\frac{A}{2T}\right)} \, , \nonumber\\
 \overline{N_2} = N -2\overline{N_2} = N\left[ 1-2\exp{\left(-\frac{A}{2T}\right)}\right] \, .
\end{eqnarray}

\begin{figure} [H]
\begin{center}
\includegraphics*[width=0.5\columnwidth]{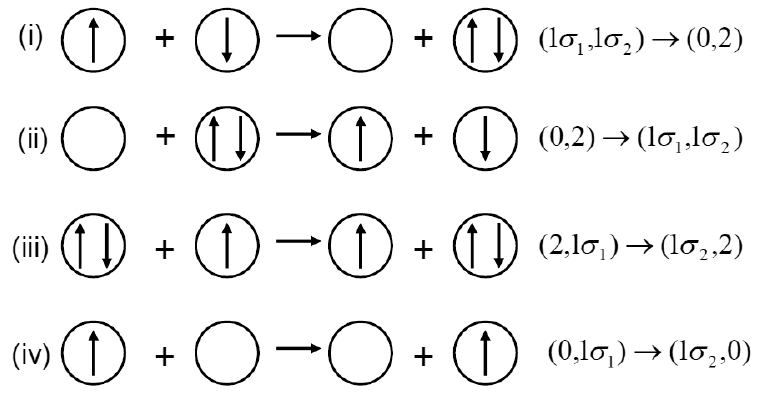} \end{center}
\caption{\label{tunnel-proc} Elementary tunneling processes \cite{RakhmanovPRB2001resistivity}. The arrows and $\sigma_i$ denote spin projections.}
\end{figure}

The electron tunneling implies one of the four possible events (see Fig.~\ref{tunnel-proc}) (i) two droplets with one electron are converted into the droplet with two electrons and an empty droplet, (ii) process, which is inverse to the previous one, (iii) droplets with one and two electrons exchange their positions, (iv) the same for the droplet with one electron and an empty droplet. The corresponding total current density is given by the sum of the contributions of all these processes $j =j_1+j_2+j_3+j_4$. It was shown in \cite{RakhmanovPRB2001resistivity} that all these processes equally contribute to the total electrical current.

Let us consider, for example, the contribution of the first two processes
\begin{equation}\label{j12}
j_{1,2} = e n_{1,2}\langle \sum_{i}{v_{1,2}^i}\rangle = e n_{1,2}\langle \sum_{i}{\frac{r^i\cos{\theta^i}}{\tau_{1,2}(r^i, \theta^i)}}\rangle \, ,
\end{equation}
In Eq.~(\ref{j12}), $n_{1,2} = N_{1,2}/V_S$ denote the densities of one-electron and two-electron droplets, $\langle...\rangle$ corresponds to the statistical and time averages, $\langle v^i\rangle$  is the average velocity of electrons along the electric field ${\mathbf E}$, $r^i$ is the distance between the droplets involved in these processes, $\theta^i$ is the angle between vector ${\mathbf E}$ and the direction of the electron motion, and  $\tau_{1,2}(r^i, \theta^i)$ are characteristic times associated with these processes. The  conventional expressions for $\tau_{1,2}$  (which are derived in \cite{RakhmanovPRB2001resistivity} using the detailed balance equation) have the following form
\begin{equation}\label{tau12}
\tau_{1,2} = \omega_0^{-1}{1,2}\exp{\left(\frac{r}{l} \pm \frac{A}{2T} - \frac{eEr\cos{\theta}}{T}\right) } \, ,
\end{equation}
where  $l$ is the tunneling length and $\omega_0$ is a characteristic magnon (or depolarization) frequency.
Let us stress that when we are far from the percolation threshold, the average in (\ref{j12}) is reduced to the space average for velocity $v_i$  multiplied by the number of droplets, to which the hopping is allowed. These numbers are equal to $N_1$ for process (i) and  respectively to $N_2$ for  process (ii). If we assume that the external electric field is small, $eEl/T \ll 1$  and moreover, the repulsion energy is large, $A\gg T$, then in the first order in $E$, we have
\begin{equation}\label{average-v}
\langle \sum_{i}{v_{1,2}^i}\rangle = \frac{eE\omega_0}{T}N_{1,2}\exp{\left(\mp\frac{A}{2T}\right)}\langle r^2\cos^2{\theta} \exp{\left(-\frac{r}{l}\right)}\rangle_V \, ,
\end{equation}
where $\langle ...\rangle_V$ denotes the average over the sample volume
\begin{equation}\label{averaging}
\langle ...\rangle_V = V_S^{-1}\int{...d^3{\mathbf r}} \, .
\end{equation}
Performing the integration, we get
\begin{equation}\label{integrated-j12}
j_{1,2} = \frac{32\pi e^2E\omega_0l^5n_{1,2}^2}{T}\exp{\left(\mp\frac{A}{2T}\right)} \, .
\end{equation}
Analogously, it is possible to demonstrate rather straightforwardly that for other tunneling processes, the corresponding current densities read
\begin{equation}\label{integrated-j34}
j_{3,2} = \frac{32\pi e^2E\omega_0l^5n_1n_2}{T} \, .
\end{equation}
Comparing Eqs.~(\ref{integrated-j34}) and (\ref{integrated-j34}), we can see that in the last one, the exponential factor is missing and the factors $n_{1,2}^2$ are replaced by $n_1n_2$.

From Eqs.~(\ref{N-aver}), (\ref{integrated-j12}), and (\ref{integrated-j34}), we can obtain the expression for dc conductivity

\begin{equation}\label{aver-cond}
\sigma = 1/\rho = \frac{128\pi e^2n^2\omega_0l^5}{T}\exp{\left(-\frac{A}{2T}\right)}k  \,
\end{equation}

For small droplets described by Eq.(2), we can show that only one-electron droplets are stable, while the empty droplets and two-electron droplets decay. The decay time of the empty droplets is governed by the inverse depolarization frequency. At the same, time an empty droplet can get an electron from the one-electron or two-electron droplets, located nearby. This process involves a time scale, which can be estimated as (see [37])
in agreement with Eqs.~(\ref{rho}) and (\ref{factorB}). For small droplets described by  Eq.~(\ref{Rpol}), we can show that, in fact, only one-electron droplets are stable, whereas empty droplets and two-electron droplets decay. The decay time of empty droplets is governed by the inverse depolarization frequency $1/\omega_0$. At the same time, an empty droplet can get an electron from one-electron or two-electron droplets, located nearby. This process involves a time scale $\tau_0$, which can be estimated as (see  \cite{RakhmanovPRB2001resistivity})
\begin{equation}\label{tau0}
\tau_0 = \frac{\exp{\left(-\frac{A}{2T}\right)}}{8\pi\omega_0l^5n} \, .
\end{equation}
The similar estimate is valid for an electron, which leaves a two-electron droplet. The validity of this picture is determined by the inequality $\tau_0 \ll\omega_0^{-1}$. Thus, our approach requires inequality $A \gg T$  and not too small droplet density $n$.

\subsubsection{Comparison with experimental data} \label{exp_rho}

\begin{figure}[H] \centering
\subfigure[]
{\includegraphics[width=0.38\columnwidth]{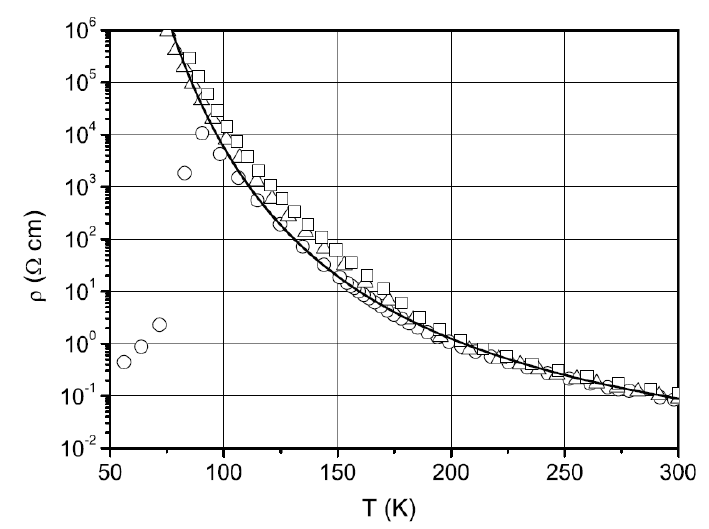}}
 \subfigure[]
{\includegraphics[width=0.38\columnwidth]{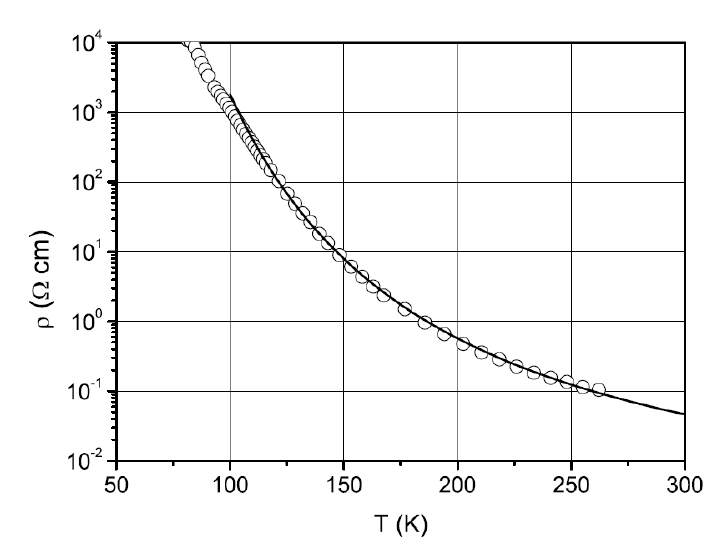}}
 \subfigure[]
{\includegraphics[width=0.38\columnwidth]{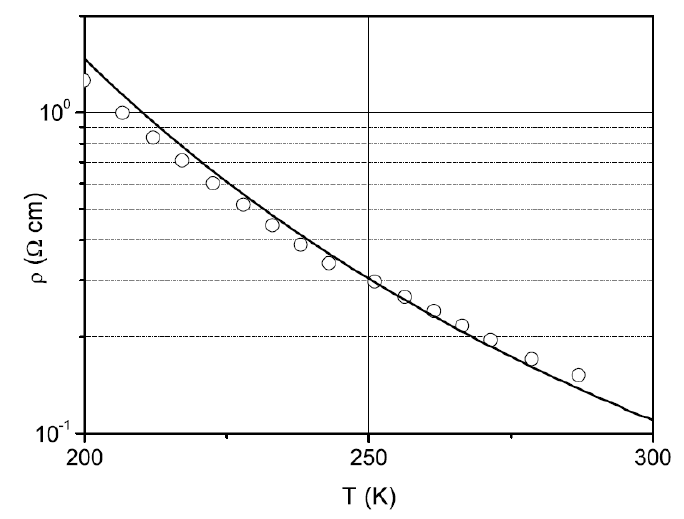}}
 \subfigure[]
{\includegraphics[width=0.38\columnwidth]{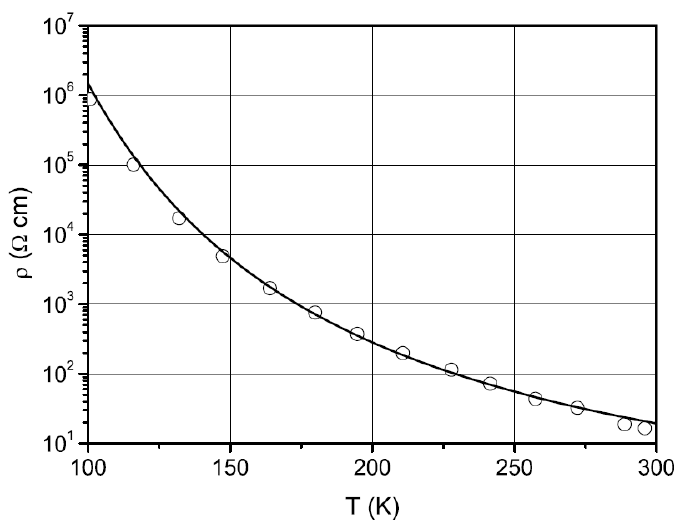}}
\caption{Temperature dependence of electrical resistivity for (a) (La$_{1-y}$Pr$_y$)$_{0.7}$ Ca$_{0.3}$MnO$_3$ with $^{16}$O $\rightarrow$ $^{18}$O oxygen isotope substitution for $y=1$ (squares) and $y=0.75$ (triangles); (b) Pr$_{0.71}$Ca$_{0.29}$MnO$_3$; (c) a layered manganite with the chemical composition (La$_{0.4}$Pr$_{0.6}$)$_{1.2}$ Sr$_{1.8}$Mn$_2$O$_7$; and (d) La$_{0.8}$Mg$_{0.2}$MnO$_3$ \cite{kugel2004characteristics}.} \label{rho-exper}
%\end{centering}
\end{figure}

In Fig.~\ref{rho-exper}, we present the plots for resistivity $\rho(T)$ in six families of manganites using the experimental data reported in \cite{BabushkinaFTT2003,BabushkinaPRB2003,WagnerEPL2002,ZhaoJPCM2001,ZhaoPRB2002}. As can be seen from these data, the samples differ in chemical composition, the type of the crystal structure, the magnitude  and the low temperature behavior of the resistivity (some of the samples behave as metals, some as the insulators, whereas the resistivity at a given temperature varies more than in two orders of magnitude). However, in the high-temperature range above the Curie temperature, the resistivity exhibits an identical behavior for all the samples governed by the universal relation (\ref{aver-cond}). This relation is typical of many tunneling systems \cite{MottBook1971} and is represented by solid curve in Fig.~\ref{rho-exper}.

Using the experimental data and Eq.~(\ref{aver-cond}), we can determine the value of the Coulomb barrier energy $A$ and make a reasonable estimate for the characteristic depolarization frequency $\omega_0  \simeq 10^{13}-10^{14}$ Hz. We can also determine the important factor $l^5n^2k$ which enters Eq.~(\ref{factorB}) for the temperature-independent coefficient $B$ (see Table~\ref{table-Coulomb}).

\begin{table} [H]
\caption{\label{table-Coulomb} Coulomb energy $A$, resistivity $\rho$ at 200 K, and product $l^5n^2k$ for several families of manganites~\cite{kugel2004characteristics}}
\begin{center}
%\begin{ruledtabular}
\begin{tabular}{ccccc}
 \hline
 Samples&$A$ (K) &$\rho$(200 K) &$l^5n^2k$ & Data source\\
&&$\Omega\cdot$cm&cm$^{-1}$\\
\hline
La$_{1-y}$Pr$_y$)$_{0.7}$ Ca$_{0.3}$MnO$_3$&3650&1.25&$2\times10^5$&Fig.~\ref{rho-exper}a (Babushkina et al., 2003 \cite{BabushkinaFTT2003})\\
Pr$_{0.71}$Ca$_{0.29}$MnO$_3$&3500&0.57&$3\times10^5$&Fig.~\ref{rho-exper}b (Fisher et al., 2003 \cite{BabushkinaPRB2003})\\
(La$_{0.4}$Pr$_{0.6}$)$_{1.2}$ Sr$_{1.8}$Mn$_2$O$_7$&3600&1.5&$1.5\times10^5$&Fig.~\ref{rho-exper}c (Wagner et al., 2002 \cite{WagnerEPL2002})\\
La$_{0.8}$Mg$_{0.2}$MnO$_3$&3700&283&$1\times10^3$&Fig.~\ref{rho-exper}d
(Zhao et al., 2001 \cite{ZhaoJPCM2001})\\
\hline
\end{tabular}
%\end{ruledtabular}
\end{center}
% \footnotetext[1]{nominal Eu content (of 0.7)} \footnotetext[2]{actual Eu
% content (of 1)}
\end{table}

\subsubsection{Magnetoresistance of phase-separated manganites}
\label{MR-PS}

Now we can proceed to the analysis of the magnetoresistance in phase-separated manganites. Using Eq.~(\ref{Zeeman}) for the effective Heisenberg exchange, we can obtain the following expression for the droplet radius in experimentally accessible magnetic fields
\begin{equation}\label{Rpol_H}
R_{pol}(H) = R_{pol}(0)(1 + bH) \, ,
\end{equation}
where in the 3D case, the coefficient $b$ is
\begin{equation}\label{coef-b}
b = \frac{g\mu_B}{5J_{ff}S} \, .
\end{equation}
For the corresponding decrease in the Coulomb barrier height $A$, we find
\begin{equation}\label{A(H)}
A(H) = A(0)(1 - bH) \, .
\end{equation}
Thus, using eq/~(\ref{rho}), we find that the magnetoresistance in the nanoscale phase-separated state is negative and can be written as
\begin{equation}\label{MR-PS1}
MR = 1 - \exp{\left(\frac{AbH}{2T}\right)} \, .
\end{equation}
It is shown in \cite{kugel2004characteristics} that such exponential behavior really takes place in strong magnetic fields $H>10$ Tesla. Since the  gyromagnetic ratio $g \sim 10$ is typical of manganites, this behavior corresponds  just to the magnetic field range where the argument of exponential in Eq.~(\ref{MR-PS1}) becomes larger than unity. At the same time, at low and moderate magnetic fields, the magnetoresistance is quadratic in the field (in agreement with the textbooks on solid state physics \cite{AbrikosovBook1988}) and is given by the relation
\begin{equation}\label{MR-lowH}
MR \propto -H^2/T^n \, ,
\end{equation}
where the exponent $n$ in Eq.~(\ref{MR-lowH}) is  either 1, 2, or 5 depending on the range of magnetic fields under consideration, the value of magnetic anisotropy, and so on~\cite{SboychakovJETP2002tunMR,SboychakovJMMM2003,kugel2004characteristics}. The magnetoresistance in this case is determined  by the effects of the spin-dependent tunneling. The value $n=5$ is typical of the magnetoresistance in manganites in the range of low magnetic fields and high temperatures. In this case, we have~\cite{kugel2004characteristics}
\begin{equation}\label{MR-T5}
|MR| = 5\times10^{-3}\frac{\mu_B^3S^5J_{FM}^2N_{eff}^3Z^2g^3H_a}{T^5}H^2 = 5\times10^{-3}\frac{M_0^3J_{FM}^2Z^2S^2H_a}{T^5}H^2\, ,
\end{equation}
where we introduced the magnetic moment of a droplet
\begin{equation}\label{M0}
M_0 = g\mu_BSN_{eff}\, .
\end{equation}

In Eq.~(\ref{MR-T5}), $N_{eff}$ is the number of Mn atoms inside the droplet, $S = 2$ is the total spin of Mn$^{3+}$ ions in manganites under the hole doping, and $Z$ is the number of nearest neighbors of a Mn ion. Let us stress that for 3D cubic manganites with $Z =6$, $H_a$ is the effective magnetic anisotropy field of a droplet. At the same time, an effective FM exchange integral $J_{FM}$ is specified by the well-known relationship of the molecular field theory  \cite{SmartBook1966}
\begin{equation}\label{T-Curie}
T_C = S(S+1)ZJ_{FM}/3 \, .
\end{equation}

\begin{figure}[H] \centering
\subfigure[]
{\includegraphics[width=0.38\columnwidth]{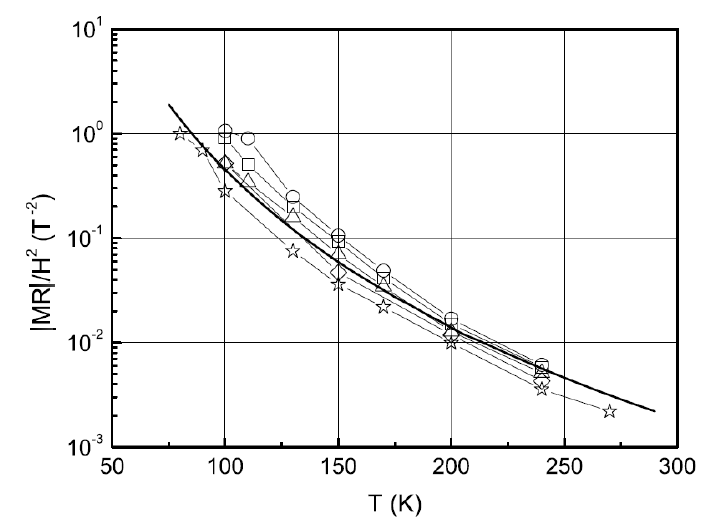}}
 \subfigure[]
{\includegraphics[width=0.38\columnwidth]{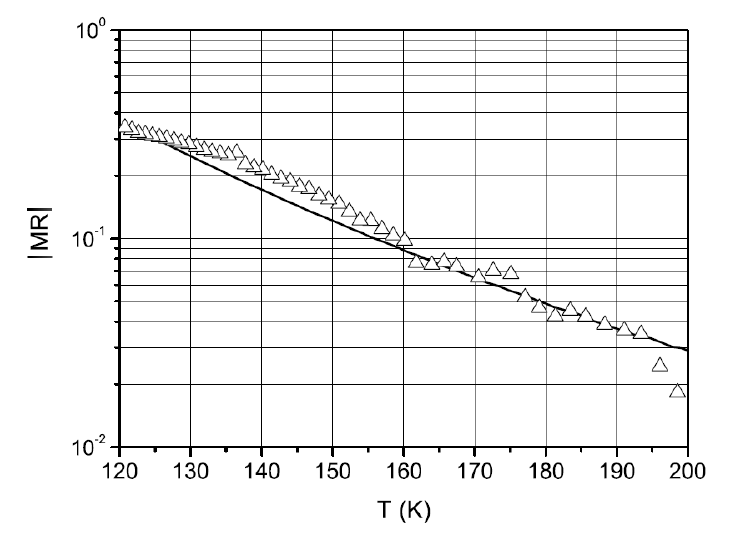}}
 \subfigure[]
{\includegraphics[width=0.38\columnwidth]{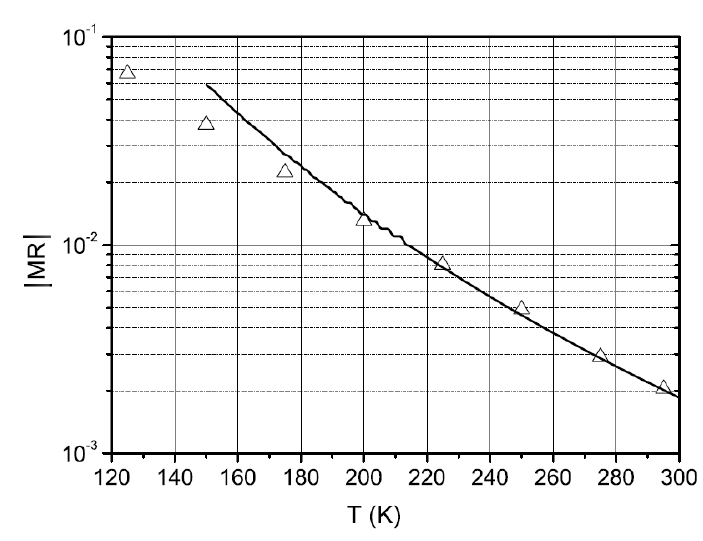}}
 \subfigure[]
{\includegraphics[width=0.38\columnwidth]{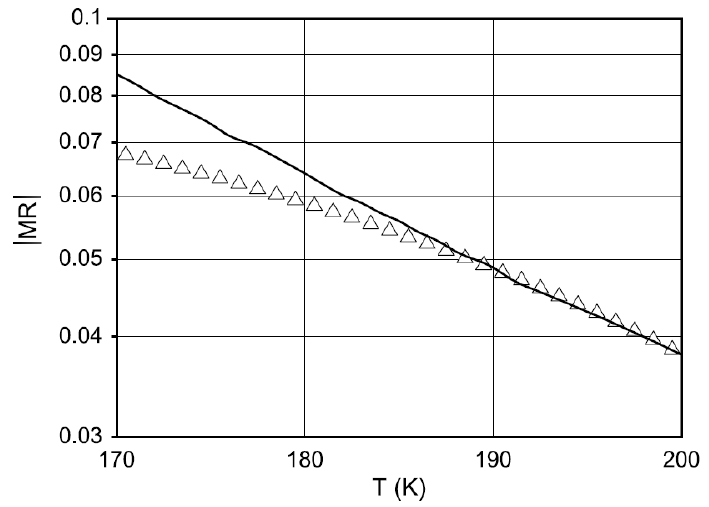}}
\caption{Temperature dependence of the absolute value of magnetoresistance. Solid curve show the results of the theoretical calculations using relation (\ref{MR-T5}), $|MR|\propto 1/T^5$ . (a) $|MR|/H^2$  ratio for (La$_{1-y}$Pr$_y$)$_{0.7}$ Ca$_{0.3}$MnO$_3$  at $y=1$ and $y=0.75$; (b) Pr$_{0.71}$Ca$_{0.29}$MnO$_3$ at $H = 2$ T; (c) layered manganite (La$_{0.4}$Pr$_{0.6}$)$_{1.2}$ Sr$_{1.8}$Mn$_2$O$_7$  at $H = 1$ T; and (d) La$_{0.8}$Mg$_{0.2}$MnO$_3$ for $H = 1.5$ T \cite{kugel2004characteristics}.} \label{MR-exper}
%\end{centering}
\end{figure}

The behavior of the magnetoresistance given by Eq.~(\ref{MR-T5}) was also experimentally observed in different families of manganites \cite{BabushkinaFTT2003,BabushkinaPRB2003,WagnerEPL2002,
ZhaoJPCM2001,ZhaoPRB2002} (see Fig.~\ref{MR-exper}).

The solid curves in Fig.~\ref{MR-exper} were calculated for the value of $T_C = 120$ K. At the same time, the fitting parameter $N_{eff}$ was determined from the relation $k = N_{eff}x$, where $x$ is the atomic fraction of the dopant. The values of  $N_{eff}$, $x$, and $k$ are presented in Table~\ref{table-fits}.

\begin{table} [H]
\caption{\label{table-fits} Effective number $N_{eff}$ of manganese atom and the number of electrons $k$ in a droplet, and the dopant fraction $x$ for several different families of manganites, estimated using Eq.~(\ref{MR-T5}) and experimental data presented in Fig.~\ref{MR-exper}}
\begin{center}
%\begin{ruledtabular}
\begin{tabular}{ccccc}
 \hline
 Samples&$N_{eff}$&$x$&$k$ & Data source\\
\hline
La$_{1-y}$Pr$_y$)$_{0.7}$ Ca$_{0.3}$MnO$_3$ &250&0.35&75&Fig.~\ref{MR-exper}a (Babushkina et al., 2003 \cite{BabushkinaFTT2003}) \\
Pr$_{0.71}$Ca$_{0.29}$MnO$_3$&200&0.29&58&Fig.~\ref{MR-exper}b (Fisher et al., 2003 \cite{BabushkinaPRB2003})\\
(La$_{0.4}$Pr$_{0.6}$)$_{1.2}$ Sr$_{1.8}$Mn$_2$O$_7$ &250&0.4&100&Fig.~\ref{MR-exper}c (Wagner et al., 2002 \cite{WagnerEPL2002})\\
La$_{0.8}$Mg$_{0.2}$MnO$_3$ &265&0.2&53&Fig.~\ref{MR-exper}d
(Zhao et al., 2001 \cite{ZhaoJPCM2001})\\
\hline
\end{tabular}
%\end{ruledtabular}
\end{center}
% \footnotetext[1]{nominal Eu content (of 0.7)} \footnotetext[2]{actual Eu
% content (of 1)}
\end{table}

In brief, the derivation of Eq.~(\ref{MR-T5}) is as follows. The conductivity of the system is controlled by the ``spin" contribution $\Sigma(H)$ to $\sigma$, that is $\sigma(H) = \sigma_0 \langle \Sigma(H))\rangle$. Therefore, we have $|MR| = \langle\Sigma(H)\rangle - 1$. Assuming the interaction between the droplets to be negligibly small and using an expression (\ref{M0}) for the effective magnetic moment, we can represent the free energy of the droplet in a magnetic field in the following form
  \begin{equation}\label{U(H)}
  U(H) = U(0) - M(H\cos{\theta} + H_a\cos^2{\Psi})\,.
\end{equation}
In Eq.~(\ref{U(H)}), $\theta$ is the angle between the external magnetic field ${\mathbf H}$ and the magnetic moment ${\mathbf M}$, and $H_a$ stands for the field of magnetic anisotropy. Finally, $\Psi$ corresponds to the angle between the axis of the uniaxial anisotropy and ${\mathbf M}$. Let ${\mathbf H}$ be parallel to the $z$ axis and let the anisotropy axis lie in the ($x,z$) plane and have the angle $\beta$ with vector ${\mathbf H}$. In this configuration, we find
\begin{equation}\label{cosPsi}
\cos{\Psi} = \sin{\theta}\sin{\beta}\cos{\phi} + \cos{\theta}\cos{\beta} \, .
\end{equation}
Tn Eq.~(\ref{cosPsi}), $\phi$  denotes the angle between the $x$ axis and the projection of magnetic moment ${\mathbf M}$  onto the ($x,y$) plane.

In the classical limit, a given orientation of vector ${\mathbf M}$ corresponds to the probability
\begin{equation}\label{cosPsi1}
P(H, \theta, \varphi) = C(H)\exp{\frac{M(H\cos{\theta} + H_a\cos^2{\Psi(\theta,\varphi)}}{T}} \, ,
\end{equation}
where $C(H)$ is the normalization factor. The eigenstates for a conduction electron correspond to the conservation of the spin projection $s \pm 1/2$ onto the effective field direction in FM correlated region. Let the electron interact with $Z$ magnetic moments of the droplet. The energy of this interaction is $E_s = -J_{FM}SZs$. Since the value of $J_{FM}SZ$ product is according to Eq.~(\ref{MR-T5}) is of the order of the Curie temperature, $E_s$ is much larger than the energy of interaction between the electron spin and magnetic field for realistic magnetic fields $H < 100$ T.
In this situation, the direction of the effective field is parallel to the direction of the magnetic moment ${\mathbf M}$. Therefore, the probability for the projection of the electron spin to be equal to $s$ can be written as
\begin{equation}\label{spin_prob}
P_s = \frac{\exp{(-E_s/T)}}{2\cosh{(E_s/T)}} \, .
\end{equation}

In the process of the transfer from droplet 1 to droplet 2, an electron moves in the effective field, which forms an angle $\nu$ with the effective field in the initial state. This angle is naturally defined via the obvious relation $\cos{\nu} = \cos{\theta_1}\cos{\theta_2} + \sin{\theta_1}\sin{\theta_2}\cos{(\varphi_1 -\varphi_2)}$, where subscripts 1 and 2 refer to the droplet numbers. Then, the work performed in the process of electron transfer from droplet 1 to  droplet 2 is $\Delta E_s(1-\cos{\nu})$ . Accordingly, the probability of this transfer is proportional to $\exp{(-\Delta E_s/T)}$. Taking into account all the probability factors introduced above, we obtain the final expression for the ``spin contribution"
\begin{equation}\label{Sigma-H}
\langle\Sigma(H)\rangle = \int_0^{2\pi}{d\varphi_1}\int_0^{2\pi}{d\varphi_2}
\int_0^{\pi}{\sin{\theta_1}d\theta_1}\int_0^{\pi}{\sin{\theta_2}d\theta_2}
P(\theta_1, \phi_1)P(\theta_2, \phi_2)\sum_{s =\pm 1/2}{P_s\exp{\left(\frac{-\Delta E_s}{T}\right)}}  \, .
\end{equation}

The high-temperature expansion of the expression (\ref{Sigma-H}) for  $T \gg \{\mu_BgSN_{eff}H, \mu_BgSN_{eff}H_a\}$, yields formula (\ref{MR-T5}).

\subsubsection{Specific features of the magnetic susceptibility in the phase-separated state} \label{suscept}

At the high-temperature limit, the susceptibility of nanoscale phase-separated state is given by the Curie--Weiss law~\cite{kugel2004characteristics}
\begin{equation}\label{chiT}
  \chi(T) = \frac{n(\mu_BgSN_{eff})^2}{3(T -\Theta)} =  \frac{nM_0^2}{3(T -\Theta)} \, ,
\end{equation}
where  $\Theta$ is the Curie--Weiss constant.

\begin{figure}[H] \centering
\subfigure[]
{\includegraphics[width=0.32\columnwidth]{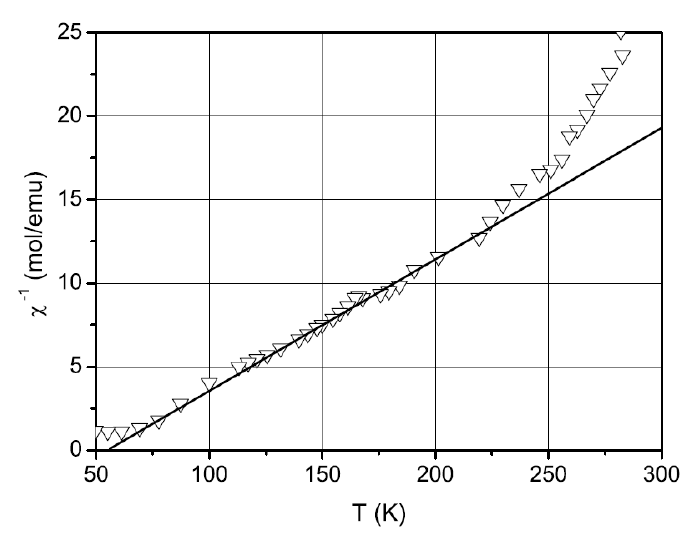}}
 \subfigure[]
{\includegraphics[width=0.32\columnwidth]{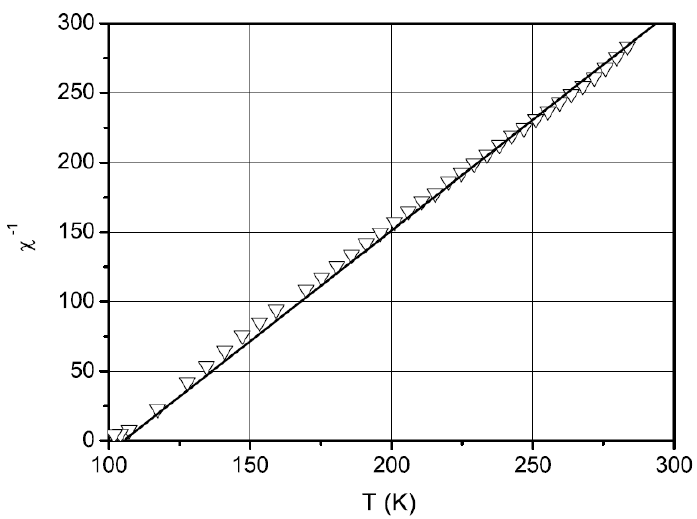}}
 \subfigure[]
{\includegraphics[width=0.32\columnwidth]{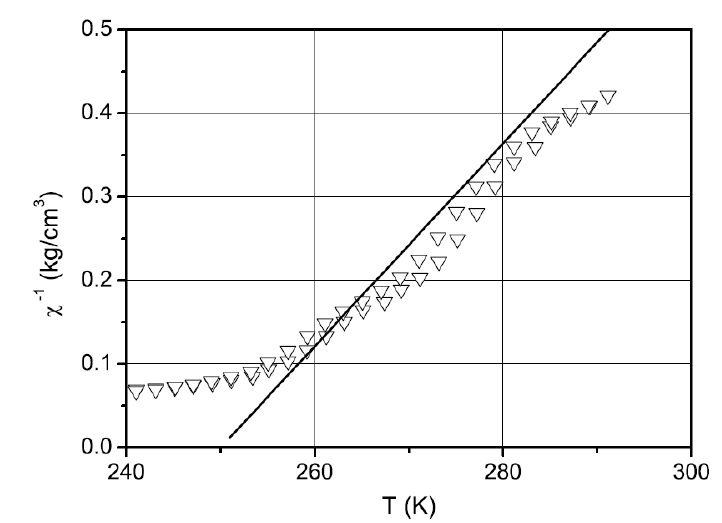}}
 \caption{Temperature dependence of the inverse magnetic susceptibility of (a) (La$_{1-y}$Pr$_y$)$_{0.7}$ Ca$_{0.3}$MnO$_3$  for $y=1$; (b) Pr$_{0.71}$Ca$_{0.29}$MnO$_3$; and (c) (La$_{0.4}$Pr$_{0.6}$)$_{1.2}$ Sr$_{1.8}$Mn$_2$O$_7$~\cite{kugel2004characteristics}.} \label{susc-exper}
%\end{centering}
\end{figure}

The linear temperature dependence for the inverse susceptibility is also confirmed in experiments  \cite{BabushkinaFTT2003,BabushkinaPRB2003,WagnerEPL2002,ZhaoJPCM2001,
ZhaoPRB2002} for several different families of manganites (see Fig.~\ref{susc-exper}, where solid curves illustrate the fitting procedure based on Eq.~(\ref{chiT})). Using these  experimental results  \cite{BabushkinaFTT2003,BabushkinaPRB2003,WagnerEPL2002,ZhaoJPCM2001,
ZhaoPRB2002} and theoretical results of Eq.~(\ref{chiT}), we can evaluate the Curie--Weiss constant $\Theta$, the density of ferrons $n$, the FM phase fraction $p = nN_{eff}a^3$  (where for all the samples, we used the value $a = 3.9$\AA   as a lattice constant), as well as the effective tunneling length $l$ (see Table~\ref{table-chi}).

\begin{table} [H]
\caption{\label{table-chi} Curie--Weiss constant $\Theta$, density of ferrons $n$, FM phase fraction $p$, and effective tunneling length $l$ in different families of manganites}
\begin{center}
%\begin{ruledtabular}
\begin{tabular}{cccccc}
 \hline
 Samples&$\Theta$, K&$n$, cm$^{-3}$&$p$&$l$, \AA& Data source\\
\hline
La$_{1-y}$Pr$_y$)$_{0.7}$ Ca$_{0.3}$MnO$_3$ &55&$1.8\times10^{18}$&0.03&24&Fig.~\ref{susc-exper}a (Babushkina et al., 2003 \cite{BabushkinaFTT2003}) \\
Pr$_{0.71}$Ca$_{0.29}$MnO$_3$
&105&$6.0\times10^{18}$&0.07&17&Fig.~\ref{susc-exper}b (Fisher et al., 2003 \cite{BabushkinaPRB2003})\\
(La$_{0.4}$Pr$_{0.6}$)$_{1.2}$ Sr$_{1.8}$Mn$_2$O$_7$ &255&$2.5\times10^{18}$&0.04&19&Fig.~\ref{susc-exper}c (Wagner et al., 2002 \cite{WagnerEPL2002})\\
La$_{0.8}$Mg$_{0.2}$MnO$_3$
&150&$0.6\times10^{18}$&0.01&14&Zhao et al., 2001 \cite{ZhaoJPCM2001}\\
\hline
\end{tabular}
%\end{ruledtabular}
\end{center}
% \footnotetext[1]{nominal Eu content (of 0.7)} \footnotetext[2]{actual Eu
% content (of 1)}
\end{table}

Let us emphasize that effectively we can regard a small FM droplet as a qubit since the direction of the magnetic moment of the droplet can be parallel or antiparallel to the easy axis. If we apply a high magnetic field, we can strongly polarize FM droplets and thus can write down the information. In our case, an entangled two-qubit state (which is very important for quantum calculations) is organized by a pair of two droplets located nearby. The droplets in the pair are connected with the tunneling bridges. Their interaction has a magneto-dipole origin. Thinking about the applications in e spintronics, we can say that the network of FM droplets or nanocomposites (multilayers) Fe/Cr,Co/Cu, which is connected by the tunneling bridges, can serve as the basic element for the memory device~\cite{KaganPrague2014}.

Returning back to qubits, note that a popular nowadays magnetic principle of their realization is based on the quantum spin tunneling in the system of large magnetic molecules ~\cite{BarbaraPhWorld1999} and thus is very similar to the spin-dependent tunneling of conduction electron between the neighboring FM droplets in our case.

In addition, there are interesting possibilities for the applications of CMR manganites in the nanoscale phase-separated state for optoelectronics and nanoplasmonics. Possible application for the fabrication of the new optical devices can be related to a strong change in the infrared behavior of the transmission coefficient in the  phase-separated state, when we apply an external magnetic field to our system. This change in the behavior gives us a possibility to construct an infrared radiation modulator based on such manganites~\cite{SukhorukovTPL2003}.

Another possible application of the phase-separated manganites in the nanoplasmonics \cite{ShalaevBook2007} was first proposed in \cite{SarychevPRL2011}. The main idea here is to construct on the surface of a bulk metal an artificial layer with  small metallic droplets embedded in the insulating charge ordered (CO) matrix. Let us consider an incident electromagnetic wave coming at the surface from the vacuum. It is very interesting then to study the angular and frequency dependence of the transmission ($T$) and reflection ($R$) coefficients in this geometry assuming that metallic droplets form a periodic superstructure within the surface layer. The reflection and transmission coefficients will have the resonance peaks of different shapes for some critical angles including the angles of total internal reflection. It is important to stress that the critical angles themselves will correspond to the resonance excitation of the surface plasmon modes here. Thus, we can calculate the spectrum of surface plasmons and find the resonance frequencies. The experimental realization of this idea can lead to the development of a new generation of the devices for Raman spectroscopy, which will have a strongly enhanced  sensitivity.

\subsubsection{$1/f$ voltage noise in the systems with phase separation} \label{noise}

The difficulties, which we have on the way of the creation of magnetic devices are mostly associated with a very large amplitude of the $1/f$ noise in the phase-separated state. In \cite{PodzorovPRB2000}, we find the following expression for the intensity of the $1/f$ noise spectrum (or the Hooge constant ) in the framework of the droplet model~\cite{RakhmanovPRB2001resistivity,RakhmanovFMM2001,SboychakovJETP2002tunMR,
SboychakovJMMM2003})
\begin{equation}\label{Hooge}
  \alpha_H = \frac{\langle \delta U^2\rangle_{\omega}V_s \omega}{U_{dc}^2} =  2\pi^2l^3 \ln^2{\left(\frac{\bar{\omega}}{\omega}\right)}\, .
\end{equation}
In Eq.~(\ref{Hooge}), $\langle\delta U^2\rangle_{\omega}$ is the spectral density of voltage fluctuations, $U_{dc}$ is the applied voltage, $\langle \delta U^2\rangle_{\omega}/U_{dc}^2 = \delta \sigma^2\rangle_{\omega}/\sigma^2$ and $\bar{\omega} = \omega_0\exp{(A/2T)}$. To derive Eq.~(\ref{Hooge}), we should use the important relation \cite{RakhmanovPRB2001resistivity}
\begin{equation}\label{fluc-sigma}
  \delta \sigma = \sigma\frac{\delta n_2}{{\bar n_2}}[1 - 2\exp{(-A/2T)}]
  \end{equation}
between conductivity fluctuations and the fluctuations of the density of two-electron droplets. We use then the general prescription from the textbooks on statistical physics (see, e.g.  \cite{DuttaRMP1981,LandauBook_StPhI1980}), which relates fluctuation spectrum to the thermal average
\begin{equation}\label{fluc-n}
  \langle \delta n_2^2\rangle_{\omega} = \langle \delta n_2^2\rangle_T\langle \sum_i{\frac{2\tau(r^i)}{1 + \omega^2\tau^2}}\rangle \, ,
  \end{equation}
where neglecting the effect of electric field on the relaxation time, we have $\tau(r) =\omega_0^{-1}\exp{(r/l- A/2T)}$. In Eq.~{\ref{fluc-n}), we calculate the sum over the pairs which include empty and two-electron droplets. Correspondingly, in Eq.~{\ref{fluc-n}), $r^i$ is the intersite distance within the pair. In our case, all the pairs contribute to the noise spectrum. That is why, in Eq.~({\ref{fluc-n}), we can safely proceed from the average to the spatial integration, in which the dominant contribution is governed by small distances
\begin{equation}\label{fluc-n-int}
  \langle \delta n_2^2\rangle_{\omega} = 8\pi{\bar n_2}\langle \delta n_2^2\rangle_T\int_0^{\infty}{\frac{\tau(r)}{1 + \omega^2\tau^2(r)}r^2dr} \, .
  \end{equation}
As a result, we obtain the final expression (\ref{Hooge}) for the noise spectrum, which in the broad range of frequencies has almost the $1/f$ form.

%and frequencies in the range 1-1000 . It should be noted that the value of the Hooge constant  is 3-5 orders of magnitude higher in phase-separated state than  in homogeneous semiconductors. Note also that the application of the external magnetic field or the proximity to the percolation threshold can lead to the additional increase of the Hooge constant[42].

Substituting in Eq.~(\ref{Hooge}) the parameters typical of manganites, we obtain $\alpha_H \simeq 10^{-16}$ cm$^3$  at temperatures 100--200 K and frequencies in the 1--1000 s$^{-1}$ range. The value of Hooge constant $\alpha_H$ in phase-separated state is is by 3--5 orders of magnitude higher than that in homogeneous semiconductors. The application of external magnetic field or the proximity to the percolation threshold can lead to an additional increase in the Hooge constant~\cite{SboychakovJPCM2003}.

\section{Some model systems. Internal structure of magnetic polarons}
 \label{ModelSys}

\subsection{FM polarons in layered manganites}
 \label{PolaronsLayeredSys}

In the previous section, we mostly considered isotropic 3D  manganites (such as La$_{1-x}$Ca$_x$MnO$_3$ or La$_{1-x}$Sr$_x$MnO$_3$), which typically have a cubic perovskite structure. In addition to this family of manganites, a serious attention %in the last 15 years
was also focused on the family of layered manganites with a structure specified by La$_{2-x}$Ca$_x$MnO$_4$, La$_{3-x}$Ca$_x$Mn$_2$O$_7$, and so on (see e.g., \cite{nagaev2001colossal,KaganUFN2001}). In the literature, these compounds are usually referred to as the Ruddelsden--Popper phases. The first member of this family La$_{2-x}$Ca$_x$MnO$_4$  structurally is close to the well-known high-$T_c$ cuprate La$_{2-x}$Cu$_x$MnO$_4$, whereas the second one La$_{3-x}$Ca$_x$Mn$_2$O$_7$ contains two neighboring MnO planes.

In the absence of doping, at low concentration of Ca or Sr, the layered manganites usually exhibit the so-called A-type AFM structure. In this structure (see Fig.~\ref{layer-str}), spins in the conducting MnO layer are aligned ferromagnetically but the FM moments of neighboring layers are antiparallel to one another. At finite doping, the physics of layered manganites is similar to the physics of high-$T_c$ compounds. Indeed, the basic magnetic and transport properties of layered manganites are determined by 2D metallic MnO layers, while the LaO and SrO layers play a role of  charge reservoir similar to that in cuprates.

\begin{figure} [H]
\begin{center}
\includegraphics*[width=0.35\columnwidth]{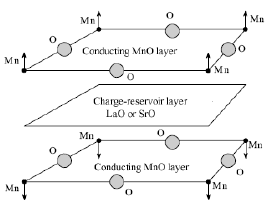} \end{center}
\caption{\label{layer-str} Spin and charge arrangement in layered manganites \cite{KaganUFN2001}.}
\end{figure}

At low doping levels, the main properties of layered manganites are captured by the anisotropic double-exchange model. The Hamiltonian of this model can be written as
\begin{equation}\label{Hamil_anisDE}
\hat{H} = -J_H\sum_{ia}{\mathbf{S}_{ia}\mathbf{\sigma}_{ia}}
- t_{\parallel}\sum_{\langle i,j\rangle a \sigma}{c_{i a \sigma}^{\dag}c_{j a \sigma}} -t_{\perp}\sum_{i, a \sigma}{c_{i a \sigma}^{\dag}c_{i a+1 \sigma}} -J_{FM}\sum_{\langle i,j\rangle a}{\mathbf{S}_{ia}\mathbf{S}_{ja}}
+J_{ff}\sum_{i a}{\mathbf{S}_{ia}\mathbf{S}_{i a+1}}  \, ,
\end{equation}
where subscript $a$ denotes the layer number, $i$ and $j$  refer to a lattice site within the layer, and $\langle i,j\rangle$ means the summation over neighboring sites. Accordingly, $t_{\parallel}$ and $t_{\perp}$ are the intralayer and interlayer hopping integrals, respectively. Finally, $J_{FM}$   describes the FM exchange between local spins inside the layer, whereas $J_{ff}$ takes into account the AFM exchange between local spins of the neighboring layers. We again consider the case of a large local spin $S \gg 1$ and work in the strong coupling limit, where the following set of inequalities is satisfied
\begin{equation}\label{chain_ineq}
J_HS \gg \{t_{\parallel}, t_{\perp}\} \gg \{J_{FM}S^2, J_{ff}S^2\}\,.
\end{equation}

We start the analysis of the different possible states in the layered manganites with Hamiltonian (\ref{Hamil_anisDE}) in the strong coupling limit (\ref{chain_ineq}) with the homogeneous classical canted state. In this state, a conduction electron travelling parallel to the layer should occupy the bottom of the conduction band. At the same time, its motion perpendicular to the layer should be an 1D analogue of the de Gennes classical canted state. Then, it is reasonable to assume that the spectrum of conduction electrons in the layered case has the form
\begin{equation}\label{anis-spect}
\varepsilon = -4t_{\parallel} - 2t_{\perp}\cos{\frac{\theta}{2}}\, ,
\end{equation}
where $\theta$  is the canting angle between the FM moments of the neighboring metallic MnO layers and $z=4$ is the number of the nearest neighbors in a MnO layer on a square lattice. The spectrum in Eq.~(\ref{anis-spect}) corresponds to the total spin at one site equal to $S_{tot} = S +1/2$ and the total spin projection $S_{tot}^z = S +1/2$. At the same time, it is possible to show (see \cite{nagaev2001colossal,KaganUFN2001,NagaevJLett1967,NagaevUFN1996,NagaevBook1983}) that unlike the isotropic situation, the second conduction band, which corresponds to the total spin projection $S_{tot}^z = S -1/2$, has the spectrum
\begin{equation}\label{anis-spect1}
\varepsilon_-= -\frac{-4t_{\parallel} - 2t_{\perp}\cos{\frac{\theta}{2}}}{2S +1} \, .
\end{equation}

Thus, $\varepsilon_-/\varepsilon \propto 1/(2S +1)\ll 1$, so we can ignore the second band and consider the classical de Gennes canted state. Minimization of the classical canting energy with respect to the parameter $\cos{(\theta/2)}$ leads for the layered case to the following result
\begin{equation}\label{cos-theta}
 \cos{\frac{\theta}{2}}= \frac{t_{\perp}n}{2J_{ff}S^2} \, ,
\end{equation}
where $n$ is the number of electrons per site. It is possible to demonstrate that the compressibility of the canted state ($\kappa^{-1} = d^2E/dn^2$) is negative. Moreover, the nanoscale phase-separated state with small FM droplets inside AFM matrix is more favorable in energy. %Let us find out the size and the structure of the polaron in this case.

We will assume that ferrons have an ellipsoidal shape (Fig.~\ref{el1ip_drop}). Similarly to the isotropic case, the energy of a FM polaron can be written as
\begin{equation}\label{anisEpol}
E_{pol} = E_0 -4t_{\parallel}n - t_{\perp}n\left( 2-\frac{\pi^2a^2}{R_{\perp}^2}\right) +2J_{ff}S^2n\Omega  \, ,
\end{equation}
where $\Omega$  is the volume of FM droplet, $R_{\perp}$  is the droplet radius in the direction perpendicular to the layers, and $E_0$  is given by
\begin{equation}\label{anisE0}
E_0 = -J_{ff}S^2 -2J_{FM}S^2  \, .
\end{equation}
The volume of ellipsoidal droplet is
\begin{equation}\label{vol_ellip}
\Omega_{ell} = \frac{4}{3}\pi \frac{R_{\perp}^2 R_{\parallel}}{a^3} \, .
\end{equation}
It is possible to show that in the anisotropic double-exchange model, the optimum energy of the system corresponds to
\begin{equation}\label{Rparal}
R_{\parallel} = R_{\perp}\sqrt{\frac{t_{\parallel}}{t_{\perp}}} \gg  R_{\perp} \, .
\end{equation}

\begin{figure}[H]
\begin{center}
\includegraphics*[width=0.2\columnwidth]{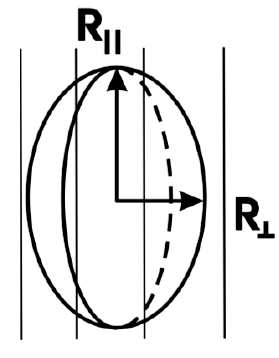} \end{center}
\caption{\label{el1ip_drop} FM droplet of the ellipsoidal shape in layered manganites \cite{KaganUFN2001}.}
\end{figure}

Relation (\ref{Rparal}) is in agreement with the neutron scattering experiments of Hennion et al. \cite{HennionPRL1998}. Using this relation, we can write the expression for the droplet volume in the following form
\begin{equation}\label{vol_ellip_t}
\Omega_{ell} = \frac{4}{3}\pi \left(\frac{R_{\perp}}{a}\right)^3\frac{t_{\parallel}}{t_{\perp}} \, .
\end{equation}
The minimization of the droplet energy (\ref{anisEpol}) with respect to parameter $R_{\perp}$  yields
\begin{equation}\label{Rperp}
R_{\perp} = a \left(\frac{\pi t_{\perp}^2}{4J_{ff}S^2t_{\parallel}}\right)^{1/5} \, .
\end{equation}
Correspondingly, the optimum volume of an ellipsoidal droplet is
\begin{equation}\label{vol_ellip_opt}
\Omega_{ell} = \frac{4}{3}\pi \left(\frac{\pi t_{\perp}^2}{4J_{ff}S^2t_{\parallel}}\right)^{3/5} \frac{t_{\parallel}}{t_{\perp}} \, .
\end{equation}

It is interesting to compare the energies of the droplets of  different shapes and thus to get the feeling what shape of the droplet is the optimum one for layered manganites. Let us illustrate this point considering also the droplets of the cylindrical shape  \cite{KaganUFN2001,NagaevUFN1996}.

The volume of a cylindrical droplet is (see Fig.~\ref{cyl-drop})
\begin{equation}\label{vol_cyl}
\Omega_{cyl} = \frac{\pi \rho^2L}{a^3} \, .
\end{equation}
The energy of cylindrical droplet is given by
\begin{equation}\label{anisEpol-cyl}
E_{pol} = E_0 -t_{\parallel}n\left (4 -\frac{9\pi^2a^2}{16\rho^2}\right) - t_{\perp}n\left( 2-\frac{\pi^2a^2}{L^2}\right) +2J_{ff}S^2n\Omega  \, ,
\end{equation}
where $E_0$ is defined in (\ref{anisE0}).

\begin{figure}[H]
\begin{center}
\includegraphics*[width=0.25\columnwidth]{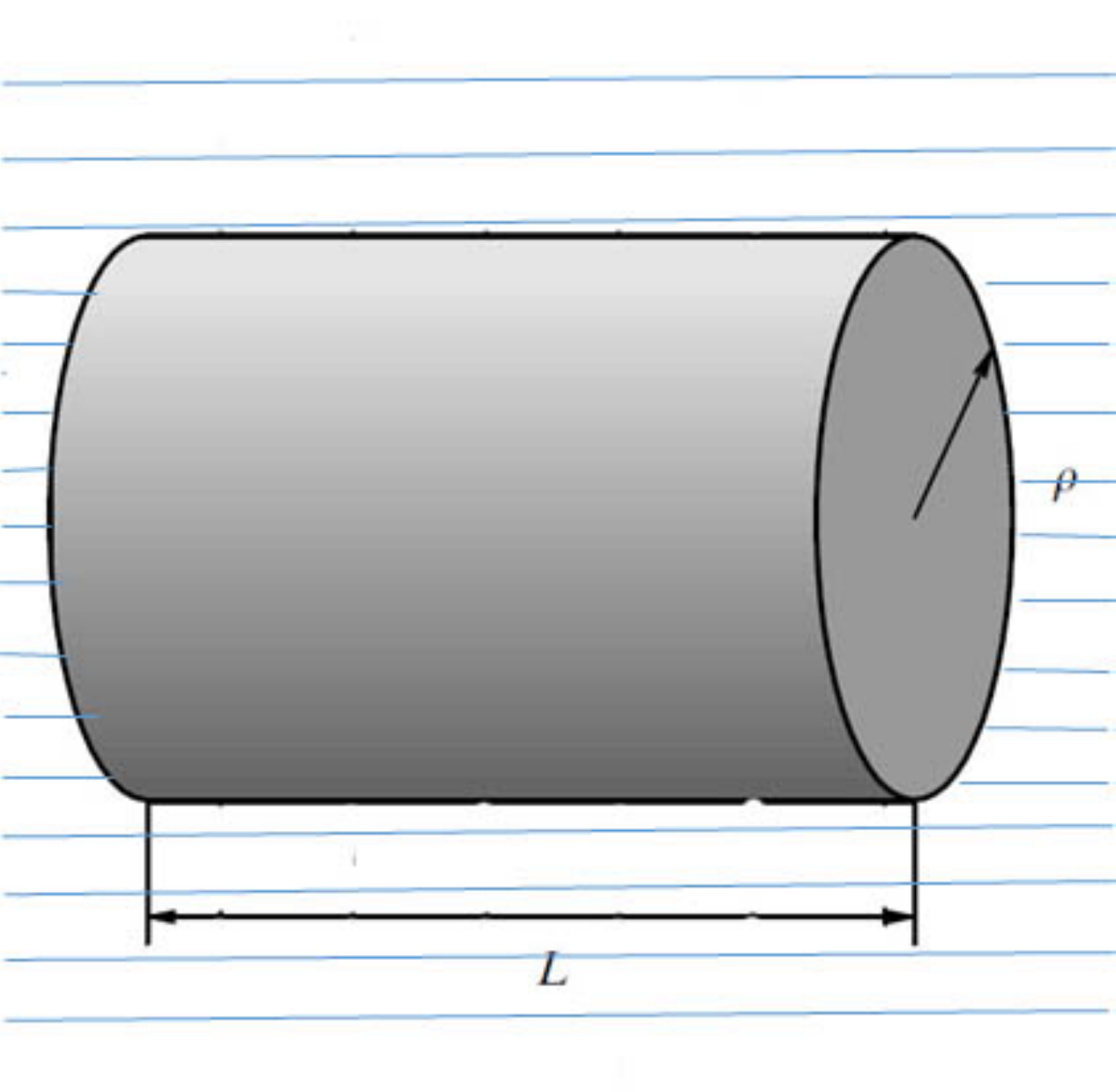} \end{center}
\caption{\label{cyl-drop} FM droplet of the cylindrical shape in layered manganites \cite{KaganUFN2001,NagaevUFN1996}.}
\end{figure}

The coefficient $9\pi^2/16$ in the expression for the energy shift from the bottom of the band (which is just $-4t_{\perp}n$) is determined by a zero of the Bessel function $J_0 =0$  for $x \approx 3\pi/4$, where we have introduced the dimensionless variable $x =\rho/a$. It corresponds to the energy of the lowest bound state of an itinerant  electron in the 2D cylindrically symmetric potential well.

The minimization of the droplet energy in Eq.~(\ref{anisEpol-cyl}) with respect to both $\rho$ and $L$ yields
\begin{eqnarray}\label{rhoL-cyl}
\rho &\approx& 0.53L\sqrt{\frac{t_{\parallel}}{t_{\perp}}}\, ,
 \nonumber \\
  L &\approx& 0.98a\left(\frac{4\pi t_{\perp}^2}{t_{\parallel}J_{ff}S^2}\right)^{1/5} \, .
\end{eqnarray}
As a result, the optimum volume of the cylindrical droplet reads
\begin{equation}\label{Omega-cyl}
\Omega_{cyl} \approx \pi\frac{L^2}{a^3}\frac{t_{\parallel}}{4t_{\perp}} \approx 0.26\pi\left(\frac{4\pi t_{\perp}^2}{t_{\parallel}J_{ff}S^2}\right)^{3/5}\frac{t_{\parallel}}{t_{\perp}} \, .
\end{equation}
The direct comparison of the optimum energies of ellipsoidal and cylindrical droplets shows that
\begin{equation}\label{ell-cyl-ratio}
\frac{E_{ell.pol} - \tilde{E}_0}{E_{cyl.pol} - \tilde{E}_0} \approx \frac{\Omega_{ell}}{\Omega_{cyl}} \approx 0.96 \, ,
\end{equation}
where $\tilde{E}_0 = E_0 - 4t_{\perp}n -2t_{\parallel}n$.

Hence, we come to the conclusion that the optimum shape of the droplet for layered manganites is an ellipsoidal one. Physically, this is quite reasonable since the optimum shape of FM polaron (its ``ground state" to some extent) repeats the form of the electronic spectrum in the layered case, which is elliptically symmetrical and is given by $\varepsilon(p) \propto t_{\parallel}p_{\parallel}^2 t_{\perp}p_{\perp}^2$.  Of course, after the dilatation transformation of the momenta $p_{\parallel} = \tilde{p}_{perp}\frac{}{}\sqrt{\frac{t_{\parallel}}{t_{\perp}}}$ to the new reference frame, an electronic spectrum becomes spherical. In the layered case, the percolation threshold, which corresponds to the overlap of the droplets, takes place at the critical density
\begin{equation}\label{crit-dens1}
\tilde{n}_c = \frac{1}{\Omega_{ell}} = \frac{3}{4\pi} \left(\frac{4J_{ff}S^2t_{\parallel}}{\pi t_{\perp}^2}\right)^{3/5} \frac{t_{\perp}}{t_{\parallel}} \, .
\end{equation}
For all densities $n < \tilde{n}_c$, the energy of the droplet state is lower than all the optimum energies of the homogeneous state.

\subsection{FM polarons in doped anisotropic AFM systems}
\label{polarons-doped-anis}

We are now ready to face a more general situation and to consider the shape and the size of FM droplets in anisotropic two-dimensional (2D) or 3D cases when the electron hopping integrals along $x$, $y$, and $z$ directions, $t_x$, $t_y$, and $t_z$, as well as the AFM exchange integrals  $J_x$, $J_y$, and $J_z$ are different. We find that in analogy with the situation in layered manganites, the most favorable shape of the FM droplet is again an ellipsoidal one. The optimum energy and the optimum volume of the droplet are expressed in terms of the universal averaged parameters $\overline{J} = (J_x + J_y + J_z)S^2$ and $t_{eff} = (t_xt_yt_z)^{1/3}$ in 3D or respectively $\overline{J} = (J_x + J_y )S^2$ and $t_{eff} = (t_xt_y)^{1/2}$   in 2D anisotropic AFM systems~\cite{KaganJPCM2006}. These results are, in particular, interesting in relation to neutron scattering experiments \cite{HennionPRL1998,HennionPRB2000,HennionNJP2005,HennionPRL2005}, which prove the existence of FM droplets of different shapes in perovskite and layered manganites. Additional experimental evidence of the existence of nanoscale FM droplets in layered manganites comes from scanning tunneling microscopy with atomic resolution \cite{RonnowNature2006}. The ferrons in low-doped layered manganites often proved to be ``platelets" characteristic of 2D rather than spherical droplets characteristic of the 3D case \cite{HennionNJP2005,HennionPRL2005}.

\subsubsection{Shape of FM droplets in anisotropic 2D case}
\label{polarons-anis2D}

In the anisotropic 2D case, the system is described by the anisotropic FM Kondo-lattice model in the double-exchange limit
\begin{equation}\label{Hamil_anis2D}
\hat{H} = -J_H\sum_{i}{\mathbf{S}_{i}\mathbf{\sigma}_{i}}
 + \sum_{\langle i,j\rangle_a}{J_a\mathbf{S}_{ia}\mathbf{S}_{ja}}
 - \sum_{\langle i,j\rangle_a \sigma}{t_ac_{i a \sigma}^{\dag}c_{j a \sigma}} \, ,
\end{equation}
where $\langle i,j\rangle_a$ denotes the neighboring sites in the square lattice along the $a$ direction, $a =\{x, y\}$, and $J_H \gg \{t_x, t_y\}\gg \{J_x, J_y\}$.

The analysis of the volume and energy of the FM droplets of  different shapes in anisotropic 2D case is very similar to the analysis of the energy and volume of the ellipsoidal and cylindrical droplets in the layered manganites. It uses again the dilatation transformation of the coordinates (which restores the cylindrical symmetry in the new reference frame in 2D) alongside with the properties of the zeroth order Bessel function $J_0$ (the value of its first zero) for the lowest binding state of an electron in the cylindrical potential well. As a result, the most energetically favorable shape of the droplet in the anisotropic 2D case is an elliptical one with the volume
\begin{equation}\label{Omega-anis2D}
\Omega = \frac{j_{0,1}}{2}\left(\frac{\pi t_{eff}}{\overline{J}}\right)^{1/2}  \, ,
\end{equation}
where $j_{0,1} = 0.2404 \approx 3\pi/4$ is the first zero of the zeroth order Bessel function $J_0(x)$ corresponding to the solution of the Schr\"{o}dinger equation in polar coordinates and  $t_{eff},  \overline{J}$ are the universal averaged parameters mentioned above.

Similarly, in the 3D anisotropic case the most favorable shape of the FM droplet is an ellipsoidal one. Its  optimum volume is
\begin{equation}\label{Omega-anis3D}
\Omega = \frac{\pi^{8/5}2^{1/5}}{3}\left(\frac{t_{eff}}{\overline{J}}\right)^{3/5}  \, ,
\end{equation}
and is again determined by the ratio $t_{eff}/\overline{J}$ of the universal averaged parameters.

\subsubsection{FM droplets in the case of frustrated triangular lattice}
\label{polarons-triang2D}

Just in the same manner, we can find the volume and the most favorable shape of the FM droplet on the 2D frustrated triangular lattice (with the same magnitude of AFM exchanges and hopping integrals along the sides of triangle) for the planar spin configuration. Similarly to the situation described by  Eq.~(\ref{Omega-anis2D}), the most favorable shape of the ferron corresponds to the circle with an area \cite{KaganOgarJPCM2008}
\begin{equation}\label{Omega-triang2D}
\Omega = j_{0,1}\left(\frac{\pi t}{3J}\right)^{1/2}  \, .
\end{equation}
%Note that the difference of the Schr\"{oe}dinger equation on the triangular lattice from the corresponding equation on the square lattice is related to the factor 3/4  at $t\Delta_r$.

\subsection{Bound magnetic polarons with extended spin distortions}
\label{extended_spin-dist}

In Section~\ref{MagPolaronTrans}, we have mentioned an old idea, which belongs to de Gennes about the magnetic polarons of a different type, which resemble the magnetic impurity and can produce a long-range tail of spin distortions due e.g., to the magnetodipole interaction. This type of bound magnetic polarons was introduced in \cite{OgarkovPRB2006,KugelOgarPhB2008,KaganOgarJPCM2008}, based on the  model for an electron, which is bound at a donor impurity by the Coulomb attractive potential. The Hamiltonian of such a model has the following form
\begin{equation}\label{Hamil-bound-ferron}
\hat{H} = \hat{H}_{el} +J_{ff}\sum_{\langle i,j\rangle}{[\mathbf{S}_i\mathbf{S}_j + S^2]} - K\sum_{i}{[(S_i^x)^2 -S^2]} \, ,
\end{equation}
\begin{equation}\label{Hamil-el-bound-ferron}
\hat{H}_{el} = -t\sum_{\langle i,j\rangle \sigma}{c_{i  \sigma}^{\dag}c_{j \sigma}} - -J_H\sum_{i}{\mathbf{S}_{i}\mathbf{\sigma}_{i}}  -V \sum_{i,\sigma}{\frac{c_{i  \sigma}^{\dag}c_{i \sigma}}{|\mathbf{n} -\mathbf{n}_0|}} \, ,
\end{equation}
where $K$ is the constant of magnetic anisotropy. The last term in Eq.~(\ref{Hamil-el-bound-ferron}) describes the Coulomb interaction between conduction electrons and the impurity ion located in the middle of the main diagonal of the unit cell (e.g., in the 3D case between the lattice sites $\mathbf{n} = (0,0,0)$ and $\mathbf{n} = (1,1,1)$, i.e. $\mathbf{n}_0 = (1/2,1/2,1/2)$ and similarly, $\mathbf{n}_0 = (1/2,1/2)$ in the 2D case, see Fig.~\ref{unit-cell}.

\begin{figure}[H]
\begin{center}
\includegraphics*[width=0.2\columnwidth]{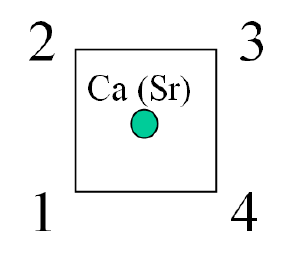} \end{center}
\caption{\label{unit-cell} Nonmagnetic Ca(Sr) donor impurity is located in the middle of the quadratic unit cell in 2D  \cite{KaganBookSpringer}.}
\end{figure}

We work in the limit of strong Coulomb interaction and strong Hund's rule coupling
\begin{equation}\label{hierarchy}
V \gg J_HS \gg t \gg J_{ff}S^2, KS^2 \, .
\end{equation}
In this range of parameters, the radius of electron localization is of the order of interatomic distance $a$. The magnetic distortions outside the electron localization region (the ferron core), as it was shown in \cite{OgarkovPRB2006,KugelOgarPhB2008,KaganOgarJPCM2008}, decrease with the distance $r$  as $1/r^4$ and $1/r^2$  in the 3D cubic and 2D square lattices, respectively. At the same time, they decay even slower, as $1/r$  in the 2D frustrated triangular lattice (for the planar spin configuration).

In both 3D and 2D cases, these ``coated" magnetic polarons correspond to the ground state of the system at very small doping levels. At the same time, in the 1D case for the AFM chain of interacting neighboring spins with donor impurities located just in the middle (between the nearest neighbor spins ) the situation is quite different. Here,  as it was shown in \cite{GonzalezPRB2004,SboychakovGonzPRB2005,KugelMakhach2005}, the ``coated" magnetic polarons with slowly decaying tails of spin distortions (shown e.g., in Fig.~\ref{coated-1D}) correspond usually to the excited state of the system, while the ground state corresponds to the more standard (``bare" or rigid) magnetic polarons with very rapidly decreasing tails of spin distortions. In other words, we can say that in rigid ferrons, the intermediate region where the canting angle $\nu$  changes from zero (FM domain) to $\pi$ (AFM domain) is narrow (of the order of the interatomic distance $a$), so the radius of magnetic polaron $R_{pol}$ is a well-defined quantity. Moreover, the tails of magnetic distortions exponentially decrease for the distances $r > R_{pol}$ (outside the ferron).

\begin{figure}[H]
\begin{center}
\includegraphics*[width=0.4\columnwidth]{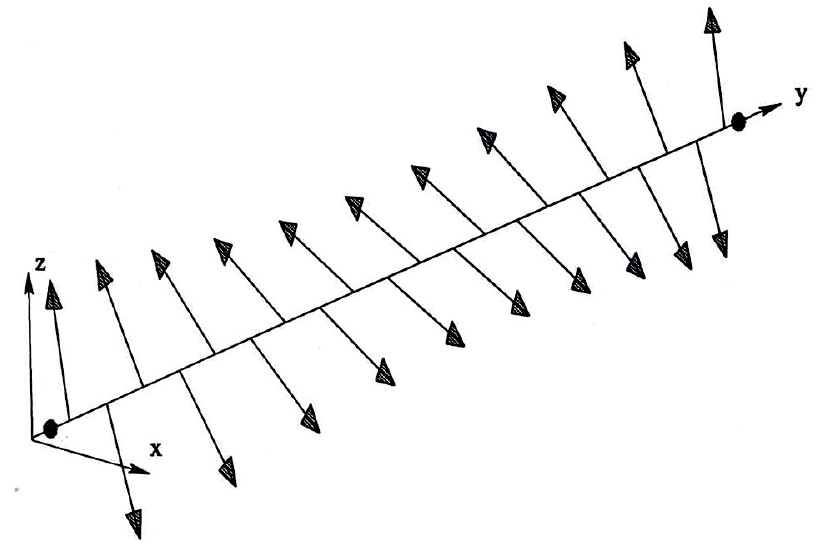} \end{center}
\caption{\label{coated-1D}  Magnetic structure with slowly decaying spin distortions between two donor impurities corresponding to the coated magnetic polaron in the 1D AFM chain~\cite{KugelMakhach2005}.}
\end{figure}

Let us now focus on in the 3D and 2D cases. At low electron or hole doping ($n \ll 1$ or $x \ll 1$), we can consider a one-particle problem. In the limit of strong Hund’s rule coupling $J_H \rightarrow\infty$, the spin $\mathbf{\sigma}_n$ of a conduction electron at site $n$ should be parallel to the local spin $\mathbf{S}_n$. If $S \gg 1$, it is sufficient to consider the local spin as a classical vector, i.e., to represent it in the 3D case as \cite{OgarkovPRB2006,KugelOgarPhB2008,KaganOgarJPCM2008}
\begin{equation}\label{S-3D}
\mathbf{S}_n =S(\sin{\theta_n}\cos{\varphi_n}, \sin{\theta_n}\sin{\varphi_n}, \cos{\theta_n}) = S\mathbf{e}_n \, ,
\end{equation}
where $\mathbf{e}_n$ is the unit vector, $\theta_n$ and $\varphi_n$ are polar and azimuthal angles of $\mathbf{e}_n$.

For the planar structure, when all local spins lie in ($x,y$) plane, we have $\theta_n = \pi/2$  and
\begin{equation}\label{S-2D}
\mathbf{S}_n =S(\cos{\varphi_n}, \sin{\varphi_n}, 0), \,\,(-\pi < \varphi_n \leq \pi)\, .
\end{equation}

In the limit of strong electron--impurity coupling $V \rightarrow \infty$, the electron wave function $\Psi_n$ will be nonzero only at sites nearest to the impurity. Let us focus on the 2D case (the 3D case is considered in a similar way). In this case, we can assume that $\Psi_n \neq 0$  only at four sites nearest to the impurity: $\mathbf{n}_1 = (1,1)$,  $\mathbf{n}_2 = (1,0)$,  $\mathbf{n}_3 = (0,1)$, and  $\mathbf{n}_4 = (0,0)$ in the square lattice (see Fig.~\ref{unit-cell}) and similarly, $\Psi_n$ is nonzero at eight nearest sites for the 3D cubic lattice, and three nearest sites at the  frustrated triangular lattice. The wave function of the conduction electron in the explicit form can be written as  \cite{OgarkovPRB2006,KugelOgarPhB2008,KaganOgarJPCM2008}
\begin{equation}\label{wavefuncPhi}
|\Phi\rangle = \frac{1}{\sqrt{2}}\sum_{n}{\Psi_n\left(\hat{c}_{n\uparrow}^{\dag} +\exp{(-i\varphi_n)}\hat{c}_{n\downarrow}^{\dag}\right)}|0\rangle \, ,
\end{equation}
where $|0\rangle$ is a vacuum state. The energy of a conduction electron in the state $|\Phi\rangle$ is
\begin{equation}\label{E-cond-el}
E_{el}[\Phi] = \langle \Phi |\hat{H}_{el}| \Phi \rangle = -t\sum_{\langle n,m \rangle}{[T_{mn}\Psi_n^*  \Psi_m + T_{mn}^* \Psi_n \Psi_m^*]} - \frac{1}{2} J_HS -V \sum_{n}{\frac{|\Psi_n|^2}{|\mathbf{n} - \mathbf{n}_0|}} \, ,
\end{equation}
where $\sum_{n}{|\Psi_n|^2} =1$,
\begin{eqnarray}\label{hoppings}
T_{mn}&=&\frac{1}{2}\left[1+ \exp{(i(\varphi_n-\varphi_m))}\right] =\cos{\frac{\nu_{nm}}{2}}\exp{(i\omega_{nm})}\, ,\nonumber \\
\nu_{nm}&=&\varphi_n-\varphi_m\,, \,\,\, \omega_{nm} = \frac{\nu_{nm}}{2}\, .
\end{eqnarray}
In Eq.~(\ref{hoppings}), $\nu_{nm}$ is the angle between the local spins $\mathbf{S}_n$  and  $\mathbf{S}_m$ (canting angle) and $\omega_{nm}$ is the Berry phase~\cite{BerryPRSocA1984,MullerHartPRB1996}.

The electron energy in the ground state for fixed planar spin configuration (at given angles $\varphi_n$) can be found from the minimization of Eq.~(\ref{E-cond-el}) with respect to $\Psi_n$. As a result, we have
\begin{equation}\label{E-cond-el-min}
E_{el}(\varphi_n) = \frac{1}{2} J_HS -\frac{V}{b} -t\varepsilon(a_{nm}) \, ,
\end{equation}
where $b = 1/\sqrt{2}$ for the square lattice, $a_{nm} = \cos{\frac{\nu_{nm}}{2}}$, and
\begin{equation}\label{epsilon-a}
\varepsilon(a_{nm}) = \frac{1}{\sqrt{2}}\{a_{12}^2 + a_{23}^2 +a_{34}^2 + a_{41}^2 + [((a_{12} - a_{34})^2 + (a_{23} + a_{41})^2)((a_{12} + a_{34})^2 + (a_{23} - a_{41})^2)]^{1/2}\}^{1/2}  \, .
\end{equation}
Correspondingly, for the triangular lattice, we have $b = 1/\sqrt{32}$ and
\begin{equation}\label{epsilon-a-triang}
\varepsilon(a_{nm}) = g(a_{nm}) + \frac{a_{12}^2 + a_{23}^2 +a_{31}^2}{3g(a_{nm}} \, ,
\end{equation}
where
\begin{equation}\label{g-anm}
g(a_{nm})= \{a_{12}a_{23}a_{31} + \frac{1}{3\sqrt{3}}[(a_{12}^2 + a_{23}^2 +a_{31}^2)^3 -27a_{12}^2a_{23}^2a_{31}^2]^{1/2}\}^{1/3} \, .
\end{equation}

The electron energy $E_{el}$ has a minimum when all neighboring local spins surrounding an impurity are parallel (or as we will see later, almost parallel) to each other.  Hence, at this level of approximation, we have a bound magnetic polaron state, which can be described by a ferromagnetic core of radius $b$ (with four local spins at the square lattice and three local spins at the triangular one) embedded into the antiferromagnetic host. When in the next subsection, we include magnetic degrees of freedom into account, then after the minimization of the total energy, we will observe that for the most energetically favorablel ``coated" magnetic polaron, its magnetic core will not be fully polarized and the AFM  background around it will be substantially distorted on rather large distances.

\subsection{Magnetic structure of a coated ferron}
\label{coated ferron}

Magnetic contribution to the total energy of a coated ferron can be found in the general case of 3D spins by the minimization of the total energy with respect to the angles $\theta_n$ and $\varphi_n$.  To do this, it is convenient, first, to perform the following transformation of angles: $\varphi_n \rightarrow \varphi_n + \pi, \theta_n \rightarrow \pi -\theta_n$ for one of the sublattices of the cubic (or square) lattice. For planar spin configuration in the ($x,y$) plane, after the transformation, $\theta_n$  is still equal to $\pi/2$. For the 2D triangular lattice and planar spin configuration, the corresponding transformation reads $\varphi_n \rightarrow \varphi_n \pm 2\pi/3$  for two of the three  magnetic sublattices. As a result of this transformation, the AFM order becomes FM and \emph{vice versa}. Such a transformation allows us to work with continuously changing orientation of spins outside the ferron core. Moreover, all the angles belong to the range from 0 to  $\pi/2$  for cubic and square lattices and to the range from 0 to  $2\pi/3$ for the triangular lattice.The total energy is
\begin{equation}\label{Etot}
E = E_{el} + E_{mag} \, ,
\end{equation}
where $E_{el}$ is given by Eq.~(\ref{E-cond-el-min}) and
\begin{equation}\label{Emag}
E_{mag} = J_{ff}S^2\sum_{\langle n,m\rangle \sigma}{(1 -\cos{\nu_{nm}})}  -KS^2\sum_{n}{(\sin^2{\theta_n}\cos^2{\varphi_n} -1)} \, .
\end{equation}
After this transformation, we should replace $a_{nm}$ in Eqs.~(\ref{epsilon-a})--(\ref{g-anm}) by $\alpha_{nm} = \sin{\frac{\nu_{nm}}{2}}$.

Minimization of the total energy with respect to the angles $\theta_n$ and $\varphi_n$  yields two possible types of the solutions for the bound magnetic polaron. The first one corresponds to the magnetic polaron with completely polarized spins inside its core, which is embedded in the purely AFM background. The total magnetic moment of the polaron is parallel to the easy axis. We refer this trivial solution as a ``bare" magnetic polaron. Its energy for the 2D square lattice is
\begin{equation}\label{Ebare2D}
E_{pol}^0 = -2t +16J_{ff}S^2 \, ,
\end{equation}
and
\begin{equation}\label{Ebare3D}
E_{pol}^0 = -3t +48J_{ff}S^2 \, ,
\end{equation}
for the 3D cubic lattice  \cite{OgarkovPRB2006,KugelOgarPhB2008,KaganOgarJPCM2008}. The corresponding expression for the energy of a ``bare" magnetic polaron on the 2D triangular lattice is more cumbersome. Its explicit form is given in \cite{KaganOgarJPCM2008}.

There exists another solution corresponding to the magnetic polaron state with magnetic moment perpendicular to the easy axis. We call this solution a ``coated" magnetic polaron. Let us consider the planar spin configuration in the ($x,y$) plane on the 2D square lattice. Then, from the symmetry requirements, it is clear that the angles $\varphi_n, \,\, n = 1, ..., 4$ for the local spins inside the core of magnetic polaron should satisfy the conditions
\begin{equation}\label{phi-square}
\varphi_1 = \varphi_3 = \varphi_0, \,\, \varphi_2 = \varphi_4 = -\varphi_0 \, ,
\end{equation}
where $0<\varphi_0<\leq \pi/2$ (see Fig.~\ref{spin-sym})

\begin{figure}[H]
\begin{center}
\includegraphics*[width=0.25\columnwidth]{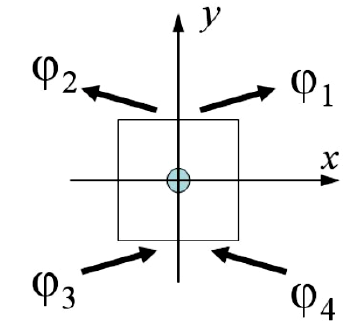} \end{center}
\caption{\label{spin-sym}  Spin symmetry inside the ferron after the lattice transformation~\cite{KaganBookSpringer}.}
\end{figure}

For the 3D cubic lattice and the 2D triangular lattice, the angles $\varphi_n$  satisfy, respectively  the conditions
\begin{eqnarray}\label{phi-cube_triang}
1 = \varphi_3 = \varphi_5 =\varphi_7 = \varphi_0, \,\, \varphi_2 = \varphi_4 = \varphi_6 = \varphi_8 = -\varphi_0 \, , \textrm{for the 3D cubic lattice}, \nonumber \\
\varphi_1 = 0, \, \varphi_2 = -\varphi_3 = \varphi_0, \,  0 < \varphi_0 \leq 2\pi/3\, , \textrm{for the 2D triangular lattice} \, . \end{eqnarray}
Minimization of the total energy of a coated magnetic polaron with respect to $\varphi_n$  yields the following set of nonlinear equations
\begin{equation}\label{nonlinear_eqns}
\sum_{\mathbf{\Delta}}{\sin{(\varphi_{\mathbf{n}+\mathbf{\Delta}} -\varphi_{\mathbf{n}})}} -\frac{\kappa}{2}\sin{2\varphi_0} = \frac{t}{2J_{ff}S^2} \sum_{i}{\delta_{\mathbf{nn}_i}(-1)^i \cos{\varphi_i}} \, ,
\end{equation}
where $\mathbf{n} = (n_x, n_y), \, \kappa = 2K/J_{ff}$, $\delta_{\mathbf{nm}}$  is the Kronecker symbol, and $\mathbf{\Delta}$  takes the values ($\pm1, 0$) and (0, $\pm1$).

Equations~(\ref{nonlinear_eqns}) with conditions~(\ref{phi-square}) were solved numerically for the cluster containing $40\times40$  sites. The further growth in the number of sites in the cluster does not change the obtained results. The initial angle  $\varphi_0$ was also found. The calculated magnetic structure is shown in Fig.~\ref{coated-ferron}.

\begin{figure}[H]
\begin{center}
\includegraphics*[width=0.7\columnwidth]{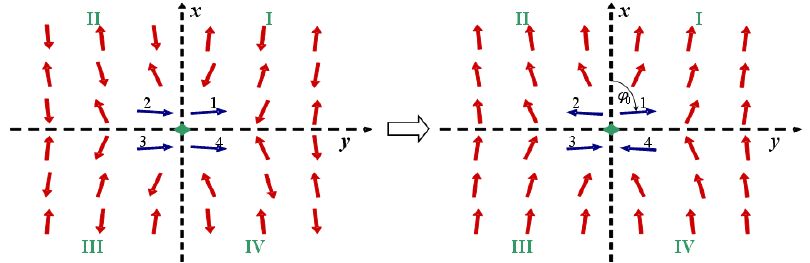} \end{center}
\caption{\label{coated-ferron}  ``Coated" magnetic polaron on the square lattice before (left panel) and after (right panel) transformation of angles in one of sublattices. The magnetic structure is calculated by solving Eq.~(\ref{nonlinear_eqns}) at $t/J_{ff}S^2 = 50$ and $\kappa =5\times 10^{-3}$. At these values of parameters, we have $\varphi_0 = 85^{\circ}$  \cite{OgarkovPRB2006}.}
\end{figure}

We see from this figure that in contrast to the ``bare" ferron, the ``coated" magnetic polaron produces spin distortions of the AFM background outside the region of electron localization (the core region). The ``coated" magnetic polaron has a magnetization lower than its saturation value ($\varphi_0 < \pi/2$ ). Moreover the ``coat" has a magnetic moment opposite to that of the core (sites 1--4 in Fig.~\ref{coated-ferron} for the square lattice). To obtain analytical estimations for the spatial distribution of the spin distortions, it is important to find an approximate solution of Eq.~(\ref{nonlinear_eqns}) in the continuum limit. To do this, we should treat the angles $\varphi_n$  as the values of the continuous function $\varphi (\mathbf{r})$  at points  $\mathbf{r} =\mathbf{n} - \mathbf{n}_0$. Assuming that outside the magnetic polaron, the following condition for spatial variations of the angles $|\varphi_{\mathbf{n}+\mathbf{\Delta}} -\varphi_{\mathbf{n}}| \ll 1$ is met, we can expand $\varphi (\mathbf{r}+\mathbf{\Delta})$ in the Taylor series up to the second order in $\mathbf{\Delta}$
\begin{equation}\label{phiTaylor}
\varphi (\mathbf{r}+\mathbf{\Delta}) \approx \varphi (\mathbf{r})+\Delta^{\alpha}d_{\alpha}\varphi (\mathbf{r}) + \frac{1}{2} \Delta^{\alpha}\Delta^{\beta}d_{\alpha}d_{\beta}\varphi (\mathbf{r}) \, .
\end{equation}
Substituting this expansion in Eq.~(\ref{nonlinear_eqns}), we find that the function $\varphi(\mathbf{r})$ outside the magnetic polaron should satisfy the 2D sine-Gordon equation
\begin{equation}\label{sine-Gordon}
\Delta \varphi - \frac{\kappa}{2}\sin{2\varphi} = 0  \, .
\end{equation}
In the range of the parameters under study, $K < J_{ff}$, that is, $\kappa < 1$, we can linearize this equation. As a result we obtain
\begin{equation}\label{sine-Gordon-lin}
\Delta\varphi - \kappa \varphi = 0  \, .
\end{equation}
This equation should be solved with the boundary conditions at infinity $\varphi(\mathbf{r})_{\mathbf{r} \rightarrow \infty} \rightarrow 0$  and with some boundary conditions at the surface of the magnetic polaron. We model the magnetic polaron by a circle of radius $R_{pol} = a/\sqrt{2}$, where $a$  is the intersite distance, and choose the Dirichlet boundary conditions (see Fig.~\ref{Dirich-fig})
\begin{equation}\label{Dirichlet}
\varphi(\mathbf{r})|_{r=b} = \tilde{\varphi}(\zeta)  \, ,
\end{equation}
where we introduced the polar coordinates ($r, \zeta$)  in the ($x, y$) plane.
\begin{figure}[H]
\begin{center}
\includegraphics*[width=0.2\columnwidth]{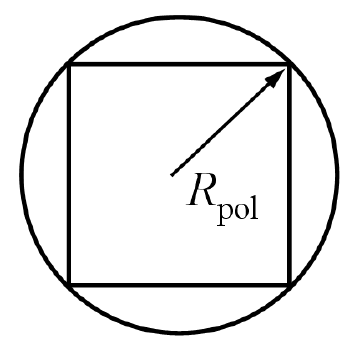} \end{center}
\caption{\label{Dirich-fig}  Dirichlet boundary conditions at the surface of magnetic polaron for the 2D square lattice. We model the core of magnetic polaron by the circle of radius $R_{pol} = a/\sqrt{2}$  \cite{KaganBookSpringer}.}
\end{figure}
The function $\tilde{\varphi}(\zeta)$ can be found in the following way. For the square lattice, $\tilde{\varphi}(\zeta)$  should meet symmetry conditions (\ref{phi-square}) at the points $\zeta_i = \pi(2i-1)/4$
\begin{equation}\label{phi-i}
\tilde{\varphi}(\zeta) = \varphi_i\, ,\,\,\,i = 1, ..., 4 \, .
\end{equation}
Since the function $\tilde{\varphi}(\zeta)$ is periodic, it can be expanded into Fourier series
\begin{equation}\label{phiFourier}
\tilde{\varphi}(\zeta) = \sum_{m=0}^{\infty}{[a_m\cos{m \zeta}+ b_m\sin{m \zeta}]} \, .
\end{equation}
In Eq.~(\ref{phiFourier}), we neglect the terms with $m > 2$ , which allows us to keep the minimum number of terms to satisfy conditions (\ref{phi-square}). It follows from Eq.~(\ref{phi-i}) that $a_0 = a_1 =a_2 = b_1 = 0$. Finally, we obtain
\begin{equation}\label{phi-fin}
\tilde{\varphi}(\zeta) = \varphi_0\sin{2\zeta} \, .
\end{equation}

The solution of Eq.(\ref{sine-Gordon-lin}) with boundary condition (\ref{Dirichlet}) yields
\begin{equation}\label{phi-solution}
\varphi(\mathbf{r}) =  \frac{\varphi_0}{K_2(R_{pol}/r_0)}K_2\left(\frac{r}{r_0}\right) \sin{2\zeta}\, , \,\,\, r_0 = \frac{a}{\sqrt{\kappa}}\, ,
\end{equation}
where $K_2(x)$ is the Macdonald function of the second order. Within the $|\mathbf{r}| < r_0$ range, the function $\varphi(\mathbf{r})$ behaves as $R_{pol}^2/r^2$, whereas at large distances it decreases exponentially, $\varphi(\mathbf{r}) \propto \exp{(-r/r_0)}$. Thus, $R_{pol}$ and $r_0$ play role of the size of the core and the radius of the ``coat", respectively. Moreover, in the case of small anisotropy ($\kappa \ll 1$), the distance, where function $\varphi(\mathbf{r})$ has a power-law behavior, can be large enough in agrement with the de Gennes conjecture~\cite{deGennesPR1960}. The function $\varphi(\mathbf{r})$ at $\zeta = \pi/4$ and the numerical results for $\varphi_{n, n}$  are plotted in Fig.~\ref{phi-vs-r}.

\begin{figure}[H]
\begin{center}
\includegraphics*[width=0.5\columnwidth]{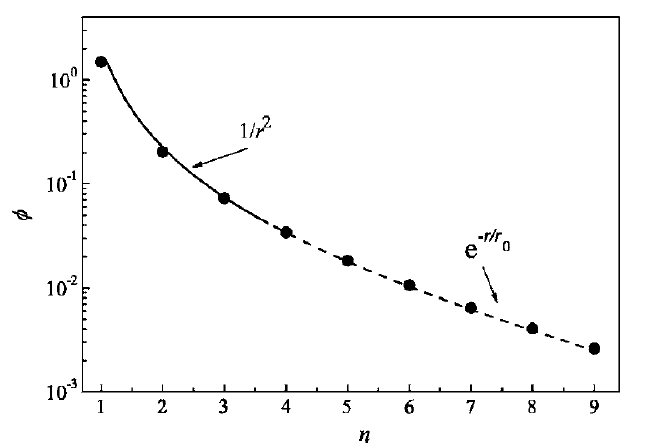} \end{center}
\caption{\label{phi-vs-r}  Decrease with the distance of an angle  $\varphi(\mathbf{r})$ between misaligned spins and easy axis for a coated magnetic polaron on 2D square lattice~\cite{OgarkovPRB2006}. The circles correspond to the numerical solution of the discrete problem on a square cluster at $t/J_{eff}S^2 = 50$  and $\kappa =5\times10^{-2}$  ($r_0 = 4.5a$). The solid curve corresponds to analytical solution (\ref{phi-solution}) in the continuum limit. At small distances, $\varphi(\mathbf{r})$  decreases proportionally to $1/r^2$, while at large distances $r \gg r_0$, it decreases exponentially.}
\end{figure}

For the 3D cubic lattice, the similar analytical solution of Eq.~(\ref{sine-Gordon-lin}) with symmetry conditions (\ref{phi-cube_triang}), the Dirichlet boundary conditions, and a spherical core of magnetic polaron with the radius $R_{pol} = a\sqrt{3}/2$ yields
\begin{equation}\label{phi-solution3D}
\varphi(x,y,z) =  \varphi_0\frac{R_3(r/r_0)}{R_3(R_{pol}/r_0)}\frac{3\sqrt{3}xyz}{r^3} \, ,
\end{equation}
where $r_0 =a/\sqrt{\kappa}$ and
\begin{equation}\label{R3}
R_3(\rho) = \left(1+\rho +\frac{2\rho^2}{5} + \frac{\rho^3}{15}\right)\frac{e^{-\rho}}{\rho^4}\, , \quad \rho = \frac{r}{r_0} \, .
\end{equation}
It follows from Eqs.~(\ref{phi-solution3D}) and (\ref{R3}) that $\varphi(\mathbf{r})$ decreases as $(R_{pol}/r)^4$  at $r < r_0$. At distances $r$  exceeding the characteristic radius $r_0$, the spin distortions decay exponentially $\varphi(\mathbf{r})\propto \exp{(-r/r_0)}$ as in the 2D case.

Finally, for the 2D triangular lattice an analytical solution of Eq.~(\ref{sine-Gordon-lin}) with symmetry conditions (\ref{phi-cube_triang}) and the Dirichlet boundary conditions for polar coordinates and a circular core of magnetic polaron with the radius $R_{pol} = a/\sqrt{3}$ is given by~\cite{KaganOgarJPCM2008}
\begin{equation}\label{phi-solution_triang}
\varphi(\mathbf{r}) =  \frac{2\varphi_0}{\sqrt{3}}\frac{K_1(r/r_0^*)}{K_1(R_{pol}/r_0^*)}\sin{\zeta}\, ,
\end{equation}
where $r^* = a\sqrt{4\kappa/3} = r_0\sqrt{3}/2$ and $K_1(x)$ is the Macdonald function of the first order. Within the $r < r_0^*$ range, the function $\varphi(\mathbf{r})$ slowly decreases as $R_{pol}/r$.. At the same time, at large distances, it decreases exponentially again, $\varphi(\mathbf{r})\propto \exp{(-r/r_0^*)}$ \cite{KaganOgarJPCM2008}.

Let us compare now the energies of the ``coated" magnetic polarons with the ``bare" ones. For the 2D square lattice, the energy of the ``bare" magnetic polaron is given by Eq.~(\ref{Ebare2D}) and the energy of the ``coated" magnetic polaron is given by Eqs.~(\ref{Etot}) and (\ref{Emag}), where $\theta_n = \pi/2$  and $\varphi_n$ are the solutions of Eqs.(\ref{nonlinear_eqns}). In the continuum approximation, we expand $\cos{\nu_{nm}}$ and $\cos^2{\varphi_n}$ in the Taylor series up to the second order outside the ferron core. Changing the summation by the integration over $r$, we can represent the total energy of the ``coated" magnetic polaron as
\begin{equation}\label{Etot-coated}
E =E_{el} +E_M =\left[-\frac{1}{2}J_HS - V\sqrt{2}- 2t\sin{\varphi_0} \right] +8 J_{ff}S^2 \left( 1+ \frac{\kappa}{4} \right)\sin^2{\varphi_0} +\frac{J_{ff}S^2}{2}\int_{r\geq r_0}{d^2r[(\nabla\varphi)^2 + \kappa\varphi^2]} \, ,
\end{equation}
In Eq.~(\ref{Etot-coated}), the last two terms correspond to the magnetic part of the total energy. The first one of them comes from the summation over spins in the core of magnetic polaron. Substituting the solution for the function $\varphi(\mathbf{r})$  from Eq.~(\ref{phi-solution}) into Eq.~(\ref{Etot-coated}) and performing the integration, we get
\begin{equation}\label{Epol-coated}
E_{pol}  = 8 J_{ff}S^2 \left( 1+ \frac{\kappa}{4} \right)\sin^2{\varphi_0}+ \frac{J_{ff}S^2\varphi_0^2}{2}I(R_{pol}/r_0) -2t\sin{\varphi_0} \, ,
\end{equation}
where
\begin{equation}\label{functionI}
I(\rho_0) = 2\pi \left[1 +\frac{\rho_0K_1(\rho_0)}{2K_2(\rho_0)} \right]
\end{equation}
and $\rho_0 = R_{pol}/r_0$.

Since $R_{pol} < r_0$, the value of the function $I(\rho_0)$ is close to $2\pi$. The optimum angle $\varphi_0$ is found by the minimization of the energy in Eq.~(\ref{Epol-coated}) and is given by
\begin{equation}\label{Epol-minimization}
\cos{\varphi_0} - \frac{4J_{ff}S^2}{t} \left( 1+ \frac{\kappa}{4} \right)\sin{2\varphi_0} - \frac{J_{ff}S^2\varphi_0^2}{2t}I(R_{pol}/r_0) = 0 \, .
\end{equation}
Hence,
\begin{equation}\label{phi-optim}
\varphi_0 \approx \frac{\pi}{2}\left[1 - \frac{\pi J_{ff}S^2}{t}\right] \, .
\end{equation}
At $J_{ff}S^2/t = 50$ correspnfing to Fig.~\ref{phi-vs-r}, estimate~(\ref{phi-optim}) gives us just  the required value $\varphi_0 \approx 85^{\circ}$. The dependence of energy difference between ``coated" and ``bare" magnetic polarons $\Delta E = E_{pol} -E_{pol}^0$ at the 2D square lattice  on the parameter of anisotropy $\kappa$  is shown in Fig.~\ref{coated-vs-bare}. As it follows from this figure, ``coated" polarons are more favorable if the anisotropy parameter $\kappa$is not too large. Otherwise, ``bare" magnetic polarons are more favorable.

\begin{figure}[H]
\begin{center}
\includegraphics*[width=0.5\columnwidth]{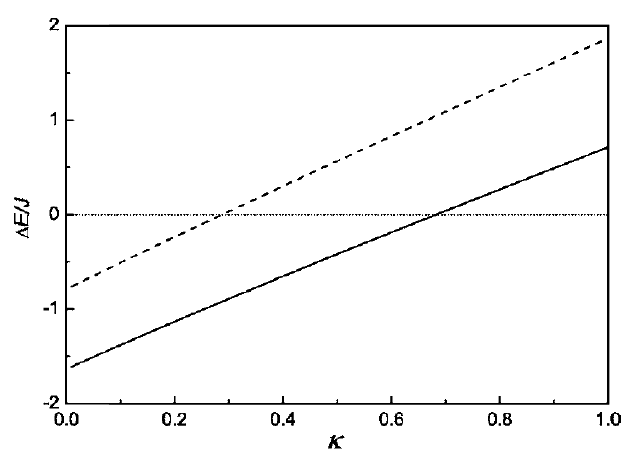} \end{center}
\caption{\label{coated-vs-bare}  Energy difference  between ``coated" and ``bare" magnetic polarons {\emph vs} $\kappa =2K/J_{jj}$ calculated at $J_{ff}S^2/t = 50$. Solid curve corresponds to the numerical calculations , whereas dashed curve is calculated using Eq.~(\ref{Epol-coated})
\cite{OgarkovPRB2006} }
\end{figure}

It can be shown in the same manner that for 3D cubic lattice and 2D triangular lattice, the energies of the ``coated" magnetic polarons for not very high values of the magnetic anisotropy can be  also lower then the energies of the corresponding ``bare" magnetic polarons~\cite{OgarkovPRB2006,KugelOgarPhB2008,KaganOgarJPCM2008}.

\section{Charge ordering and phase separation}
 \label{ChargeOrder}

\subsection{Introduction}
 \label{IntroChargeOrder}

In this section, we present the results of our group  \cite{KaganUFN2001,KaganBookSpringer,KaganJETP2001,KaganFNT2001,
KaganEfrJETPL2011}, as well as of other groups  \cite{CastellaniPRL1995,CastellaniJPhChSol1998,PeraliPRB1996,
LorenzanaPRB_II_2001} on the charge ordering (CO) and formation of nanoscale metallic droplets in CO insulating matrices in framework of the Verwey \cite{VerweyNature1939,VerweyPhys1941} and Shubin--Vonsovsky models \cite{ShubVonsPrRoySoc1934}. We will also consider ferromagnetic metallic droplets, which arise in CO and antiferromagnetic insulating matrices in more extended models  \cite{KaganUFN2001,KaganBookSpringer,KaganJETP2001,
KaganFNT2001} combining the double exchange magnetic interactions with the strong intersite Coulomb interactions. Finally, we will comment on
the large-scale phase separation and formation of stripes and zig-zag structures typical of these models and confirmed by different experimental techniques including the electron diffraction and dark-image electron microscopy \cite{MoritomoPRB1999,MoriNature1998,UeharaNature1999}.

Historically, the problem of CO in magnetic oxides has attracted attention of theorists since the discovery of the Verwey localization transition in magnetite Fe$_3$O$_4$ at the end of the 1930s \cite{VerweyNature1939}. The early theoretical description of this phenomenon is given, for example, in  \cite{KhomskiiPrepr1969}. At the end of 1990s, this problem was reexamined in a number of papers in relation to the colossal magnetoresistance (CMR) phenomena in manganites and related compounds  \cite{MutouPRL1999,vdBrinkPRL1999,JackeliPRB2000}. Mechanisms stabilizing the CO state can be different: the Coulomb repulsion of charge carriers (the energy minimization requires keeping the charge carriers as far away as possible from each other, similarly to the Wigner crystallization) or the electron--lattice interaction leading to the effective repulsion of electrons at the nearest-neighbor sites \cite{NewnsAdvPh1987,ColemanPRB1987,TsvelikBook2007,VarmaPhB2006} if the electron bandwidth is sufficiently small. In the opposite case, large electron kinetic energy stabilizes the homogeneous metallic state. In actual materials, in contrast to the Wigner crystallization, the underlying lattice periodicity determines the preferential type of CO. Thus, in the simplest bipartite lattice, to which belong, for example, both cubic (La$_{1-x}$Sr$_x$MnO$_3$) and  layered (La$_{2-x}$Sr$_x$MnO$_4$, La$_{2-2x}$Sr$_{1+2x}$Mn$_2$O$_7$) hole-doped manganites, the optimum conditions for the formation of CO state  exist for $x = 1/2$. At this value of $x$, the concentrations of Mn$^{3+}$ and Mn$^{4+}$ ions are equal and the simple checkerboard arrangement is possible (see Fig.~\ref{checker_struct}). A remarkable experimental fact is that even at $x \neq 1/2$ (in underdoped manganites with $x < 1/2$), only the simplest version of CO is experimentally observed with the alternating checkerboard structure of the occupied and empty sites in the basal plane. In other words, this structure corresponds to the doubling of the unit cell, whereas the more complicated structures with a longer period (or even incommensurate structures) do not actually appear in this case.

\begin{figure} [H]
\begin{center}
\includegraphics*[width=0.25\columnwidth]{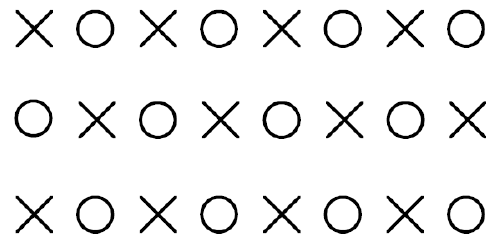}
\end{center}
\caption{\label{checker_struct} Verwey type of charge ordering at the doping level $x = 1/2$ with the checkerboard distribution of  Mn$^{3+}$ (crosses) and  Mn$^{4+}$ (open circles) ions, which here play the role of electrons and holes, respectively~\cite{KaganBookSpringer}.}
\end{figure}

 Then, a natural question arises, namely, how the extra or missing electrons can be redistributed for an arbitrary doping level such that the superstructure remains the same as for $x = 1/2$? To answer this question, the experimentalists introduced the concept of the incipient CO state corresponding to the distortion of a long-range charge ordering by microscopic metallic clusters  \cite{MoriNature1998,UeharaNature1999,ArulrajPRB1998}. In fact, the existence of this state implies a kind of phase separation. In this section, we show that the interplay between the CO and the tendency toward the phase separation plays a very important role in the physics of manganites and other related compounds (magnetite  \cite{VerweyNature1939}, cobaltites  \cite{MoritomoPRB1998}, and nickelates \cite{AlonsoPRL1999}).

\subsection{The simplest model for charge ordering}
 \label{CO_simple_model}

 We start our theoretical considerations of the CO and phase separation with the simplest Verwey model, which includes the Coulomb repulsion of electrons at the neighboring sites and the intersite electron hopping. The Hamiltonian of the Verwey model is
\begin{equation}\label{Hamil_Verwey}
\hat{H}' = -t\sum_{<i,j>}{c_i^{\dag}c_j}+V\sum_{<i,j>}{n_in_j}-\mu \sum_i{n_i}  \, ,
\end{equation}
where $t$ is the hopping integral, $V$ is the nearest-neighbor Coulomb repulsion (a similar nearest-neighbor repulsion can be obtained via the interaction with the breathing type optical phonons), $\mu$ is the chemical potential, $c_i^{\dag}$ and $c_j$  are respectively the electron creation and annihilation operators at site $i$, $n_i = c_i^{\dag}c_i$. The symbol $\langle i,j \rangle$  denotes as usual the summation over the nearest-neighbor sites. In Eq.~\eqref{Hamil_Verwey}, we omit spin and orbital indices for simplicity, thus considering one orbital for spinless electrons and emphasizing the most robust effects related to the nearest-neighbor Coulomb repulsion. We also assume that the double occupancy does not occur in this model because of the strong onsite repulsion $U \to \infty$ between electrons.

The explicit inclusion of the arbitrary (either weak or strong) onsite Hubbard repulsion $U$  in the model together with an account taken of spin degrees of freedom transforms the Verwey model \cite{VerweyNature1939} to the Shubin--Vonsovsky model  \cite{ShubVonsPrRoySoc1934} (or, using a more common term, the extended Hubbard model) with the Hamiltonian  \cite{KaganEfrJETPL2011}
\begin{equation} \label{ShubVons_model}
\hat{H}' =-t\sum_{<i,j> \sigma}{c_{i\sigma}^{\dag}c_{j\sigma}}+U\sum_i{n_{i\uparrow} n_{i\downarrow}}+\frac{V}{2}\sum_{<i,j>}{n_in_j}-\mu \sum_{i\sigma} {n_{i\sigma}}\, ,
\end{equation}
where $n_{i,\sigma} = c_{i\sigma}^{\dag}c_{i\sigma}$ is the electron density at site $i$ with spin projection $\sigma = |\uparrow, \downarrow \rangle$, and  $n_i = \sum_{\sigma}{n_{i\sigma}}$. We will show later in this section (see also  \cite{KaganBookSpringer,KaganEfrJETPL2011}) that in the strong coupling limit of the Shubin--Vonsovsky model ($U \gg V \gg  W = 2zt$ and densities $ n\neq 1/2$), we obtain the same results for the radius of the nanoscale metallic droplets in the CO matrix as in  the strong-coupling limit of the Verwey model ($V \gg W$) after the evident substitution $ V \to V/2$.

In the Shubin--Vonsovsky model, the Verwey localization transition at densities $n\to 1/2$ is an additional one to the more  standard Mott--Hubbard localization transition \cite{HubbardPrRoySocA1963,MottBookMIT1990} at $U \gg W$  and $n \to 1$.

In the main part of this section (see Eqs.~\eqref{Hamil_Verwey} and \eqref{ShubVons_model}), we usually speak about electrons. However, in application to manganites, for example, we mostly have in mind underdoped systems like La$_{1-x}$Sr$_x$MnO$_3$  with the hole doping level lower than 1/2 ($x < 1/2$). For them, we must therefore think in terms of holes. All the theoretical treatment definitely remains the same (from the very beginning, we could define the operators in the Hamiltonian of the Verwey model \eqref{Hamil_Verwey} as the operators of holes). We hope that this does not lead to any misunderstanding.

In what follows, we consider the simplest case of square (2D) and cubic (3D) lattices, where the simple two-sublattice ordering occurs (in manganites for $x = 1/2$). This is the case for layered manganites, whereas in 3D perovskite manganites, this ordering occurs as we already mentioned only in the basal plane: the ordering is not checkerboard {``in-phase") along the $c$-axis direction. A more complicated model is apparently needed to account for this behavior.

If the bandwidth $W$ is even smaller than the Coulomb repulsion between electrons located at more distant sites ($V \gg V_2 \gg V_3 \gg \dots \gg V_m \gg W$ , where $V_m\sum_i{n_in_{m+i}}$  accounts for the Coulomb interaction between electrons at a distance $ma$ apart), then the variation of the charge carrier density will give rise to the whole cascade of localization phase transitions, which differ in their ground state structure. For instance, if $V_2 \gg W$ is included to the model, the system will exhibit an additional phase transition to a localized state at electron densities $n \to 1/3$. A more general localization criterion for narrow-band systems ($\mathbf{a}\triangledown V(\mathbf{r}) \gg W$ was originally proposed in \cite{KaganMaksJETP1984,
KaganMaksJETP1985}.

Let us return to the simplest model with only the nearest-neighbor Coulomb repulsion (Eq.~\eqref{Hamil_Verwey}). For densities $n =1/2$,  the Verwey model was analyzed in many papers \cite{KhomskiiPrepr1969,MutouPRL1999,vdBrinkPRL1999,JirakJMMM1985,PietigPRL1999}). In this section, we follow the treatment of \cite{KhomskiiPrepr1969} (see also  \cite{KaganJETP2001,KaganFNT2001,KaganBookSpringer}). As mentioned above, the Coulomb repulsion stabilizes the CO in the form of a checkerboard arrangement of the occupied and the empty sites, whereas the first term (band energy) opposes this tendency. At arbitrary values of the electron density $n$, we first consider a homogeneous CO solution and use the same ansatz as in \cite{KhomskiiPrepr1969}, namely,
\begin{equation}\label{anzatz}
n_i = n[1 +(-1)^i\tau]\, .
\end{equation}
This expression implies doubling the lattice periodicity with the local densities
\begin{equation}\label{doubling}
n_1 = n(1 + \tau), \quad  n_2 = n(1 - \tau) \, .
\end{equation}
at the neighboring sites. At $n = 1/2$ for a general form of the electron dispersion without nesting, the CO state exists only for a sufficiently strong repulsion $V > 2t$ \cite{KhomskiiPrepr1969}. The order parameter is $\tau < 1$  for finite $V/2t$ , so the average electron density  differs from zero or unity even at $T =0$.

We use the coupled Green's function approach  \cite{KhomskiiPrepr1969}), which yields  \cite{KaganJETP2001,KaganFNT2001}
 \begin{eqnarray} \label{coupledGreen}
(E+\mu )G_1-t_kG_2-zVn(1-\tau)G_1&=&1/2\pi \, ,
\nonumber \\
(E+\mu )G_2-t_kG_1-zVn(1+\tau)G_2&=&0 \, .
\end{eqnarray}
In Eq.~\eqref{coupledGreen}, $G_1$ and $G_2$ are the Fourier transforms of the normal lattice Green functions $G_{ij} = \langle \langle c_ic_j^{\dag} \rangle \rangle$
for the sites $i$ and $j$  belonging respectively to the same or different sublattices, $z$ is the number of nearest neighbors, and $t_k$ is the Fourier transform of the hopping matrix element. In deriving Eq.~\eqref{coupledGreen}, we performed a mean-field decoupling and replaced the averages $\langle c_i^{\dag}c_i \rangle$  by the onsite densities  from Eq.~\eqref{anzatz}. The solution of Eq.~\eqref{coupledGreen} gives the following spectrum
\begin{equation} \label{COspectrum}
E+\mu =Vnz \pm \sqrt{(Vn\tau z)^2+t_k^2}=Vnz \pm \omega_k\, .
\end{equation}
The spectrum defined by Eq.~\eqref{COspectrum} resembles the spectrum of superconductor and, hence, the first term under the square root is analogous to the superconducting gap squared. In other words, we can introduce the CO gap by the formula
\begin{equation} \label{COgap}
\Delta = Vn\tau z\, .
\end{equation}
It depends on the density not only explicitly, but also via the density dependence of $\tau$. We thus obtain
\begin{equation} \label{omegaCO}
\omega_k = \sqrt{\Delta^2 + t_k^2}\, .
\end{equation}

Using Eqs.~\eqref{coupledGreen}, \eqref{COgap}, and \eqref{omegaCO}, we can  find the Green's functions
\begin{eqnarray} \label{Green_func}
G_1 &=& \frac{A_k}{E+\mu -Vnz-\omega_k+i0}+\frac{B_k}{E+\mu -Vnz+\omega_k+i0} \, ,
\nonumber \\
G_2 &=& \frac{t_k}{4\pi\omega_k}\left [\frac{1}{E+\mu -Vnz-\omega_k+i0}+\frac{1}{E+\mu -Vnz+\omega_k+i0}\right] \, .
\end{eqnarray}
where
\begin{equation} \label{AandB}
A_k = \frac{1}{4\pi} \left(1 -\frac{\Delta}{\omega_k}\right), \quad B_k = \frac{1}{4\pi} \left(1 +\frac{\Delta}{\omega_k}\right) \, .
\end{equation}

After a standard Wick transformation $E +i0 \to iE$  in the expression for $G_i$, we find the densities in the following form
\begin{eqnarray} \label{densities}
n_1 &=& n(1+\tau)=\int{\left[\left(1-\frac{\Delta}{\omega_k}\right)
f_F(\omega_k-\mu+Vnz)+\left(1+\frac{\Delta}{\omega_k}\right)
f_F(-\omega_k-\mu+Vnz)\right]\frac{d^D\mathbf{k}}{2\Omega_{BZ}}} \, ,
\nonumber \\
n_2 &=& n(1-\tau)=\int{\left[\left(1+\frac{\Delta}{\omega_k}\right)
f_F(\omega_k-\mu+Vnz)+\left(1-\frac{\Delta}{\omega_k}\right)
f_F(-\omega_k-\mu+Vnz)\right]\frac{d^D\mathbf{k}}{2\Omega_{BZ}}}\, ,
\end{eqnarray}
where
\begin{equation*}
f_F(y)=\frac{1}{e^{y/T}+1}
\end{equation*}
is the Fermi distribution function and $\Omega_{BZ}$  is the volume of the first Brillouin zone.
Adding and subtracting the equations for $n_1$  and $n_2$, we obtain the final system of equations for $n$ and $\mu$
\begin{eqnarray} \label{n-mu}
n &=& \int{\left[f_F(\omega_k-\mu+Vnz) +
f_F(-\omega_k-\mu+Vnz)\right]\frac{d^D\mathbf{k}}{2\Omega_{BZ}}} \, ,
\nonumber \\
1 &=& Vz\int{\frac{1}{\omega_k}\left[f_F(-\omega_k-\mu+Vnz)-
f_F(\omega_k-\mu+Vnz)\right]\frac{d^D\mathbf{k}}{2\Omega_{BZ}}}\, .
\end{eqnarray}
For low temperatures ($T \to 0$ ) and $n \leq 1/2$, it is reasonable to assume the filling of only the lowest levels (negative $\mu-Vnz$). Therefore,
\begin{equation*}
f_F(\omega_k-\mu+Vnz)=0
\end{equation*}
and
\begin{equation*}
f_F(-\omega_k-\mu+Vnz)= \theta(-\omega_k-\mu+Vnz)
\end{equation*}
is the step function.

It can be shown that for $n=1/2$, the system of equations \eqref{n-mu} yields identical results for all $-\Delta \le \mu -Vnz\le \Delta$. This means that $n=1/2$ is a point of indifferent equilibrium. For infinitely small deviations from $n=1/2$, i.e. for densities $n=1/2 - 0$, the chemical potential should be defined as
\begin{equation}\label{mu_defin}
\mu =-\Delta +\frac{Vz}{2}=\frac{Vz}{2}(1-\tau)\, .
\end{equation}

If we consider the strong coupling $V \gg 2t$ and assume a constant density of states inside the band, then for $n =1/2$, we can find~\cite{KaganJETP2001,KaganFNT2001}
\begin{equation}\label{tau _simp_cube}
\tau =1-\frac{2W^2}{3V^2z^2}
\end{equation}
with $z =6$ for the simple cubic lattice and, therefore,
\begin{equation}\label{mu _simp_cube}
\mu =\frac{W^2}{3Vz}\, .
\end{equation}

For the density $n =1/2$, the CO gap $\Delta$ appears for an arbitrary interaction strength $V$. This is due to the existence of nesting in our simple model. In the weak coupling case $V \ll 2t$  and with perfect nesting, we have
\begin{equation}\label{Delta_weak_coupl}
\Delta \propto W\exp{\left(-\frac{W}{Vz}\right)}\, ,
\end{equation}
and $\tau$ is exponentially small. For $V \gg W =2tz$, or  accordingly for $V \gg 2t$, it follows that $\Delta \approx Vz/2$ and  $\tau \to 1$ (Eq.~\eqref{COgap}). In the general case, when there is no nesting, the CO exists only if the interaction strength $V$ exceeds a certain critical value of the order of the bandwidth $W$ \cite{KhomskiiPrepr1969}. In what follows, we restrict ourselves to the physically more instructive strong-coupling case $V \gg 2t$.

For the constant density of states, the integrals in Eq.~\eqref{n-mu} can be taken explicitly and the system of equations \eqref{n-mu} can be easily solved for arbitrary $n$. However, in the strong-coupling case $V \gg 2t$ and for small density deviations from 1/2 ($\delta \ll 1$), the results are not very sensitive to the form of the electron dispersion. %That is why, we do not need to solve the system of equations \eqref{n-mu} exactly.

When $n = 1/2 -\delta$, the chemical potential is $\mu = \mu(\delta,\tau)$, and we have two coupled equations for $\mu$ and $\tau$. As a result, we obtain~\cite{KaganJETP2001}
\begin{equation}\label{mu-delta}
\mu(\delta) \approx Vnz(1-\tau)-\frac{4W^2}{Vz}\delta^2 \approx \frac{W^2}{3Vz}+\frac{4W^2}{3Vz}\delta +O(\delta^2)\, .
\end{equation}
The energy of CO state is therefore given by
\begin{equation}\label{energyCO}
E_{CO}(\delta) = E_{CO}(0)-\frac{W^2}{3Vz}\delta -\frac{2W^2}{3Vz}\delta^2+O(\delta^2) \, ,
\end{equation}
where
\begin{equation}\label{energyCO-zero}
E_{CO}(0) = -\frac{W^2}{6Vz}
\end{equation}
is the energy precisely corresponding to the density $n =1/2$. Moreover, $|E_{CO}(0)| \ll W$. At the same time, the CO gap $\Delta$  is given by
\begin{equation}\label{COgap-delta}
\Delta \approx \frac{Vz}{2}\left[1-2\delta -\frac{2W^2}{3V^2z^2}(1+4\delta)\right]\, .
\end{equation}
The dependence of the chemical potential $\mu$ and the total energy $E$ on $\delta$ in Eqs.~\eqref{mu-delta} and \eqref{energyCO} is in fact related to the linear decrease of the CO gap $\delta$  with the deviation from half-filling.

For $n > 1/2$, the energy of the CO state starts to increase rapidly due to a large contribution of the Coulomb repulsion (the upper Verwey band is partially filled at $n > 1/2$). For $n > 1/2$, in contrast to the case of $n < 1/2$, each extra electron placed into the checkerboard CO state necessarily has the occupied nearest-neighbor sites, increasing the total energy by $Vz|\delta|$. For $|\delta| = n -1/2 > 0$, we then have
\begin{equation}\label{energyCO_larger_onehalf}
E_{CO}(\delta) = E_{CO}(0)+\left(Vz - \frac{W^2}{3Vz}\right)|\delta| -\frac{2W^2}{3Vz}\delta^2+O(\delta^3) \, .
\end{equation}
Accordingly, the chemical potential is given by
\begin{equation}\label{mu-delta_larger_onehalf}
\mu(\delta) = Vz-\frac{W^2}{3Vz}-\frac{4W^2}{3Vz}|\delta| +O(\delta^2)\, .
\end{equation}
It undergoes a stepwise change equal to $Vz$ at $ \tau \to 1$. The gap $\Delta$ is symmetric for $n > 1/2$ and is given by
\begin{equation}\label{Delta-delta_larger_onehalf}
\Delta \approx \frac{Vz}{2}\left[1-2|\delta| -\frac{2W^2}{3V^2z^2}(1+4|\delta|)\right]\, .
\end{equation}

We could make the entire picture symmetric with respect to $n = 1/2$ by shifting all the one-electron energy levels and the chemical potential by $Vx/2$, i.e. defining $\mu' = \mu -Vz/2$. In terms of $\mu'$, Eqs.~\eqref{mu-delta} and \eqref{mu-delta_larger_onehalf} can be written as
\begin{eqnarray} \label{mu-delta_prime}
\mu' &=& -\frac{Vz}{2}+\frac{W^2}{3Vz}+\frac{4W^2}{3Vz}\delta, \quad n < 1/2 \, ,
\nonumber \\
\mu' &=& \frac{Vz}{2}-\frac{W^2}{3Vz}-\frac{4W^2}{3Vz}|\delta|, \quad n > 1/2 \, .
\end{eqnarray}

Similarly to the situation in semiconductors, we have $\mu' = 0$ precisely at the point $n = 1/2$, which means that the chemical potential lies in the middle of the band gap (see Fig.~\ref{Verwey_bands}). At densities $n = 1/2 - 0$ , the chemical potential $\mu' = -Vz/2$ coincides with the upper edge of the filled Verwey band. We get just the same band structure in the strong-coupling limit of the Shubin--Vonsovsky model~\cite{KaganBookSpringer}.

\begin{figure} [H]
\begin{center}
\includegraphics*[width=0.25\columnwidth]{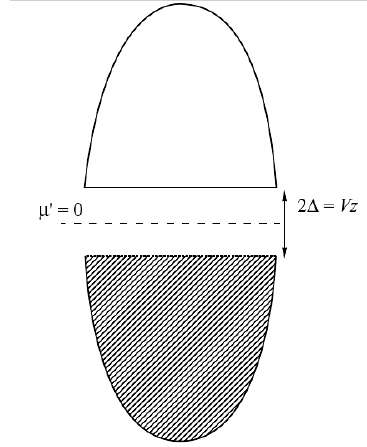}
\end{center}
\caption{\label{Verwey_bands} Band structure of the Verwey and Shubin--Vonsovsky models at $n =1/2$  \cite{KaganJETP2001,KaganEfrJETPL2011}. The lower Verwey band is completely filled. The upper Verwey band is empty. The chemical potential $\mu' = 0$ lies in the middle of the gap with the width of $2\Delta$.}
\end{figure}

\subsection{Phase separation in the Verwey model}
 \label{PS_Verwey}

 We now check the stability of the CO state in the Verwey model. At the densities close to $n = 1/2$, the dependence of the energy on the charge density has the form illustrated in Fig.~\ref{COenergy-n}.

\begin{figure} [H]
\begin{center}
\includegraphics*[width=0.4\columnwidth]{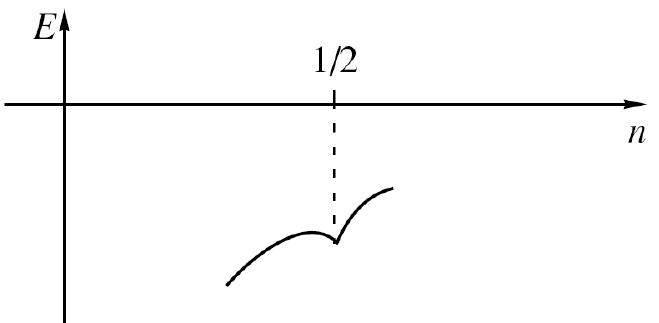}
\end{center}
\caption{\label{COenergy-n} Energy of the CO state versus charge density for $n \to 1/2$. The energy has a kink at $n= 1/2$  \cite{KaganJETP2001}.}
\end{figure}

Figure~\ref{COenergy-n} clearly demonstrates an instability of the CO state. Indeed, the most remarkable implication of Eqs.~\eqref{mu-delta}--\eqref{mu-delta_prime} is that the compressibility $\kappa$ of the homogeneous CO state is negative for the densities different from 1/2 ,
\begin{equation}\label{kappa_inverse}
\kappa^{-1} \propto \frac{d\mu}{dn} = \frac{d^2E}{d\delta^2} = -\frac{4W^2}{3Vz} <0 \, ,
\end{equation}
where $\delta =1/2 - n$. As we already mentioned, this is a manifestation of the tendency toward the phase separation characteristic of the CO system with $\delta \neq 0$. The presence of a kink in the $E_{CO}(\delta)$ curve determined by  Eqs.~\eqref{energyCO} and \eqref{energyCO_larger_onehalf} implies that one of the states. into which the system might separate, would correspond to the checkerboard CO state with $n = 1/2$, whereas the other would have a certain density $n'$ smaller or larger than $n$. This conclusion (based on very general considerations  \cite{LandauBook_StPhI1980}) resembles that discussed in  \cite{KaganEPJB1999,vdBrinkPRL1999,ArovasPRB1998}, although the detailed physical mechanism is different. The possibility of the phase separation in model \eqref{Hamil_Verwey} away from half-filling was also reported earlier in  for the infinite-dimensional case. In what follows, we focus on the situation with $n < 1/2$. The case when $n > 1/2$  apparently has certain special properties -- the existence of stripe phases etc.  \cite{MoriNature1998,UeharaNature1999}, the detailed origin of which should
will be analyzed separately.

It is easy to understand the physics of phase separation in our case. As follows from Eq.~\eqref{COgap-delta}, the CO gap decreases linearly with the deviation from the half-filling. Correspondingly, the energy of the homogeneous CO state rapidly increases, and it is more favorable to ``extract" extra holes from the CO state, putting them into one part of the sample, while creating the ``pure" checkerboard CO state in the other part. The energy loss due to this redistribution of holes is overcompensated by the gain provided by the better CO.

However, the hole-rich regions would not be completely ``empty" similarly to pores or voids (clusters of vacancies) in crystals: we can gain extra energy by ``dissolving" a certain amount of electrons there. In doing this, we decrease the band energy of the electrons due to their delocalization. Thus, the second phase would be a metallic one. The simplest state of this kind is a homogeneous metal with the electron density $n_{met}$. This density, as well as the relative volumes of the metallic and CO phases, can be easily calculated by minimizing the total energy of the system. The energy of the metallic part of the sample, $E_{met}$,  in the 3D case is given by
\begin{equation}\label{Emet}
E_{met} = -tzn_{met}+ct(n_{met})^{5/3}+\frac{Vz}{2}(n_{met})^2 \, ,
\end{equation}
where $c$ is a constant.

Minimizing Eq.~\eqref{Emet} with respect to $n_{met}$, we find the equilibrium electron density in the metallic phase. For the  strong-coupling case $V \gg 2t$, we obtain the following simple expression (neglecting the relatively small correction provided by the term which contains $n_{met}^{5/3}$)
\begin{equation}\label{n_met-zero}
n_{met}^{(0)} = t/V \, .
\end{equation}
The optimum energy of the metallic phase is given by \begin{equation}\label{Emet-ECO}
\left(E_{met} \approx -\frac{t^2z}{2V} = -\frac{W^2}{8Vz}\right) > \left(E_{CO} = -\frac{W^2}{6Vz}\right)\, ,
\end{equation}
and it is larger than the energy of CO state.

\subsection{Phase separation into two large clusters (complete phase separation)}
 \label{complete_PS}

In accordance with the general treatment in statistical physics, the system with the density $n_{met}^{(0)} < n < 1/2$  separates fully on two large clusters (Fig.~\ref{completePS}): a metallic one with the density $n_{met}^{(0)}$  and a CO one with the density 1/2. For arbitrary $n$, the relative volumes $\Omega_{met}$ and $\Omega_{CO}$ of these clusters can be found from the Maxwell construction
\begin{equation}\label{Maxwell_constr}
\frac{\Omega_{met}}{\Omega_{CO}} = \frac{1/2-n}{n-n_{met}^{(0)}}\, ,
\end{equation}
which implies that the metallic cluster occupies the part $\Omega_{met}$ of the total volume $\Omega = \Omega_{met} +\Omega_{CO}$ given by
\begin{equation}\label{Omega_met}
\frac{\Omega_{met}}{\Omega} = \frac{1/2-n}{1/2-n_{met}^{(0)}} = \frac{\delta}{\delta_0}\, ,
\end{equation}
where we introduced deviations from the half-filling, $\delta = 1/2-n$ and $\delta_0 = 1/2-n_{met}^{(0)}$.

Accordingly, the energy of the fully separated state $E_{comp.sep.} =E_{met}(\Omega_{met}/\Omega)+E_{CO}(\Omega_{CO}/\Omega)$ takes the form
\begin{equation}\label{E_fully_sep}
 E_{comp.sep.} \approx -\frac{W^2}{8Vz}\frac{\delta}{\delta_0}-\frac{W^2}{6Vz}
 (1-\frac{\delta}{\delta_0})\, .
\end{equation}
The metallic cluster will occupy the entire sample volume when the total electron density $n$ is less than $n_{met}^{(0)}$, while CO cluster occupies all the volume for $n =1/2$.

\begin{figure} [H]
\begin{center}
\includegraphics*[width=0.4\columnwidth]{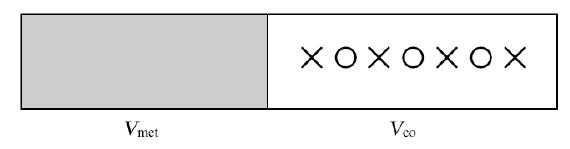}
\end{center}
\caption{\label{completePS} Complete phase separation into two large clusters (metallic and CO ones)  \cite{KaganUFN2001}. }
\end{figure}

\subsection[Nanoscale phase-separated state within the charge-ordered insulating host]{Nanoscale phase-separated state with small metallic droplets within the charge-ordered insulating host material}
%\addcontentsline{toc}{subsection}{%
  %Nanoscale phase-separated state with small metallic droplets \\ \protect\hspace*{-\cftchapnumwidth}within the charge-ordered insulating host material}
 \label{nanoscPS_CO-met}

We now consider a different possibility for the phase-separated state, namely the possibility, which is closer to the nanoscale phase separation in manganites and magnetic semiconductors, described in Section~\ref{MagPolaronTrans}. To be more specific, we will analyze metallic droplets of a the small radius in the CO insulating host. The expression for the energy of this state in the 3D cubic case can be written in the form \cite{KaganUFN2001,KaganJETP2001}
\begin{equation} \label{Edrop-CO}
E_{drop} = -tn_{drop}\left(z-\frac{\pi^2a^2}{R^2}\right)-
\frac{W^2}{6Vz}\left[1-\frac{4}{3}\pi\left(\frac{R}{a}\right)^3n_{drop}\right]\, .
 \end{equation}
It corresponds to the situation illustrated in Fig.~\ref{nanscalPS-CO} and describes the process, in which small metallic droplets with one conduction electron are formed in the CO matrix.
\begin{figure}[H]
\begin{center}
\includegraphics*[width=0.4\columnwidth]{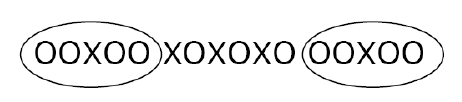}
\end{center}
\caption{\label{nanscalPS-CO} Simplest nanoscale phase separation scenario involving the formation of small metallic droplets with one conduction electron embedded into a CO host~\cite{KaganUFN2001,KaganEfrJETPL2011} .}
\end{figure}

In  expression \eqref{Edrop-CO}, similarly to the situation considered in Section~\ref{MagPolaronTrans}, $R$ is the radius of a droplet, and $a$ is the interatomic spacing. The first term in Eq.~\eqref{Edrop-CO} describes the kinetic energy gain due to the localization of an electron inside the droplet. The second term is the energy of CO in the insulating region of the sample. The parameter $n_{drop}$ in Eq.~\eqref{Edrop-CO} is the concentration of metallic droplets. Similarly to the situation of FM metallic droplets in AFM matrix, a surface energy term can be incorporated into the total energy of the system. This term, however, is of the order of $W^2R^2/V$ for a metallic droplet in a CO matrix, so that for $R \gg a$, it is always small compared to the bulk energy term (which is proportional to $R^3$ in the 3D case) and it will be neglected in the following discussion. The minimization of energy in \eqref{Edrop-CO} with respect to the droplet radius $R$ results in the following condition
\begin{equation} \label{R-optimCO}
\frac{R}{a} \propto \left(\frac{V}{t}\right)^{1/5}\, .
\end{equation}

Therefore, at the critical concentration of droplets
\begin{equation} \label{n_crit_dropCO}
n_{drop}^{(c)} = \frac{3}{4\pi}\left(\frac{a}{R}\right)^3 \propto \left(\frac{t}{V}\right)^{3/5}
\end{equation}
the metallic droplets begin to overlap and the sample as a whole undergoes a transition to the  metallic state. Expression \eqref{Edrop-CO} is in fact the analogue of the Nagaoka theorem  \cite{NagaokaPR1966} for the Verwey model. Similarly, in the 2D case, the exponent 1/5 in Eq.~\eqref{R-optimCO} for the droplet radius is replaced by 1/4  and the exponent 3/5 in the expression for the critical density in Eq.~\eqref{n_crit_dropCO} is replaced by 1/2  \cite{KaganBookSpringer,KaganEfrJETPL2011}.

Just the same expression for the critical density corresponding to the phase separation was obtained in  \cite{KaganEfrJETPL2011} for the strong-coupling limit of the Shubin--Vonsovsky model, whereas for low electron densities $n \ll 1$ the ground state of the model corresponds to the triplet $p$-wave superconductivity, which is governed by the Kohn--Luttinger mechanism \cite{KohnLattPRL1965,KaganKorovUFN2015,KaganEfrJETPL2011}. The qualitative phase diagram for the phase-separation regions in the strong-coupling limit of the 2D Shubin--Vonsovsky model is presented in Fig.~\ref{phdiagPS-CO}.

\begin{figure} [H]
\begin{center}
\includegraphics*[width=0.4\columnwidth]{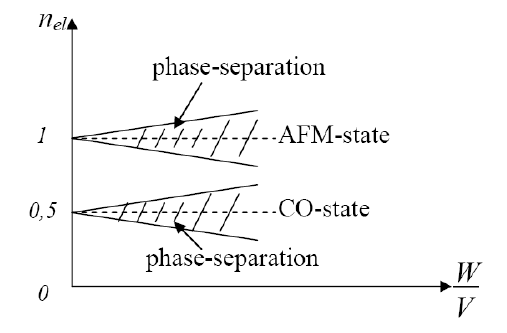}
\end{center}
\caption{\label{phdiagPS-CO} Qualitative phase diagram of the nanoscale phase-separated state in the strong-coupling limit of the 2D Shubin--Vonsovsky model \cite{KaganBookSpringer}.}
\end{figure}

If we use Eq.~\eqref{n_crit_dropCO} for the critical droplet concentration, the total energy \eqref{Edrop-CO} of the system can be rewritten as
\begin{equation} \label{EdropCO_optim}
E_{drop}=-tn_{drop}\left[z-\tilde{c}(n_{drop}^{(c)})^{2/3}\right]-
\frac{W^2}{6Vz}\left[1-\frac{n_{drop}}{n_{drop}^{(c)}}\right]\, ,
\end{equation}
where $\tilde{c}$ is a numerical coefficient of the order of unity.

The ratio $n_{drop}/n_{drop}^{(c)}$ in Eq.~\eqref{EdropCO_optim} describes the relative volume $\Omega_{drop}/\Omega$ occupied by droplets. This volume can be found from the condition for the conservation of the total number of particles in the system, which reads
\begin{equation} \label{consN_drop}
N = \frac{1}{2}(\Omega -\Omega_{drop})+n_{drop}\Omega_{drop}=n\Omega\, ,
\end{equation}
where $n$ is the average density of the conduction electrons. From condition \eqref{consN_drop}, the required relation (which resembles relation \eqref{Omega_met} for complete phase separation) follows immediately and has the form
\begin{equation} \label{Omega_drop}
\frac{\Omega_{drop}}{\Omega} = \frac{1/2-n}{1/2-n_{drop}^{(c)}} = \frac{\delta}{\delta_c} \, ,
\end{equation}
where again  $\delta =1/2-n$ and $\delta_c = 1/2-n_{drop}^{(c)}$ . From Eq.~\eqref{consN_drop}, it follows that $n_{drop}=0$  at $n = 1/2$. Accordingly, at the critical density $n_{drop} = n_{drop}^{(c)}$, we have $\delta =\delta_{c} = 1/2 -n_{drop}^{(c)}$. Finally, the energy of the phase-separated state can be written as
\begin{equation} \label{Edrop_fin}
E_{drop}=-t\frac{\delta}{\delta_c}n_{drop}^{(c)}[z-\tilde{c}(n_{drop}^{(c)})^{2/3}]
-\frac{W^2}{6Vz}\left(1-\frac{\delta}{\delta_c}\right) \approx -\frac{W}{2}\frac{\delta}{\delta_c}n_{drop}^{(c)}-\frac{W^2}{6Vz}
\left(1-\frac{\delta}{\delta_c}\right) \, .
\end{equation}
Comparison of  energy \eqref{Edrop_fin} with that of the CO state given by \eqref{energyCO} shows that for the densities $n_{drop}^{(c)} < n < 1/2$, we have
\begin{equation} \label{Edrop-ECO}
E_{drop}-E_{CO} \approx -\frac{W}{2}\frac{\delta}{\delta_c}n_{drop}^{(c)} < 0\, .
\end{equation}

Thus, a nanoscale phase-separated state with small metallic droplets inside a CO matrix is more favorable than a homogeneous CO state. The direct comparison of the energy for nanoscale phase-separated state \eqref{Edrop_fin} and the energy for the complete separation into two large clusters \eqref{E_fully_sep} shows that for $n_{met}^{(0)} < n_{drop}^{(c)}$,  we have $E_{drop} < E_{comp.sep}$ and the nanoscale phase separation is more favorable than complete one, as well. Physically, it is related to a kind of Nagaoka theorem for the Verwey model. The point is that even when metallic droplets overlap, each conduction electron occupies a sphere of the radius $R/a \sim (V/t)^{1/5}$ and thus effectively it is located at the larger distance from the other electrons than it would be in the case of complete phase separation. Therefore, at the complete phase separation, the energy of Coulomb interactions between electrons turns out to be higher than for metallic droplets of a small radius.

In addition to the one-electron metallic droplets, a nanoscale phase separation scenario of another type can be organized \cite{KaganUFN2001,KaganBookSpringer} (see Fig.~\ref{undermeltedCO}). In this scenario, a metallic droplet can be formed by replacing one electron with a hole in the center of the droplet. Note, however, that the energy of such an undermelted CO state (a resonance valence bond or RVB state for the Verwey model) is much more difficult to evaluate than that for a one-electron droplet, and this problem is not considered in the present review.

\begin{figure} [H]
\begin{center}
\includegraphics*[width=0.4\columnwidth]{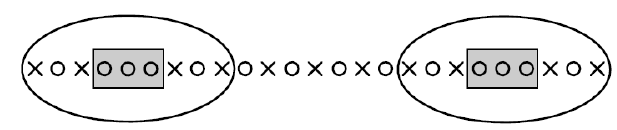}
\end{center}
\caption{\label{undermeltedCO} Nanoscale phase-separation scenario with an undermelted  CO  state inside a metallic droplet embedded into the CO matrix \cite{KaganUFN2001}.}
\end{figure}

Concluding this subsection, we can summarize the main results as follows. In the strong-coupling limit, for $V \gg 2t$ and $(t/V)^{3/5} <n < 1/2$, the 3D Verwey model is unstable toward the nanoscale phase separation. Moreover, in the simplest case, the most energetically beneficial scenario of the phase separation corresponds to small one-electron metallic droplets inside the CO insulating host. Similar results for the nanoscale phase separation, we get in the strong-coupling limits of the 2D Verwey model \cite{KaganBookSpringer} and the 2D Shubin--Vonsovsky model \cite{ShubVonsPrRoySoc1934} for the densities $(t/V)^{1/2} <n < 1/2$.

\subsection[Phase separation in the extended double exchange model]{Phase separation in the extended double exchange model with the nearest-neighbor Coulomb interaction}
 \label{extendedDE-nnCoulomb}

 Let us now return to the basic double exchange model (or the ferromagnetic Kondo-lattice model) considered in Section~\ref{MagPolaronTrans} (see Eq.~\eqref{Hexch} there) and add the term with nearest-neighbor Coulomb interaction to this model. The Hamiltonian of this extended model reads
\begin{equation} \label{DE-Coulomb-Hamilt}
\hat{H}' =-J_H\sum_i{\mathbf{S}_i\bm{\sigma_i}-t\sum_{<i,j>\sigma}
{Pc_{i\sigma}^{\dag}c_{j\sigma}}P}+J_{ff}\sum_{<i,j>}{\mathbf{S}_i\mathbf{S}_j}
+V\sum_{<i,j>}{n_in_j}-\mu \sum_{i\sigma}{n_{i\sigma}}\, .
\end{equation}

It is physically reasonable to consider this model in the strong-coupling limit
\begin{equation} \label{strong-coupl_inequal}
J_HS > V > W > J_{ff}S^2\, .
\end{equation}
In the absence of the Coulomb term it is exactly the conventional double exchange model.

Let us consider the possibility of the nanoscale phase separation in the extended model \eqref{DE-Coulomb-Hamilt}. The simplest scenario of the phase separation in this model corresponds in analogy with Section~\ref{MagPolaronTrans} and previous subsection to small FM metallic droplets (with one conduction electron inside) embedded into the AFM CO insulating host  \cite{KaganUFN2001,KaganEPJB1999,
KaganKhomPhB2000} (see Fig.~\ref{FM-AFMCO_drop}).
\begin{figure} [H]
\begin{center}
\includegraphics*[width=0.4\columnwidth]{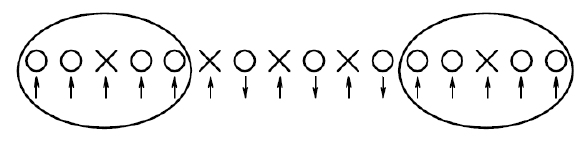}
\end{center}
\caption{\label{FM-AFMCO_drop} Formation of metallic FM droplets of small radius embedded into the AFM CO insulating host material ] \cite{KaganUFN2001}.}
\end{figure}

The energy of this state in the 3D case is given by the expression
\begin{eqnarray} \label{energyFM-AFMCO_drop}
E&=&-tn_{drop}\left(z-\frac{\pi^2a^2}{R^2}\right)+\frac{1}{2}zJ_{ff}S^2
\frac{4}{3}\pi\left(\frac{R}{a}\right)^3n_{drop}
\nonumber \\
&&-\frac{1}{2}zJ_{ff}S^2
\left[1-\frac{4}{3}\pi\left(\frac{R}{a}\right)^3n_{drop}\right]-
\frac{W^2}{6Vz}\left[1-\frac{4}{3}\pi \left(\frac{R}{a}\right)^3n_{drop}\right]\, .
\end{eqnarray}
The first three terms in Eq.~\eqref{energyFM-AFMCO_drop} are in fact identical to the FM polaron energy in the double exchange model (see Eq.~\eqref{Epol} in Section~\ref{MagPolaronTrans}) with the electron density  replaced by the droplet density $n_{drop}$. At the same time, the last term in Eq.~\eqref{energyFM-AFMCO_drop} is identical to the second term in Eq.~\eqref{Edrop-CO} corresponding to the energy of a homogeneous CO Verwey state. Minimization of the droplet energy with respect to radius $R$ yields
\begin{equation} \label{R_opt_FM-AFMCO}
\frac{R}{a} \propto \left(\frac{t}{V}+\frac{J_{ff}S^2}{t}\right)^{-1/5}\, .
\end{equation}
For $t/V \ll J_{ff}S^2/t$, we obtain $R/a \propto (t/J_{ff}S^2)^{1/5}$ and we reproduce the double exchange result  for the metallic droplet radius. In the opposite limit $t/V \gg J_{ff}S^2/t$, we have $R/a \propto (V/t)^{1/5}$, and we arrive at the result of the Verwey model \eqref{R-optimCO}. Then, we have
\begin{equation} \label{n_drop_FM-AFMCO_crit}
n_{drop}^{(c)} \propto \left(\frac{t}{V}+\frac{J_{ff}S^2}{t}\right)^{3/5}\, .
\end{equation}
Naturally, the droplet concentration in the expression for the  energy of the phase-separated state \eqref{energyFM-AFMCO_drop} is again given by
\begin{equation} \label{n_drop_FM-AFMCO}
n_{drop} = n_{drop}^{(c)}\frac{\delta}{\delta_c}
\end{equation}
with $n_{drop}^{(c)}$ from Eq.~\eqref{n_drop_FM-AFMCO_crit}. Physically, the minimization of the energy in \eqref{energyFM-AFMCO_drop} with respect to the droplet radius implies that there is only one conduction electron inside a metallic droplet and that this electron is surrounded by FM ordered local spins. At the same time, outside the droplets, we have a checkerboard CO arrangement of conduction electrons as well as AFM arrangement of the local spins.

The latter result illustrates the main difference between the phase-separated states that are obtained in the extended model \eqref{DE-Coulomb-Hamilt} at densities $n \to 0$ and $n \to 1/2$. The point is that in the low-density regime for $ n \ll n_c \propto \left(\frac{J_{ff}S^2}{t}\right)^{3/5}$, the mean distance between conduction electrons $a/n^{1/3}$  in the 3D case far exceeds the polaronic radius $R$, which in its turn is larger than the intersite distance $a$ ($a/n^{1/3} \gg R \gg a$) (see Fig.~\ref{distr_cond_el_PS}).

\begin{figure} [H]
\begin{center}
\includegraphics*[width=0.4\columnwidth]{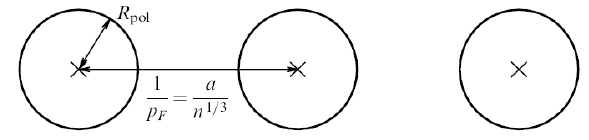}
\end{center}
\caption{\label{distr_cond_el_PS} Distribution of conduction electrons in a nanoscale phase-separated state with FM polarons embedded into the AFM host \cite{KaganUFN2001}.}
\end{figure}

Thus, for FM polarons (ferrons) with one conduction electron inside the ferron the inclusion to the model of even strong Coulomb repulsion between electrons at the neighboring sites does not lead to a charge redistribution. Therefore, upon the inclusion of the Coulomb interaction, both the energy of the phase-separated state with FM polarons inside AFM matrix and the energy of the homogeneous canted state (see Eq.~(6) in Section~\ref{MagPolaronTrans}) acquires only a Hartree--Fock correction term proportional to $zVn^2/2$, so that the energy difference $E_{pol} -E_{hom}$ between the state with magnetic polarons and the homogeneous state remains negative and does not change. Hence, the global minimum for the energy of the system in the extended model \eqref{DE-Coulomb-Hamilt} corresponds at low density to the phase-separated state with FM polarons embedded into the AFM host just as in the conventional double exchange model. Thus, at low average densities ($n \ll 1$), the local conduction electron density outside FM polarons is zero and the entire electric charge is contained within metallic droplets. At the same time, at densities close to 1/2, most conduction electrons are localized within the CO regions outside metallic droplets.

\subsection{Concluding remarks}
\label{conclusionCO}

Eventually, the sequence of phases with the increase in the doping level for the double exchange model with the Coulomb interaction can be described as follows.
\begin{enumerate}
  \item {For $0 < n < (J_{ff}S^2/t)^{3/5}$, the system separates into nanoscale FM metallic droplets inside the AFM insulating matrix.}
  \item {For $(J_{ff}S^2/t)^{3/5} < n < (t/V+J_{ff}S^2/t)^{3/5} <1/2$, the system is a FM metal. Of course, we should have a ``window" of parameters to satisfy the inequality in the right-hand side of this expression. In manganites, for example, $t/V \sim 1/2-1/3$, and at the same time, $J_{ff}S^2/t \sim 0.01$. Therefore, the inequality $n < (t/V+J_{ff}S^2/t)^{3/5} <1/2$ is not necessarily met. Experimental evidence indicates that the desired parameter range exists for La$_{1-x}$Ca$_x$MnO$_3$, but definitely not for Pr$_{1-x}$Ca$_x$MnO$_3$.}
  \item {Finally, for $(t/V+J_{ff}S^2/t)^{3/5} < n < 1/2$, the phase separation into nanoscale FM metallic droplets inside an AFM CO matrix occurs. It should be emphasized that an ideal AFM structure in CO matrix is obtained only accounting for the Heisenberg interaction between local spins $\mathbf{S}$. A more thorough account of the large Hund's rule exchange between a local spin $\mathbf{S}$  and spin of a conduction electron $\bm{\sigma}$  can change this naive expectation and lead to the FM order in the CO matrix. At $n =1/2$, a homogeneous CO state is realized.}
\end{enumerate}
Here, we limit ourselves to the case of $n  < 1/2$, since at higher doping levels, the model under study should take into account some additional factors, such as interactions with the lattice distortions.

We can get analogous results for the phase diagram in the 2D extended model containing the double exchange magnetic interactions and the nearest-neighbor Coulomb interaction in the strong-coupling limit.

Concluding this section, let us emphasize that the phase diagram discussed above is in a good qualitative agreement with many available experiments  \cite{HennionPRL1998,AllodiPRB1997,
 BabushkinaPRB1999,VoloshinJETPL2000,YakubovsPRB2000} and numerical calculations \cite{YunokiPRL2000} concerning the nanoscale phase separation in manganites and, in particular, with the experimental observation of the nanoscale phase separation close to the density $n =1/2$  \cite{MoritomoPRB1999} (or to the doping $x =1/2$ in hole-doped manganites). We also note that our phase diagram has certain similarities with the phase diagram obtained in \cite{BalentsPRL2000,BarzykinPRL2000} for the problem of spontaneous ferromagnetism in doped excitonic insulators.

The coexistence of FM reflections and the reflections of the CE type of the magnetic structure (typical of the CO state at $x =1/2$) were observed by neutron scattering in \cite{KajimotoPRB1999}. This coexistence provides additional arguments in favor of the proposed phase diagram for the 3D  underdoped ($x \leq 1/2$) manganites.

The long-range Coulomb interaction may modify our results, but we do not expect any qualitative changes. For realistic values of the parameters, we can conclude that the radius of FM polarons inside CO host is of the order of  10 \AA, and the excess charge contained in them is rather small for the doping close to 1/2.

Our treatment is also applicable to other systems with charge ordering such as cobaltites \cite{MoritomoPRB1998} and nickelates \cite{AlonsoPRL1999}. It would be interesting to study them for charge carrier densities different from the commensurate ``checkerboard" one. Among the important questions, which deserve further investigations, remain also the origin of the ``in-phase" ordering along the $c$ direction in perovskite manganites, as well as  the behavior of the CO state at finite temperatures.

Let us now draw attention to interesting experimental results reported in Ref.~\cite{HongNatCom2018} on the electronic phase separation in CaFe$_3$O$_5$ compound exhibiting the charge ordering, which manifests itself as the alternation of Fe$^{3+}$ and Fe$^{2+}$ sites similar to that in magnetite, Fe$_3$O$_4$. It is demonstrated that below the magnetic transition temperature at 302 K, this compound undergoes separation into two phases with different electronic and spin orders. One of them has the charge order; however, it involves not a simple alternation of charges in a checkerboard manner, but also a kind of orbital ordering leading to the formation of electronic molecules (trimerons). Another phase is characterized by the Fe$^{3+}$/ Fe$^{2+}$ charge averaging including, nevertheless, ferromagnetic chains. Lattice symmetry is unchanged but different strains from the electronic orders probably drive the phase separation. Indeed, as it is shown in in \cite{OlesArxiv2020}, the trimeron--phonon interaction is rather strong, thus determining the specific features of the ground state. However, for the phase separation in charge ordered materials, a deviation from the stoichiometry may be needed \cite{CassidyNatCom2019}. Thus, we clearly see here that the electronic phase separation reveals a complex interplay of charge, spin, orbital, and lattice degrees of freedom, which will  be discussed in the following sections. Let us mention that magnetite itself exhibits an inhomogeneous state characterized by structural, spin, and charge fluctuations even above the Verwey transition~\cite{PerversiNatCom2019}.

\section{Phase separation in strongly correlated systems with two electron bands}
\label{twobands}

As it was stated above, the phase separation in strongly correlated electron systems arises due to the competition between the contributions of kinetic and potential energies to the total energy of charge carriers. The interplay between the effects related to magnetic ordering and kinetic energy considered above is one of such possibilities. The second rather general case of this competition can be observed in multiband materials~\cite{sboychakov2007phase}. Actually, the bandwidth in a strongly correlated multiband material depends on the number of electrons in the band. The competition between the phases with more localized and more itinerant electrons can give rise to the phase separation. The phase separation in the classical Falicov--Kimball model~\cite{FalKimbPRL1969} is the most evident and simple illustration for this. This model is often used as a toy model for heavy-fermion compounds. It describes the system of itinerant and localized electrons with a strong on-site Coulomb repulsion. The numerical simulations for the Falicov--Kimball model demonstrated an inhomogeneous charge density distribution at some relation between the itinerant electron bandwidth and the distance between the localized level and the bottom of conduction band~\cite{FreericksPRB1999,FreericksPRL2002,MaskaPSS2005pattern}. The competition between the metallicity and localization in a similar system with magnetic interactions was studied in Refs.~\cite{KugelPRL2005,SboychakovPRB2006jahn,KugeJMMMl2007electronic} with an emphasis on the phase diagram of magnetic oxides such as manganites. The system with a band and localized level is a limiting case of a more common case of two bands having different width.

\subsection{Phase separation in the two-band Hubbard model}\label{Falicov-Kimball}

\subsubsection{General equations}\label{Falicov-KimballGE}

Here, we use a two-band Hubbard model for the description of a strongly correlated electron system with two types of charge carriers. We demonstrate that the phase separation in this system arises even without any ordering if the ratio of the bandwidths exceeds some threshold value.

We consider the system with two bands $a$ and $b$. Let the first band, $a$, be wider than the second one, $b$. In the Hubbard approach, the Hamiltonian of the system can be written in the form~\cite{sboychakov2007phase}:
\begin{equation}\label{HubbardTwoB}
\hat{H} =-\sum_{\langle \mathbf{ij}\rangle\alpha,\sigma}t^\alpha a^\dag_{\mathbf{i}\alpha\sigma}a_{\mathbf{j}\alpha\sigma}-
\epsilon\sum_{\mathbf{i}\sigma}n_{\mathbf{i}b\sigma}
  -\mu\sum_{\mathbf{i}\alpha\sigma}n_{\mathbf{i}\alpha\sigma}
  +\frac{1}{2}\sum_{\mathbf{i}\alpha\sigma}U^\alpha n_{\mathbf{i}\alpha\sigma}n_{\mathbf{i}\alpha\bar{\sigma}}
  +\frac{U'}{2}\sum_{\mathbf{i}\alpha\sigma'} n_{\mathbf{i}\alpha\sigma}n_{\mathbf{i}\bar{\alpha}\sigma'}.
\end{equation}
Here, $a^\dag_{\mathbf{i}\alpha\sigma}$ and $a_{\mathbf{i}\alpha\sigma}$ are the creation and annihilation operators for electrons corresponding to bands $\alpha =a,b$ at site $\mathbf{i}$ with spin projection $\sigma$, and $n_{\mathbf{i}\alpha\sigma}=a^\dag_{\mathbf{i}
\alpha\sigma}a_{\mathbf{i}\alpha\sigma}$. The first term in the right-hand side of Eq.~(\ref{HubbardTwoB}) corresponds to the kinetic energy of the conduction electrons in bands $a$ ($b$) with the hopping integral $t_a$ ($t_b$). We ignore the interband hopping. The second term describes the shift $\epsilon$ of the center of band $b$ with respect to the center of band $a$. The last two terms describe the on-site Coulomb repulsion of two electrons either in the same state $\alpha$ with the Coulomb energy $U^\alpha$ or in the different states $U'$. The assumption of the strong electron correlations means that $U^\alpha,U'\gg t^\alpha,\epsilon$. The total number $n$ of electrons per site is a sum of electrons in the $a$ and $b$ states, $n=n_a+n_b$. Below, for definiteness sake, we consider the case $n\leq 1$.

We define the single-particle and two-particle Green's functions as
\begin{equation}\label{GreenFs}
G_{\alpha\sigma}\left(\mathbf{j}-\mathbf{j}_0,t-t_0\right)=-i\langle \hat{T}a_{\mathbf{j}\alpha\sigma}(t)a^\dag_{\mathbf{j}_0\alpha\sigma}(t_0)\rangle,\quad
G_{\alpha\sigma\,\beta\sigma'}\left(\mathbf{j}-\mathbf{j}_0,t-t_0\right)=-i\langle \hat{T}a_{\mathbf{j}\alpha\sigma}(t)n_{\mathbf{j}\beta\sigma'}(t)
a^\dag_{\mathbf{j}_0\alpha\sigma}(t_0)\rangle,
\end{equation}
where $\hat{T}$ is the time-ordering operator. The equation of motion for the single-particle Green's function with Hamiltonian \ref{HubbardTwoB}) can be written as
\begin{equation}\label{Green1Eq}
\left(\!i\frac{\partial}{\partial t}+\mu+\epsilon^\alpha\!\right)\!G_{\alpha\sigma}\left(\mathbf{j}\!-
\!\mathbf{j}_0,t-t_0\right)\!=\!
\delta_{\mathbf{ij}}\delta(t-t_0)-t^\alpha\!\sum_\mathbf{\Delta}
\!G_{\alpha\sigma}\!\left(\mathbf{j}\!-\!\mathbf{j}_0\!+
\!\mathbf{\Delta},t-t_0\right)
+U^\alpha G_{\alpha\sigma\,\alpha\bar{\sigma}}+
U'\!\sum_{\sigma'}G_{\alpha\sigma\,\bar{\alpha}\sigma'}\, ,
\end{equation}
where $\epsilon^\alpha=0$ for $\alpha =a$ and $\epsilon^\alpha=\epsilon$ for $\alpha =b$, the summation in the second term in the right-hand side of Eq.~(\ref{Green1Eq}) is performed over the sites nearest to $\mathbf{j}$, and $\mathbf{\Delta}$ are the vectors connecting site $\mathbf{j}$ with its nearest neighbors. Equation~(\ref{Green1Eq}) includes the two-particle Green's functions. Then, we should write the equations of motion for these functions, which will include the next-order Green's functions, etc.

Then, it is necessary to cut such an infinite chain of equations; this problem was treated in Ref.~\cite{sboychakov2007phase} within a standard mean-field approach. In the limit of strong Coulomb repulsion, the presence of two electrons at the same site is unfavorable, and the two-particle Green's function is of the order of $1/U$, where $U\sim U^\alpha ,U'$. In the equations of motion for the two-particle Green's functions, the three-particle terms are of the order of $1/U^2$ and were neglected. Following the Hubbard I approximation~\cite{HubbardPrRoySocA1963}, in the equations of motion for the two-particle Green's functions, the term $\langle \hat{T}a_{\mathbf{j+\Delta}\alpha\sigma}(t)
n_{\mathbf{j}\beta\sigma'}(t)a^\dag_{\mathbf{j}_0\alpha\sigma}(t_0)\rangle$ [coming from the commutator of $a_{\mathbf{j}\alpha\sigma}(t)$ with the kinetic-energy terms of Hamiltonian (\ref{HubbardTwoB})] is replaced by $\langle n_{\mathbf{j}\beta\sigma'}\rangle\langle \hat{T}a_{\mathbf{j+\Delta}\alpha\sigma}(t)a^\dag_{\mathbf{j}_0\alpha\sigma}(t_0)\rangle$. The similar decoupling in the terms coming from the commutator of $n_{\mathbf{j}\alpha\sigma}$  with the same kinetic-energy operator yields zero. As a result, the equations for the
two-particle Green's functions can be written as~\cite{sboychakov2007phase}
\begin{equation}\label{Green2Eq}
 \left(\!i\frac{\partial}{\partial t}+\mu+\epsilon^\alpha-U\!\right)\!\!\left[\!
     \begin{array}{c}
     G_{\alpha\sigma\,\alpha\bar{\sigma}}(\mathbf{j}-\mathbf{j}_0,t-t_0) \\
      G_{\alpha\sigma\,\bar{\alpha}\sigma}(\mathbf{j}-\mathbf{j}_0,t-t_0) \\
       \end{array}
  \!\right]
 \!=\!\left(
    \begin{array}{c}
      n_{\alpha\bar{\sigma}} \\
      n_{\bar{\alpha}\sigma} \\
    \end{array}
  \right)\!\!
\left[\!\delta_{\mathbf{jj}_0}\delta(t-t_0)-
t^\alpha\sum_{\mathbf{\Delta}}G_{\alpha\sigma}(\mathbf{j}\!-
\!\mathbf{j}_0\!+\!\mathbf{\Delta},t-t_0)\!\right],
\end{equation}
where $n_{\alpha\sigma}=\langle n_{\mathbf{j}\alpha\sigma}\rangle$ is the average number of electrons per site in the state ($\alpha,\sigma$) and we put for brevity $U^\alpha,U'=U$. The Hubbard I approximation is an appropriate method to find out the main features of the electron band structure, which should be confirmed by comparison with experiments and numerical results~\cite{FuldeBook2002,Ovchin_book2004}. The detailed analysis of the applicability of the Hubbard I approximation to the problems of phase separation is given in Ref.~\cite{SboychakovPhB2013}.

Equations (\ref{Green1Eq}) and (\ref{Green2Eq}) can be solved in a conventional manner~\cite{HubbardPrRoySocA1963} by passing from the time--space to the frequency--momentum representation. Eliminating the two-particle Green's functions, we can calculate the single-particle Green's functions, which have poles corresponding to the Hubbard subbands for bands $a$ and $b$. The splitting between these subbands is determined by the on-site Coulomb repulsion $U$ of electrons belonging to the same bands. The on-site Coulomb repulsion of electrons from different bands gives rise to the correlation between the fillings of $a$ and $b$ bands~\cite{sboychakov2007phase,KugelPRL2005,SboychakovPRB2006jahn}. In the case when the total number of the electrons per site is $n<1$, the upper Hubbard subbands are empty, and one can proceed to the limit $U\rightarrow\infty$. As a result, we derive
\begin{equation}\label{G1abs}
G_{\alpha\sigma}(\mathbf{k},\omega)=\frac{g_{\alpha\sigma}}{\omega+
\mu+\epsilon^\alpha-g_{\alpha\sigma}w_\alpha\varsigma(\mathbf{k})},
\end{equation}
where $w_\alpha=zt^\alpha$, $z$ is the number of nearest neighbors
\begin{equation}\label{G1notations}
g_{\alpha\sigma}=1-\sum_{\sigma'}n_{\bar{\alpha}\sigma'}
-n_{\alpha\bar{\sigma}},\quad \varsigma(\mathbf{k})=\frac{1}{z}\sum_\mathbf{\Delta}e^{i\mathbf{k\Delta}}.
\end{equation}
The spectral function $\varsigma(\mathbf{k})$ depends on the lattice symmetry. For example, in the case of a simple cubic lattice, we have $\varsigma(\mathbf{k})=\left[\cos{k_xd}+\cos{k_yd}+\cos{k_zd}\right]/3$, where $d$ is the lattice constant.

In the the main approximation in $1/U$ considered here, the magnetic ordering does not appear. Then, one can put $n_{\alpha\downarrow}=n_{\alpha\uparrow}\equiv n_{\alpha}/2$ and the filling of each band $g_\alpha \equiv (g_{\alpha\downarrow}+g_{\alpha\uparrow})/2$. It follows from the expression for the density of states $\rho_\alpha(E)=-\pi^{-1}\textrm{Im}\int{G_\alpha(\mathbf{k},E+i0)
d^3\mathbf{k}/(2\pi)^3}$ that~\cite{sboychakov2007phase}
\begin{equation}\label{nab}
n_\alpha=2g_\alpha n_0\left(\frac{\mu+\epsilon^\alpha}{g_\alpha w_\alpha}\right),
\end{equation}
where
\begin{equation}\label{n0}
n_0(\varepsilon)=\int_{-1}^\varepsilon{dE\int{\frac{d^3\mathbf{k}}
{(2\pi)^3}\delta[E-\varsigma(\mathbf{k})]}}.
\end{equation}
The chemical potential $\mu$ in Eq.~(\ref{nab}) can be determined from the equality $n=n_a+n_b$.

Let the energy difference $\epsilon$ between centers of $a$ and $b$ bands be of the order of the characteristic width of the $b$ band, $w_b$. In this case, there exist only $a$ electrons at low doping until the chemical potential reaches the bottom of the $b$ band, $\mu=-\epsilon -w_b$, at some electron density $n_c$. If $n>n_c$, the $b$ electrons appear in the system, and the effective width of the $a$ band, $W_a =2w_ag_a(n_a,n_b)$, starts to decrease. In other words, at $n=n_c$ the topology of the Fermi surface transforms since the number of the electronic bands at the Fermi surface changes from one to two. Such a singularity is also referred to as the topological Lifshitz transition or so called 2.5 order transition~\cite{Lifshitz1960anomalies}. The plots of $n_a$, $n_b$, and the effective bandwidths as functions of $n$ are shown in Figs.~\ref{fig1FinWidth} and \ref{fig2FinWidth} for the simple cubic lattice.
\begin{figure}[H] \centering
  \includegraphics[width=0.6\columnwidth]{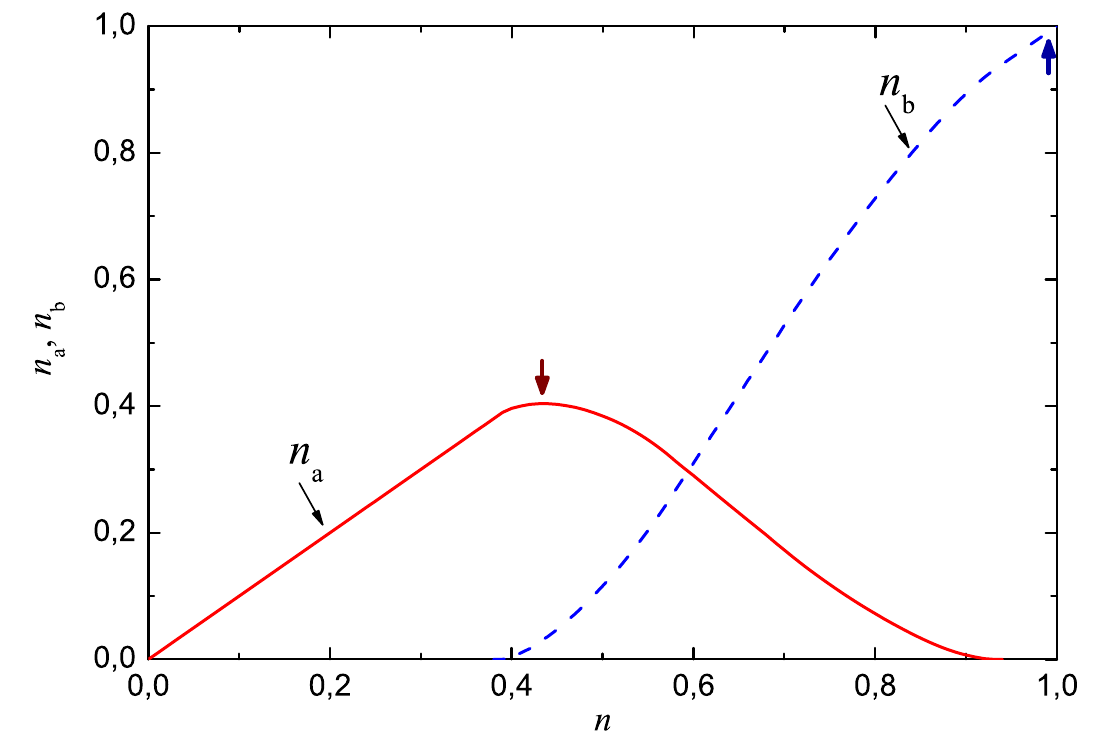}\\
  \caption{ \label{FigN} (Color online) Electron densities
$n_a$ (solid line) and $n_b$ (dashed line) vs the total
number of charge carriers $n$; $w_b/w_a=0.2$, $\epsilon/w_a=0.12$.
Vertical arrows show the values of $n_a$ and $n_b$ corresponding
to the inhomogeneous state~\cite{sboychakov2007phase}/
  }\label{fig1FinWidth}
\end{figure}

\begin{figure}[H] \centering
  \includegraphics[width=0.6\columnwidth]{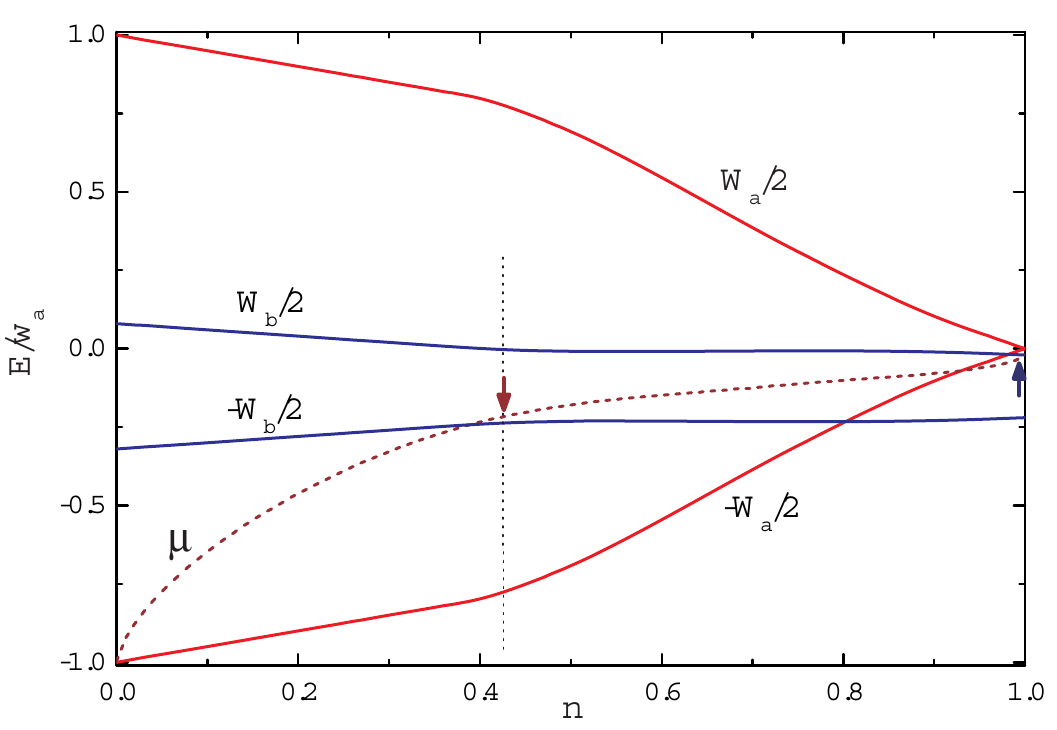}\\
  \caption{ (Color online) Effective bandwidths $W_\alpha=2w_\alpha g_\alpha$ vs $n$. The dashed curve is the chemical potential $\mu$. The values of the parameters are the same as in Fig.~\ref{fig1FinWidth}. Vertical arrows show the values of $n_a$ and $n_b$ in the inhomogeneous state~\cite{sboychakov2007phase}.
  }\label{fig2FinWidth}
\end{figure}

The energy of the system in the homogeneous state, $E_{\textrm{hom}}$, is the sum of electron energies in all filled bands. Similar to Eq.~(\ref{nab}), we get~\cite{sboychakov2007phase}
\begin{equation}\label{Ehom}
 E_{\textrm{hom}}=2\sum_\alpha g^2_\alpha w_\alpha\varepsilon_0\left(\frac{\mu+\epsilon^\alpha}{g_\alpha w_\alpha}\right),
\end{equation}
where
\begin{equation}\label{Eps0}
\varepsilon_0(\varepsilon)=\int_{-1}^\varepsilon{\!\!\!E\;dE\!
\int{\frac{d^3\mathbf{k}}{(2\pi)^3}\delta[E-\varsigma(\mathbf{k})]}}.
\end{equation}
The $E_{\textrm{hom}}(n)$ function is plotted in Fig.~\ref{fig3FinWidth} by the solid line.

\begin{figure}[H] \centering
  \includegraphics[width=0.6\columnwidth]{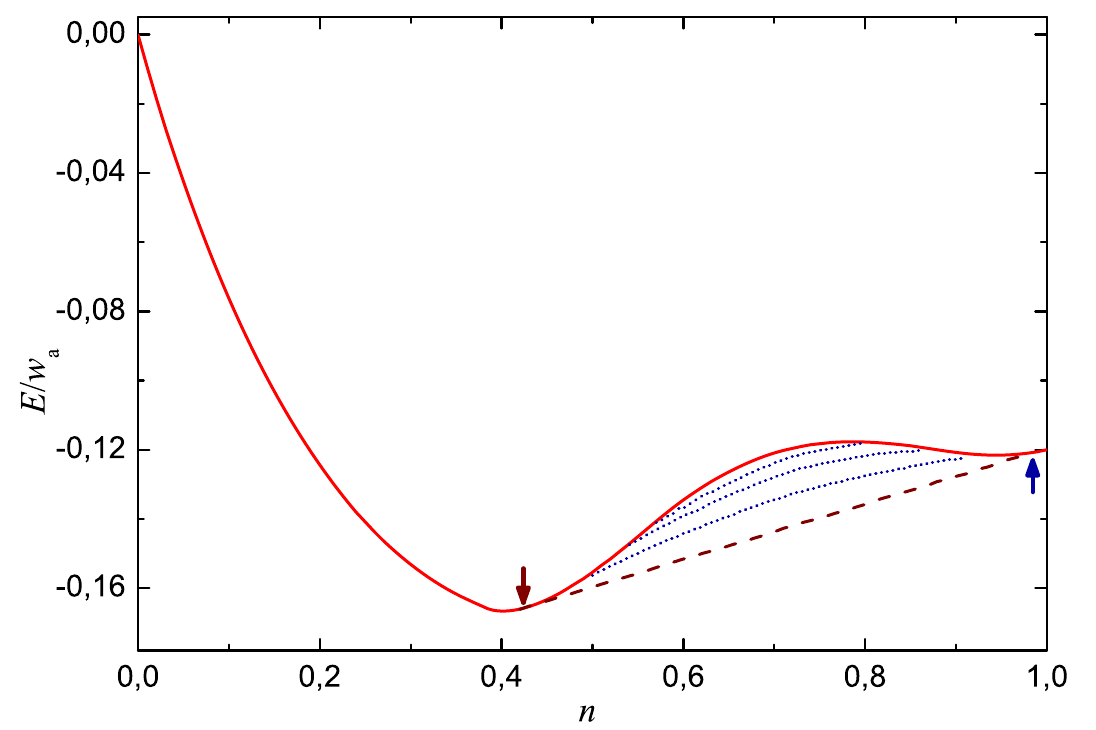}\\
  \caption{(Color online) Energy of the system vs doping level $n$. The solid curve corresponds to the homogeneous state, whereas the dashed curve is the energy of the phase-separated state without taking into account electrostatic and surface contributions to the total energy. Dotted curves are the energies of the inhomogeneous state at $V_0 /w_a$=0.1, 0.05, and 0.01 from top to bottom, respectively, (see the text below). Here, $w_b/w_a=0.2$ and $\epsilon/w_a=0.12$. Vertical arrows show the values of $n_a$ and $n_b$ in the inhomogeneous state~\cite{sboychakov2007phase}
  }\label{fig3FinWidth}
\end{figure}

\subsubsection{Phase separation}\label{Falicov-KimballPS}

As one can see in Fig.~\ref{fig3FinWidth}, the energy of the homogeneous state, $E_{\textrm{hom}}(n)$, has two minima. Thus, we can expect that a system forms two phases with different electron densities. However, the phase separation may be hindered by the increase of the energy due to surface effects and a charge redistribution. However, at first, we ignore these. We consider two phases, I (low charge carrier density per site $n_1$ and volume fraction $p$) and II (high charge carrier density $n_2$ and volume fraction $1-p$). We seek a minimum of the system energy
\begin{equation}\label{EpsNwoB}
E^0_{\textrm{ps}}=pE_{\textrm{hom}}(n_1)+(1-p)E_{\textrm{hom}}(n_2)
\end{equation}
under the condition of the conservation of the number of charge carriers
\begin{equation}\label{chacons}
n=pn_1+(1-p)n_2.
\end{equation}

The calculated energy of the system with simple cubic lattice in the phase-separated state is shown in Fig.~\ref{fig3FinWidth} by the dashed curve~\cite{SboychakovPRB2006jahn}. The phase separation exists within the range of $n$ where both types of charge carriers coexist in the homogeneous state. The densities of charge carriers in each phase vary slowly with $n$, remaining close to certain optimal values for each phase: $n_1\approx n_a \approx 0.5$ and $n_2\approx n_b\approx 1$.  Phase II can be considered as a Mott--Hubbard insulator since the corresponding lower Hubbard subband is almost completely filled. Therefore, we can conclude that the system may become separated into metallic and insulating phases in a certain parameter range. The above discussion demonstrates that the width and filling of one band depend on the width and filling of the other band.

The phase separation can be favorable only if the bands are appreciably different. As we know, the negative compressibility $\chi=\partial^2E/\partial n^2<0$ is a signature of the possible phase separation. The numerical analysis of the considered particular model in Ref.~\cite{sboychakov2007phase} demonstrates that $\partial^2E/\partial n^2<0$ in a certain range of $n$ if $w_b/w_a<0.4$. Naturally, this threshold value of thebandwidth ratio may be different for different systems. The region of the phase separation in the ($\epsilon/w_a,1-n$) plane for the model under discussion is shown in Fig.~\ref{figTwBPD}.

\begin{figure}[H] \centering
  \includegraphics[width=0.6\columnwidth]{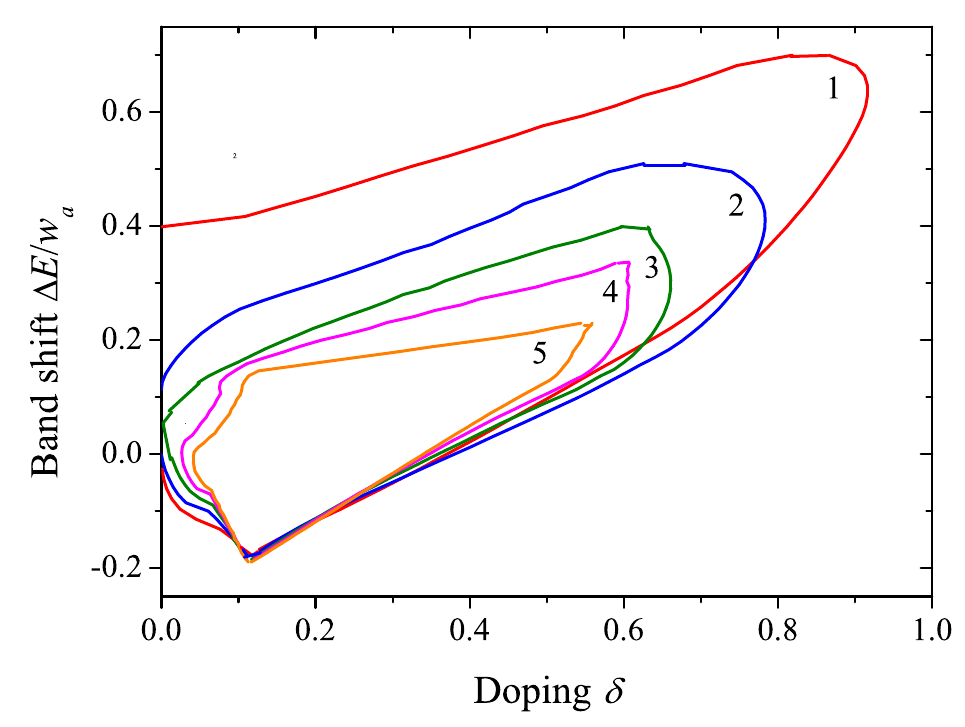}\\
  \caption{ (Color online) Phase separation regions (value $\Delta E$ in the figure corresponds to $\epsilon$ in the text and $\delta=1-n$) for different values of the bandwidth ratio: $w_b/w_a = 0.1$ (curve 1), 0.2 (2), 0.3 (3), 0.35 (4), and 0.4 (5). If $w_b/w_a > 0.4$, the phase separation is unfavorable (long-range Coulomb interaction is neglected, $V=0$)~\cite{KugelSuST2008}.
  }\label{figTwBPD}
\end{figure}

\subsection{Role of the interband hybridization} \label{hybridization}

In the previous discussion of the two-band Hubbard model, we did not take into account a possibility of the interband charge transfer or the hybridization between bands $a$ and $b$. Such a hybridization may be especially relevant in the case of cuprate superconductors. Cuprate superconductors can be considered as typical examples of the strongly correlated electron systems. Effects occurring in them are usually treated in the framework of the multiband Hubbard model or its generalizations. In such situations, we are usually dealing with the hybridization of oxygen $p$ orbitals and copper $d$ orbitals. Then, the $a$ and $b$ bands under discussion can be treated as $p$ and $d$ bands of cuprates. Thus, we can  generalize Hamiltonian (\ref{HubbardTwoB}) taking the kinetic energy term in the form
\begin{equation}\label{hybr-term}
-\!\!\!\sum_{\langle\mathbf{ij}\rangle\alpha\beta\sigma}t^{\alpha\beta}
\left(a^{\dag}_{\mathbf{i}\alpha\sigma}a_{\mathbf{j}\beta\sigma}+h.c.\right).
\end{equation}
Here, the notation is the same as in Eq.(\ref{HubbardTwoB}), and $\alpha$ and $\beta$ mean indices of $a$ and $b$ (or $p$ and $d$) bands. Hence, we have three different hopping integrals $t_{pp}$, $t_{dd}$, and $t_{pd}$ (later on we also use notation $w_{pp}=zt_{pp}$, $w_{dd}=zt_{dd}$, and $w_{pd}=zt_{pd}$). The detailed analysis based on such two-band Hubbard Hamiltonian with the interband hybridization is given in Ref.~\cite{sboychakov2008mechanism}. The problem is considered following the same lines as those for the two-band Hubbard model without hybridization studied in the previous subsections. The corresponding results can be summarized as follows. First, we can diagonalize the kinetic energy part of the Hamiltonian. Solving the corresponding eigenvalue problem, we obtain the energy spectrum of charge carriers in two new unhybridized bands (labelled by $j=\pm 1$). In contrast to $p$ and $d$ holes with short lifetime due to interband hoping, the new quasiparticles have a longer lifetime due to scattering by, e.g., phonons and impurities. We denote the lower band as $j=1$ and the upper one as $j=-1$. The densities of states (DOS) for these bands are plotted in Fig.~\ref{hybr-DOS} in the case of the simple cubic lattice. Again, each band is split in two Hubbard sub-bands due to the on-site Coulomb repulsion. So, in the two-band Hubbard model, we have four bands, two lower and two upper bands separated by the gap of the order of $U$. As above, we can consider here the doping range $n<1$, for which we can limit ourselves to the case $U \rightarrow \infty$, when the upper Hubbard sub-bands are empty and do not contribute to the total electron energy.

Two qualitatively different quasiparticle spectra are shown in the insets of Figs.~\ref{hybr-DOS}(a) and \ref{hybr-DOS}(b). If the hybridization is small [see inset of Figs.~\ref{hybr-DOS}(a)], the anticrossing of two bands corresponds to a metallic behavior for any doping. When doping increases, the chemical potential $\mu$ [shown by the dotted (green) line] shifts upward: at low doping, we have one metallic band and one empty band; then two metallic bands; with further increase of the doping we have one metallic and one filled band. For larger interband hybridization, when $t^{pd}>(t^{pp}t^{dd})^{1/2}$ [see inset of Figs.~\ref{hybr-DOS}(b)], there exists a transition to an insulator at some doping level. Indeed, for a certain doping, $\mu$ is located in the gap between the bands. Nearby the anticrossing point of the two bands, the quasiparticle spectrum becomes flattened. This could be related to the nesting of the Fermi surface. As a result, the DOS exhibits peaks at the energies corresponding to the anticrossing points (Fig.~\ref{hybr-DOS}). The optimum doping for superconductivity corresponds to the case when $\mu$ is close to the energy, where these peaks are observed.

\begin{figure}[H]
\begin{center}
\includegraphics*[width=0.5\columnwidth]{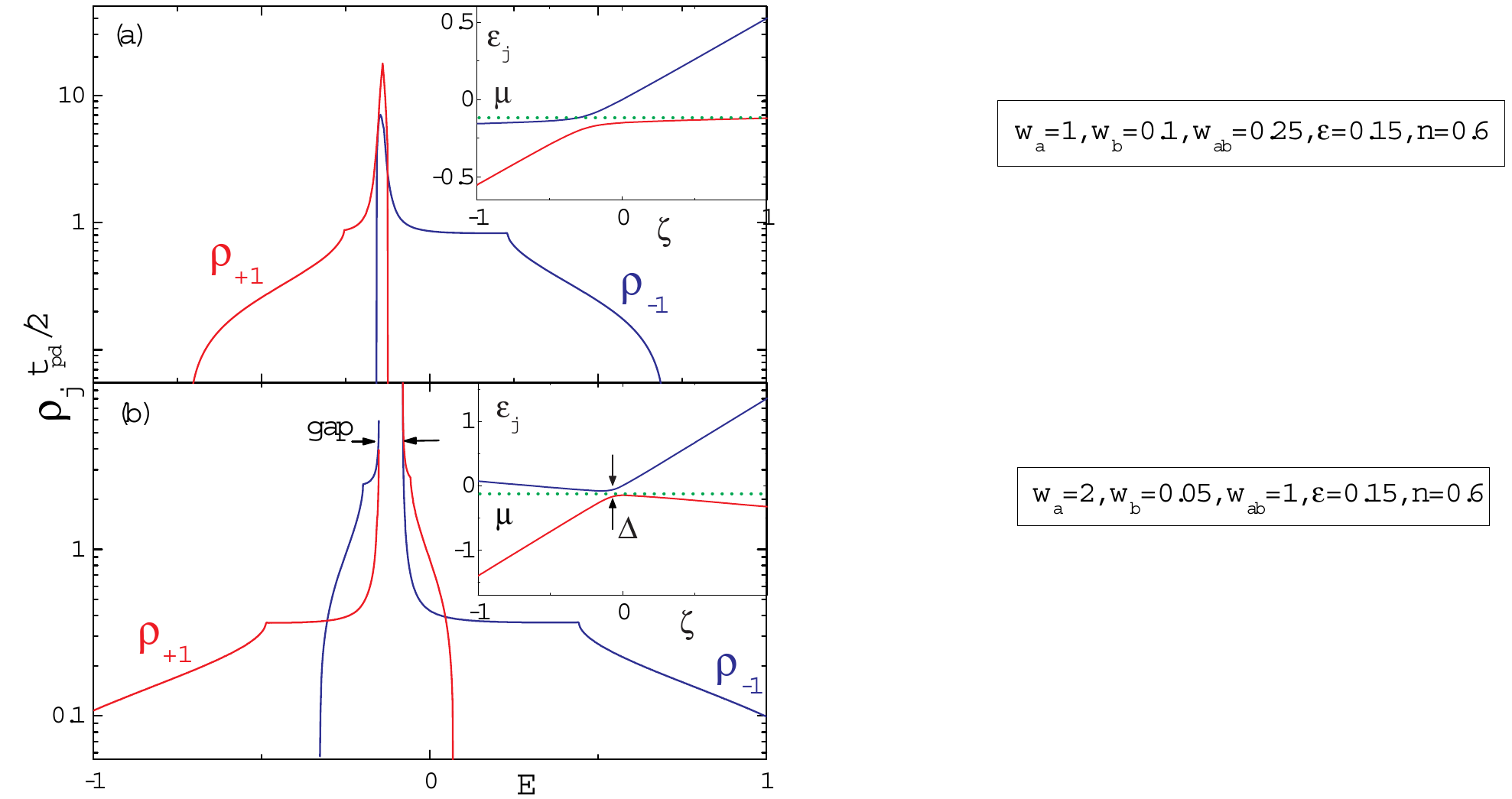}
\end{center} \caption{\label{hybr-DOS} (Color online) Density of states $\rho_{\pm 1}$ versus energy $E$ for quasiparticles (holes) located in bands
$j=\pm 1$ for two different types of spectra, shown in the corresponding insets. The gapless spectrum shown in inset (a) was calculated at $n=0.6$, $\varepsilon_d=0.15$~eV, $w_{pp}=1$~eV, $w_{dd}=0.1$~eV, and relatively small interband hybridization $w_{pd}=0.25$~eV. The gapped spectrum in inset (b) was calculated for the same parameters but for stronger interband
hybridization $w_{pd}=1$~eV. In case (b), the transition to the
insulating state occurs at a certain doping level for which the
chemical potential $\mu$ (shown by the green dotted line) is
located inside the gap. In both cases, a large peak in the DOS is
observed at energies corresponding to the anticrossing of the
bands, where a significant flattening of the Fermi surface (see
insets) takes place. The quasiparticle energy spectra are shown in
the insets as a function of the variable $\zeta$, since
$\varepsilon_j$ depends on the crystal momentum $\textbf{k}$ only
via $\zeta(\textbf{k})$~\cite{sboychakov2008mechanism}. } \end{figure}

The energy of the homogeneous state versus doping $n$ is shown in the inset of Fig.~\ref{hybrPS}(a). The curvature of $E_{\textrm{hom}}(n)$ is negative between the two marked points $n_1$ and $n_2$. This indicates the instability of the homogeneous state with respect to the separation into two phases with the hole densities $n_1$ and $n_2$, when $n_1<n<n_2$. The energy $E_{\textrm{ps}}(n)=pE_{\textrm{hom}}(n_1)+(1-p)E_{\textrm{hom}}(n_2)$ of the phase-separated state is lower than the energy $E_{\textrm{hom}}(n)$ of the homogeneous phase in this doping range [see dashed line in the inset of Fig.~\ref{hybrPS}(a)]. In a wide parameter range, the hole density $n_2$ turns out to be near the optimum value for superconductivity. A typical dependence of the phase concentration $p$ versus doping is shown in the inset of Fig.~\ref{hybrPS}(b). At a low doping level, the system is in a homogeneous state ($p=1$). If $n>n_1$, the sample becomes segregated into droplets with two different hole densities, $n_1$ and $n_2$. With the further increasing in $n$, the relative concentration $p$ of the phase with lower hole content $n_1$ decreases almost linearly, and when $n>n_2$, this phase disappears and the system becomes homogeneous again. Further analysis shows that the phase-separated state occurs in the parameter range, where $w_{pd}<w_{pp}$. If $w_{pd}\gg w_{pp}$, the two-band Hubbard Hamiltonian reduces to an effective single-band model and the discussed mechanism for the phase separation in the system does not work.

\begin{figure}[H]
\begin{center}
\includegraphics*[width=0.5\columnwidth]{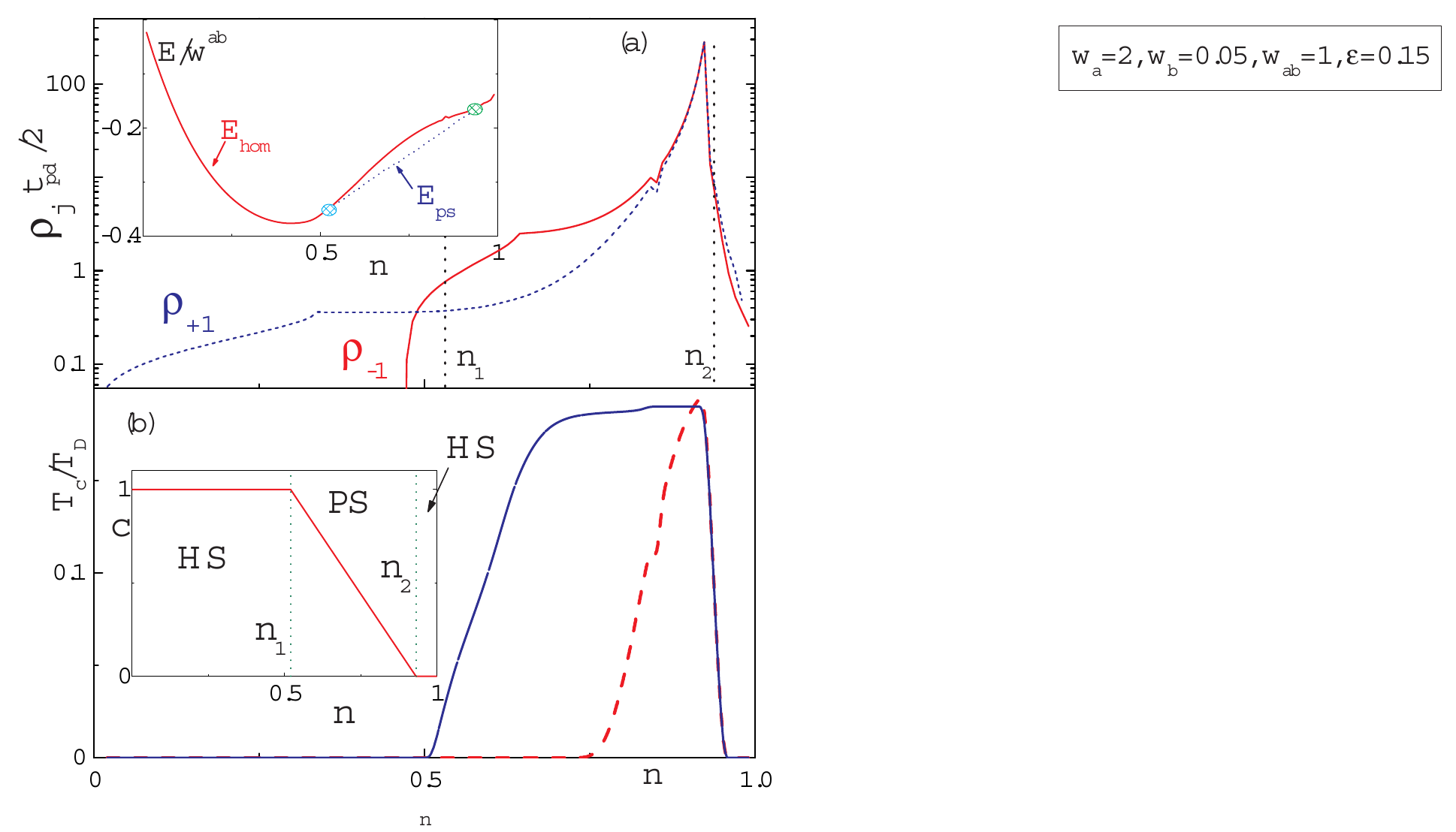}
\end{center} \caption{\label{hybrPS}(Color online) (a) Density of states at the Fermi level versus doping in the two-band Hubbard model. Inset of panel (a): energy of the homogeneous (red solid line) and the phase-separated (blue dashed line) states. The dependence of the energy in the homogeneous state, $E_{\textrm{hom}}$, on $n$ has a negative curvature if $n_1<n<n_2$. In this range of doping, the phase separated-state becomes more favorable since its energy, $E_{\textrm{ps}}$, is lower than $E_{\textrm{hom}}$. The values $n_{1,2}$ are indicated in the main panel of (a) by black vertical dotted lines. The point $n_2$ is near the peak in the DOS. (b) Critical temperature, $T_c$, of the superconducting transition versus doping level $n$ for the homogeneous (red dashed line) and phase-separated (blue solid line) states calculated using Eq.~(\ref{Tc}). The $T_c(n)$ of the homogeneous state decreases fast when the doping $n$ deviates from its optimum value about $n_2$. In contrast to this, $T_c(n)$ in the phase separated state is a broad function and exhibits a plateau within the $n_1<n<n_2$ range. The inset of panel (b) demonstrates the dependence of concentration of the phase with lower hole content on $n$ (the notation $c$ in the figure corresponds to the notation $p$ in the text). The regions of homogeneous (HS) and the phase-separated (PS) states are indicated in the inset. Here, we use the following parameters: $\varepsilon_d=0.2$~eV, $w_{pp}=1$~eV,
$w_{dd}=0.3$~eV, and $w_{pd}=0.7$~eV ~\cite{sboychakov2008mechanism}. }
\end{figure}

The phase separation leads to a redistribution of the charge carriers and charge neutrality breaking. The structure of the phase-separated state can be either ordered (e.g., checkerboard structure) or random (i.e., randomly distributed droplets of one phase within a matrix of the other phase). The charge redistribution and finite droplet size give rise to the additional size-dependent terms in Eq.~(\ref{EpsNwoB}). For the random phase-separated state, the droplet size $D$ is determined by the competition between the Coulomb and surface energies. This problem will be discussed in detail below (in subsection~\ref{CoulombSurface}). Following the approach used in that subsection, we can estimate $D$ in the range from three to six lattice constants $d$ when $w_{pd}\sim w_{pp} \sim1$~eV, in agreement with the experimental data reported for the cuprates, e.g, in Ref.~\cite{LangNature2002}, where spatial variations of the DOS and superconducting gap were measured using STM.

Let us discuss the possible effect of the considered type of the phase separation on the critical temperature $T_c$ of the superconducting transition. We do not focus on any particular mechanism of superconductivity in cuprates. We should only relate somehow the DOS to $T_c$. Then, to illustrate the possible effects, we use the simplest BCS formula for $T_c$ accounting for the Coulomb repulsion~\cite{deGennes_book}
\begin{equation}\label{Tc}
T_c=T_D\exp\left[-\;1/\left(\rho_jV_p-\nu_c\right)\right],
\end{equation}
where $T_D$ is the Debye temperature, $V_p$ is the BCS-type electron--phonon coupling constant,
\begin{equation}\label{Qu}
\nu_c=\rho_jV_c/\left[1+\rho_jV_c\ln\left(\mu/k_BT_D\right)\right]
\end{equation}
is the Coulomb pseudopotential, and $V_c\sim U^p\approx 5$~eV is the Coulomb matrix element. The possible applicability of Eq.~(\ref{Tc}) in the case of high-$T_c$ cuprates is discussed in Ref.~\cite{MaksimovUFN2007}.

In the approach under discussion, the homogeneous superconducting state can appear in different bands depending on doping, see Fig.~\ref{hybrPS}: in the band $j=1$ for low doping and in the band $j=-1$ for higher doping. The dependence of $T_c$ on doping $n$ for the homogeneous state is shown by the dashed line in Fig.~\ref{hybrPS}(b). For the chosen range of parameters, $T_c$ is nonzero only for high doping level and achieves a maximum at the optimum doping, which corresponds to the value of chemical potential near the anticrossing point (see insets of Fig.~\ref{hybr-DOS}). Away from the optimum doping, $T_c(n)$ decreases fast with $n$. In the phase-separated state, one of the phases retains the optimum hole density (which is about $n_2$) within a wide interval of doping. This results in a in a plateau of $T_c(n)$ [see blue solid line in Fig.~\ref{hybrPS}(b)]. This can provide a possible explanation for the observed dependence of $T_c$ on the hole doping in cuprates (see Refs.~\cite{MaksimovUFN2007,MaksimovUFN2004} and references therein).

The above picture relates $T_c$ to the variation of the density of states and doping. It can be qualitatively applicable for other models of the superconductivity. The relation between DOS and $T_c$ is widely discussed in literature in connection with many other models of superconductivity (see, e.g., Refs.~\cite{DagottoPRL1995,FeinerPRL1996}).

\subsection{Several remarks: beyond the Hubbard I approximation}
\label{beyondHubbard}

In our analysis of the electronic phase separation, we were mostly dealing with an approach based on the Hubbard I approximation. As we have already mentioned, Hubbard I IS an appropriate method to find out the main features of the electron band structure capturing the essential physics related to electron correlations. Moreover, it is a good starting point for revealing the tendency to electronic phase separation in various systems and finding out the structure of corresponding phase diagrams, especially in the limit of strong correlations.

However, Hubbard I misses some important effects such as interband quasiparticle peak clearly seen if we use more refined approaches such as the dynamic mean-field theory (DMFT). The detailed comparison of the results obtained within the Hubbard I and DMFT approaches has been undertaken in Ref.~\cite{SboychakovPhB2013}. The spinless Falicov--Kimball model for the simple cubic lattice and a more general model, in which heavier charge carriers correspond to a band with finite width $W$, were analyzed. The Green's function for itinerant charge carriers, the evolution of the system with doping, and the possibility of the phase separation were studied. It was shown that both approximations lead to nearly identical results for the Falicov--Kimball model, i.e. in the narrow-band limit, $W \rightarrow 0$. For finite $W$ and in the strong correlation limit, $U \gg W$, , both approximations do not contradict each other giving qualitatively similar results in the spectral properties of itinerant charge carriers and the behavior of the system with doping. Both approximations predict an instability of the homogeneous state toward the phase separation in a wide range of model parameters. It was also shown that the system remains unstable toward the phase separation even for finite bandwidth $W$ for heavy charge carriers. This suggests that the phase separation phenomenon is the inherent feature of Hubbard-like models with wide and narrow bands, and not an artefact of this or that approximation scheme.

Moreover, the Hubbard I approximation was used for the direct evaluation of the equal-time anomalous two-electron propagator for the Hubbard model on a two-dimensional square lattice \cite{RozhkovJPCM2011}. The results were compared with the quantum Monte Carlo calculations reported in Ref.~\cite{AimiJPSJp2007} in the limit of strong electron–electron interaction. The Hubbard I predictions appear to be in a good qualitative agreement with the Monte Carlo results. In particular, $d$-wave type correlations decay as $cr^{-3}$ (``free-electron'' behavior), if the separation $r$ exceeds 2--3 lattice constants. However, the Hubbard I approximation underestimates the coefficient $c$ by a factor of about 3. This demonstrates that the Hubbard I approximation, despite its simplicity and artefacts, captures the qualitative behavior of the two-particle propagator for the Hubbard model, at least, for moderate values of $r$.

Note, however, that using the Hubbard I approximation, we can miss additional mechanisms, which can also lead to the phase separation. Quite a vivid example is the so-called charge-transfer instability, which was widely discussed in the context of cuprate superconductors~\cite{GriliPRL1991,GriliIJMPB1991,BangPRB1991,RaimondiPRB1993}. The charge-transfer instability manifests itself in the vanishing with  doping of the lowest excitonic particle-hole excitation. This leads to the diverging compressibility giving rise to a phase separation. A thorough analysis of this effect undertaken in Ref.~\cite{RaimondiPRB1993} shows that it has rather fundamental nature and is related  to the violation of the Landau stability criterion for the Fermi liquid (so called Pomeranchuk instability). In this analysis, the used slave-boson technique \cite{KotliarPRL1986} appears to be very fruitful since it reveals nontrivial electron correlation effects beyond the Hubbard I approximation. In this connection, let us mention a series of papers of Irkhin et al. \cite{Igoshev_JMMM2015,IgoshevJPCM2015spiral,TimirgazinJPCM2016,IgoshevJMMM2018}, where the slave-boson technique has been successfully implemented for finding out the phase-separation regions for different magnetic transitions in strongly correlated electron systems.

\subsection{Electron--lattice interaction and the phase separation} \label{el-lattice}

Electron--lattice coupling is frequently incorporated in the multiband Hubbard model to describe electronic structure of cuprates and manganites.  Indeed, such compounds contain Jahn--Teller ions (copper or manganese) and this implies the crucial role of this coupling. Even at the initial stages of the studies of the colossal magnetoresistance effect, it was shown that both the strong electron correlations and electron--lattice interactions are fundamental ingredients for any adequate interpretation of this effect~\cite{MillisPRL1995,MillisPRL1996,MillisNature1998,MillisJAP1998}. Moreover,  the electron--lattice coupling can lead to quite unusual forms of the phase separation (see, e.g., Ref.~\cite{ramakrishnan2004theory}).

In Ref.~\cite{sboychakov2010effect}, the effect of this coupling on the phase separation in the two-band Hubbard model, Eq.~(\ref{HubbardTwoB}), was considered. The interaction of electrons with the lattice distortions results, first, in the change of the hopping probability and, second, in the relative shifts of the electron bands. If the electron--lattice interaction is strong enough, then there appears a competition between states with different values of strains and the transition between these states can occur in a jump-like manner. It was also demonstrated that the electron--lattice interaction produces a pronounced effect on the conditions of the electronic phase separation since it affects the value of the bandwidth ratio and the relative positions of the bands.

The electron--lattice interaction can be chosen in the following
form~\cite{sboychakov2010effect}
\begin{eqnarray}\label{e-p} H_{el-latt}=\sum_{\langle
\mathbf{i}\mathbf{j}\rangle,\sigma}\left(\lambda\,
a^\dag_{\mathbf{i}\sigma}a_{\mathbf{j}\sigma}u_\mathbf{j}+\lambda'
b^\dag_{\mathbf{i}\sigma}b_{\mathbf{j}\sigma}u_\mathbf{j}+h.c.\right)
+\sum_{\mathbf{i},\sigma}\left(\lambda_a
u_{\mathbf{i}}n^a_{\mathbf{i}\sigma}+
\lambda_bu_{\mathbf{i}}n^b_{\mathbf{i}\sigma}\right),
\end{eqnarray} where $\lambda$, $\lambda'$ and $\lambda_{a,b}$ are
corresponding constants of the electron--lattice interaction and
$u_\mathbf{i}$ are the local distortions of the unit cell containing site
$\mathbf{i}$. The first sum in this equation describes the change of the hopping probabilities due to lattice distortions affecting both intersite distance and the form of electron wave functions. The second sum corresponds to the change in the electron--ion interaction induced by strain. Thus, we take into account the effect of lattice
strains both on the on-site electron energy and intersite charge transfer.

The elastic energy of the system depending on distortions at different sites can be written as
\begin{equation}\label{el} {\cal{F}}_{\rm
elast}=\frac{K}{2}\sum_{\mathbf{i}}u^2_{\mathbf{i}},
\end{equation}
where $K$ is the elastic modulus.

To find a self-consistent solution to the problem in the adiabatic approximation, we first perform averaging of the Hamiltonian over the electronic degrees of freedom. From the condition of energy minimum with respect to strains, $\partial{\langle H(u_\mathbf{i})\rangle}/\partial u_\mathbf{i}=0$, we obtain for the lattice local distortions
\begin{equation}\label{ui}
\bar{u}_\mathbf{i}=-\frac{\lambda z\langle
a^\dag_{\mathbf{i}\sigma}a_{\mathbf{j}\sigma}\rangle+\lambda'z\langle
b^\dag_{\mathbf{i}\sigma}b_{\mathbf{j}\sigma}\rangle+\lambda_a\langle
n_\mathbf{i}^a\rangle+\lambda_b\langle n_\mathbf{i}^b\rangle}{K},
\end{equation}
where $\mathbf{i}$ and $\mathbf{j}$ are the nearest-neighbor sites. Using Eq.~(\ref{ui}), we can present the effective electron Hamiltonian as
\begin{eqnarray}\label{Heff1}
H_{\rm eff}=&-&\sum_{\langle \mathbf{i}\mathbf{j}\rangle,\sigma}
\left[\left(t^a-\lambda
\bar{u}_\mathbf{i}\right)a^\dag_{\mathbf{i}\sigma}a_{\mathbf{j}\sigma}
+
\left(t^b-\lambda'\bar{u}_\mathbf{i}\right)
b^\dag_{\mathbf{i}\sigma}
b_{\mathbf{j}\sigma}+h.c.\right]
-\nonumber \sum_{\mathbf{i},\sigma}\left[\Delta
E+(\lambda_a-\lambda_b
)\bar{u}_\mathbf{i}\right]n_{\mathbf{i}\sigma}^b\nonumber \\
&+&H_U-\sum_{\mathbf{i},\sigma}(\mu-\lambda_a\bar{u}_\mathbf{i})
\left(n^a_{\mathbf{i}\sigma}+n^b_{\mathbf{i}\sigma}\right)+
\frac{K}{2}\sum_\mathbf{i}\bar{u}^2_\mathbf{i},
\end{eqnarray}
where $H_U$ includes all the terms responsible for the on-site electron--electron interaction [see Eq.~(\ref{HubbardTwoB})].

Hamiltonian (\ref{Heff1}) clearly demonstrates that the  effect
of electron--lattice interaction is twofold. This interaction in its
nondiagonal terms changes the effective bandwidth, whereas in
diagonal terms, it shifts the positions of the bands and the
chemical potential. In the previous subsections, we have demonstrated that the qualitative features of the phase diagram for the
two-band Hubbard model are mainly determined by two dimensionless
parameters: the ratio of the bandwidths and the relative positions
of the bands. Thus, to construct a minimal model capturing the
main physical effects of electron--lattice interaction, it is
sufficient to keep only $\lambda$ and $\lambda_b$. In addition, we
can put $\Delta E=0$ to emphasize the effect of band shift related
only to the electron--lattice interaction. As a result, we get
\begin{equation}\label{ui1} \bar{u}_\mathbf{i}=-\frac{\lambda
z\langle a^\dag_{\mathbf{i}\sigma}a_{\mathbf{j}\sigma}\rangle
+\lambda_b\langle n_\mathbf{i}^b\rangle}{K}
\end{equation}
and
\begin{eqnarray}\label{Heff2} \nonumber
H_{\rm eff}&=&-\sum_{\langle
\mathbf{i}\mathbf{j}\rangle,\sigma} \left[\left(t^a-\lambda
\bar{u}_\mathbf{i}\right)a^\dag_{\mathbf{i}\sigma}a_{\mathbf{j}\sigma}
+t^b b^\dag_{\mathbf{i}\sigma}b_{\mathbf{j}\sigma}+h.c.\right]+H_U\\
&+&\sum_{\mathbf{i},\sigma}\lambda_b
\bar{u}_\mathbf{i}n_{\mathbf{i}\sigma}^b
-\sum_{\mathbf{i},\sigma}\mu
\left(n^a_{\mathbf{i}\sigma}+n^b_{\mathbf{i}\sigma}\right)+
\frac{K}{2}\sum_\mathbf{i}\bar{u}^2_\mathbf{i}.\,\,
\end{eqnarray}

Starting from Hamiltonian (\ref{Heff2}), we can point out  the
main qualitative effects of the electron--lattice interaction in
the two-band model. In the absence of doping, $n=0$, bands $a$ and
$b$ are empty and their centers coincide. With the growth of $n$,
the wider band $a$ begins to be filled up from the bottom. The band
filling is accompanied by strain $\bar{u}_{\mathbf{i}}=-\lambda
z\langle a^\dag_{\mathbf{i}\sigma}a_{\mathbf{j}\sigma}\rangle/K$.
The average $\langle a^\dag_{\mathbf{i}\sigma}a_{\mathbf{j}\sigma}\rangle$ is proportional to the hopping probability and thus is positive. The
strain $\bar{u}_{\mathbf{i}}$ is positive if $\lambda<0$ and
negative if $\lambda>0$. From Eq.~(\ref{Heff2}), it is easy to see
that at any sign of $\lambda$ the bandwidth increases with the
strain. At a certain doping level, the chemical potential attains
the bottom of the narrower band $b$ and this second band starts to
be filled up. In this range of doping, we have two types of the
electrons and the energy of the system depends on $n$ in a rather
complicated manner due to electron--electron correlations. Such situation is favorable for the phase separation even in the absence of the electron--lattice interaction. However, the characteristic feature of the system under study  is the dependence of the effective band shift $\Delta E_{\rm eff}=-\lambda_b\bar{u}_{\mathbf{i}}$ on the strain and, hence, on
the doping. The sign of the shift depends on the signs of $\lambda$ and $\lambda_b$. A simple analysis of Eqs.~(\ref{ui1}) and (\ref{Heff2}) shows that for the same signs of $\lambda$ and $\lambda_b$ the value of $\Delta E_{\rm eff}$ is negative and the sign of $\bar{u}_{\mathbf{i}}$
remains the same at any $n$. If the signs of $\lambda$ and
$\lambda_b$ are different, the strain can change its sign at some
doping level. The change of the sign of the strain results in the
change of the sign of the effective band shift $\Delta E_{\rm
eff}$.

The dependence of $\Delta E_{\rm eff}$ on doping can give rise to
a more sophisticated behavior of the system. If at some doping
level $n^*$, the narrower band crosses the bottom
of the wider band, that is, $\lambda_b^2n^*/K > zt_a$, then, it
could be favorable to have almost all electrons in the $b$ band.
So, there appears a competition between two states with
different values of the strain. It suggests the possibility of a
transition between these two states, which can have a jump-like
form. The corresponding results are illustrated in Figs. \ref{FigJumpU} and \ref{FigJumpE}.

\begin{figure}[tbh] \begin{center}
\includegraphics[width=0.6\columnwidth]{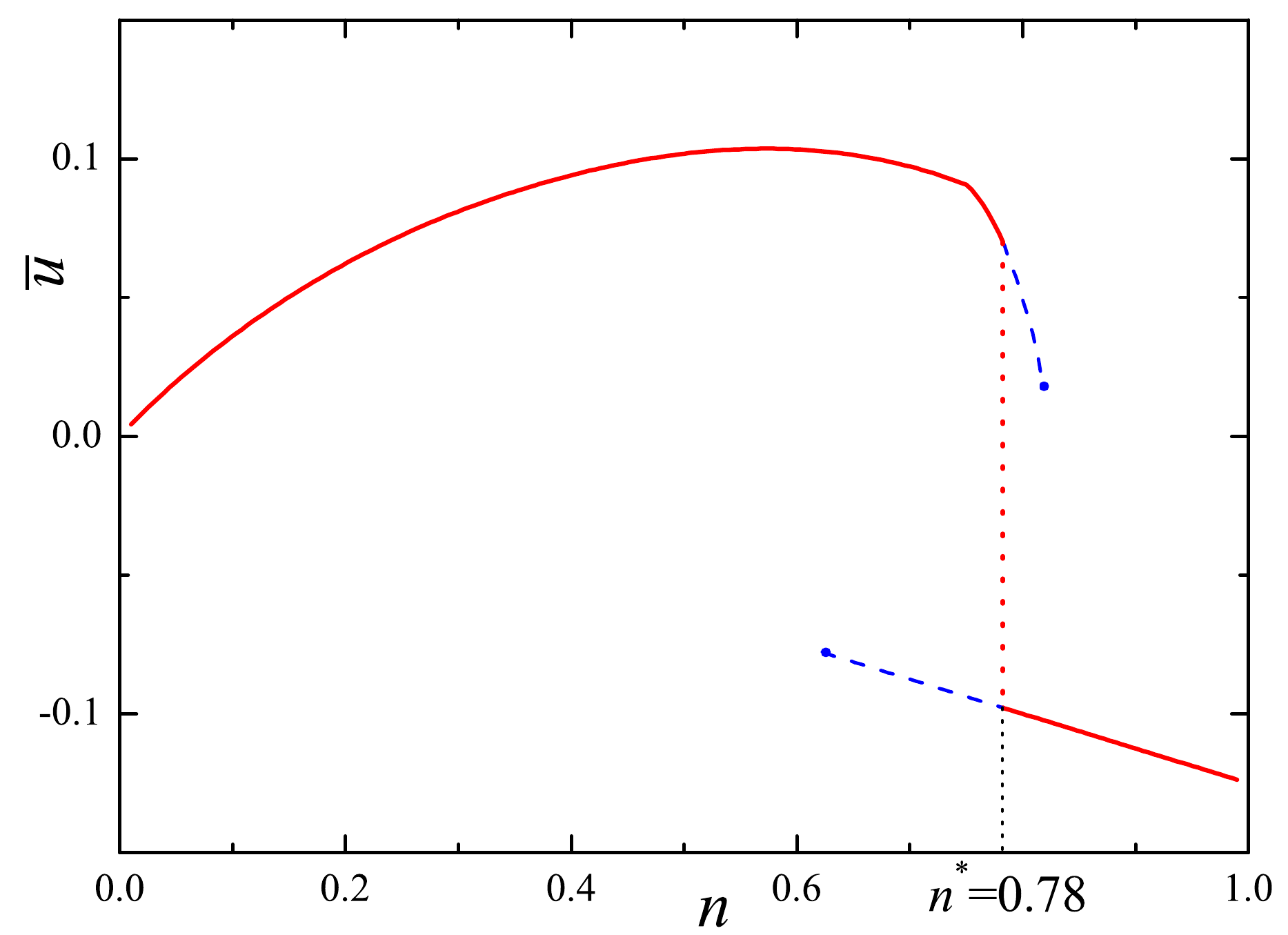} \end{center}
\caption{\label{FigJumpU} (Color online)
Lattice strain as a function of doping. The parameters
are $\lambda/w_a=-1.2$, $\lambda_b/w_a=2$, $K=16w_a$
and $w_b/w_a=0.25$. Jump-like transition between two
states with different values of lattice distortions occurs
at $n=n^*$. Solid (red) lines correspond to the states
with minimum energy, whereas dashed (blue) lines correspond
to metastable states.~\cite{sboychakov2010effect}}
\end{figure}

\begin{figure}[tbh] \begin{center}
\includegraphics[width=0.6\columnwidth]{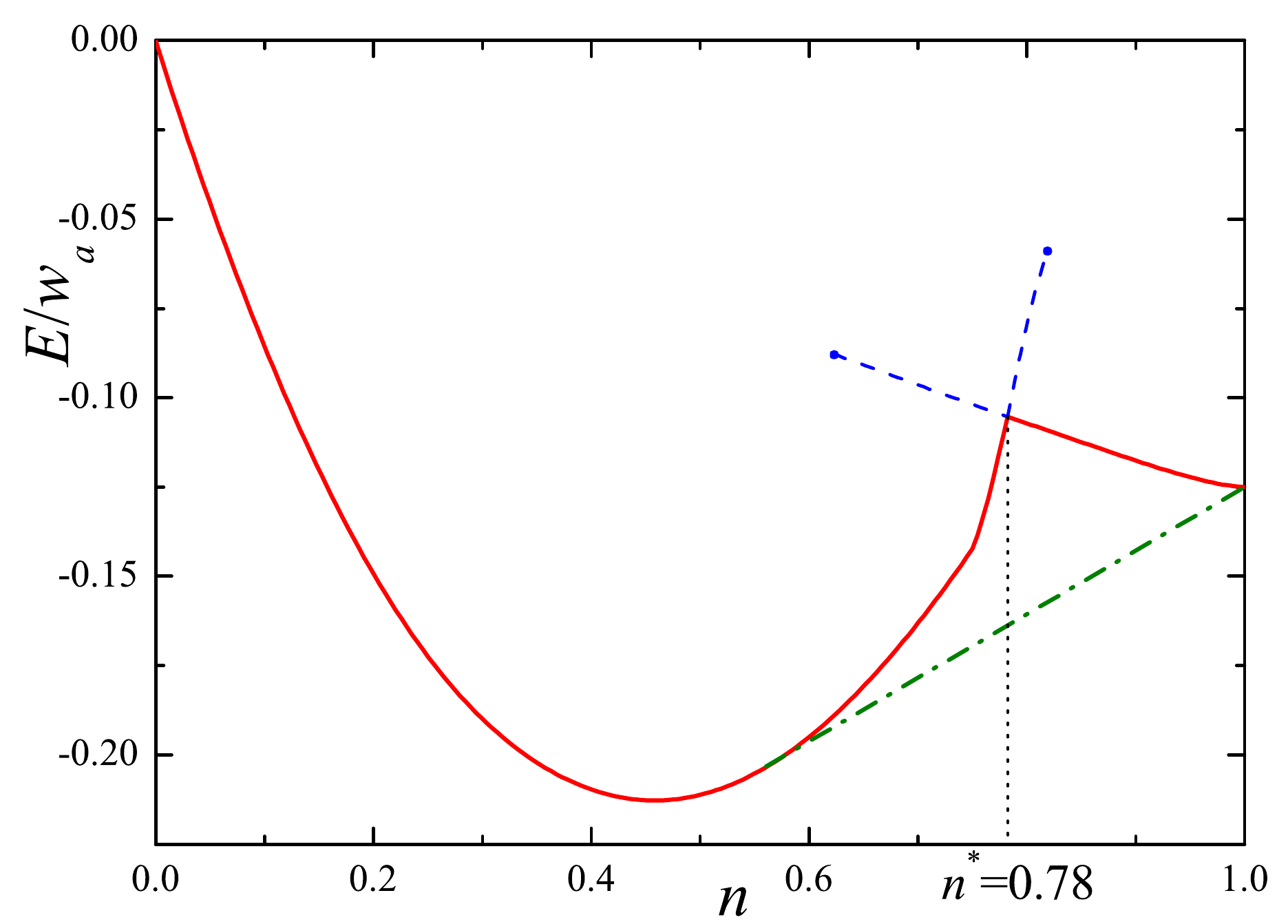} \end{center}
\caption{\label{FigJumpE} (Color online) Energy of the system versus doping level $n$. Solid (red) curve corresponds to the homogeneous state, whereas the dot-dash (green) curve is the energy of the phase-separated state.
The dashed (blue) curves correspond to the energies of metastable states. The parameters are the same as in Fig.~\ref{FigJumpU}. The kink in the curve $E_{\rm{hom}}(n)$ corresponds to the jump-like change in the lattice distortion.~\cite{sboychakov2010effect}}
\end{figure}

\subsection{Characteristic size and geometry of inhomogeneities}\label{CoulombSurface}

The phase separation leads to the violation of charge neutrality ($n_1\neq n_2$). Therefore, we should take into account the electrostatic contribution, $E_C$, to the total energy of the phase-separated state. The value $E_C$ evidently depends on the size and shape of inhomogeneities. The second contribution $E_S$ to the total energy depending on the size of inhomogeneities is related to the interface between two phases. Therefore, the interplay between $E_C$ and $E_S$ determines the equilibrium sizes and geometry of the inhomogeneities. The simplest is a droplet-like geometry of the inhomogeneities. However, the long-range Coulomb interaction can give rise to a more sophisticated geometry of phase separation (stripes, layers, rods, etc.)~\cite{KugelSuST2008,low1994study,LorenzanaPRB_I_2001,
LorenzanaEPL2002,OrtixLorDiCastroPRL2008}.

In Refs.~\cite{KugelSuST2008,LorenzanaPRB_I_2001,LorenzanaEPL2002}, the Coulomb contribution to the total energy was calculated for different shapes of the inhomogeneities: spherical ($D = 3$), cylindrical ($D = 2$), and layered ($D = 1$). The calculations were performed using the Wigner--Seitz approximation, that is, the system is divided into cells, where an internal region (core) containing phase $a$ is surrounded by a shell of phase $b$, if the concentration $p$ of phase $a$ is lower than 0.5. When $p > 0.5$, the core contains the phase $b$ and the shell contains the phase $a$. The relative volumes of the core and shell are determined by the charge conservation condition Eq.~(\ref{chacons}). It was assumed that the characteristic sizes of inhomogeneities are larger than the lattice constant $d$ and the macroscopic approach can be applied to calculate the electrostatic energy. The interaction between neutral cells is neglected. In so doing, the the electrostatic energy, $E_C$, for the inhomogeneities with dimensionality $D$ can be expressed as~\cite{KugelSuST2008,LorenzanaPRB_I_2001,LorenzanaEPL2002}
\begin{equation}\label{CoulEn}
E_C=V(n_1-n_2)^2\left(\frac{R_S}{d}\right)^{\!2}u_D(p),
\end{equation}
where $V=e^2/d\varepsilon$ is the characteristic Coulomb interaction, $\varepsilon$ is the average permittivity, $R_s$ is the size of the core (which is the radius of the internal sphere or cylinder in $3D$ or $2D$ cases, respectively, and the half-width of the internal layer in the $1D$ case), and $u_D(p)$ is a dimensionless function. When $p < 0.5$, we have
\begin{eqnarray}\label{uD}
\nonumber
u_3 &=&\frac{4\pi p}{5}\left(2-3p^{1/3}+p\right),\quad D=3\,\, \textrm{(droplets)}, \\
u_2 &=&  \pi p\left(-\ln{p}+p-1\right),\,\,\,\quad D=2\,\, \textrm{(rods)},\\
u_1 &=& \frac{4\pi}{3}\left(1-p\right)^2,\qquad\qquad\quad D=1\,\, \textrm{(planes or layers)}.
\end{eqnarray}
In the case $p > 0.5$, we should replace $n_1\Leftrightarrow n_2$ and $p\Leftrightarrow 1-p$ in all equations.

The second term in the total energy depending on the size of the inhomogeneities, $E_S$, is related to the boundary between two phases. It comes from the size quantization and depends on the electron densities in both phases. In Ref.~\cite{sboychakov2007phase}, the energy $E_S$ was calculated using the perturbative approach proposed by R. Balian and C. Bloch~\cite{BalianBlochAnnPh1970}. If $R_s > d$, the surface tension $\sigma(n_1,n_2)$ does not depend on the shape of inhomogeneity.  Then, we can write the surface energy per unit cell as $E_S=\sigma(n_1,n_2)S_Dd^3/V_D$, where $S_D$ is the surface area of the core and $V_D$ is the total volume of the Wigner--Seitz cell. It is convenient to write
\begin{equation}\label{surfEn}
E_S=\sigma(n_1,n_2)pDd^2\frac{d}{R_s}.
\end{equation}
Minimization of the sum $E_{CS}=E_S+E_C$ with respect to $R_s$ gives
\begin{equation}\label{ECSD}
E_{CS}=3\left[\frac{d^2\sigma(n_1,n_2)|n_1-n_2|}{2}pD\sqrt{Vu_D}\right]^{2/3}.
\end{equation}
The equations used for the numerical calculation of $\sigma(n_1,n_2)$ can be found in the Appendix of Ref.~\cite{sboychakov2007phase}. We do not rewrite them here. Note only, that the function $\sigma(n_1,n_2)$ varies rather slowly with the change of its arguments.

\begin{figure}[tbh] \centering
\includegraphics[width=0.6\columnwidth]{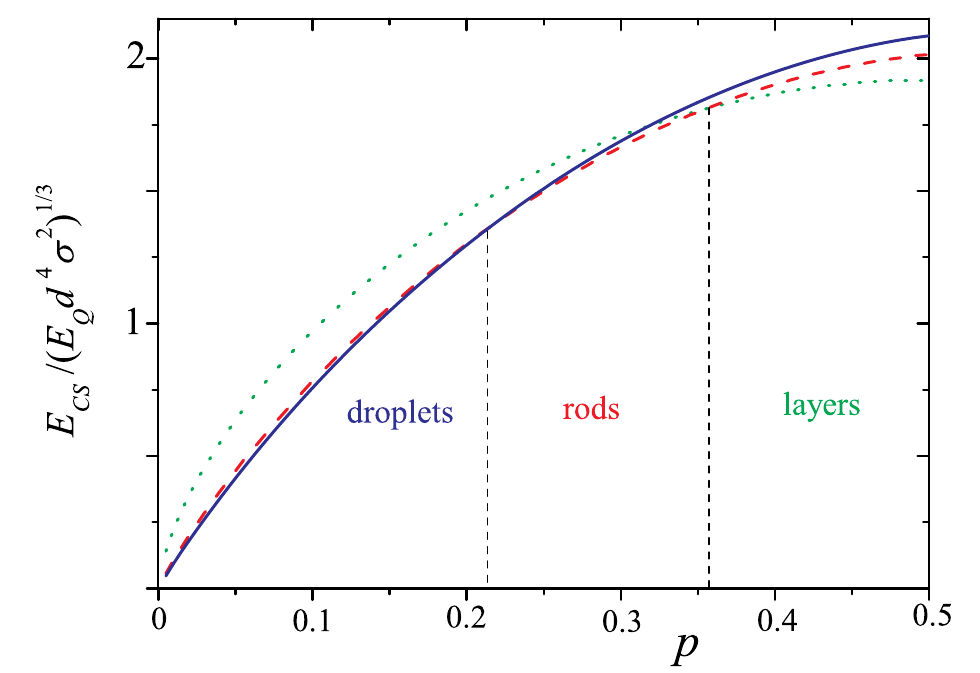}\\
\caption{ (Color online) Normalized energy of an inhomogeneity as a function of the phase content $p$ at $p < 0.5$. The picture for $p > 0.5$ is a mirror reflection of this figure with respect to the vertical line $p = 0.5$, $E_Q=V(n_1-n_2)^2$. Solid (blue), dashed (red), and dotted (green) lines correspond to droplets, rods, and planes or layers, respectively. ~\cite{KugelSuST2008}
}\label{figGeom}
\end{figure}

The behavior of energy $E_{CS}$ as a function of $p$ is illustrated in Fig.~\ref{figGeom}. We see that the droplet geometry is the most favorable at small $p$ ($p < p_1 = 0.215$). At intermediate values of $p$ ($p_1 < p < p_2 = 0.355$), we have rods, which are transformed to layers at $p > p_2$. However, the different geometries are close in energy. Thus, we can use $E_{CS}$ calculated, e.g., for droplets in different estimates as it was done in Fig.~\ref{fig3FinWidth} for calculation of the energy of the phase-separated phase. Note also, that the presented derivation of $E_{CS}$ does not depend on the particular electron Hamiltonian and can be applied in different cases.

The electroneutrality condition (\ref{chacons}) determines the relationship between the phase content and the number of electrons per site $p = (n-n_2)/(n_1-n_2)$. In the main approximation, the values $n_{1,2}$ are independent of $n$. So, the phase content is almost a linear function of $n$ or doping $\delta=1-n$. Then, the inhomogeneities evolve with doping as follows. Phase separation arises if the doping exceeds some value $\delta=\delta^*(w_a/w_b,\epsilon)>1-n_2$. If $\delta^*<\delta_1=1-n_1p_1-n_2(1-p_1)$, the phase separation appears in the form of the nanoscale droplets of the phase $b$ in $a$ host. With the increase of doping, the geometry changes to the rod-like [$\delta_1<\delta<\delta_2=1-n_1p_2-n_2(1-p_2)$], and, finally, to the stripe-like layered configuration when $\delta>\delta_2$. If the value of $\delta^*$ is larger than $\delta_1$ and/or $\delta_2$, droplet-like or even rod-like structure can be missed.

\subsection{Electronic phase separation in high-$T_c$ superconducting cuprates and pnictides}

The solution of the problem of high-$T_c$ superconductivity in cuprates and pnictides requires a correct description of the normal state where spin, charge, orbital, and lattice degrees of freedoms compete, with the formation of nanoscale puddles of spin density wave stripes, puddles of charge density wave stripes, and/or puddles of ordered mobile oxygen interstitials. In this way, it was  suggested that the minimal model to capture the essential physics of high-temperature superconductors needs to take into account both the presence of two electronic components with different orbital symmetry~\cite{BednMulRMP1988,BianconiSSC1987,BianconiPRB1988} and a nanoscale phase separation~\cite{muller2005superconductivity,
KresinPhRep2006,BianconiNatPh2013,BianconiScSciTech2015,YukalovPRB2004}. There is now an experimental evidence that the high temperature superconductivity in cuprates and pnictides emerges in the proximity of a topological Lifshitz transition~\cite{LiuPRB2011importance,LeboeufPRB2011lifshitz,
LaliberteNatCom2011fermi}. In Ref.~\cite{innocenti2010resonant}, a theory for the high-temperature superconductivity was suggested based on the shape resonance near the Lifshitz transition between a BCS-like superconducting gap and a second gap in the Bose--Einstein condensate.

In Refs.~\cite{KugelSuST2008,BianconiScSciTech2015,
sboychakov2008mechanism,kugel2008model}, a model with two-band Hubbard Hamiltonian (\ref{HubbardTwoB}) was proposed as a version of minimal model for the description of the normal state in high-$T_c$ cuprate superconductors. This approach naturally provides a reasonable theoretical background for the phase diagram region where the nanoscale phase separation emerges near the topological Lifshitz transition.  The detailed analysis of experimental evidence on the electronic phase separation in cuprate superconductors is given in review article \cite{LiarokapisCondMat2019}.

The phase separation in pnictides will be discussed in subsection~\ref{imperfect_pnictides}. In pnictides, the electron--electron interaction is not so strong as in cuprates and an important feature of these systems is an imperfect nesting of the Fermi surface.

\subsection{Phase separation in two-dimensional electron gas} \label{2DEG}

The phase separation is a common feature in two-dimensional electron gas (2DEG) that forms at the interface of two insulators or semiconductors~\cite{SpivakPRB2003,ScopignoPRL2016} and in highly crystalline ultrathin films, like ZrNCl~\cite{DeziPRB2018}. In general, the inhomogeneous states in the 2DEG arises due to electrostatic interaction between electrons in the gas with substrate. However, the details of this interaction are different for different heterostructures with 2DEG. We briefly consider here two typical cases. First, the 2DEG that forms at the interface of two insulating oxides, like LaAlO$_3$/SrTiO$_3$ (or LaTiO$_3$/SrTiO$_3$) hereafter referred to as LXO/STO (and, possibly, in the ultrathin films of ZrNCl, TiSe$_2$, MoS$_2$, etc). Second, the 2DEG at the interface of two semiconductors GaAs/AlGaAs. Both systems exhibit a number of nontrivial phenomena, e.g., a gate-tunable metal--to--superconductor transition~\cite{ReyrenScience2007,CamjayiNatPhys2008}, inhomogeneous magnetic response~\cite{AriandoNatComm2011,TenehPRL2012,PudalovJETPL2020}, and very intriguing inhomogeneous superconductivity~\cite{DeziPRB2018,SminkPRB2018,SinghNatMat2019}.

The physics of 2DEG in LXO/STO structures has recently attracted considerable interest (see, e.g., a review~\cite{PaiRPP2018}), and the electronic phase separation is a key phenomena in that system~\cite{ScopignoPRL2016}. This interest is highly driven by experimental discovery of an unusual inhomogeneous superconducting states in these systems~\cite{SminkPRB2018,SinghNatMat2019}. Now, two mechanisms of the phase separation are debated for LXO/STO, both associated with a high dielectric constant of SrTiO$_3$~\cite{BovenziJPCS2019}.

According to a detailed investigation of the stability of 2DEG in LXO/STO~\cite{ScopignoPRL2016}, the fluctuation of the electron density in 2DEG gives rise to a high polarization of the SrTiO$_3$ substrate. As a result, electrons are captured in the self-consistent potential well, which depth in LXO/STO is sufficient to form quantum bound states. Naturally, the compressibility of such a system is negative and the phase separation occurs in 2DEG. The second mechanism is more tricky but also rather natural. The inhomogeneous distribution of electron density is accompanied by an inhomogeneous distribution of the electric field confining the electrons at the interface. In turn, this inhomogeneous transverse electric field (with large gradients) induces an inhomogeneous Rashba spin--orbit coupling giving rise to capturing of the electrons in 2DEG by a self-consistent potential and, again, to the arising quantum bound states, negative compressibility, and the phase separation~\cite{BovenziJPCS2019}.

The existence of the inhomogeneity and/or multi-component phases in LXO/STO systems is reliably confirmed by a number of experiments. In particular, the magnetotransport measurements reveal the presence of high- and low-mobility charge carriers in LTO/STO, and the superconductivity develops, when high-mobility charge carriers arise in 2DEG~\cite{BiscarasPRL2012,BellPRL2009}. Inhomogeneities are observed in magnetic experiments~\cite{AriandoNatComm2011,LiNatPhys2011,BertNatPhys2011} and in the tunneling spectra~\cite{RisticEPL2011}. Moreover, a percolative insulator-to-metal transition with the increase of the gate-tuned electron density, which is commonly observed in 2DEG, has a natural explanation in the framework of the picture with the electronic phase separation.

In the 2DEG in GaAs/AlGaAs heterostructures, the electronic phase separation can be also attributed to an electrostatic electron--substrate interaction but the nature of this interaction is different as compared to LXO/STO. It is commonly accepted that the inhomogeneous state in GaAs/AlGaAs arises due to disordered electrostatic potential produced by donor (or acceptor) atoms in the substrate~\cite{ShiPRL2002,FoglerPRB2004,TripathiPRB2006,TripathiPRB2011,TripathiPRB2012}. The gate voltage rips out the electrons (or holes) from the substrate and tune the electron (hole) density in the 2DEG. This electron redistribution produces a disorder of the charge density in the substrate. Interplay of the kinetic energy of the 2D Fermi gas, disordered potential, and Coulomb repulsion between charge carriers in the 2DEG gives rise to formation of the self-consistent electron (hole) droplets. This mechanism is studied in detail in the cited papers. However, in the case of GaAs/AlGaAs, we deal with an initially inhomogeneous system, which is out of the scope of the present review.

\subsection{Electronic phase separation in magnetic oxides with Jahn--Teller ions}

As we have already mentioned above, the phase separation phenomena play a fundamental role in the physics of magnetic oxides, especially manganites and related compounds. The most frequently discussed type of the phase separation in them is the formation of nanoscale inhomogeneities such as ferromagnetic (FM) metallic droplets in an insulating antiferromagnetic (AF) matrix arising due to self-trapping of charge carriers~\cite{DagottoBook2003,KaganUFN2001}. The analysis of the experimental data, especially on the magnetotransport~\cite{kugel2004characteristics} and voltage noise~\cite{RakhmanovPRB2001resistivity}, shows that such type of the phase separation is indeed a common feature of these materials.

The model for the phase separation in  manganites and related compounds was suggested in Refs.~\cite{KugelPRL2005,SboychakovPRB2006jahn,KugeJMMMl2007electronic}. First, we should take into account the Jahn--Teller (JT) nature of the magnetic ions, which could give rise to the localization of charge carriers at the lattice distortions. This situation is typical of manganites and could be described in terms of the Kondo-lattice model in the double exchange limit with account taken for the JT distortions and the superexchange interaction between the localized electrons. The intraatomic exchange coupling is assumed to be large enough to align the spins of $e_g$ electrons in a magnetic ion parallel to spin $\mathbf{S}$ of core ($t_{2g}$) electrons. That is, we assume the Hund's rule coupling constant to be large compared to hopping integrals. The effective electron hopping integral for neighboring sites depends on an angle $\nu$ between the directions of spins $\mathbf{S}$ at these sites: $t_{\textrm{eff}} = t \cos{\!\left(\nu/2\right)}$~\cite{deGennesPR1960}. The JT effect leads to the splitting of the double-degenerate $e_g$ level. Then, following Ref.~\cite{ramakrishnan2004theory}, we can divide $e_g$ electrons into two groups: ``localized'' ($l$, with hoping integral $t^l\ll t$) producing the maximum splitting of the $e_g$ level and ``itinerant'' (b) electrons with non-zero hopping integrals $t$, leading to smaller distortions of MnO$_6$ octahedra. Thus, we come to the two-band picture, which reduces to the Falicov--Kimbal model in the limit $t^l\rightarrow 0$. Really, this is only some ``minimal model'' describing the magnetic phases and the effect of the phase separation in the manganites and related systems.

The corresponding effective Hamiltonian can be written as~\cite{KugelPRL2005,SboychakovPRB2006jahn,KugeJMMMl2007electronic}
\begin{equation}\label{HubbardTwo_Mang}
\hat{H} =-t\sum_{\langle \mathbf{ij}\rangle} b^\dag_{\mathbf{i}}b_{\mathbf{j}}\cos{\left(\frac{\nu_{\mathbf{ij}}}{2}\right)}-
\epsilon_{\textrm{JT}}\sum_{\mathbf{i}}n_{\mathbf{i}l}
-\mu\sum_{\mathbf{i}}\left(n_{\mathbf{i}b}+n_{\mathbf{i}l}\right)
  +U\sum_{\mathbf{i}}n_{\mathbf{i}b}n_{\mathbf{i}l}
  +J\sum_{\langle \mathbf{ij}\rangle}\cos{\left(\nu_{\mathbf{ij}}\right)}.
\end{equation}
Here $n_{\mathbf{i}b}=b^\dag_\mathbf{i}b_\mathbf{i}$, $n_{\mathbf{i}l}=l^\dag_\mathbf{i}l_\mathbf{i}$, are the number operators of $b$ and $l$ electrons, $b^\dag_\mathbf{i}$, $b_\mathbf{i}$, $l^\dag_\mathbf{i}$, and $l_\mathbf{i}$ are creation and annihilation operators of the localized and itinerant electrons at site $\mathbf{i}$, respectively, $\nu_{\mathbf{ij}}$ is the angle between the localized spins at sites $\mathbf{i}$ and $\mathbf{j}$, $\epsilon_{\textrm{JT}}$ is the energy gain due to electron localization at the  Jahn--Teller distortions. The last term in the Hamiltonian (\ref{HubbardTwo_Mang}) is the AFM ($J>0$) exchange interaction between local spins. The number of localized, $n_l$, and band, $n_b$, electrons per lattice site obeys an obvious relation $n_b+n_l =1-x$, where $x$ is the number of doped charge carriers per site. The half-filling, $n_b+n_l =1$, corresponds to the undoped system. In the considered approximation the charge carriers are ``spinless'', since their spins are aligned along the direction of the local spins $\mathbf{S}$.

Hamiltonian (\ref{HubbardTwo_Mang}) was studied in Refs.~\cite{KugelPRL2005,SboychakovPRB2006jahn} in the framework of the mean-field approximation within entire range of doping, $0\leq x\leq 1$. A similar model has been studied numerically in Refs.~\cite{ShenoyPRL2007,ShenoyPRB2009}. Depending on parameters of the model, doping and temperature, it can describe either homogeneous (FM, AFM, or canted) or inhomogeneous (phase-separated) states. The phase-separated state turns out to be favorable in a large part of the phase diagram. The relative number of different kinds of charge carriers depends on the doping level, temperature, and magnetic field. In Ref.~\cite{SboychakovPRB2006jahn} it was taken into account that the hopping integral $t$ in Eq.~(\ref{HubbardTwo_Mang}) is temperature dependent due to the polaron band narrowing~\cite{AlexandrovBook1995_polarons}: $t(T) =t(0)\exp{\left[-2\lambda^2/\left(e^{\hbar\Omega/T}-1\right)\right]}$, where $\lambda$ is the dimensionless electron--phonon coupling constant and $\Omega$ is the characteristic phonon frequency. The energy spectrum of band electrons was calculated based the Green's function technique using the Hubbard I decoupling similar to that described in subsection~\ref{Falicov-Kimball} of this review.

\begin{figure}[tbh] \centering
\includegraphics[width=0.6\columnwidth]{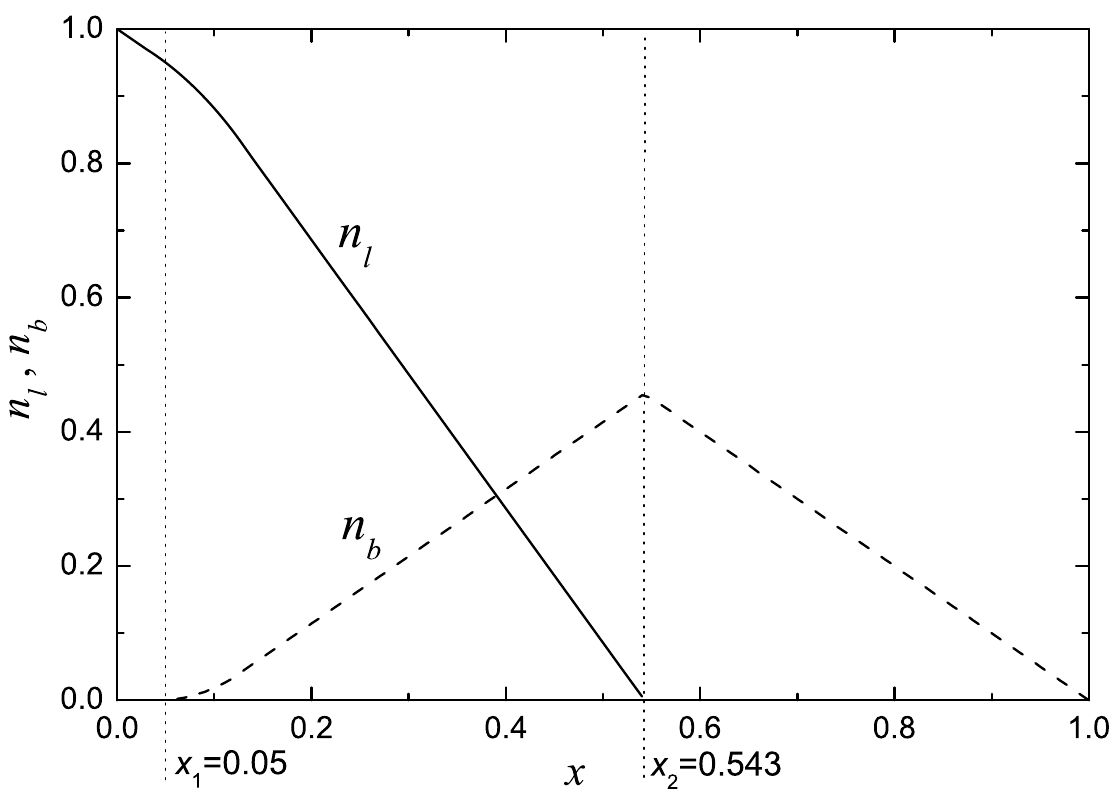}\\
\caption{Dependence of the number of localized $n_l$ (solid line) and band $n_b$ (dashed line) electrons on doping $x$ at  $\epsilon_{\textrm{JT}}/t(0)=1.8$.~\cite{SboychakovPRB2006jahn}
}\label{Fig_charge_carriers}
\end{figure}

The number of band and localized electrons depends on the relative positions of $\mu$ and $\varepsilon_{\textrm{JT}}$. If $\mu <\varepsilon_{\textrm{JT}}$, then $n_l =0$ and $n_b = 1 - x$. With the growth of $n_b$, the chemical potential attains the value of $-\varepsilon_{\textrm{JT}}$ and the further growth of the number of band electrons becomes unfavorable in energy. The chemical potential $\mu$ becomes pinned at the level $-\varepsilon_{\textrm{JT}}$ and the localized electrons arise in the system. As a result, there appear two critical values of $x$ (each depends on temperature), $x_1$ and $x_2$. When $x < x_1$, we have $n_b=0$, while $n_l=0$ if $x > x_2$. The behavior of $n_l$ and $n_b$ as functions of $x$ is illustrated in Fig.~\ref{Fig_charge_carriers}. Depending on the doping level and temperature, the system may be in the homogeneous ferromagnetic (FM), AFM ($n_b = 0$), and canted ($0 < \nu < \pi)$ states, as well as in two different kinds of paramagnetic (PM) states (PM insulating phase, $n_b = 0$, and PM metallic phase, $n_b \neq 0$). The AFM state is favorable at small $x$, whereas the canted state appears close to $x = 1$.

\begin{figure}[H] \centering
\includegraphics[width=0.6\columnwidth]{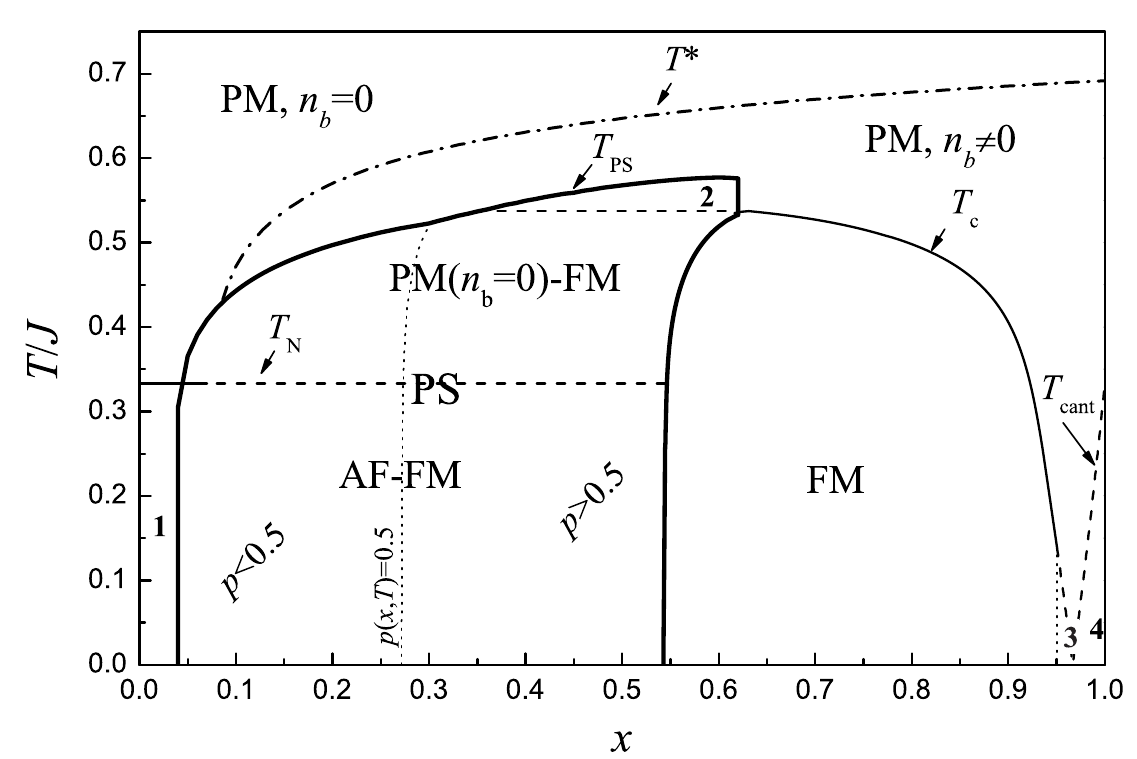}\\
\caption{Phase diagram of the model at $\epsilon_{\textrm{JT}}/t(0)=1.8$, $J/t(0) = 0.36$, $\lambda = 10$, $\Omega/t(0) = 0.18$, and $V/t(0) = 3$. AF means AFM state, FM -- ferromagnetic state, PM -- paramagnetic state, PS -- phase-separated state. The numbers denote: 1 -- homogeneous AFM phase; 2 -- the mixture of two PM states with $n_b \neq 0$ and $n_b = 0$; 3 and 4 -- homogeneous canted states with $\cos{\nu} > 0$ and $\cos{\nu} < 0$, respectively. In the picture $p$ is the volume content of the metallic phase, $T_\textrm{N}$ is the N\'eel temperature, $T_\textrm{C}$ is the Curie temperature, $T_{\textrm{PS}}$ is the transition temperature between the PS and homogeneous states, $T^*$ is the transition temperature between the PM metallic and insulating phases, and $T_{\textrm{cant}}$ is the transition temperature between the canted and PM phases.~\cite{SboychakovPRB2006jahn}
}\label{Fig_Phase_Diagram}
\end{figure}

The phase-separated states now can be considered based on the above analysis of the homogeneous phases of the model. According to Refs.~\cite{KugelPRL2005,SboychakovPRB2006jahn}, in the phase-separated state, the system consists of the metallic ($n_b\neq 0$) and insulating ($n_b=0$) phases. The metallic phase can be either FM or PM while insulating phase is AFM or PM. As it was done in the previous analysis, when calculating the free energy of the phase-separated state, we should take into account the Coulomb contribution due to the redistribution of the charge density, Eq.~(\ref{CoulEn}), and the correction to the electron density of states related to the finite size of inhomogeneities, Eq.~(\ref{surfEn}), and find the minimum of their sum, Eq.~(\ref{ECSD}). As it was discussed above, the geometry of the inhomogeneous regions is different, however, the energies of these configurations is close to each other, Fig.~\ref{figGeom}. Thus, to derive the phase diagram of the model, we can consider only the inhomogeneities of one type, e.g., in the form of spherical droplets.

In Fig.~\ref{Fig_Phase_Diagram}, we present the phase diagram of the model in the ($x,T$) plane from Ref.~\cite{SboychakovPRB2006jahn}. The phase diagram is very rich. The details of this diagram are described in the figure caption. Some phases in the diagram can disappear if we choose different model parameters. The obtained results demonstrate that in the considered model the phase separation exists in a wide range of intermediate doping values and disappears at low and high doping levels. These predictions are in agreement with the general features of the experimentally found phase diagrams of manganites~\cite{DagottoBook2003,DagottoPhysRep2001,nagaev2001colossal}. For the characteristic values of parameters for manganites, the FM droplet includes 10--30 unit cells.

The applied magnetic field $\mathbf{H}$ can affect significantly the phase diagram of the systems under study. This effect was considered in Ref.~\cite{SboychakovPRB2006jahn} with taking into account the interaction of the field $\mathbf{H}$ with the local spins $\mathbf{S}$ in the classical limit. Thus, in the presence of the external magnetic field, we should add the term $\mu_\textrm{B}g\sum_\mathbf{i}{\mathbf{S}_\mathbf{i}\mathbf{H}}$ in the Hamiltonian, where $\mu_\textrm{B}$ is the Bohr magneton and $g$ is the Land\'{e} factor. The applied magnetic field evidently favors the FM ordering and, consequently, the number of itinerate electrons increases.

As it is seen from the phase diagram in Fig.~\ref{Fig_Phase_Diagram}, the metal--insulator transition can take place at some characteristic values of the doping $x$ corresponding to the crossover between different types of the phase separation. Such a transition can be induced by changing temperature or magnetic field and is of a percolation type. This transition can be related to the colossal magnetoresistance effect~\cite{KugelPRL2005,SboychakovPRB2006jahn}.

\subsection[Electron polaron effect and anomalous resistivity in the two-band Hubbard model]{Electron polaron effect and anomalous resistivity in the two-band Hubbard model with one narrow band}
  \label{ElectPolEff}
\subsubsection{Introductory remarks}
 \label{IntroElPolEff}

Here, we present the results on the origin of a heavy mass, marginal Fermi-liquid behavior~\cite{VarmaPRL1989} and the anomalous resistivity characteristics in the 3D and 2D (or layered) strongly correlated electron systems generally described by the two-band Hubbard model with one narrow band (the details see in Refs.~\cite{KaganCzJPh1996,KaganJETP2011,KaganValFNT2011,KaganSchoenBook2012,
KaganValJScNM2012}). This model contains  very rich and elegant physics closely related to the problems discussed above in this section. It describes adequately different systems such as uranium-based and Yb-based heavy-fermion (HF) compounds and possibly also some other novel superconductors and transition-metal systems with orbital degeneracy such as complex magnetic oxides (possibly including layered manganites) in an optimally doped case. Moreover, the model contains such highly nontrivial effects as electron polaron effect (EPE) already in the homogeneous case, as well as the tendency toward the phase separation in the case of a strong mismatch between the densities of heavy and light components~\cite{KaganJETP2011,sboychakov2007phase,RozhkovPhRep2016}.

 In uranium-based HF compounds, we are usually dealing with the mixed valence limit~\cite{NewnsAdvPh1987,ColemanPRB1987,TsvelikBook2007,
 VarmaPhB2006,FuldeBook2002,KeiterGrewe1981} with a strong hybridization between heavy ($f$ electrons or  $f$-$d$ electrons) and light ($s$-$p$ electrons) components. The origin of a heavy mass, $m^*_h \gtrsim 200m_e$, for $f$ electrons in uranium-based HF is possibly very different from the physics of the cerium-based HF, where the Kondo effect %(or more generally the physics of the Kondo-lattice model)
 is dominant~\cite{VarmaPRB1976,HewsonBook1993,WilsonRMP1975,
 AndersonPRB1970,NozieresJLTP1974}. According to the ideas of  Refs.~\cite{KaganProkJETP1986,KaganProkJETP1987} on the level of the two-particle hybridization, it is just sufficiently strong interband Hubbard interaction, which leads to an additional enhancement of the heavy-electron mass due to the EPE. Physically, the EPE is related to a nonadiabatic part of the many-body wave function describing a heavy electron and a cloud of virtual electron--hole pairs of light charge carriers. These pairs are mixed with the wave function of heavy electrons but do not follow it, when a heavy electron tunnels from one unit cell to a neighboring one. It is shown in  Refs.~\cite{KaganProkJETP1986,KaganProkJETP1987} that in the unitary limit of the strong Hubbard interaction between heavy and light electrons, the effective heavy mass can reach the values $m^*_h/m_L \sim (m_h/m_L)^2$, and if we start from the ratio $m_h/m_L \sim 10$ of the bare masses of heavy and light electrons, for example, at the level of LDA approximation, we could eventually achieve the effective mass as high as $m^*_h \sim 100m_L$, which is typical of uranium-based HF compounds. A similar effect can also be described using the strong one-particle hybridization between heavy and light bands~\cite{KaganProkJETP1986,KaganProkJETP1987}.

A natural question arises, whether the two-band Hubbard model with one narrow band is a simple toy model to observe a non-Fermi liquid behavior and  well-known marginal Fermi liquid behavior, in particular~\cite{VarmaPRL1989}. We recall that in the marginal Fermi liquid (MFL) theory, the quasiparticles are strongly damped (Im$\varepsilon \sim$ Re$\varepsilon \sim T$). According to  \cite{VarmaPRL1989}, a strong damping $\gamma \sim T$ of quasiparticles (instead of the standard damping $\gamma \sim T^2/\varepsilon_F$  for a Landau-type Fermi liquid) can explain numerous experiments in high-$T_c$ cuprates including the linear resistivity $R(T) \sim T$  for $T > T_c$  at optimum doping levels. The MFL picture was also proposed to describe the properties of UPt$_3$ doped by Pd including specific heat measurements~\cite{KimPRB1992}. The two-band Hubbard model with one narrow band is a natural generalization of the well-known Falicov--Kimball model~\cite{FalKimbPRL1969} discussed above, but contains much richer physics. %due to a finite width of the heavy-electron band (instead of the localized level), which allows for an interesting dynamics of the heavy component.

\subsubsection{Electron polaron effect and the tendency toward phase separation}
 \label{TwoBandElPolEff}

The Hamiltonian of two-band Hubbard model already discussed in in previous subsections can be written as
\begin{eqnarray}\label{Hamil_Hub_twoband}
\hat{H}' &=& -t_h\sum_{\langle i,j\rangle \sigma}{a_{i  \sigma}^{\dag}a_{j  \sigma}} - t_L\sum_{\langle i,j\rangle \sigma}{b_{i  \sigma}^{\dag}
ba_{j  \sigma}}-\varepsilon_0\sum_{i \sigma}{n_{i  \sigma}^h} -\mu\sum_{i \sigma}{(n_{i  \sigma}^L + n_{i  \sigma}^h)}
\nonumber \\
  &+&U_{hh}\sum_i{n_{ih}^{\uparrow}n_{ih}^{\downarrow}} +U_{LL}\sum_i{n_{iL}^{\uparrow}n_{iL}^{\downarrow}}
+\frac{U_{hL}}{2}\sum_i{n_{ih}n_{iL}}  \, ,
\end{eqnarray}
where $U_{LL}$ and $U_{hh}$ describe the intraband onsite Coulomb repulsion of heavy and light electrons, respectively, $U_{hL}$ is the corresponding interband onsite repulsion of light and heavy electrons, $t_h$ and $t_L$ are the hopping integrals for heavy and light electrons, $n_{ih}^{\sigma} = a_{i  \sigma}^{\dag}a_{i \sigma}$ and $n_{iL}^{\sigma} = b_{i  \sigma}^{\dag}b_{i \sigma}$ are the densities of heavy and light electrons at site $i$ with spin projection $\sigma$, and $\mu$ is the chemical potential. The energy $-\varepsilon_0$ corresponds to the center of heavy band, and the difference $\Delta$ between the band bottoms is
\begin{equation}\label{Delta}
\Delta = -\varepsilon_0 + \frac {W_L -W_h}{2} = E_{min}^h - E_{min}^L \, .
\end{equation}

It is the most general form of the two-band model for the mixed valence situation. In manganites, we deal with the two-band degenerate Hubbard model, which can be reduced within some range of parameters region to the orbital  $t$--$J$ model and describes the physics of orbital ordering  together with the beautiful physics of orbital ferrons in the nanoscale phase-separated state (see Section~\ref{Orbitals} and Ref.~\cite{KugSboKhomPRB2008orbitals}).

Then, it is convenient to pass to the momentum representation of Hamiltonian~\eqref{Hamil_Hub_twoband}. As a result, we find~\cite{KaganBookSpringer,KaganJETP2011}
\begin{eqnarray}\label{Hamil_Hub_twoband-momspace}
\hat{H}' &=& \sum_{\mathbf{p}\sigma}{\varepsilon_h (\mathbf{p})a_{\mathbf{p}  \sigma}^{\dag}a_{\mathbf{p}  \sigma}} + \sum_{\mathbf{p}\sigma}{\varepsilon_L (\mathbf{p})b_{\mathbf{p} \sigma}^{\dag}b_{\mathbf{p}  \sigma}} + U_{hh}\sum_{\mathbf{pp'q}}{a_{\mathbf{p}\uparrow}^{\dag}a_{\mathbf{p'}  \downarrow}^{\dag}a_{\mathbf{p}-\mathbf{q}\downarrow}a_{\mathbf{p'}+\mathbf{q}\uparrow}}
\nonumber \\
&+&U_{LL}\sum_{\mathbf{pp'q}}{b_{\mathbf{p}\uparrow}^{\dag}b_{\mathbf{p'}  \downarrow}^{\dag}b_{\mathbf{p}-\mathbf{q}\downarrow}b_{\mathbf{p'}+\mathbf{q}\uparrow}}
+\frac{U_{hL}}{2}\sum_{\mathbf{pp'q}\sigma \sigma'}{a_{\mathbf{p} \sigma}^{\dag}b_{\mathbf{p'}\sigma'}^{\dag} b_{\mathbf{p}-\mathbf{q}\sigma'}a_{\mathbf{p'}+\mathbf{q} \sigma}}
  \, ,
\end{eqnarray}
where in D dimensions for the hypercubic lattice, the quasiparticle energies for heavy and light bands are
\begin{eqnarray*} \label{epsilon-h}
\varepsilon_h(\mathbf{p}) = -2t_h\sum_{\alpha=1}^{D}{\cos{(p_{\alpha}a)}} -\varepsilon_0 - \mu
\nonumber \\
%\begin{equation*} \label{epsilon-h1}
\varepsilon_L (\mathbf{p}) = -2t_L\sum_{\alpha=1}^D{\cos{(p_{\alpha}a)}} - \mu \, ,
\end{eqnarray*}
and $p_{\alpha} = \{p_x, p_y, \ldots\}$ are the Cartesian projections of  momentum. For low densities of heavy and light components $n_{tot}a^D = (n_h + n_L)a^D \ll 1$, the quasiparticle spectra are
\begin{eqnarray} \label{epsilon-h-L}
\varepsilon_h(\mathbf{p})
\approx  -\frac{W_h}{2} + 2t_h (pa)^2  -\varepsilon_0 - \mu \, ,
\nonumber \\
\varepsilon_L (\mathbf{p}) \approx -\frac{W_L                             }{2} + 2t_L (pa)^2 - \mu \,.
\end{eqnarray}
%In this section, we will work not with the dimensional charge carrier density $n$, so that the product $na^D$ is a dimensionless parameter in the D-dimensional case.
We include the chemical potential $\mu$ in the definition of $\varepsilon_h (\mathbf{p})$ and $\varepsilon_L (\mathbf{p})$ in Eq.~\eqref{epsilon-h-L}. In  Eq.~\eqref{epsilon-h-L}, $W_h = 4Dt_h$ and $W_L = 4Dt_L$ are the bandwidths of heavy and light electrons for the D-dimensional hypercubic lattice and $a$ is an intersite distance (see Fig.~\ref{two-band-str}).

\begin{figure} [H]
\begin{center}
\includegraphics*[width=0.5\columnwidth]{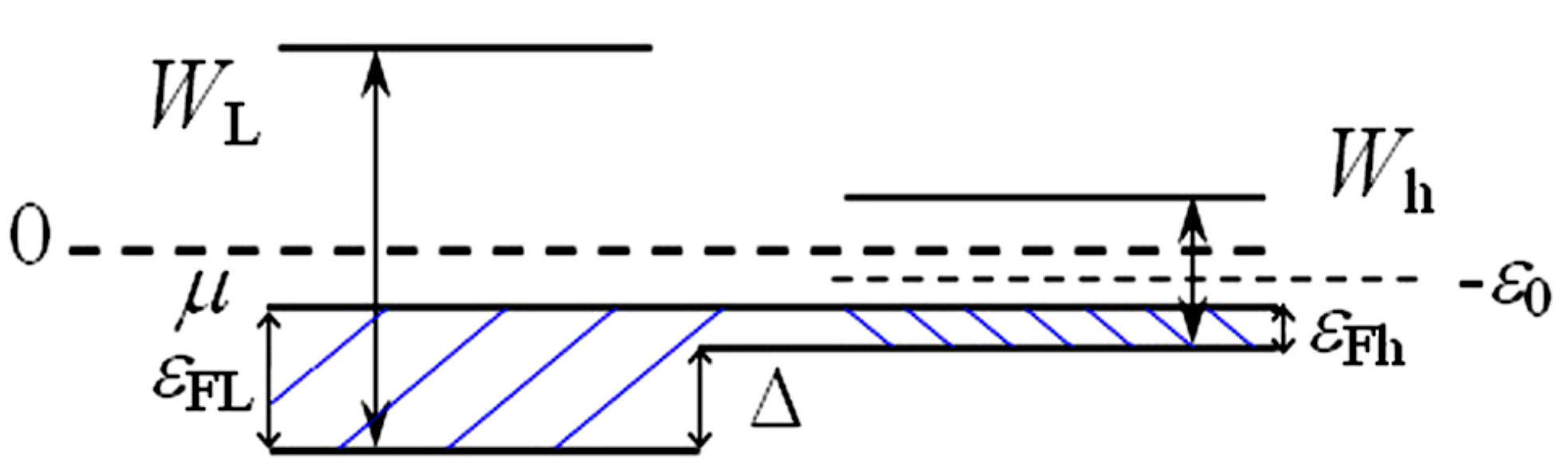}
\end{center}
\caption{\label{two-band-str} Band structure in the two-band model with one narrow band. $W_h$ and $W_L$ are the bandwidths of heavy and light electrons, $\varepsilon_{Fh}$ and $\varepsilon_{FL}$ are the Fermi energies, and $\Delta = -\varepsilon_0 + (W_L - W_h)/2 = E_{min}^h - E_{min}^L$ is the energy difference between the bottoms of heavy and light bands, while $-\varepsilon_0$ is the center of heavy band. The center of light band corresponds to zero energy. $\mu$ is the chemical potential~\cite{KaganJETP2011}.}
\end{figure}

For the further calculations, it is necessary to introduce the bare masses of heavy and light components
\begin{equation}\label{bare_masses}
m_h = \frac{1}{2t_ha^2}, \quad m_L = \frac{1}{2t_La^2}\, ,
\end{equation}
and the Fermi energies
\begin{equation}\label{Fermi_en}
\varepsilon_{Fh} = \frac{p_{Fh}^2}{2m_h} = \frac{W_h}{2}+\mu +\varepsilon_0, \quad \varepsilon_{FL} = \frac{p_{FL}^2}{2m_L} = \frac{W_L}{2}+\mu \, .
\end{equation}
Uing Eqs.~\eqref{bare_masses} and \eqref{Fermi_en}, we finally find the quasiparticle spectra at low energies,
%$T \rightarrow 0$
\begin{equation}\label{quasip_sp}
\varepsilon_h(p) = \frac{p^2}{2m_h} - \varepsilon_{Fh}, \quad \varepsilon_L(p) = \frac{p^2}{2m_L} - \varepsilon_{FL}\, .
\end{equation}
In deriving expressions \eqref{bare_masses}--\eqref{quasip_sp}, we implicitly assume that the difference in energy $\Delta$ between the bottoms of the bands is not too large, and, hence, the parabolic approximation for the spectra of both bands is valid. We note that there is no one-particle hybridization in Hamiltonians~\eqref{Hamil_Hub_twoband} and \eqref{Hamil_Hub_twoband-momspace}, but there exists the strong two-particle interaction $\frac{U_{hL}}{2}\sum_i{n_i^hn_i^L}$.

We assume that $m_h \gg m_L$, and therefore
\begin{equation}\label{ratio_bwth}
W_h/W_L = m_L/m_h \ll 1 \, .
\end{equation}
We will work in the most instructive strong-coupling limit assuming that
\begin{equation}\label{hierar_par}
U_{hh} \sim U_{LL} \sim U_{hL} \gg W_L \gg W_h\, .
\end{equation}
$U_{hL}$ can be large enough since actually the light particles are subjected to a strong scattering by heavy ones as if by a quasiresonance level. Finally, for the homogeneous state, we consider the simplest case  when the electron densities for both bands are of the same order $n_h ~\sim n_L$ (a strong mismatch between the densities, which leads to the instability of the homogeneous case, will be considered separately). In the parabolic approximation, the density has the form $n = \frac{p_F^3}{3\pi^2}$ in 3D and $n = \frac{p_F^2}{2\pi}$ in 2D for both spin projections.

Let us say a few words about the possible nature of the narrow band near the Fermi level. In the systems with magnetic impurities, the many-body effects give rise to the well-known Abrikosov--Suhl resonance \cite{AbrikosovPhys1965,SuhlPR1965}, which can lead to an instability of the Fermi liquid state \cite{ColemanPhB1994}. A thorough analysis of this situation is given in Ref.~\cite{MillisLeePRB1987} by the slave-boson treatment of the periodic Anderson model.

If we assume that the narrow band near the Fermi level is itself the Abrikosov--Suhl resonance and the EPE acts on top of the effective model, arising, say from the periodic Anderson model, the effective width of the hybridization peak according to e.g. to book of P. Fulde \cite{FuldeBook2002} is given by the expression
\begin{equation} \label{Fulde_gamma}
\Gamma^* = \pi N_L(0)(V^*)^2 = \pi N_L(0)V^2(1-n_f),
\end{equation}
where $N_L(0)$ is the density of states in the broad conduction band (light band in our notation), $V$ is the hybridization matrix element, and $n_f$ is the number of $f$ electrons per site.

If on top of this model, we will add the Falicov--Kimball type Coulomb interaction between heavy electrons (originally inside the hybridization peak) and light electrons, which is precisely our interband Hubbard interaction $U_{hL}n_hn_L$ ; then, we will get an additional narrowing of the resonance width to the values $\Gamma^{\ast \ast} = \Gamma^{\ast}(\Gamma^{\ast}/W_L)^{b/(1-b)}$,
where $W_L \sim \varepsilon_{FL}$ is the width of the light band and $b = 2N_L^2(0) U_{hL}^2$ is the exponent characterizing EPE.

Thus, EPE identically acts on the width of the heavy band $W_h^*$ in the two-band Hubbard model and on the effective width of the resonance peak $\Gamma^{**}$ (located near the Fermi level of the conduction band in mixed-valence limit) in the extended periodic Anderson model with the Falicov--Kimball term (without any double counting of these two effects)}. This result has been obtained in the important paper \cite{KaganProkJETP1987} (see also the first paper on this topic \cite{KaganProkJETP1986}). It was shown that in the two-band Hubbard model with one narrow band, we can completely neglect the one-particle hybridization in favor of the two-particle one (the interband Hubbard interaction).

As we have already mentioned above, the EPE and dressing of a heavy particle by the cloud of virtual electron--hole pairs of light particles is due to the nonadiabatic part of the heavy-particle many-body wave function at the tunnelling of the heavy particle between neighboring unit cells in the presence of Fermi liquid formed by light particles. The energy range for the nonadiabaticity was thoroughly estimated for different cases in the framework of the density matrix approach in the first fundamental works on this problem~\cite{KaganProkJETP1986,KaganProkJETP1987}. It was demonstrated that the energy range for the nonadiabaticity corresponds to that between $\hbar/\tau$  and $\varepsilon_{FL}$ or $W_L$, where $\tau$ is the effective  tunneling time for the heavy particle (see \cite{KaganProkJETP1986}).
For a more complicated problem of many heavy particles, the lower limit of nonadiabaticity is obtained in \cite{KaganProkJETP1987}. At low temperatures, it corresponds to the substitution of $\hbar/\tau$ by the heavy particle bandwidth $W_h$.

Now, let us discuss the electron polaron effect (EPE), having in mind the nonadiabatic part of the many-particle wave function, which describes a heavy particle dressed by a cloud of virtual electron--hole pairs of light particles. Nonadiabaticity of the cloud within some energy range manifests itself, when the heavy particle moves from one elementary cell to a neighboring one. Formally, EPE is related to interband Coulomb repulsion $U_{hL}$. In the second order of perturbation theory, we can write (neglecting the momentum dependence of self-energy $\Sigma_{hL}(\omega, {\mathbf q})$)~\cite{KaganJETP2011}.

\begin{equation} \label{Z_heavy}
m_h^*/m_h   \approx  Z_h^{-1} \approx 1 + b\ln{\frac{m_h}{m_L}}\, .
\end{equation}
Here, $b = 2f_0^2$ and $Z_h^{-1} = 1 -  \frac{\partial\Sigma_{hL}(\omega, {\mathbf q})}{\partial \omega}|_{\omega \rightarrow 0}$, where $Z_h$ is the $Z$-factor characterizing the renormalization of the heavy particle mass and $\Sigma_{hL}$ is the self-energy corresponding to scattering of a light particle by a heavy one. In the case under study characterized by the low charge carrier density and strong onsite Coulomb repulsion $U_{hL}$, we have in 3D, $f_0 = 2dp_F/\pi$, which is  the Galitskii gas parameter~\cite{GalitskiiJETP1958}, whereas in 2D -- $f_0 = -1/(2\ln dp_F)$ is the Bloom gas parameter~\cite{BloomPRB1975}. Here, $d$ is the intersite distance and $p_F$ is the Fermi momentum.

The exponent characterizing the electron polaron effect can be evaluated based on the nonadiabatic part of the many-particle wave function, which describes a heavy particle dressed by a cloud of electron--hole pairs of light particles \cite{KaganProkJETP1986,KaganProkJETP1987}. The evaluation yields
\begin{equation} \label{mass_ratio_b}
\frac{m_h^*}{m_h} \propto Z_h^{-1} = \left(\frac{m_h}{m_L}\right)^{b/(1-b)} \, ,
\end{equation}
For $b =1/2$ or, equivalently, for $f_0 =1/2$ (as for the coupling constant of the screened Coulomb interaction in the RPA scheme), we are in the so-called unitary limit. In this limit, according to Refs.~\cite{KaganProkJETP1986,KaganProkJETP1987}, the polaron exponent is
\begin{equation} \label{unitary_lim_b}
\frac{b}{1-b} = 1\, ,
\end{equation}
and hence
\begin{equation} \label{mass_ratio_unitary}
\frac{m_h^*}{m_h} = \frac{m_h}{m_L} \, ,
\end{equation}
or, equivalently
\begin{equation} \label{mass_ratio_unitary1}
\frac{m_h^*}{m_L} = \left(\frac{m_h}{m_L}\right)^2\, .
\end{equation}

Thus, starting from the ratio $m_h/m_L \sim 10$ of the bare masses (obtained, for instance, using the local density approximation(LDA)), in the unitary limit, we eventually obtain $m_h/m_L \sim 100$ (due to the many-body electron polaron effect), which is a typical ratio for uranium-based heavy-fermion systems.

Let us consider one more mechanism of the mass enhancement. The EPE is related  to the $Z$-factor of heavy particles. However, in the 3D case, the momentum dependence of the heavy--light self-energy also becomes important
As a result, we obtain much more dramatic mass enhancement than that in the pure electron polaron effect (which gives only $m_h/m_h^* \approx 1-2f_0^2\ln{(m_h/m_L)}$ due to the $Z$-factor of a heavy particle). The full expression for the effective heavy mass in the second order of the perturbation theory taking into account both the frequency and momentum dependence of the self-energy $\Sigma_{hL}$ yields~\cite{KaganValJScNM2012}
\begin{equation} \label{eff_heavy_mass}
\frac{m_h^*}{m_h}=1+b\ln{\frac{m_h}{m_L}}+\frac{b}{18}\frac{m_hn_h}{m_Ln_L},
\quad b=2f_0^2 \, .
\end{equation}
This expression involves an additional term, which is linear in the bare mass ratio $m_h/m_L$.
If as in LDA, we have $m_h/m_L \sim 10$, then for the large density mismatch between heavy and light particles, $n_h \ge 5n_L$, this linear term in Eq.~\eqref{eff_heavy_mass} becomes dominant over the  contribution of electron polaron effect (which is proportional only to $\ln{(m_h/m_L)}$).

It is very interesting to emphasize that in the 3D case, the same parameter $b\frac{m_hn_h}{m_Ln_L} \ge 1$ (which governs the dominance of the last term in Eq.~\eqref{eff_heavy_mass} for the effective heavy mass) is responsible for the tendency toward the phase separation in the two-band Hubbard model. It leads to the negative partial compressibility for larger densities of heavy particles $n_h \sim n_C \le 1$ and a large mismatch $n_h \gg n_L$ in the electron densities of the bands
\begin{equation} \label{inv_kappa}
\kappa_{hh}^{-1}\propto c_h^2 \propto (n_h/m_h)(\partial \mu _h/\partial n_h) < 0 \, ,
\end{equation}
where $\mu _h$  and $c_h$ are the chemical potential and sound velocity for heavy particles, respectively. This result is in qualitative agreement with the predictions of the mean-field type variational analysis~\cite{sboychakov2007phase}. In the 2D case, the contributions coming from the momentum dependence of the self-energy to $m_h^*$ and  $\mu _h$ are absent and thus, there is no tendency toward the phase separation in this case. Moreover, the EPE is governed only by the $Z$-factor (by the frequency dependence of the self-energy $\Sigma_{hL}$.

Here, we would like to emphasize that in the main approximation, EPE plays in a concert with the strong phonon polaron effect (PPE), which  has its own specific features in strongly correlated electron system (see, e.g., \cite{GrilliPRB1994}).  Namely two exponents for PPE and EPE should be multiplied by each other to result in the effective width of the heavy band $W_h^*$. This is also the case in manganites. More complicated nonlinear amplification effects due to the coupling between PPE and EPE arise when the renormalized width of the heavy band becomes smaller than the Debye temperature.

\subsubsection{Temperature dependence of the electrical resistivity}
 \label{rho_vs_T}
%\subsubsection{Imaginary parts of the self-energies for low temperatures, $T < W_h$}
%\label{Im_low_T}

The exact solution of coupled kinetic equations with an account taken of the Umklapp processes yields for $p_{Fh} \sim p_{FL} \sim p_F  \sim 1/d$ and low temperature $T < W_h^* < W_L$ gives the following estimate for the inverse scattering times
\begin{equation}\label{tauL-hL}
1/\tau_L \sim 1/\tau_{Lh} \sim f_0^2\frac{T^2}{ W_h^*} \frac{m_h}{m_L}\, , \quad
1/\tau_L \sim 1/\tau_{hL} \sim f_0^2\frac{T^2}{ W_h^*}\, .
\end{equation}
This behavior corresponds to Landau-type Fermi-liquid picture. Accordingly, for the conductivities in the Drude approximation, we have
\begin{equation} \label{sigmahL}
\sigma_{hL} \sim \sigma_{Lh} \sim \frac{\sigma_{\min}}{b} \left(\frac{W_h^*}{T} \right)^2 \, .
\end{equation}
at low temperatures $T < W_h^*$.

Thus, the electrical resistivity has the form
\begin{equation} \label{resistivv_total}
R=\frac{1}{\sigma_L + \sigma_h}  = \frac{b}{\sigma_{\min}}\left(\frac{W_h^*}{T}\right)^2\, ,
\end{equation}
where $\sigma_{\min} =e^2p_F/\hbar$ is the minimal Mott--Regel conductivity in 3D.

At high temperatures $T > W_h^*$, the inverse scattering times satisfy the relations
\begin{equation}\label{tauL-hL_highT}
1/\tau_L \sim 1/\tau_{Lh} \sim bW_L, \quad
1/\tau_L \sim 1/\tau_{hL} \sim bT\, .
\end{equation}
Thus, the heavy component exhibits a marginal behavior (heavy electrons are moving diffusively in the surrounding of light electrons). However, light electrons are scattered by heavy ones as if by static impurities, and thus light component is nonmarginal. Correspondingly, for the conductivities, we can write
\begin{eqnarray} \label{sigma_h-L}
\sigma_L \sim \sigma_{Lh} &=& \frac{n_Le^2\tau_L}{m_L} \approx \frac{n_Le^2\tau_{Lh}}{m_L} \sim \frac{\sigma_{\min}}{b}\, ,
\nonumber \\
\sigma_h \sim \sigma_{hL} &\sim & \frac{\sigma_{\min}}{b}\left(\frac{W_h^*}{T}\right)^2 \,.
\end{eqnarray}
with an account taken of the Einstein relation $\partial n_h/\partial T \sim n_h/T$ at high temperatures $T > W_h^*$. Hence the resistivity
\begin{equation} \label{resistiv_total}
R=\frac{1}{\sigma_L + \sigma_h}  \sim \frac{b}{\sigma_{\min}}\frac{1}{\left[1+
\left(\frac{W_h^*}{T}\right)^2\right]}
\end{equation}
tends to the saturation in the 3D case (see Fig.~\ref{resist_two-band}). This behavior of $R(T)$ is typical of some uranium-based heavy-fermion compounds, such as UNi$_2$AL$_3$.

\begin{figure} [H]
\begin{center}
\includegraphics*[width=0.4\columnwidth]{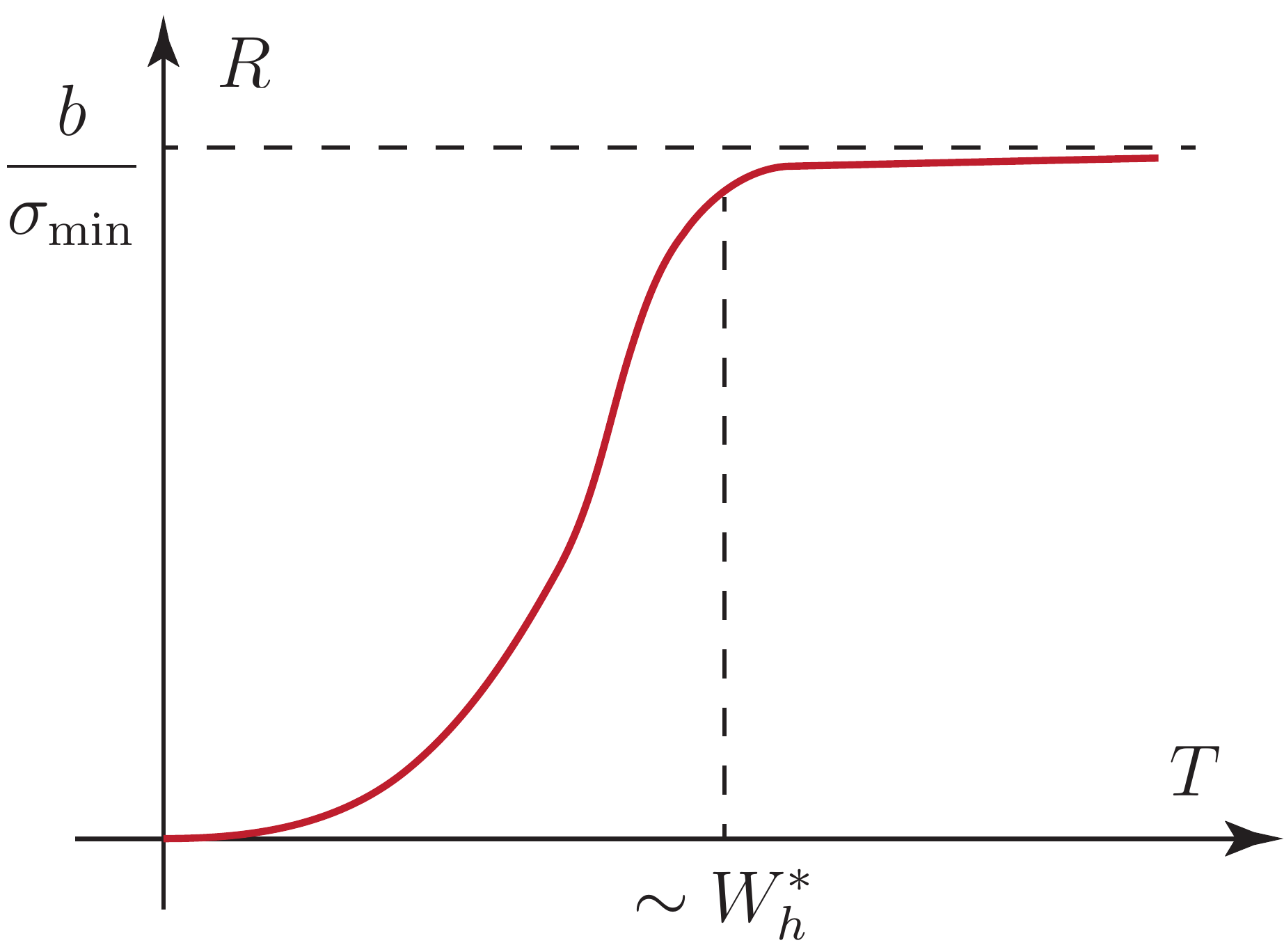}
\end{center}
\caption{\label{resist_two-band} Electrical resistivity $R(T)$ in the two-band Hubbard model with one narrow band in 3D~\cite{KaganJETP2011}.}
\end{figure}

In the 2D case, we should take into account the weak-localization corrections to the classical Drude formula for conductivity of the light band~\cite{AltArCh1985}. These corrections related to the quantum-mechanical interference (backscattering effects) have the form
\begin{equation} \label{sigma_loc}
\sigma_L^{loc} = \frac{\sigma_{\min}}{b}\left(1-b\ln{\frac{\tau_{\varphi}}{\tau}}\right)\,.
\end{equation}
Here, $\sigma_{\min} = e^2/\hbar$ is the Mott--Regel minimal conductivity in 2D, $\tau_{\varphi}$ is the inelastic (decoherence) time for light electrons, $\tau = \tau_{Lh}$, and $1/\tau_{Lh} \sim bW_L$ as in 3D. Then, we have $1/\tau_{\varphi} =b^2T$ corresponding to the weak-localization effect in ``dirty" metals in 2D (the electron--electron scattering time becomes marginal in the dirty limit when a light electron scatters many times by heavy electrons between two subsequent scattering events for light electrons, see Fig.~\ref{mult_scatter}).

\begin{figure} [H]
\begin{center}
\includegraphics*[width=0.3\columnwidth]{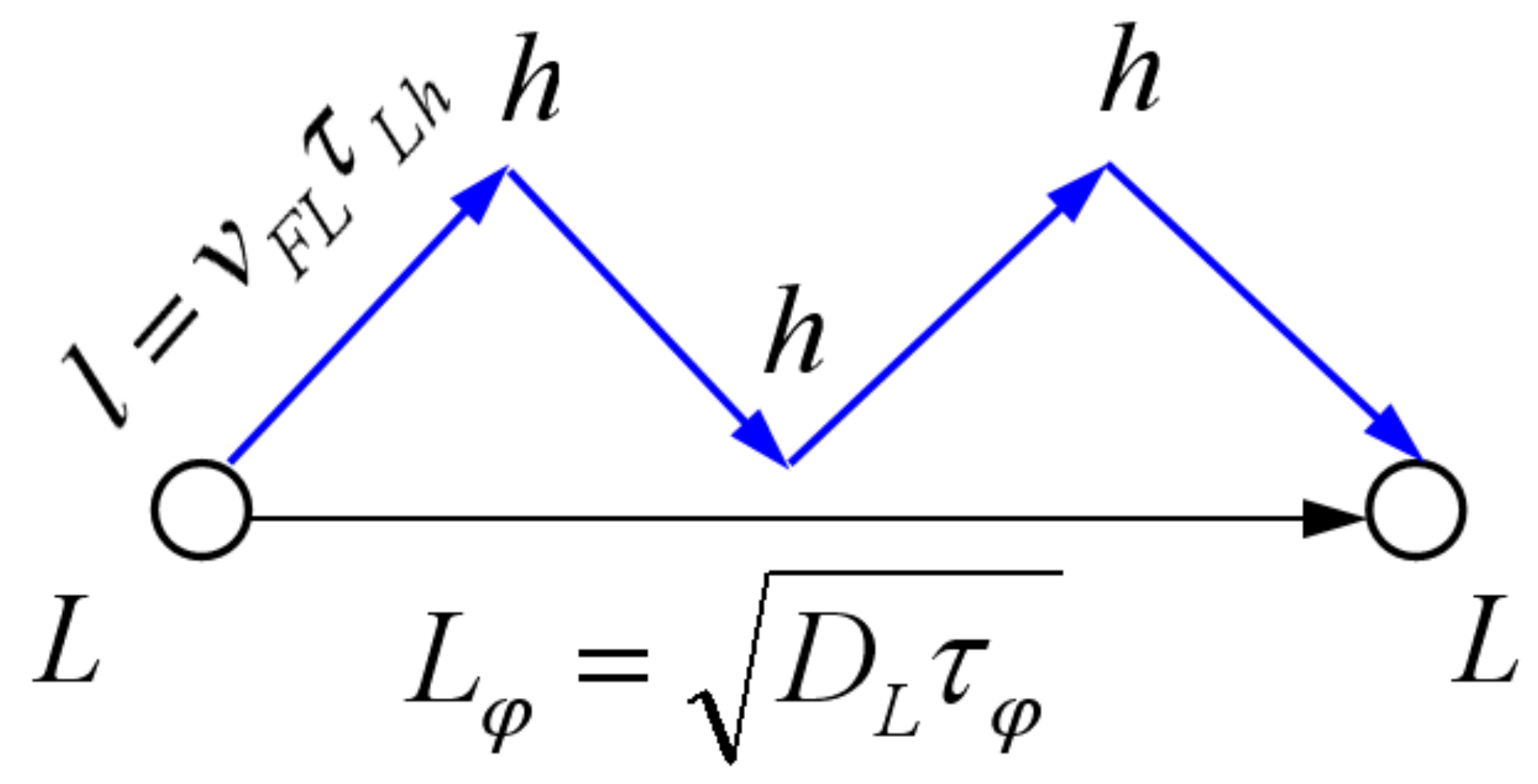}
\end{center}
\caption{\label{mult_scatter} Multiple scattering of light electrons by heavy ones in the period of time between the scattering of a light electron by another light particles. $L_{\varphi}$ is the characteristic diffusion length, $l$ is the elastic length, $D_L$ and $v_{FL}$ are the diffusion coefficient and the Fermi velocity for light electrons, respectively, and $\tau_{Lh}$ and $\tau_{\varphi}$ are respectively the elastic time for scattering of light electrons by heavy ones and the inelastic (decoherence) time~\cite{KaganJETP2011}.}
\end{figure}

Thus, in the 2D case, the light component has a tendency toward localization for $bT \geq W_h^*$.  Moreover, an additional narrowing
of the heavy band and additional localization in the light band are governed at  $bT \sim W_h^*$ by the same parameter $b\ln\left(m_h/m_L\right) \geq 1$.

As a result, instead of initially expected  marginal Fermi-liquid behavior at high temperatures $T  >W_h^*$, we obtain in 2D an  even more interesting behavior of resistivity $R  = 1/(\sigma_L + \sigma_h)$, where $\sigma_h$ is given by the same expression~\eqref{sigma_h-L} as in the 3D case. Namely, $R(T )$ in 2D has a peak and then a localization tail at higher temperatures (see Fig.~\ref{resist_2D}). Such  shape of  $R(T )$ resembles that observed in optimally doped layered manganites.

\begin{figure} [H]
\begin{center}
\includegraphics*[width=0.3\columnwidth]{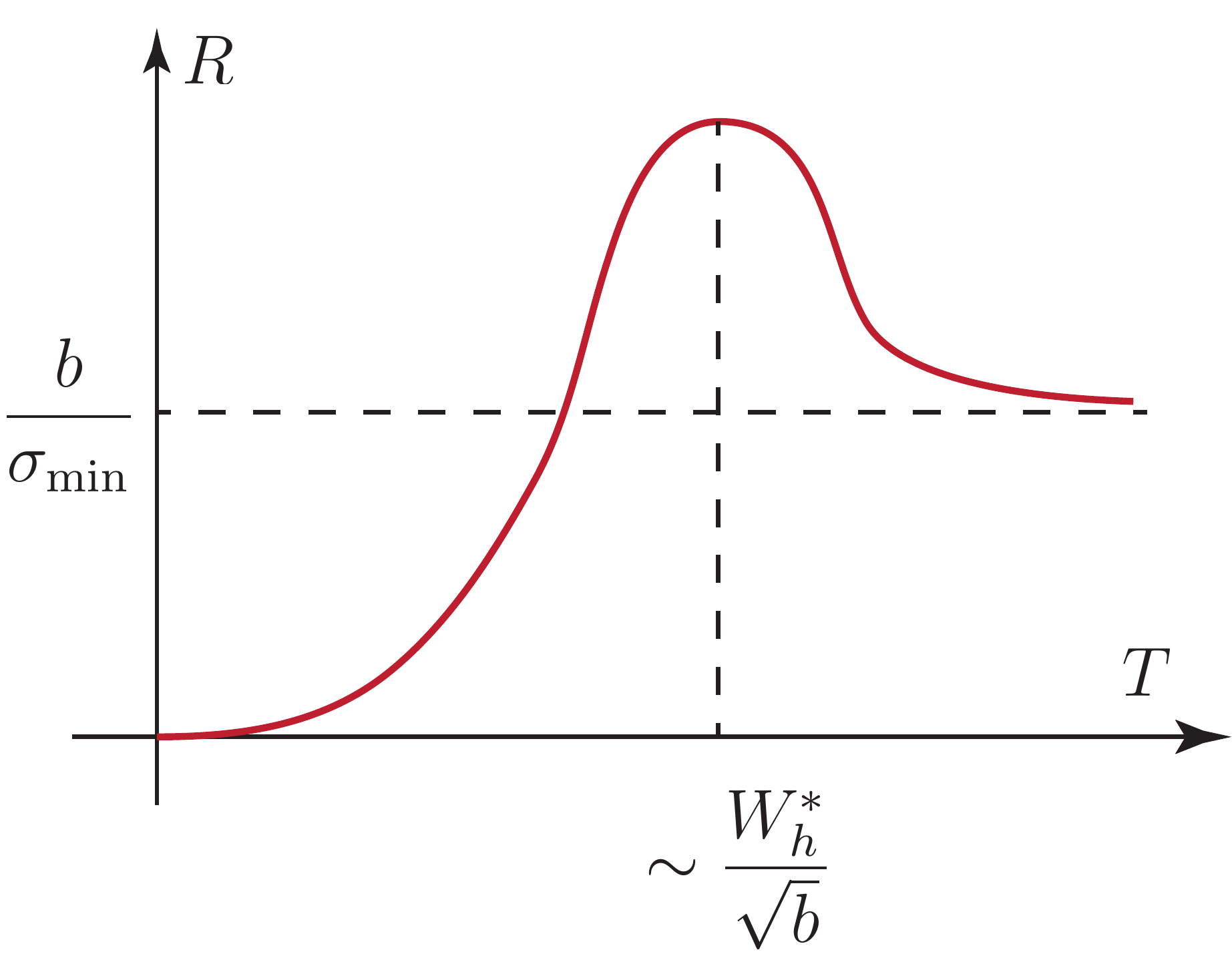}
\end{center}
\caption{\label{resist_2D} Temperature dependence of electrical resistivity $R(T)$ for the 2D two-band Hubbard model with one narrow band. The resistivity exhibits a peak and a localization tail at higher temperatures $T >W_h^*$~\cite{KaganJETP2011}.}
\end{figure}

Note, however, that in the case of layered manganites, the approach related to FM polarons, which is discussed in Section~\ref{MagPolaronTrans},  seems to be more consistent with the scanning-tunneling experiments \cite{RonnowNature2006} and with the confinement of electronic motion to two dimensions, Nevertheless, we cannot exclude that for different types of layered materials, the explanation based on the localization tail in the homogeneous state of the two-band model is valid in the regime of a dynamically destroyed (e.g., by temperature) heavy band.

At the end of this subsection, let us briefly mention the possibility of superconducting pairing in the two-band Hubbard model with one narrow band. At low electron densities, the leading mechanism of superconductivity in the two-band Hubbard model corresponds to the triplet $p$-wave pairing and is governed, especially in 2D, by the pairing of heavy electrons via polarization of light ones~\cite{KaganBookSpringer,KaganJETP2011,
KaganValFNT2011,KaganValJScNM2012,KaganKorovUFN2015,KaganJETPL2016, KaganChubJETPL1988,BaranovIJMPB1992,EfremovJETP2000,KaganPLA1991,
BaranovJETP1992,KaganDisser1994}) in the framework of the enhanced Kohn--Luttinger mechanism~\cite{KohnLattPRL1965}. The corresponding critical temperature $T_{c1}$ for such $p$-wave superconductors is higher in the 2D case. Moreover, it depends nonmonotonically on the relative doping of the bands $n_h/n_L$ and has a broad and pronounced maximum at $n_h/n_L = 4$ in 2D, where it can reach the experimentally feasible values realistic for layered ruthenates Sr$_2$RuO$_4$ ~\cite{RiceJPCMSr2RuO4,MaenoPhToday2001}, as well as for layered dichalcogenides CuS$_2$, CuSe$_2$ and semimetallic superlattices InAs-GaSb, PbTe-SnTe with the spatially separated bands belonging to different layers~\cite{MuraseSurSci1986}. In the 3D case, this mechanism can adequately describe the superconductivity in uranium-based heavy-fermion compounds such as U$_{1-x}$Th$_x$Be$_{13}$~\cite{KromerPRL1998,SigristRMP1991}. For typical values of $\varepsilon _{Fh}^* \sim$ 30--50 K, the critical temperatures $T_{c1}$ can achieve 1--5 K, which is quite nice. Moreover, as it was shown in Refs.~\cite{KaganValJScNM2012,KaganPLA1991,BaranovJETP1992}, the two superconducting gaps for heavy and light electrons are opened simultaneously below this temperature.

In conclusion, let us stress again that in 2D, where only the usual electron polaron effect (related to the $Z$-factor) contributes to the mass enhancement of heavy electrons, the restrictions on the stability of the homogeneous phase are much milder than in 3D, where a tendency toward the phase separation exists at a large mismatch between the charge carrier densities in different bands.

\section{Spin-state transitions and magnetic instabilities}
 \label{SpinState}
\subsection{Introduction}\label{Intr}

In this section, we consider specific features of the electronic phase separation, which can be observed in the systems with transition-metal ions exhibiting different multiplet states. The main emphasis will be put here on oxides with Co$^{3+}$ ions. This ion (as well as Fe$^{2+}$) can be characterized either by a low-spin (LS) state with $S$=0 ($t_{2g}^6$), intermediate-spin (IS) state, $S=1$ ($t_{2g}^5e_g^1$), or high-spin (HS) state ($t_{2g}^4e_g^2$) with $S=2$, see e.g. Ref.~\cite{GoodenoughPR1967}. The characteristic energies of these states usually do not differ much, thus the crossover between such states is quite possible, and it is usually referred to as the spin-state transition (SST). A popular compound with SST is LaCoO$_3$~\cite{GoodenoughPR1967,TokuraPRB1996,TokuraPRB1997,KorotinPRB1996}. The ions with different spin states can form a regular array~\cite{DoumercJSSCh2001,KhomLowPRBi2004}. Therefore, spin (or, in general, multiplet) ionic configuration can be treated as another degree of freedom in addition to charge, orbital, and spin ones typical of transition metal compounds. In the previous sections, we have demonstrated that under doping, the electronic phase separation can arise owing to the competition between the delocalization of charge carriers related to their kinetic energy and the possible formation of some localized magnetic or orbital structure. Therefore, in the materials exhibiting SST, the tendency to form some spin-state array can limit the motion of charge carriers appearing due to the doping. Here, again, we can locally modify the regular background to have an additional gain in kinetic energy. We know that in the framework of both the two-band double exchange
model~\cite{AndersonHasePR1955,deGennesPR1960} and  single-band Hubbard model \cite{BulKhomJETP1967,BulNagKhomJETP1968} an antiferromagnetic background hinders the motion of doped holes, whereas they can freely move on the ferromagnetic background. As it was discussed above, this favors the formation of an inhomogeneous state of the spin-polaron type~\cite{NagaevJLett1967,KasuyaSSC1970,BulNagKhomJETP1968}.

It is also well known and is discussed in detail in the next section that the orbital ordering can also suppress the motion of charge carriers and this favors a local modification of orbitals giving rise to orbital polarons~\cite{KugelPRB2008orbitalPS,
KhaliullinPRB1999,MiKhoSawPRB2000,MiKhoSawPRB2001}. In the case of SST, such interplay between the localization and delocalization can be related to the so called spin blockade ~\cite{MaignanPRL2004_spblock}. Indeed, if we add some electrons to the material with Co$^{3+}$ in the low-spin state ($S=0$), we could get the high-spin ($S=3/2$) Co$^{2+}$ .  However, if one electron move between LS Co$^{3+}$ and HS Co$^{2+}$ , the eventual configuration will include Co$^{3+}$ and Co$^{2+}$ in IS states rather than in the initial states. Indeed, the one-electron hopping leads to the $\pm 1/2$ spin change, while here the starting spin states differ by 3/2. Hence, the situation becomes similar to that discussed for the Hubbard model~\cite{BulNagKhomJETP1968}, namely, an electron can only move creating wrong spin states. This leads to the energy loss and electron localization.

Such energy loss can be minimized by local changes in the underlying spin states, i.e. by the phase separation, which has been actually reported for such cobaltites as La$_{1-x}$Sr$_x$CoO$_3$~\cite{CaciuffoPRB1999,LoshkarevaPRB2003,
PhelanPRL2006,LeightonPRB2006,LeightonPRB2007}, In particular, the magnetic measurements~\cite{TokuraPRB1996} at low hole doping ($< 1 \% $ of Sr) that the magnetic moment per doped
hole (per Sr) far exceeds that of LS Co$^{4+}$ with $S=1/2$. Actually, we have magnetic impurities with unusually large spin $S=5-10$, implying the formation of magnetic polaron owing to a transfer of some of Co$^{3+}$ ions to a magnetic state. It turns out to be possible to find out the size and shape of such magnetic inhomogeneities through the use of neutron scattering, supplemented by the ESR and NMR~\cite{PodlesnyakPRL2008}.

Below, we formulate a model allowing us to demonstrate an instability of inhomogeneous state in the systems with SST like doped cobaltites. Our discussion is based mostly on the results of Ref.~\cite{SboychakovPRB2009}.

\subsection{Spin states of cobalt ions}\label{SpinSt}

As a typical example of materials with the spin-state transitions, we will first study hole-doped perovskites of LaCoO$_3$ type. A detailed analysis of the characteristic features of such materials with the reference to the corresponding experimental data can be found in review articles~\cite{IvanovaUFN2009,DudnikovJETPL2016}. Cobalt ions in these compounds are located within oxygen octahedra. In the octahedral coordination, the $d$ levels of transition-metal ions are split into a lower $t_{2g}$ level and the higher $e_g$ one. In the case of hole doping, we are dealing with Co$^{3+}$ and Co$^{4+}$ ions, which have $3d^6$ and $3d^6$ configurations, respectively. As we have mentioned above, three low-energy states are possible for such ions, namely, low-spin (LS), intermediate-spin (IS), and high-spin (HS) states.

The spin states of Co$^{3+}$ and Co$^{4+}$ ions and their energies are illustrated in Table~\ref{Table}. For Co$^{3+}$, the lowest energy $E_0$ corresponds to the LS state ($S=0$), with fully occupied $t_{2g}$ level and empty $e_g$ one. In the IS state ($S=1$), one electron is promoted to the $e_g$ level leaving five electrons at the $t_{2g}$ level. In the HS state ($S=2$), one more electron appears at the $e_g$ level, and so there are four $t_{2g}$ and two $e_g$ electrons. We denote energy gap between $t_{2g}$ and  $e_g$ as $\Delta$ and the Hund's rule coupling constant as $J_H$. Thus, we have $E_{IS}^{(3+)} = E_0 + \Delta -J_H$ and $E_{HS}^{(3+)} = E_0 + 2\Delta -4J_H$. For the Co$^{4+}$  ion with five $3d$ electrons, these electrons can be distributed between the corresponding states in a similar way, with the only difference that the number of $t_{2g}$ electrons is smaller (one electron is missing), see Table~\ref{Table}. Therefore, the spins for LS, IS, and HS are equal to 1/2, 3/2, and 5/2, respectively, and the corresponding energies inthe case of Co$^{4+}$ ion are $E_{LS}^{(4+)} = E_1$, $E_{IS}^{(4+)} = E_1 + \Delta -2J_H$, and $E_{HS}^{(4+)} = E_1 + 2\Delta -6J_H$. As we demonstrate below, the specific values $E_0$ and $E_1$ do not produce much effect on the obtained results.

\begin{table}[tbp] \begin{center}
\includegraphics[width=0.6\columnwidth]{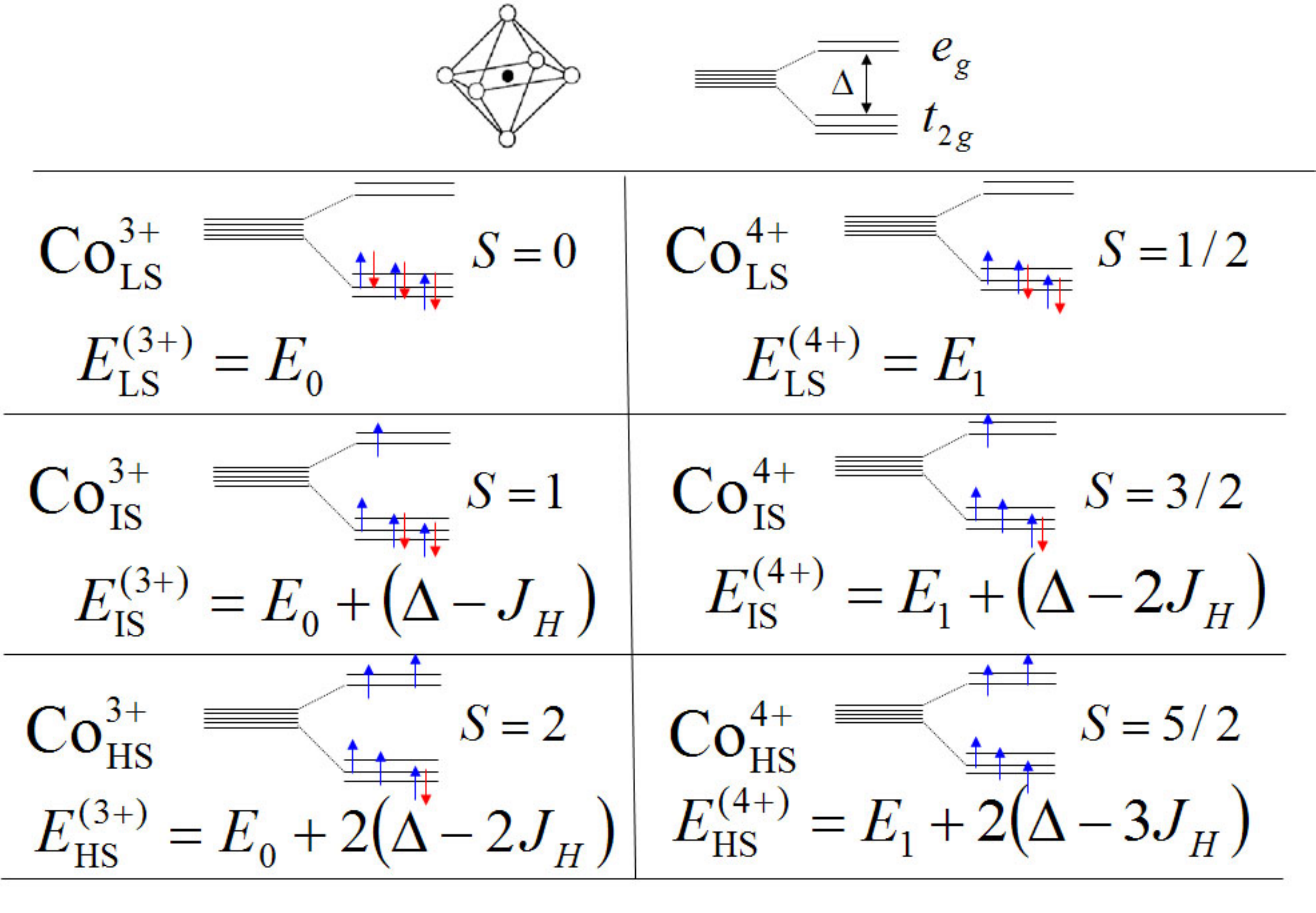}
\end{center}
\caption{(Color online) Energy levels and spin states of Co ions~\cite{SboychakovPRB2009}.
}
 \label{Table}
 \end{table}

As we can see in Table~\ref{Table}, the ground state of isolated cobalt ions is determined by the ratio of $\Delta$ and $J_H$. In particular, at $\Delta > 3J_H$, the LS state corresponds to the lowest energy for both Co$^{3+}$ and Co$^{4+}$ ions. Within the $2J_H < \Delta < 3J_H$ range, we have LS Co$^{3+}$ and HS  Co$^{4+}$. At large Hund's rule coupling constant, $2J_H > \Delta$, both ions have the HS ground state. This means that for noninteracting cobalt ions, the IS ground state is not favorable.

However, the possibility of charge transfer between cobalt ions can qualitatively change the aforementioned set of ground states. To make the further treatment clearer, let us emphasize the main effect, namely, the electron hopping between the $e_g$ states. Indeed, in cobaltites, the probability of electron hopping between the $t_{2g}$ states is usually quite low and can be neglected without any loss in generality. We also do not take into account the states with more than six electrons per site since the strong on-site Coulomb repulsion of electrons suppresses their existence. Finally, the spin blockade mentioned above prevents the electron transfer  corresponding to the spin changes exceeding one half.

Therefore, we are dealing below with two main charge transfer channels typical of doped cobaltites, namely, the electron hoppings  between the IS Co$^{3+}$ and LS Co$^{4+}$ and and those between the HS Co$^{3+}$ and IS Co$^{4+}$. Actually, to attain the maximum gain in the kinetic energy, the ground state should involve the IS cobalt ions. In particular, if the ground state for isolated cobalt ions contains LS Co$^{4+}$ and LS Co$^{3+}$ ($\Delta > 3J_H$), we can excite some LS Co$^{3+}$ ions to the IS state. Then the energy loss related to $\Delta$ could be compensated by the gain in kinetic  energy due to the charge transfer between IS Co$^{3+}$ and LS Co$^{4+}$ states. In the opposite limit $\Delta < 2J_H$, favoring the high-spin states of cobalt ions, another charge transfer channel becomes operational, namely the hopping between HS Co$^{3+}$ and IS Co$^{4+}$. Within the intermediate parameter range, $2J_H < \Delta < 3J_H$, the the charge transfer electron hopping can occur if we simultaneously excite both Co$^{3+}$ and Co$^{4+}$. However the excitation of  two ions is clearly less probable. In the next subsection, we start from the $\Delta >3J_H$ case at different doping levels, which is more often corresponds to the situation in actual cobaltites. After that, we perform the same analysis for the case of $\Delta < 2J_H$. In both ranges of parameters, we draw a special attention to the on the possible electronic phase separation and eventually determine the phase diagram in the $\Delta/J_H-$doping plane. $\Delta$ is closely related to the ionic radius $r_A$ of rare earth R in RCoO$_3$ perovskite cobaltites and increases when $r_A$ becomes smaller.

\subsection{Effects of electron delocalization and spin-state transitions: LS--LS ground state for isolated ions} \label{LS-LS}

We start with the case of $\Delta > 3J_H$. As it was mentioned above, it is characteristic of hole-doped La$_{1-x}$Sr$_x$CoO$_3$ perovskite cobaltites. Since $\Delta$ is growing with the rare earth ionic radius, this case in also typical for some other doped RCoO$_3$ compounds~\cite{LorenzPRBr2005}. In such cobaltites, Co$^{3+}$ and Co$^{4+}$ ions occupy the sites of simple cubic lattice. If these ions do not interact, they have the LS ground state and the doping level $x$ correspond to the relative number of Co$^{4+}$. If some LS Co$^{3+}$ ions are transferred to the IS state, this results to the kinetic energy gain owing to the charge transfer from IS Co$^{3+}$ to LS Co$^{4+}$.

Let us formulate a mathematical approach allowing us to justify the semiqualitative arguments put forward above. As a vacuum state, we choose the LS state of Co$^{3+}$ on and introduce creation
operators $a_{\bf{n}}^{\dag}$ and $c_{\bf{n}}^{\dag}$ for an
electron at the $e_g$ level and a hole at the $t_{2g}$ level,
respectively, at site $\bf{n}$. Their action onto the vacuum state  can be represented as
\begin{eqnarray}\label{states}
|0\rangle &=&|\textrm{Co}_{LS}^{3+}\rangle, \quad E^{(vac)} = E_0, \nonumber \\
a_{\bf{n}}^{\dag}|0\rangle &=&|\textrm{Co}^{2+}\rangle, \quad E^{(2+)} = U', \nonumber \\
c_{\bf{n}}^{\dag}|0\rangle &=&|\textrm{Co}_{LS}^{4+}\rangle, \quad E^{(h)} = E_1.
\end{eqnarray}

The IS state of Co$^{3+}$ ions is writhen in the form
\begin{equation}\label{ISstate} |\textrm{Co}_{IS}^{3+}\rangle =
c_{\bf{n}}^{\dag}a_{\bf{n}}^{\dag}|0\rangle, \quad E_{IS}^{(3+)} =
E_0 + \Delta -J_H = E_2. \end{equation}

Then, the single-site Hamiltonian involving all low-energy states have the form
\begin{equation}\label{H_onsite1}
H_{\bf{n}} =  E_0 + (E_1-E_0)n_{\bf{n}}^h +(U'-E_0)n_{\bf{n}}^e
+ \left[(E_2-E_0)-(E_1-E_0)-(U'-E_0)\right]n_{\bf{n}}^hn_{\bf{n}}^e\, ,
\end{equation}
where $n^e_{{\bf n}}= a^{\dag}_{{\bf n}}a_{{\bf n}}$ and $n^h_{{\bf n}}= c^{\dag}_{{\bf n}}c_{{\bf n}}$ are the
operators corresponding to the numbers of electrons at $e_g$ levels
and holes at $t_{2g}$ levels, respectively. Hamiltonian can be rewritten as
\eqref{H_onsite1} in a more compact form, we have
\begin{equation}\label{H_onsite2}
H_{\bf{n}} =  [E_0 + (E_1-E_0)(n_{\bf{n}}^h -n_{\bf{n}}^e)]
+ (\Delta-J_H)n_{\bf{n}}^e +Un_{\bf{n}}^e(1-n_{\bf{n}}^h)\, ,
\end{equation}
where $U=U'+E_1-\Delta+J_H - 2E_0$. After that , we can write the Hamilltonian for the lattice as a whole taking also into account the intersite electron transfer
\begin{equation}\label{H}
 H = \sum_{\bf{n}}[E_0 +
(E_1-E_0-\mu)(n_{\bf{n}}^h -n_{\bf{n}}^e)]
+ \Delta_1\sum_{{\bf n}}n^e_{{\bf n}}+
U\sum_{{\bf n}}n^e_{{\bf n}}(1-n^h_{{\bf n}}) -t\sum_{\langle{\bf n}{\bf m}\rangle}
\left(a^{\dag}_{{\bf n}}a_{{\bf m}}+h.c.\right)\, ,
\end{equation}
where $\Delta_1 =\Delta - J_H$ and $\langle{\bf n}{\bf m}\rangle$ denotes the summation over the nearest neighbor sites.

In the $t$ term of Hamiltonian~\eqref{H}, we have left only the most important electron transfer process, namely, the hopping between the occupied $e_{g}$ level of IS Co$^{3+}$ to the empty $e_{g}$ level of LS Co$^{4+}$. In addition, we assume that due to the Hund's rule coupling, the spin of an itinerant $e_g$ electron has the same direction as the total spin of core $t_{2g}$ electrons. Thus, we are dealing with a kind of ferromagnetic ground state and may not use the spin index of electron operators. Here, we do not take into account additional effects related to the degeneracy of $e_{g}$ states, which can lead only to unnecessary complications. Some of such effects in cobaltites are addressed in  \cite{KorotinPRB1996,PandeyPRB2008a,PandeyPRB2009b,LuoJSSC2009} using {\it ab initio}  band structure calculations and the analysis of X-ray photoemission spectra.

Hamiltonian~\eqref{H} turns out to be similar to that of the Falicov--Kimball model~\cite{FalKimbPRL1969}, which we have discussed in the previous section. Here, we have again the usual competition between the electron localization and itineracy: the LS cobalt ions play the role of localized states, whereas electrons promoted to the IS state become itinerant. As it was shown above, such compeition in the framework of the Falicov--Kimball model leads to the phase separation. Note also that the atomic-scale charge and spin inhomogeneities related to electronic phase separation were
revealed by the numerical solutions for small
clusters~\cite{KocharianPLA2009}. Below, using the techniques reported in Refs.~\cite{KugelPRL2005,SboychakovPRB2006jahn}, we analyze the specific features of systems with the spin-state transitions.

We introduce the average numbers of $e_g$ electrons and $t_{2g}$ holes per site $\langle n^e_{{\bf n}}\rangle = n^e$ and $\langle n^h_{{\bf n}}\rangle = n^h$ ($n^h - n^e = x$). We have in mind the average number of electrons appearing at initially empty $e_g$ levels owing to the LS--IS transitions.

The corresponding energies are
\begin{equation}\label{E}
E^{(1)} = E_0(1-x)+ E_1x + \langle H_1 \rangle /N  \, ,
\end{equation}
 where
\begin{equation}\label{H1}
H_1= \Delta_1\sum_{{\bf n}}n^e_{{\bf n}}+
U\sum_{{\bf n}}n^e_{{\bf n}}(1-n^h_{{\bf n}})
-t\sum_{\langle{\bf n}{\bf m}\rangle}\left(
a^{\dag}_{{\bf n}}a_{{\bf m}}+h.c.\right)\, .
\end{equation}

To each value of $n^e$, we can put into the correspondence some homogeneous state and calculate the energy spectrum using the Hubbard I approximation~\cite{HubbardPrRoySocA1963} for  the one-electron Green's function $G^e({\bf n,n}_0;\,t-t_0)=-i\langle
Ta_{{\bf n}}(t)a^{\dag}_{{\bf n}_0}(t_0)\rangle$ for the
$e_g$ electrons appearing in the IS state. Here, we have a clear analogy to the band $b$ electrons discussed in the previous section (see also Refs.~\cite{KugelPRL2005} and \cite{ShenoyPRL2007}). This Green's function has the following form in the frequency--momentum representation
\begin{equation}\label{Gappr}
G^e({\bf k},\omega)=-\frac{\omega+\mu -
\Delta_1-Un^h}{\left(\omega+\mu - \Delta_1-E_1({\bf
k})\right)\left(\omega+\mu-\Delta_1-E_2({\bf k})\right)}\,,
\end{equation}
where
\begin{equation}\label{E12} E_{1,2}({\bf
k})=\frac{U+\varepsilon({\bf
k})}{2}\mp\sqrt{\left(\frac{U-\varepsilon({\bf
k})}{2}\right)^2+U\varepsilon({\bf k})(1-n^h)}\,,
\end{equation}
In these expressions, $\varepsilon({\bf k})$ is the dispersion law in the absence of electron correlations,  $U=0$. We use the simplest tight-binding form of $\varepsilon({\bf k})$ without involving  orbital effects and the characteristic form of the hopping integrals for $e_g$ electrons. Thus, we use  $\varepsilon({\bf k})=-2t(\cos k_x+\cos k_y+\cos k_z)$ for the simple cubic lattice. Below, we consider the case of of strongly correlated electrons, $U/t\gg1$. In this case, the Hubbard I approximation is an adequate one capturing the essential physics related to electron correlations and agrees well both with experiment and numerical
calculations~\cite{FuldeBook2002,Ovchin_book2004}.

Using Eq.~\eqref{Gappr} for the Green's function, we calculate the
densities of $e_g$ charge carriers. The electron and hole densities, $n^e$ and $n^h$, are calculated based on Eq.~\eqref{Gappr} in the $U\to\infty$ limit. As a result, we find the total energy as function of doping $x$. These calculations are similar to those performed in
Ref.~\cite{KugelPRL2005} and discussed in detail in Section~\ref{twobands}. The $b$ electrons in
Ref.~\cite{KugelPRL2005} correspond to our $e_g$ electrons at IS
Co$^{3+}$ ions, whereas the number of localized $l$ electrons in
Ref.~\cite{KugelPRL2005}, $n_l$, corresponds to the number of LS
Co$^{3+}$ ions, $1-n^h$. Using this similarity, we could make a
direct mapping between the two systems. However, the specific feature of the systems with spin-state transitions is the existence of  another homogeneous state in addition to that considered in Ref.~\cite{KugelPRL2005}. In this state, we have all IS Co$^{3+}$ ions ($n^h=1$, $n^e=1-x$). In the usual conduction band picture, such state corresponds to an empty localized level located under the Fermi level of the conduction band. Such situation is rather exotic and hardly possible. In our case, the state corresponding to the localized level is determined by the existence of LS Co$^{3+}$ ions, and this analogy fails in the absence of such LS Co$^{3+}$ ions (the localized level simply disappears).

We denote the state similar to that discussed in Ref.~\cite{KugelPRL2005} (the state with coexisting LS and IS Co$^{3+}$ ions) as a type 1 state, and the state without LS Co$^{3+}$ as type 2 state. In Fig.~\ref{FigE_LS}, we plot the energies of such states versus the doping level $x$ at $\Delta_1/zt=0.2$ ($z=6$ is the number of nearest neighbors). The type 2 state corresponds to the  ground state at $x>x_2$. At $x>x_3$ both states, 1 and 2, are equivalent. At $x<x_1$, the number of electrons appearing at the $e_g$ level vanishes ($n^e=0$), see Fig.~\ref{FigN_LS}.

\begin{figure} [H]
\begin{center}
\includegraphics*[width=0.5\columnwidth]{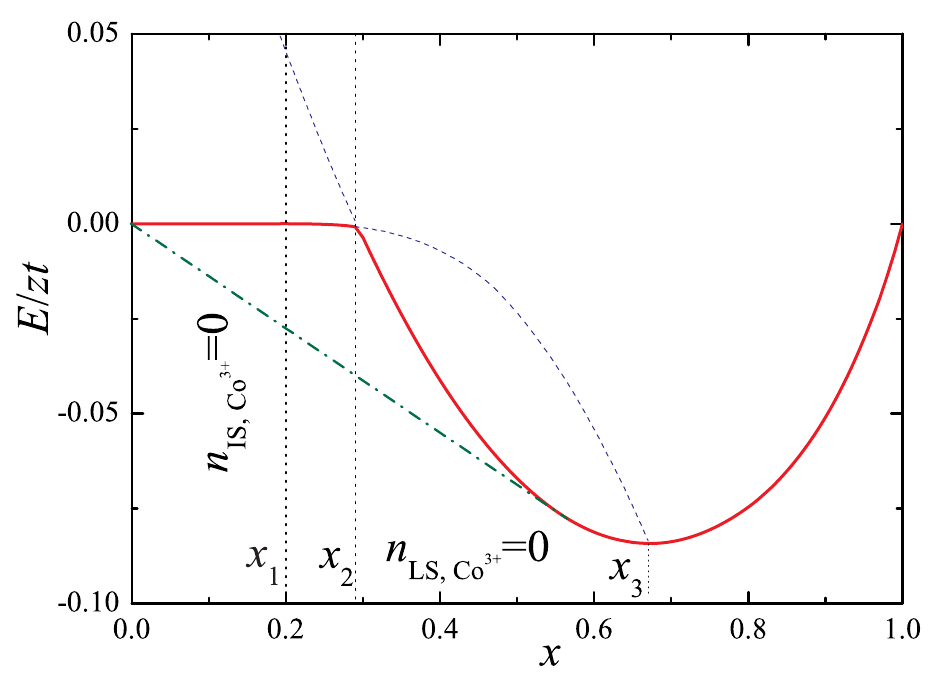}
\end{center} \caption{\label{FigE_LS} (Color online)
Energies of type 1 and 2 states versus doping $x$. The red solid curve indicates homogeneous state with the lowest energy. The blue
dashed lines denote the states with higher energies (see the text). The green dot-and-dash line shows the energy of an inhomogeneous phase-separated state. $\Delta_1/zt=0.2$ ~\cite{SboychakovPRB2009}.}
\end{figure}

At $x>x_1$, we observe a gradual increase in the number of $e_g$ electrons. In the usual Falicov--Kimball model, this increase persists up to $x=x_3$ when all Co$^{3+}$ ions transfer to the IS state. In our case, there appears the type 2 state, and at $x=x_2$, the system undergoes a jumplike transition to this state. The ground state energy $E$ for the homogeneous system as function of $x$ is shown in Fig.~\ref{FigE_LS} by the red solid curve. At the same time, this figure clearly demonstrates that within the $0<x\lesssim x_3$ doping range, we should expect the formation of inhomogeneous state, involving the domains with $n^h=1$ and $n^e=0$. In Fig.~\ref{FigE_LS}, the energy of such phase-separated state is shown  by the green dot-and-dash line.

In Fig.~\ref{FigN_LS}, we illustrate the changes with doping $x$ in the densities of IS Co$^{3+}$ ($n^e$) and LS Co$^{3+}$ ($1-n^h$) ions shown by blue dot-and-dash and red solid curves, respectively. At $x_2$, $n^e$ undergoes a jump related to the transition to the type 2 state. The plots of $n^e$ and $1-n^h$ at $x>x_2$ corresponding to the Falicov--Kimball model (without type 2 state) are drawn by thin dashed blue and red lines, respectively.

\begin{figure} [H]
\begin{center}
\includegraphics*[width=0.5\columnwidth]{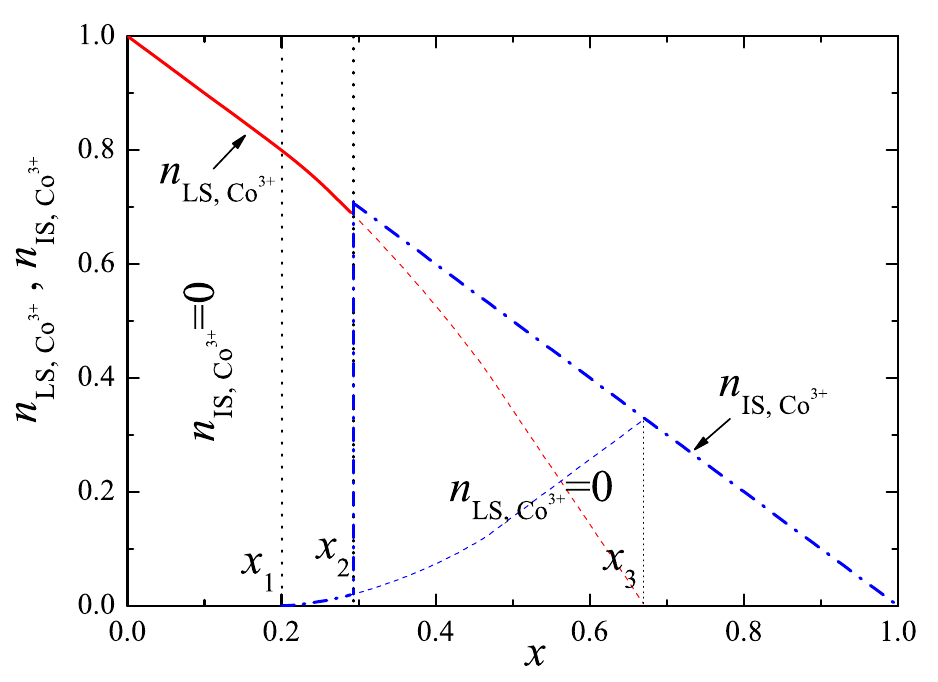} \end{center}
\caption{\label{FigN_LS} (Color online) The densities of IS Co$^{3+}$ ions ($n^e=n_{\text{IS, Co}^{3+}}$) and LS Co$^{3+}$ ions
($1-n^h=n_{\text{LS, Co}^{3+}}$) versus doping $x$ at $\Delta_1/zt=0.2$ \cite{SboychakovPRB2009}. At $x_1 < x < x_2$, both IS an LS states of Co$^{3+}$ exist simultaneously and  at $x =x_2$, we have a jump to the type 2 state with only IS Co$^{3+}$. The plots of $n_{\text{IS, Co}^{3+}}$ and $n_{\text{LS, Co}^{3+}}$ corresponding to the Falicov--Kimball model (without type 2 state) and discussed in Ref.~\cite{KugelPRL2005} are drawn by thin dashed lines.}
\end{figure}

If the number of electrons promoted to the IS state is small, $n^e\ll 1$, the approximate expression for the total energy $E$ in the case of the Fermi surface has the form
\begin{equation}\label{Etot0} E\simeq
\Delta_1n^e-tzn^en^h+
\frac{3t}{5}\left(36\pi^4n^h\right)^{1/3}(n^e)^{5/3}\,.
\end{equation}
The actual value of $n^e$ is determined by the minimization of energy~\eqref{Etot0} under condition  $n^h = x + n^e$. A nonzero value of $n^e$ can be attained only at $\Delta_1 < tzx$. Hence, at $\Delta_1/tz > 1$, IS Co$^{3+}$ ions do not arise within the whole doping range.

Since the spins of $e_g$ electrons are parallel to the total spin of $t_{2g}$ electrons due to the Hund's rule coupling, the changes in $n^e$ with doping $x$ are closely related to those in the magnetic moment of Co ions. The LS Co$^{3+}$ ions are characterized by zero magnetic moment, $S=0$, and the doping gives rise to LS Co$^{4+}$ ions with $S=1/2$ and at the same time, some LS Co$^{3+}$ ions becomes transferred to the IS state with $S=1$. Therefore, the plots shown in Fig.~\ref{FigN_LS} manifest themselves in the behavior of magnetic moments of cobalt ions illustrated in  Fig.~\ref{FigM_xLS}. In particular, the jump in $n^e$ is associated in the jump of the magnetic moment, if we consider the homogeneous states. In the phase-separated state, the behavior of the magnetic moment is different: the magnetic moment per Co$^{4+}$ ion does not change. Indeed, both the relative amount of the IS Co$^{3+}$ phase and the density of Co$^{4+}$ ions are proportional to $x$. In the case of phase separation, the magnetic moment is related to such $x$, at which the green dot-and-dash line in Fig.~\ref{FigE_LS} touches the plot of energy corresponding to the homogeneous state. In Fig.~\ref{FigM_xLS}, we can see the magnitude of the magnetic moment jump in the homogeneous state and the value of magnetic moment per Co$^{4+}$ in the phase-separated state depend on the hopping integral $t$. Namely, the increase in $t$ by a factor of two results in a much larger growth of two aforementioned parameters.

Let us emphasize here that in our analysis of magnetic moments, we imply that in the phase-separated state, the size of inhomogeneities exceeds the lattice constant. This is true for relatively high doping levels. At small $x$,  Co$^{4+}$ ion could act separately forming a kind of a small droplets, or in other words, spin-state polarons with only one Co$^{4+}$ ion surrounded by IS Co$^{3+}$. For such polaron, the effective spin value should be higher than that given by our quasimacroscopic approach. Such situation has been indeed reported  in Ref.~\cite{PodlesnyakPRL2008} for La$_{1-x}$Sr$_x$CoO$_3$ at quite small values of $x$, where the existence of spin polarons with the effective spin of  $13\mu_B$ has been suggested. We argue that the magnetic moment per Co$^{4+}$ should become smaller at higher doping levels $x$. The exact calculations for small clusters~\cite{KocharianPRB2008} provide additional evidence that within a relevant range of parameters, the effective magnetic moment can achieve the saturation even for atomic-size clusters of different shapes.
\begin{figure}[!hbt]\centering
   \subfigure[]{
      \includegraphics[width=0.5\columnwidth]{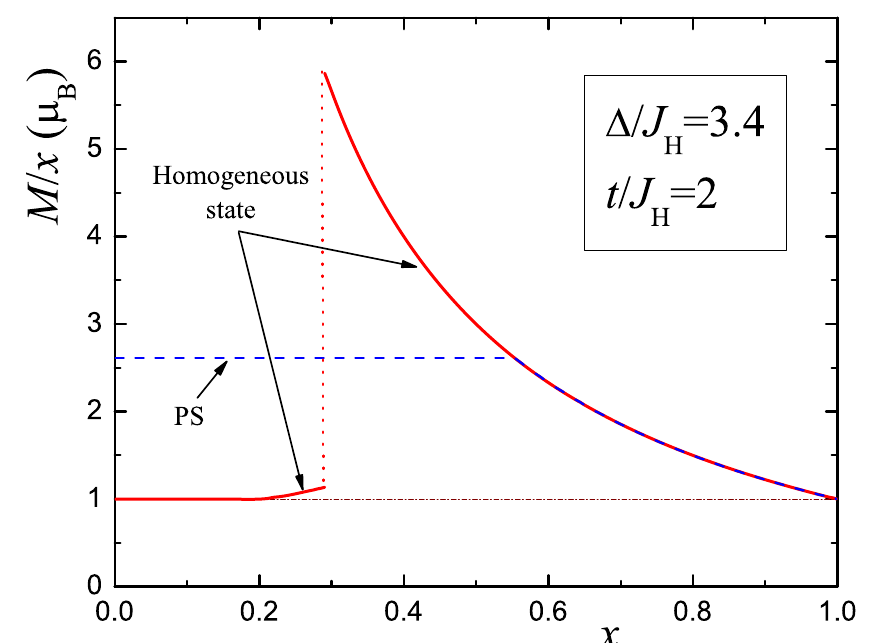}}
   \subfigure[]{
      \includegraphics[width=0.5\columnwidth]{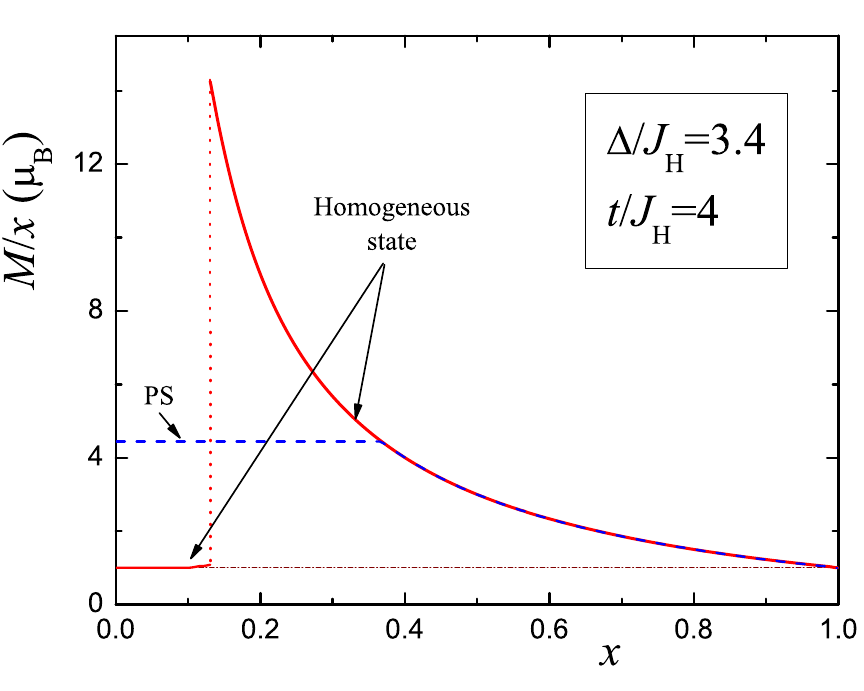}}
\caption{(Color online) The dependence of effective magnetic moment $M$ per Co$^{4+}$ ion on doping level $x$ at two values of the hopping integral $t$: (a) $t/J_H=2$; (b) $t/J_H=4$. The magnetic moment for the homogeneous states is plotted by red curves. The blue dashed line illustrates the behavior of $M$ corresponding to the phase-separated state~\cite{SboychakovPRB2009}.} \label{FigM_xLS} \end{figure}

Summarizing the above discussion, we can say that  the spin-state transitions in cobaltites with the hole doping can be be analyzed in the framework of the Falicov--Kimball type model treating an interplay of localized and itinerant electron states. In addition to the effects described by the similar model in the case of manganites~\cite{KugelPRL2005,SboychakovPRB2006jahn}, we reveal the possibility of a jumplike transition to the state, in which all charge carriers become delocalized. In respect to cobaltites, this means than all Co$^{3+}$ ions turn out to be in the intermediate-spin state. Moreover, before attaining such homogeneous state with itinerant electrons, the system has the doping range favorable for the phase separation, in which IS Co$^{3+}$ ions are located within nanoscale inhomogeneities. This picture is in an agreement with experimental data on La$_{1-x}$Sr$_x$CoO$_3$ system~\cite{CaciuffoPRB1999,LoshkarevaPRB2003, PhelanPRL2006,
LeightonPRB2006,LeightonPRB2007,PodlesnyakPRL2008}.

\subsection{The case of HS--HS ground state for isolated ions} \label{HS-HS}

As we have mentioned in the beginning of this section, a considerable energy gain can be also achieved at $\Delta < 2J_H$, corresponding to the high-spin state for  Co$^{3+}$ ions. In this case, the electron hopping  between Co ions becomes possible if Co$^{4+}$  passes to the IS state by promoting a hole to the $e_g$ level.

Thus, we can have the electron transfer from HS Co$^{3+}$
to IS Co$^{4+}$, or equivalently the transfer if a hole from IS Co$^{4+}$ to HS Co$^{3+}$. Here, the relevant vacuum state is the HS state of Co$^{4+}$, and we can write the corresponding relations by the analogy with  \eqref{states} and \eqref{ISstate} \begin{eqnarray}\label{statesHS} |0\rangle
&=&|\textrm{Co}_{HS}^{4+}\rangle, \quad \tilde{E}^{(vac)} = E_1
+2\Delta -6J_H =\tilde{E}_0,
\nonumber \\
\tilde{c}_{\bf{n}}^{\dag}|0\rangle &=&|\textrm{Co}^{5+}\rangle, \quad E^{(5+)} = \tilde{U}',
\nonumber \\
\tilde{a}_{\bf{n}}^{\dag}|0\rangle &=&|\textrm{Co}_{HS}^{3+}\rangle,
\quad E^{(e)} = E_0+2\Delta-4J_H = \tilde{E}_1, \nonumber \\
\tilde{a}_{\bf{n}}^{\dag}\tilde{c}_{\bf{n}}^{\dag}|0\rangle &=&|\textrm{Co}_{IS}^{4+}\rangle,
\, E_{IS}^{(4+)} = E_0 + \Delta -2J_H = \tilde{E}_2.
\end{eqnarray}

By making the substitution $E_0, E_1, E_2, U \rightarrow \tilde{E}_0, \tilde{E}_1, \tilde{E}_2, \tilde{U}$ and also $n^e_{{\bf n}} \rightarrow \tilde{n}^h_{{\bf n}}, n^h_{{\bf n}} \rightarrow \tilde{n}^e_{{\bf n}}$ in \eqref{H_onsite1} and \eqref{H_onsite2}, we obtain the single-site Hamiltonian. Then, instead of  Hamiltonian~\eqref{H}, we get
 \begin{equation}\label{H_HS}
H =\sum_{\bf{n}}[\tilde{E}_0 + (\tilde{E}_0-\tilde{E}_1-\mu)
(\tilde{n}_{\bf{n}}^e -\tilde{n}_{\bf{n}}^h)] + +\Delta_2\sum_{{\bf n}}\tilde{n}^h_{{\bf n}}+
\tilde{U}\sum_{{\bf n}}\tilde{n}^h_{{\bf n}}(1-\tilde{n}^n_{{\bf n}}) -t\sum_{\langle{\bf n}{\bf m}\rangle}\left(
\tilde{c}^{\dag}_{{\bf n}}\tilde{c}_{{\bf m}}+h.c.\right)\, .
\end{equation}
In this Hamiltonian, $\Delta_2 = 4J_H -\Delta$ is the difference between energies of IS and HS of Co$^{4+}$ ions,
$\tilde{c}^{\dag}_{{\bf n}}$, $\tilde{c}_{{\bf n}}$ are the creation
and annihilation operators describing a hole exited to the $e_{g}$
level of IS Co$^{4+}$ ion located at site $\bf n$, $\tilde{n}^h_{{\bf n}}= \tilde{c}^{\dag}_{{\bf n}}\tilde{c}_{{\bf n}}$, and
$\tilde{n}^e_{{\bf n}} = \tilde{a}^{\dag}_{{\bf n}}\tilde{a}_{{\bf
n}}$ is the operator corresponding to the number ($0$ or $1$) of
additional $t_{2g}$ electrons located at site ${\bf n}$
($\tilde{a}^{\dag}_{{\bf n}}$ and $\tilde{a}_{{\bf n}}$ are the creation and annihilation operators for such electrons). The relation between the average have now the form $\tilde{n}^e - \tilde{n}^h = 1-x$.

Instead of expression \eqref{E}, we have now for the energy per site  \begin{equation}\label{E2}
E^{(2)} = E_0(1-x)+ E_1x + \langle H_2 \rangle /N  \, ,
\end{equation}
where
\begin{equation}\label{H1a}
H_2= \sum_{{\bf n}}\left(2\Delta - 6J_H +
2J_H(\tilde{n}_{\bf{n}}^e -
\tilde{n}_{\bf{n}}^h)\right)+ \Delta_2\sum_{{\bf n}}\tilde{n}_{\bf{n}}^h  +
\tilde{U}\sum_{{\bf n}}\tilde{n}^h_{{\bf n}}
(1-\tilde{n}^n_{{\bf n}})-t\sum_{\langle{\bf n}{\bf m}\rangle}\left(
\tilde{a}^{\dag}_{{\bf n}}\tilde{a}_{{\bf m}}+h.c.\right)\, .
\end{equation}
It is important that the energy difference $E^{(2)} - E^{(1)}$ is independent of the specific values of $E_0$ and $E_1$, and this will be useful below for the comparison of phase diagrams in different ranges of parameters.

As a result, we can see that in the case under study in this subsection, the energy of the system and charge carrier densities, $\tilde{n}^e$ and $\tilde{n}^h$, exibit the behavior analogous to that depicted in in Figs.~\ref{FigE_LS} and~\ref{FigN_LS}. Note, however, that we should make the following replacement: $n^h\to\tilde{n}^e$, $n^e\to\tilde{n}^h$, and $x \to 1-x$. This means that the densities of Co$^{3+}$ ions in IS ($n^e=n_{\text{IS,Co}^{3+}}$) and LS ($1-n^h=n_{\text{LS, Co}^{3+}}$) states are substituted by the densities of Co$^{4+}$ ions in IS ($\tilde{n}^h=n_{\text{IS, Co}^{4+}}$) and HS
($1-\tilde{n}^e=n_{\text{HS, Co}^{4+}}$) states. Such exact mapping of the LS--LS case to HS--HS one is a consequence of the parallel arrangement of the spins of charge carriers. This implies the electron--hole symmetry, that is, the similarity of an empty $e_g$ level at LS Co$^{3+}$ and the filled $e_g$ level at HS Co$^{3+}$.

\subsection{Phase diagrams}

The above analysis allows us to illustrate the  the behavior of the systems under study with the variation of doping at different values of the characteristic parameter $\Delta/J_H$ and the form of resulting phase diagrams. Such phase diagrams are quite different at different values of the hopping integral $t$, see in Fig.~\ref{PhDiaHom}, where the possible homogeneous states are depicted. In Fig.~\ref{PhDiaHom}a corresponding to small values of $t$ ($t/J_H \lesssim 1$), there are distinct regions of the phase diagram revealing the situation at $\Delta > 3J_H$ and $\Delta < 2J_H$ (see subsections \ref{LS-LS} and \ref{HS-HS}, respectively). Each of these regions exhibits the evolution of the system with doping, at which the phase involving only localized charge carries transforms to the phase, where the localized and itinerant charge carriers coexist, and finally, we have the phase with the complete delocalization of charge carriers. Between the regions corresponding to  $\Delta > 3J_H$ and $\Delta < 2J_H$, there exists  the phase with Co$^{3+}$ in LS ($S=0$) and Co$^{4+}$ in HS ($S=5/2$) states, with the charge carriers localized owing to the spin blockade~\cite{MaignanPRL2004_spblock}.

With the growth of $t$ (at $t/J_H \gtrsim 1$), see Fig.~\ref{PhDiaHom}(b), the intermediate phase shrinks and there appears the possibility of a direct spin-state transition between the phases with purely itinerant charge carriers.

\begin{figure}[H]
\centering
   \subfigure[$\quad t/J_H = 1$]{
      \includegraphics[width=0.5\columnwidth]{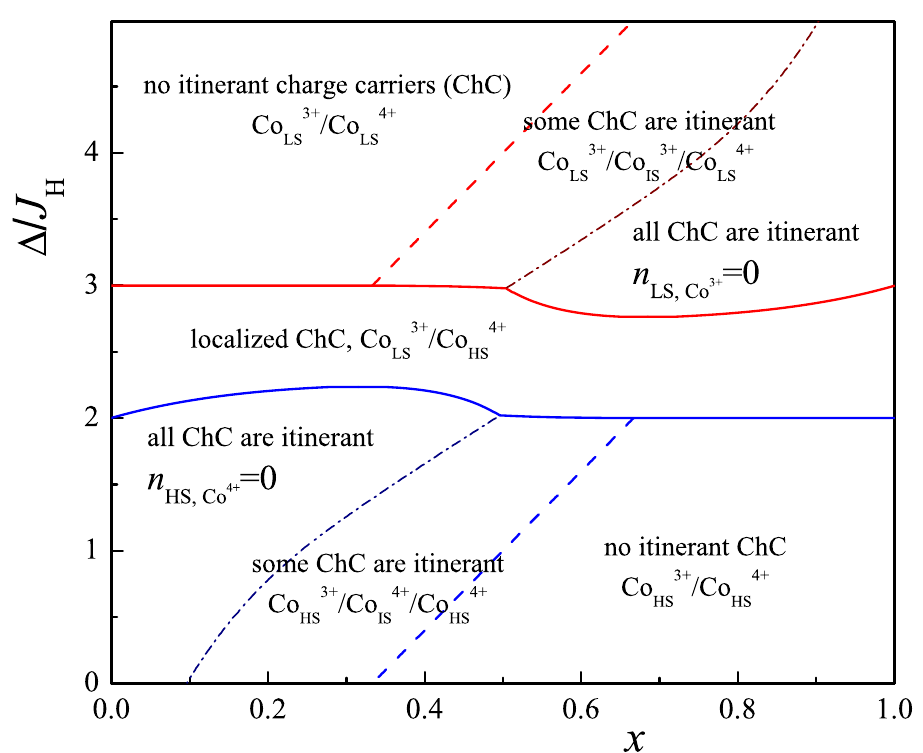}}
   \subfigure[$\quad t/J_H = 1.5$]{
      \includegraphics[width=0.5\columnwidth]{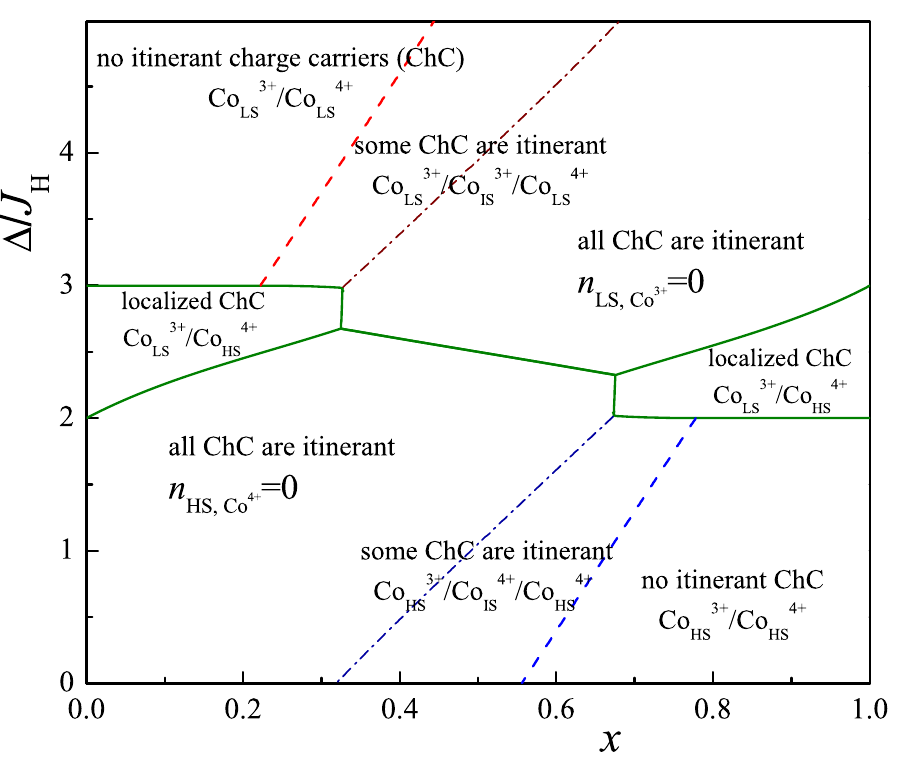}}
   \caption{(Color online) Phase diagram illustrating the range of existence for the possible homogeneous states in the system
with spin-state transitions at two characteristic values of of the hopping integral $t$: (a)$t/J_H=1$  and (b) $t/J_H=1.5$. The phase boundaries for the similar phases in the upper and lower parts of both panels are shown denoted by the same lines (solid, dashed, or dot-and-dash)~\cite{SboychakovPRB2009}. ChC is the abbreviation for charge carriers.}
\label{PhDiaHom}
 \end{figure}

\begin{figure} [H]
\centering
   \subfigure[$\quad t/J_H = 1$]{
      \includegraphics[width=0.5\columnwidth]{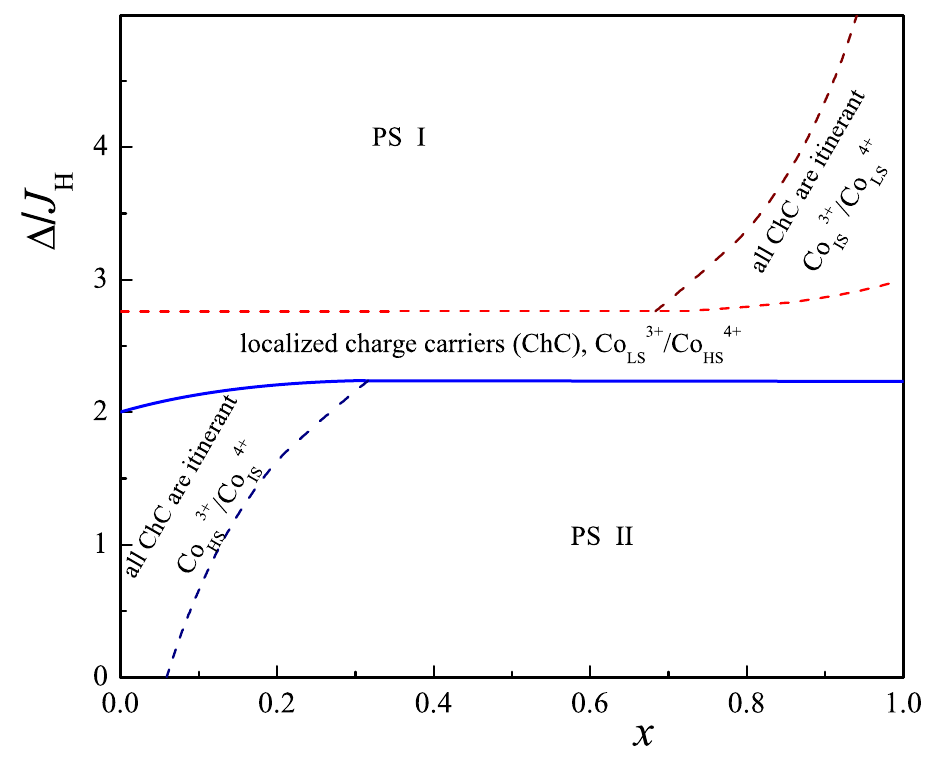}}
    \subfigure[$\quad t/J_H = 1.3$]{
      \includegraphics[width=0.5\columnwidth]{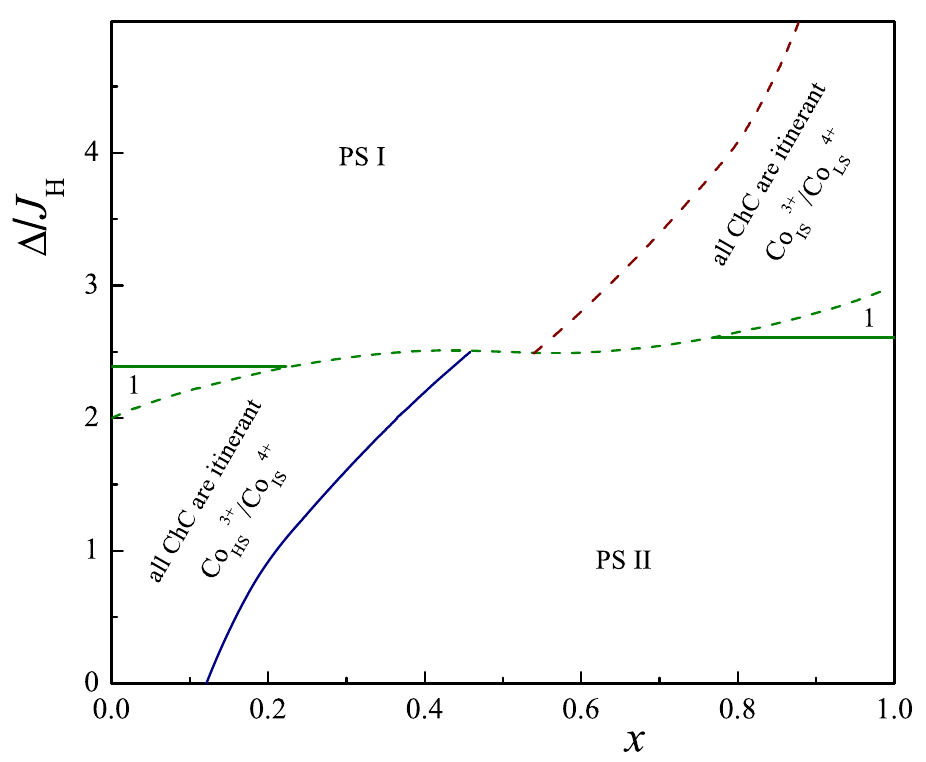}}
   \subfigure[$\quad t/J_H = 1.5$]{
      \includegraphics[width=0.5\columnwidth]{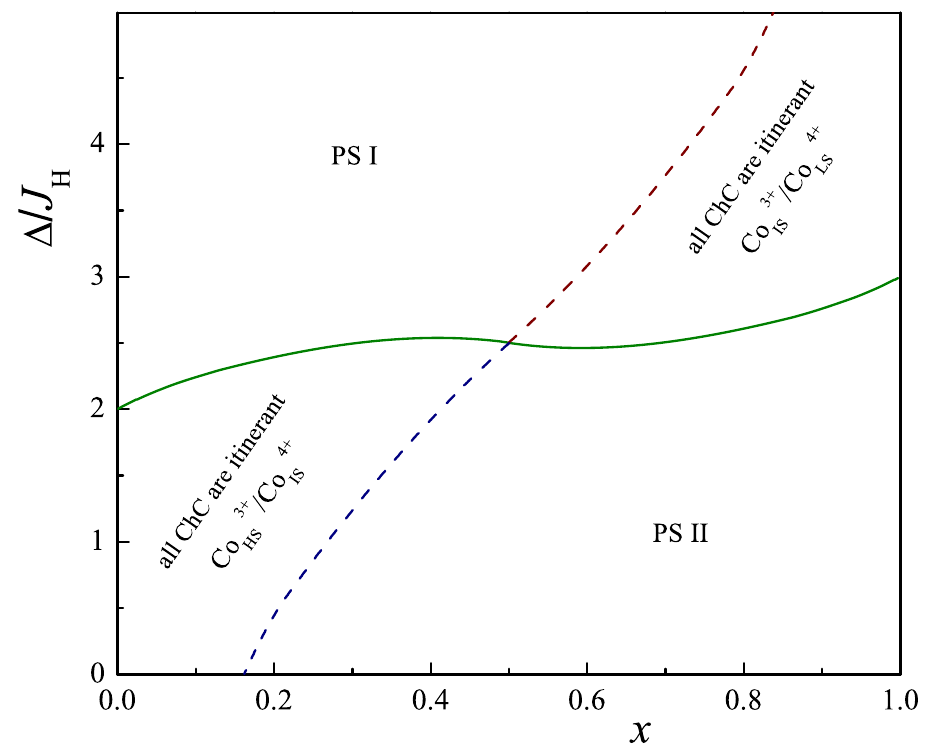}}
\caption{(Color online) Phase diagrams involving the phase-separated states in the system with spin-state transitions at at three characteristic values of of the hopping integral $t$: (a) $t/J_H=1$; (b) $t/J_H=1.3$; (c) $t/J_H=1.5$. PS I is the the state with the phase separation into the domains without itinerant charge carriers, corresponding (with LS Co$^{3+}$), and those containing purely itinerant charge carriers corresponding to IS Co$^{3+}$. PS II is the the state with the similar type of the  phase- separated, state, where the domains with and without delocalized charge
carriers correspond to IS Co$^{4+}$ and HS Co$^{4+}$, respectively. Regions 1 in panel (b) correspond to charge carriers located at HS Co$^{4+}$ and LS Co$^{3+}$, for which the electron hopping between Co ions is prohibited owing to the spin blockade~\cite{SboychakovPRB2009}.}
\label{PhDiaInhom} \end{figure}

If we take into account that the phase separation is possible, this leads to the drastic changes in the phase diagram. In Fig.~\ref{PhDiaInhom}, we present the resulting phase diagrams
at different values of parameter $t/J_H$. In comparison to the phase diagram with only homogeneous states, here the range of phase separation occupies a considerable part of the phase diagram. Within this range, we have an inhomogeneous state with intermixed domains with either localized or delocalized charge carriers. For small $t$ ($t/J_H \lesssim1$), there is again intermediate region corresponding to charge carries localized at any doping level (that is, we have with LS Co$^{3+}$ and HS Co$^{4+}$ ). An increase in $t$ leads to the shrinking of this intermediate region. At $t/J_H\cong1.24$, there appear two separated such regions [see Fig.~\ref{PhDiaInhom}(b)], and a direct boundary between PS I and PS II phases. With the further increase in $t$, these two regions of the localization are gradually vanishing and do not exist at $t/J_H > 1.44$.

Note again that the change in $\Delta$ axis can be implemented in cobaltites by varying the average ionic radius of rare-earth elements (see, e.g. Refs.~\cite{FujitaJPSJap2005}, \cite{WangPRB2006}).

Above, we have considered the $U\to\infty$ limit. This means that the spin--spin, spin--orbital, and orbital--orbital correlations between neighboring sites, which take place in two-band Hubbard Hamiltonians~\eqref{H} and~\eqref{H_HS}, are not taken into account. Such correlations can affect the actual types of magnetic and orbital orderings. However, the corresponding terms are of the order of $t^2/U\ll t,\,\Delta,\,J_H$. Hence, they are rather small and should not significantly change the phase diagrams. Using the mean-field approach, we also neglect the fluctuation effects, but it is known that such approach captures the main qualitative features of the phase diagrams.

The long-range Coulomb interaction related to the charge inhomogeneity in the phase-separated state can reduce the doping range corresponding to the phase separation and affect the size of  inhomogeneities. The minimization of both the long-range Coulomb repulsion and the surface energy provides an opportunity ro reveal the geometry of the phase-separated state (e.g., the inhomogeneities can take the shapes of droplets, alternating layers, or rods) and to determine the characteristic sizes of inhomogeneities and the percolation threshold for the doping, at which the insulator--metal
transition occurs~\cite{KugelPRL2005,LorenzanaEPL2002,KugelSuST2008}.

\subsection{Conclusions}

In this section, we have formulated a simple model capturing the main physics of strongly correlated electron systems with spin-state transitions. Based on this model, we predict the possibility of electronic phase separation for perovskite cobaltites under doping. We determined the corresponding phase diagrams exhibiting broad ranges of the phase separation. The specific shape of phase diagrams exhibits significant changes with the variation of the ratio of the electron hopping integral $t$ and and the Hund's rule coupling constant $J_H$.

We did not analyze in detail the possible structure of the
phase-separated state. However, as we have discussed in the previous sections in the case manganites, the calculations~\cite{KugelPRL2005} and numerical simulations~\cite{ShenoyPRL2007} taking into account the surface and long-range Coulomb contributions to the total energy
lead to the characteristic size of nanoscale inhomogeneities of
the order of several lattice constants. One can expect that the
inhomogeneities caused by the phase separation in cobaltites with
spin-state transitions would be of a similar scale.

Note that the existence of the electronic phase separation in the systems with the spin-state transitions is rather robust phenomenon appearing in different models. For example, the numerical studies reported in \cite{IshiharaPRB2009} revealed a possibility of the phase separation in the two-band model with the bands differing in the spin state. In such a model, the phase separation appears upon doping if the widths of these band are comparable to each other.

It is also worth noting rather active line research related to the photoinduced spin-state transitions and the formation of the corresponding inhomogeneous states, which can be revealed,for example, by the pump-probe techniques. In particular, a bound state of optically exited hole and the HS state can be created in cobaltites within the LS phase. This bound state brings about a characteristic peak structure in the optical pump-probe spectra which is distinct from the spectra in thermally excited states~\cite{IshiharaPRL2011,IshiharaEJPSpTop2013}. This brings us a clue for the new optical spin control technique.

The nanoscale phase separation in materials with spin state transitions can involve some additional degrees of freedom. The detailed studies of La$_{2-x}$Sr$_x$CoO$_4$ layered cobaltites near $x =1/3$ in the experiments including the resonant micro X-ray diffraction, neutron scattering, and muon spin rotation spectroscopy revealed nanosize undoped and hole doped islands involving the charge order correlations in addition to the magnetic ones \cite{DreesNatCom2014,GuoPRB2019}.

\section{Effects of orbital degrees of freedom, different shapes of polarons}
\label{Orbitals}
\subsection{Introduction}\label{Intr2}

In this section, we study the effects related to one more degree of freedom, namely with orbitals and their ordering. The orbital ordering (OO) plays an important role in the physics of magnetic oxides, especially of those, which contain the ions with the orbitally degenerate ground state, so called Jahn--Teller (JT) ions. The compounds with JT ions usually exhibit different kinds of superstructures (OO)~\cite{KugKhomUFN1982,KaplanVekhterBook2012}. In fact, the OO is a specific feature of magnetic insulators. The additional charge carriers introduced by doping are able to transfer an OO insulator to the metallic state. Hence, here we are dealing again with the competition between the charge carrier localization and their itinerancy, which can give rise to the phase separation, especially at low doping levels~\cite{DagottoBook2003,NagaevBook2002,KaganUFN2001}. In the analysis of phase separation in magnetic materials such as manganites, the OO is usually not addressed (except maybe the treatment of orbital and magnetic polarons~\cite{KhaliullinPRB1999,KaganUFN2001,KhalOkamotoPRL2002,
vdBKhaKho_inBook2004}). In this section, we study the role of OO for the electronic phase separation employing minimal models including itinerant charge carriers interacting with the OO superstructure. We demonstrate that at a low doping level, the orbitally ordered systems can exhibit the phase separation, which not directly related to the possible magnetic structure.

Such phase separation is treated using two approaches. In subsection~\ref{SymModel}, we base our analysis on a symmetrical model analogous to the considered above double exchange model where the orbital variables play a role of local spins. We assume that owing to the orbital degeneracy, localized electrons induce lattice distortions of JT type, thus giving rise to OO. Itinerant charge carriers appearing due to the doping move on OO background. Within the other approach (subsection~\ref{AnisModel}), we employ the $e_g$ symmetry of the wave functions for the doped charge carriers. Using both approaches, we reveal the tendency to the phase separation related to the existence of the orbital degrees of freedom. In other words, itinerant charge carriers introduced by doping stimulate the formation of inhomogeneous states with nanosize domains having the orbital order differing from that taking place in the absence of doping. In subsection~\ref{Inhom}, we analyze the possible shapes and sizes of such domains and demonstrate that at different relations between the electron hopping integral $t$ and the parameter $J$ characterizing the interaction between orbitals, such shapes can be quite different. For example, in the layered compounds, the two-dimensional case, below some critical value of of $t/J$, nearly circular domains can undergo the transformation to the needle-like ones. This is a consequence of anisotropic charge distribution characteristic of specific orbitals, which brings forward additional possibilities to the physics of phase separation.

In subsection~\ref{Model}, we are dealing with a picture more relevant to the actual JT compounds, in which the same charge carriers are involved both in OO and, after doping, in conduction. We assume that at zero doping, the orbitals are ordered due to the electron--lattice coupling. The doped charge carriers are moving on the OO background. In subsection~\ref{e_g}, we consider another model, where the same $e_g$ electrons are responsible for the band structure and for the orbital ordering.

\subsection{Symmetrical model}\label{SymModel}

Here, we analyze the situation typical of compounds JT ions having a double-degenerate ground state. Different kinds of lattice distortions (e.g. compression or stretching) lift such degeneracy making favorable two possible ground states for each ion, say, either $a$ or $b$. The $a$ and $b$ states at lattice site $\mathbf{n}$ are, describe, in fact, the occupation of electron orbitals at this site. Usually we deal with the linear combination of $a$ and $b$ orbitals, which can be expressed in terms of angle $\theta$ \begin{equation}\label{theta}
|\theta\rangle=\cos\frac{\theta}{2}|a\rangle+\sin\frac{\theta}{2}|b\rangle\,.
\end{equation}
The orbitals and the related local distortions at different sites can form regular arrays (orbital ordering). In the case of two basis states, the orbitals can be put into correspondence to some spin-1/2 variables, and the interaction between orbitals in its simplest version can be written in the form of Heisenberg Hamiltonian
\begin{equation}\label{H_Heis}
H_{\text{OO}}=J\sum_{\langle{\bf n}{\bf m}\rangle}\bm{\tau}_{\bf
n}\bm{\tau}_{\bf m}\,, \end{equation}
where
$\bm{\tau_n}=\{\tau^x_{\mathbf{n}}$, $\tau^z_{\mathbf{n}}\}$ are
the Pauli matrices, and $a$ and $b$ states of the ion ${\bf n}$
correspond to eigenvectors of operators $\tau^z_{\bf n}$, with
eigenvalues $1$ and $-1$, respectively. Hamiltonian~\eqref{H_Heis}, provides a possibility of two simplest types of orbital ordering: ferro-OO (similar orbitals at each site) and antiferro-OO (alternating orbitals at neighboring sites). At zero doping (no itinerant charge carriers), the system can have either the antiferro-OO ground state (at $J>0$) or the ferro-OO one (at $J<0$). In actual materials containing Jahn--Teller ions, orbital Hamiltonians are not so simple. Nevertheless, model~\eqref{H_Heis} appears to be sufficient for capturing main  physical effects related to the orbital ordering.

Let us now pass to the case of nonzero doping and hence of nonzero density of itinerant charge carriers in the system, $n\neq 0$. These itinerant charge carriers are moving on the background of orbitally ordered state formed by localized electrons. (Below, we demonstrate that the main qualitative results remain the same, if itinerant  charge carries such as $e_g$ electrons, are responsible both for the  orbital ordering and for the electrical conductivity). The electron hopping integrals are determined by the specific electron orbitals located at the neighboring lattice sites. The Hamiltonian describing only the electron states have the form
\begin{equation}\label{H_el}
H_{\text{el}}=-\sum_{\langle{\bf n}{\bf
m}\rangle,\alpha,\beta,\sigma}t^{\alpha\beta}\left(P_{{\bf
n}\alpha\sigma}^{\dag}a^{\dag}_{{\bf n}\alpha\sigma}a_{{\bf
m}\beta \sigma}P_{{\bf m}\beta\sigma}+h.c.\right), \end{equation}
where, $a^{\dag}_{{\bf n}\alpha\sigma}$, $a_{{\bf n}\alpha\sigma}$
are creation and annihilation operators cottespondingg to charge carriers at site ${\bf n}$ with spin projection $\sigma$ at orbital $\alpha$. Projection operators $P$ in \eqref{H_el} take into account that the  double occupancy of the lattice sites by electrons is highly unfavorable in energy due to the strong on-site Coulomb repulsion. Below, we are dealing mostly with the case $n<1$. In this section, our analysis of the electron contribution to the total energy is based on the tight-binding approximation. To avoid unnecessary complications, we illustrate the essential physics of orbitally ordered systems considering the square lattice in the two-dimensional (2D) case and simple cubic lattice in the three-dimensional (3D) case. In the tight-binding approximation, the spectrum of charge carriers has the form
\begin{equation}\label{E(k)}
E^{\alpha\beta}({\bf k})=-t^{\alpha\beta}\frac{1}{D}\sum_{i=1}^D\cos k_i
=t^{\alpha\beta}\xi({\bf k})\,,
\end{equation}
where $D$ (2 or 3) is the number of spatial dimensions. Components $k_i$ of the wave vector ${\bf k}$ are expressed in the units of inverse lattice constant $1/d$.

Within the approach adopted in this section, the charge carriers introduced by doping are built in to the JT distorted background and occupy orbital states $|a\rangle$ or $|b\rangle$. Then, their intersite charge transfer is described by three hopping integrals: $t^{aa}$, $t^{bb}$, and $t^{ab}=t^{ba}=t'$. For simplicity, we can assume that $t^{aa}=t^{bb}=t$. Such system is characterized by an interplay two contributions to the total energy: the possible gain in the kinetic energy can come from the increase in the electron bandwidth, whereas the potential energy can be reduced by the optimization of the orbital ordering.

In a more general case, when there is a superposition of $|a\rangle$ or $|b\rangle$ states at site $\mathbf{n}$  characterized by angle  $\theta$, (see,  by Eq.~\eqref{theta}), we can write the expression for the hopping integral between the sites characterized by orbital states $|\theta_1\rangle$ and $|\theta_2\rangle$ \begin{equation}\label{t_theta}
t^{\theta_1\theta_2}=t\cos{\frac{\theta_1 -\theta_2}{2}}
+t'\sin{\frac{\theta_1+\theta_2}{2}}
\end{equation}
This is in fact a generalization to the OO case of the semiclassical  expression for the effective hopping integral in the double exchange model~\cite{deGennesPR1960}.

We shall start from the homogeneous phase, in which orbitals  $|\theta_1\rangle$ and $|\theta_2\rangle$ alternate at different lattice sites ($J>0$). Within the mean-field approach, the total energy per site is given by the expression
\begin{equation}\label{Etot1}
E_{tot}(\theta_1,\theta_2)=zt^{\theta_1\theta_2}\varepsilon_0(n)
+\frac{zJ}{2}\cos(\theta_1-\theta_2)\,,\;\;\varepsilon_0(n)<0\,.
\end{equation}
Here, $z$ is the number of nearest-neighbor sites and $\varepsilon_0(n)$ is the dimensionless kinetic energy, which depends  on the type of crystal lattice. In this subsection, we consider an isotropic case, for which $t$, $t'$, and, hence, $t^{\theta_1\theta_2}$, are independent on the direction in space. For such isotropic system, $\varepsilon_0(n)$ does not depend on
$\theta_{1,2}$, and we can easily calculate the orbital structure
by minimization of total energy~\eqref{Etot1} with respect to
angles $\theta_1$ and $\theta_2$. At high doping levels, namely for
$t|\varepsilon_0(n)|>2J$, the ferro-OO state with
$\theta_1=\theta_2=\pi/2$ becomes favorable. In the opposite limiting case,
$t|\varepsilon_0(n)|<2J$, the orbital structure corresponds to
$\theta_2=\pi-\theta_1$, and
\begin{equation}\label{thetaMin}
\theta_1=\arcsin\left(\frac{t|\varepsilon_0(n)|}{2J}\right)\,,
\;\;\frac{t|\varepsilon_0(n)|}{2J}<1\,. \end{equation}
Such alternation of orbitals can be interpreted as a kind of canting for the directions of pseudospins $\tau$ in the orbital space. For the energy of such state, we have
\begin{equation}\label{Etot2}
E_{tot}=zt'\varepsilon_0(n)-\frac{zt^2}{4J}
\varepsilon_0^2(n)-\frac{zJ}{2}\,. \end{equation}
If it is possible to represent the kinetic energy as
$\varepsilon_0(n)=nf(n)$, where $f(n)$ a slowly changing function of $n$, then $E_{tot}$ can exhibit a negative curvature, at least, at small $n$, which signals on an instability of a homogeneous orbitally ordered state (negative compressibility).

We can find a specific form of $\varepsilon_0(n)$ and determine the shape of the curve depicting the behavior of the total energy with doping. For the electron energy ~\eqref{E(k)}, the density of states $\rho_0(E)$, can be written as
\begin{equation}\label{rho0}
\rho_0(E)=\!\int\!\frac{d\mathbf{k}}{(2\pi)^D}\delta(E-\xi(\mathbf{k}))=%
\!\int\limits_0^{\infty}\!\frac{ds}{\pi}\cos(Es)J_0^D\!\left(\frac{s}{D}\right)\!,
\end{equation}
where $J_0$ is the Bessel function. The expression for $\varepsilon_0(n)$ n has the form
\begin{equation}\label{epsilon}
\varepsilon_0(n)=\int_{-1}^{\mu(n)}\!\!\!\!\!dE\,E\rho_0(E)\,,
\end{equation}
where the chemical potential $\mu$ can be found based on the usual relation $n=\int_{-1}^{\mu}dE\rho_0(E)$.

At small density of itinerant charge carriers, $n\ll1$, we can write an explicit expression for $\varepsilon_0(n)$. For example, in a 2D case, we have $\varepsilon_0(n)\approx-n+\pi n^2/2$, and the corresponding expression for the total energy is
\begin{equation}\label{Etot2Da}
E_{tot}\approx-zt'n-z\left(\frac{t^2}{4J}-\frac{\pi
t'}{2}\right)n^2-\frac{zJ}{2}\,.
\end{equation}
Using Eq.~\eqref{Etot2Da}, we can find the range of parameters corresponding to the negative curvature of the energy plot, i.e. $d^2E_{tot}/dn^2 < 0$. Such situation takes place if
\begin{equation}
\frac{t}{J}>\frac{2\pi t'}{t},
\end{equation}
This means that the homogeneous state with canted orbitals is unstable with respect to the phase separation into a inhomogeneous states with ferro- and antiferro-orbital ordered domains. Here, we have a close analogy to the situation typical of the double exchange (in the simplest single-band case)~\cite{KaganEPJB1999}, which corresponds to $t'=0$. In the opposite limiting case, that is, for $2\pi t'/t>t/J$, a homogeneous state is favorable within the whole doping range -- in contrast to the usual double exchange.

A rough estimate for the phase separation range can be found if we neglect the quadratic term in Eq.~\eqref{thetaMin}, $\varepsilon_0(n)\approx-n$. Thus, we find
\begin{equation}\label{une1}
0<n\lesssim\frac{2J}{t}.
\end{equation}
Actually, the homogeneous orbitally canted state appears to be unfavorable in energy within almost the whole doping range, where the difference
$\theta_2-\theta_1=\pi-2\theta_1$ with $\theta_1$ from
Eq.~\eqref{thetaMin} does not vanish. The situation does not qualitative change, if we use more accurate calculations for $E_{tot}$ based on $\varepsilon_0(n)$ determined employing density of states found from~\eqref{rho0}. The corresponding $E_{tot}(n)$ plot in the 2D case is shown in Fig.~\ref{FigEtot2D}.

\begin{figure} [H]
\begin{center}
\includegraphics*[width=0.45\textwidth]{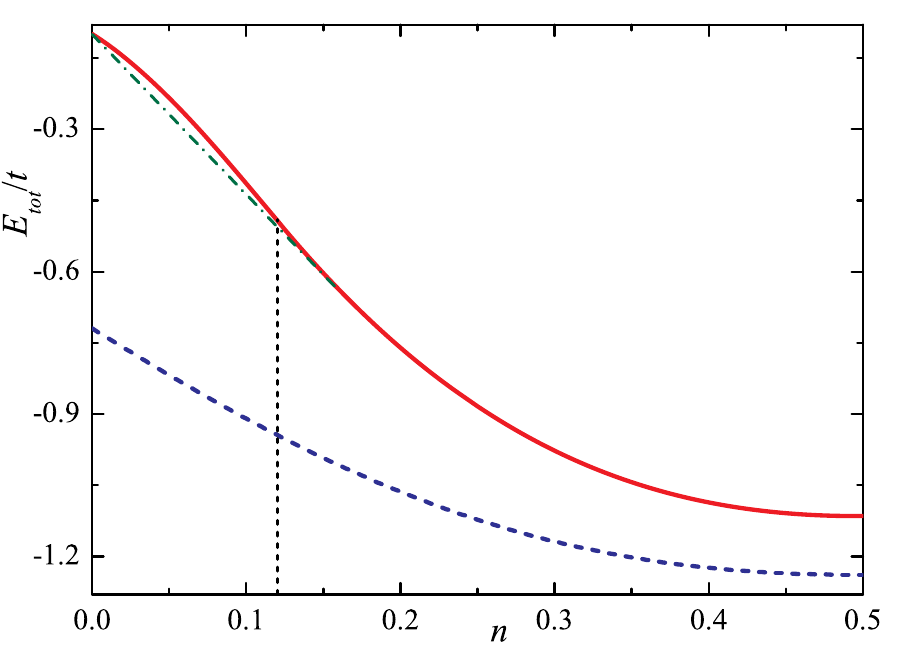} \end{center}
\caption{\label{FigEtot2D} (Color online) Total energy of homogeneous orbitally ordered state state ~\eqref{Etot2} in 2D versus doping $n$. Two types of behavior are shown: with the range corresponding to the negative curvature (red solid line, $J/t=0.05$), and without it (blue dashed line, $J/t=0.35$); $t'=0.5t$ for both curves. The range of instability for the homogeneous state (with the negative curvature) is located on the left-hand side of vertical line. The energy of the phase-separated state is drawn by the dot-and-dash line. The blue dashed curve represents the energy of the homogeneous state, which at $J/t=0.35$ is stable within the whole doping range~\cite{KugSboKhomPRB2008orbitals}.}
\end{figure}

In the 3D case, the situation is not so simple. At low doping levels, the corresponding expression for the kinetic energy has the form $\varepsilon_0(n)\approx-n+an^{5/3}$, where $$
a=\frac35\left(\frac{\pi^2}{\sqrt{6}}\right)^{2/3}\!\!\!, $$ and
the total energy is
\begin{equation}\label{Etot3Da}
E_{tot}\approx-zt'n-z\left(\frac{t^2}{4J}-\frac{a}{n^{1/3}}
\right)n^2-\frac{zJ}{2}\,.
\end{equation}
The second derivative of $E_{tot}$ is positive at $n\to0$ and becomes negative at
\begin{equation} \label{n-crit}
n_c\approx\left(\frac{5aJt'}{9t^2}\right)^3. \end{equation}
The same line of reasoning as that in the 2D case leads to the following estimate for the phase-separation range
\begin{equation}\label{une2}
n_c\lesssim n\lesssim\frac{2J}{t}. \end{equation}
According to Eq.~\eqref{n-crit}, a nonzero value of the nondiagonal hopping integral $t'$ manifests itself in a nonzero critical concentration $n_c$ for phase separation. The phase separation range obtained by the Maxwell construction is in fact broader and starts from some $n_0<n_c$. Here, the orbital state is described by a classical vector, but nevertheless we have the lower critical concentration for the canted state and hence for the phase s eparation, which in the usual double exchange model appears only, if we take into account the quantum nature of core spins~\cite{KaganEPJB1999}. Inequalities~\eqref{une1} and~\eqref{une2} are approximate ones and are valid only at quite small values of parameter  $J/t$.

\subsection{Anisotropic model} \label{AnisModel}

Let us now analyze the effects related to the specific form of $e_g$ orbitals. To make the underlying physics clearer, we discuss in this subsection the two-dimensional case (square lattice). This case is relevant, e.g., to layered cuprates, like K$_2$CuF$_4$, or manganites (La$_2$MnO$_4$ or La$_2$Mn$_2$O$_7$). Another simplification is that we choose the Heisenberg-like form~\eqref{H_Heis} for the orbital exchange. Such simplification do not lead to a loss in generality. At least, in the analysis based on the superexchange mechanism of the orbital ordering~\cite{KugKhomUFN1982} in the case of of $e_g$ orbitals demonstrates the the main physics of the phase separation discussed below remains qualitatively the same.

As we have mentioned above, any wave function can be written as a linear combination of basis orbitals. For  $e_g$ orbitals, we have $|a>=|x^2 - y^2>$ and $|b>=|2z^2 - x^2 - y^2>$ and
$|\theta>=\cos(\theta/2)|x^2-y^2>+\sin(\theta/2)|2z^2-x^2-y^2>$.
Hopping integrals $t^{\alpha\beta}$ in Eq.~\eqref{H_el} now
become anisotropic and can be presented in the matrix form
\begin{eqnarray}\label{t_xy}
(t_{x,y})^{\alpha\beta} =\frac{t_0}{4}\left(%
\begin{array}{cc}
3 & \mp\sqrt{3} \\
\mp\sqrt{3} & 1 \\
\end{array}%
\right)\,,
 \end{eqnarray}
where minus and plus signs correspond to the electron hopping along the axes $x$ and $y$, respectively.

Again, we choose the  the orbital ground state in the form of
alternating $|\theta_1\rangle$ and $|\theta_2\rangle$ orbitals.
In this case, the spectrum of charge carriers takes the form
\begin{equation}\label{specEg}
E(\mathbf{k})=-t_0\left(A_{x}(\theta_1,\theta_2)\cos
k_x+A_{y}(\theta_1,\theta_2)\cos k_y\right)\,,
\end{equation}
where \begin{equation}\label{Axy}
A_{x,y}(\theta_1,\theta_2)=\left|\cos\left(\frac{\theta_1-\theta_2}
{2}\right)+\cos\left(\frac{\theta_1+\theta_2}{2}
\pm\frac{\pi}{3}\right)\right|.
\end{equation}
The corresponding expression for the total energy is
\begin{eqnarray}\label{EtotAn}
E_{tot}(\theta_1,\theta_2)&=&t_0\left(A_{x}(\theta_1,\theta_2)
+A_{y}(\theta_1,\theta_2)\right)\times\nonumber\\%
&&\varepsilon(n;\theta_1,\theta_2)+2J\cos(\theta_1-\theta_2)\,,
\end{eqnarray}
where $\varepsilon(n;\theta_1,\theta_2)=
\int_{-1}^{\mu}dE\,E\rho(n;\theta_1,\theta_2)$,
and the electronic density of states has the form
\begin{equation}\label{rho12}
\rho(n;\theta_1,\theta_2)=\!\int\limits_0^
{\infty}\!\frac{ds}{\pi}\cos(Es)J_0\!\left(\frac{sA_x}{A_x+A_y}\right)
J_0\!\left(\frac{sA_y}{A_x+A_y}\right)\!. \end{equation}
Density of states \eqref{rho12} is the function of orbital angle $\theta_1$ and $\theta_2$ . Therefore, to determine the optimum orbital configuration, it is necessary to minimize $E_{tot}$, Eq.~\eqref{EtotAn}, with respect to angles $\theta_1$ and $\theta_2$. As a result of such minimization, we find that at low doping levels ($n$ less than some threshold value, $n_1$, which depends on parameter $J/t_0$), the ground state configuration corresponds to $\theta_{1}=0$, $\theta_{2}=\pi$. This is the homogeneous phase with alternating $|x^2-y^2>$ and $|2z^2-x^2-y^2>$ orbitals, i.e. the antiferro-OO. We do not take in to account here anharmonic terms and higher-order interactions, which favor locally elongated octahedra corresponding in our notation to the angles equal to $\theta=\pi$, $\pm 2\pi/3$, see
Refs.~\cite{KanamoriJAP1960,Khom_vdBrPRL2000}). After the minimization, the total energy takes the form
\begin{equation} E_{tot}=t_0\sqrt{3}\varepsilon_0(n)-2J\,.
\end{equation}
The plot of this energy has the positive curvature, $\partial^2E_{tot}/\partial n^2>0$, corresponding to the stability of such state.

At $n n_1$, the system passes in a jump-like manner to another state characterized by alternating orbital angles $\theta_2=-\theta_1$, where
\begin{equation}
\theta_1=\arccos\left(\frac{t_0|\varepsilon_0(n)|}{4J}\right)\,,
\end{equation}
and $E_{tot}(n)$ exhibits a kink at $n=n_1$. This canted state arising at $n>n_1$ has the energy
\begin{equation}\label{EtotEg}
E_{tot}=t_0\varepsilon_0(n)-\frac{t_0^2}{4J}\varepsilon_0^2(n)-2J\,.
\end{equation}

The angle $\theta_1$ decreases with the increase in $n$ and eventually vanishes at $n=n_2$. The value of $n=n_2$ can be found from the equation $t_0|\varepsilon_0(n_2)|/4J=1$. The resulting state, $\theta_1 = \theta_2 = 0$, is the ferro-orbital one formed by $|x^2-y^2\rangle$ orbitals. The plot total energy versus doping is presented in in Fig.~\ref{FigEtot2DEg}. In this figure, we see that the plot of energy~\eqref{EtotEg} can exhibit both positive and  negative curvature (see the inset of Fig.~\ref{FigEtot2DEg}). For the positive curvature ensuring the stability, the phase separation can nevertheless arise due to the kink within the doping range close to $n=n_1$. At the negative curvature, the phase separation naturally appears. Thus, both the negative curvature of $E_{tot}$ plot and the kink can give rise to different kinds of inhomogeneous states~\cite{OrtixLorDiCastroPRL2008}.

\begin{figure}[H]
\begin{center}
\includegraphics*[width=0.45\textwidth]{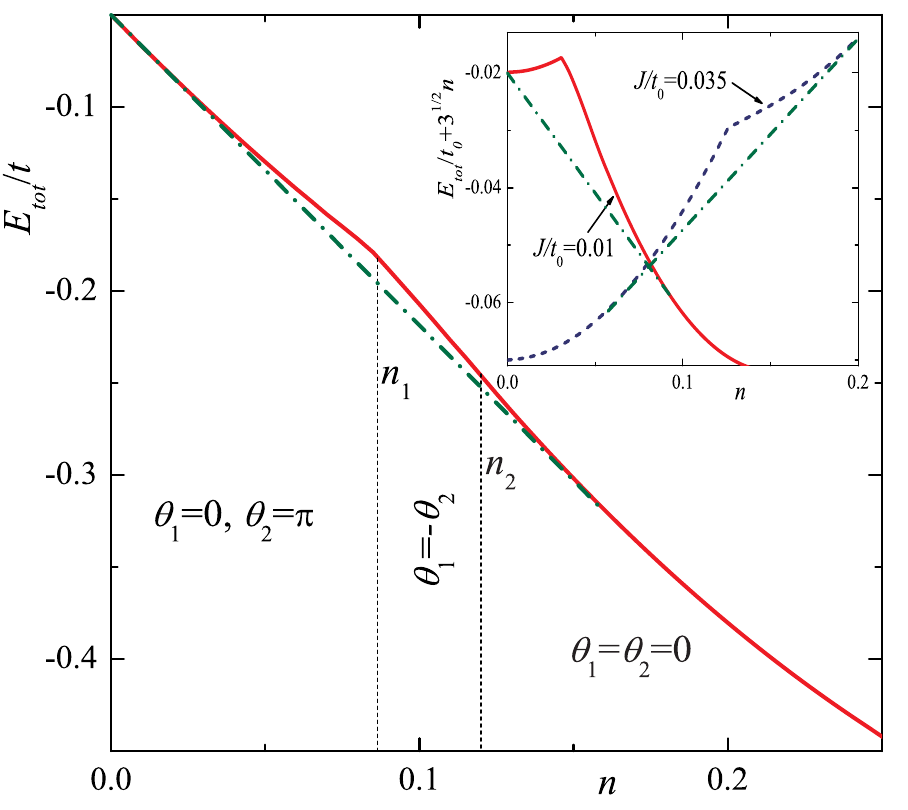} \end{center}
\caption{\label{FigEtot2DEg} (Color online) Energy  Eq.~\eqref{EtotAn} versus doping at $J/t_0=0.025$ (red solid curve). Within the doping range close $n_1\approx0.08$, the homogeneous state is unstable, and a phase separation can arise. In the inset, we plot the function $E_{tot}(n)+t_0\sqrt{3}n$ (for the better visuality, we subtract the linear term in $E_{tot}(n)$) in the vicinity of $n<n_1$ at different values of  parameter $J/t$. The red solid curve (blue dashed curve) corresponds to $J/t_0=0.01$ ($J/t_0=0.035$), and have a negative (positive) value of $\partial^2E_{tot}/\partial n^2$ in the region $n>n_1$ close to $n_1$. The phase separation appears in both cases. The dot-and-dash line is the result of the Maxwell construction~\cite{KugSboKhomPRB2008orbitals}.} \end{figure}

\subsection{Inhomogeneities in the orbitally ordered structures}\label{Inhom}

In the previous subsections, we have shown that itinerant charge carriers introduced by doping can favor the formation of charge inhomogeneities in the case of orbital ordering. Now, let us discuss possible shapes of such inhomogeneities. We base our discussion on the case of $e_g$ electrons forming a regular array at the square lattice. We assume that the host lattice is characterized by the antiferro-OO structure with alternating $|x^2 - y^2>$ and $|2z^2 - x^2 - y^2>$ orbitals considered in the previous subsection. An additional charge carrier favors the formation of another type of orbital ordering within a limited region close to it. The optimum structure can be found by minimization of the additional energy related to the doping. Without a loss in generality, we can consider alternating
$|\theta_1\rangle$ and $|\theta_2\rangle$ orbitals.

In such a situation, the energy spectrum of charge carriers is specified by Eq.~\eqref{specEg}. Since we are dealing with a limited region, we can write an effective Hamiltonian by expanding Eq.~\eqref{specEg} up to up to the second order in  $\mathbf{k}$. As a result, we have
\begin{equation}\label{Heff}
\hat{H}_{eff} = -t_0\left(A_x + A_y\right)%
+\frac{t_0}{2}\left(A_x\frac{\partial^2}{\partial x^2}
+A_y\frac{\partial^2}{\partial y^2}\right), \end{equation}
where $A_x$, $A_y$ are defined in Eq.~\eqref{Axy}. Then, we assume that the region under study is an ellipse with semiaxes $\sqrt{A_x}\rho_0$ and $\sqrt{A_y}\rho_0$ and search for the solution of the Schr\"{o}dinger
equation within this ellipse employing Hamiltonian~\eqref{Heff}.
In such elliptical droplet, the kinetic energy of a charge carrier can be written as
\begin{equation}\label{Ekin} E_{kin} = -t_0\left(A_x +
A_y\right)+\frac{t_0j_{0,1}^2}{2\rho_0^2}\,. \end{equation}
Here, $j_{0,1}\cong2.405$ is the first root of the Bessel function $J_0$.

For the orbitally ordered state, we can write the potential energy $E_{pot}$ related to the orbital ordering, which is proportional to the volume
$v$ of the elliptical droplet ($v=\pi\sqrt{A_xA_y}\rho_0^2$). It includes the energy of the new canted OO state $zJ\cos(\theta_1-\theta_2)/2$ and the loss in energy of the antiferro-OO host structure occurring due to arising droplets with the modified OO structure ($zJ/2$, $z=4$).  For a set of identical droplets, the sum of these two contributions is
\begin{equation}\label{Epot}
E_{pot}=4\pi\rho_0^2J\sqrt{A_xA_y}\cos^2
\left(\frac{\theta_1-\theta_2}{2}\right)\,. \end{equation}
Minimizing the droplet energy $E_{kin}+ E_{pot}$ with respect to
$\rho_0$, we find \begin{equation}\label{rho_0} \rho_0
=\left(\frac{t_0j_{0,1}^2}{8\pi J\sqrt{A_xA_y}
\cos^2\left(\frac{\theta_1-\theta_2}{2}\right)}\right)^{1/4}\!\!.
\end{equation} The total energy (per lattice site) then reads
\begin{eqnarray}\label{Etot(theta)}
E_{tot}&=&-2J+E(\theta_1,\,\theta_2)\,n\,,\\
E(\theta_1,\,\theta_2)&=&-2t_0(A_x+A_y)\phantom{\frac12}+j_{0,1}\left(8\pi
t_0J\sqrt{A_xA_y}\right)^{1/2}\left|\cos
\left(\frac{\theta_1-\theta_2}{2}\right)\right|\, .
\end{eqnarray}

To determine the optimum shapes of OO droplets, we perform the minimization of energy\eqref{Etot(theta)} with respect to orbital angles $\theta_1$ and $\theta_2$ using expressions \eqref{Axy}[for $A_{x,y}$. The behavior of$E(\theta_1,\,\theta_2)$ at two different values of $J/t_0$ is illustrated in Fig~\ref{FigE3D}. In this figure, we can clearly see that the function $E(\theta_1,\,\theta_2)$ exhibits several minima, and the specific angles $\theta_1$ and $\theta_2$ corresponding to the lowest energy strongly depend on the value of $J/t_0$. At small $J/t_0$ [Fig~\ref{FigE3D}(a)], the lowest energy is achieved at to $\theta_1=\theta_2=0$ (ferro-OO state with the filled $|x^2-y^2>$ orbitals). For such state, the optimum shape of droplets is a circle (see the left panel of Fig.~\ref{FigPolaron}). At $J/t_0$ exceeding a certain critical value ($J_{cr}/t_0\simeq0.0075$), the minimum
$\theta_1=\theta_2=0$ corresponds to a metastable state, and the energy
$E(\theta_1,\,\theta_2)$ attains the lowest value at four degenerate  structures: with the alternation of either $\theta_1=\pi/3$ and $\theta_2=2\pi/3$, or $\theta_1=-\pi/3$ and $\theta_2=-2\pi/3$, and the other  two minima correspond to the similar structures with the replacement $\theta_1\leftrightarrow\theta_2$ [see Fig~\ref{FigE3D}(b)]. Such four orbital arrays are the chains with alternating $|x^2-z^2>$ and $|2y^2 -x^2-z^2>$ (or $|y^2-z^2>$ and $|2x^2-y^2-z^2>$) orbitals. This, in fact, leads to the formation of nearly one-dimensional (cigar-shaped) droplets directed along $y$ or $x$ axes (right panel of Fig.~\ref{FigPolaron}).  At the further increase in $J/t_0$, the metastable $\theta_1=\theta_2=0$ array  splits into two more colicated structures with $\theta^*_1=-\theta^*_2$, in which $\theta^*_1$ can be either positive and (see Fig~\ref{FigE3D}(b)).
The inhomogeneities related to the latter arrays should have the circular shape, but the OO structure inside them is canted.

\begin{figure}[H]
\centering
   \subfigure[$J/t_0 = 0.005$]{
      \includegraphics[width=0.45\columnwidth]{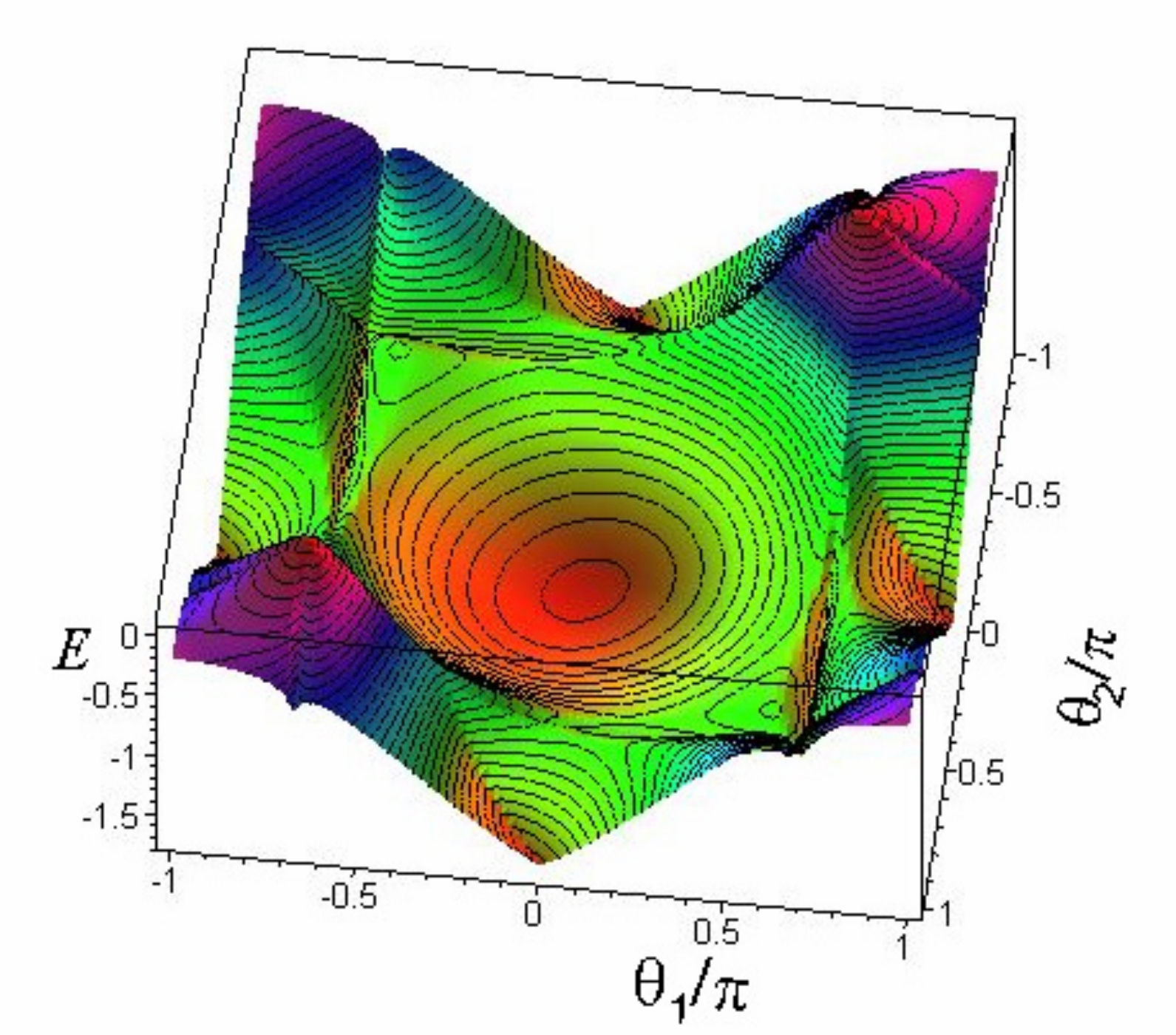}}  \subfigure[$J/t_0 = 0.02$]{
      \includegraphics[width=0.45\columnwidth]{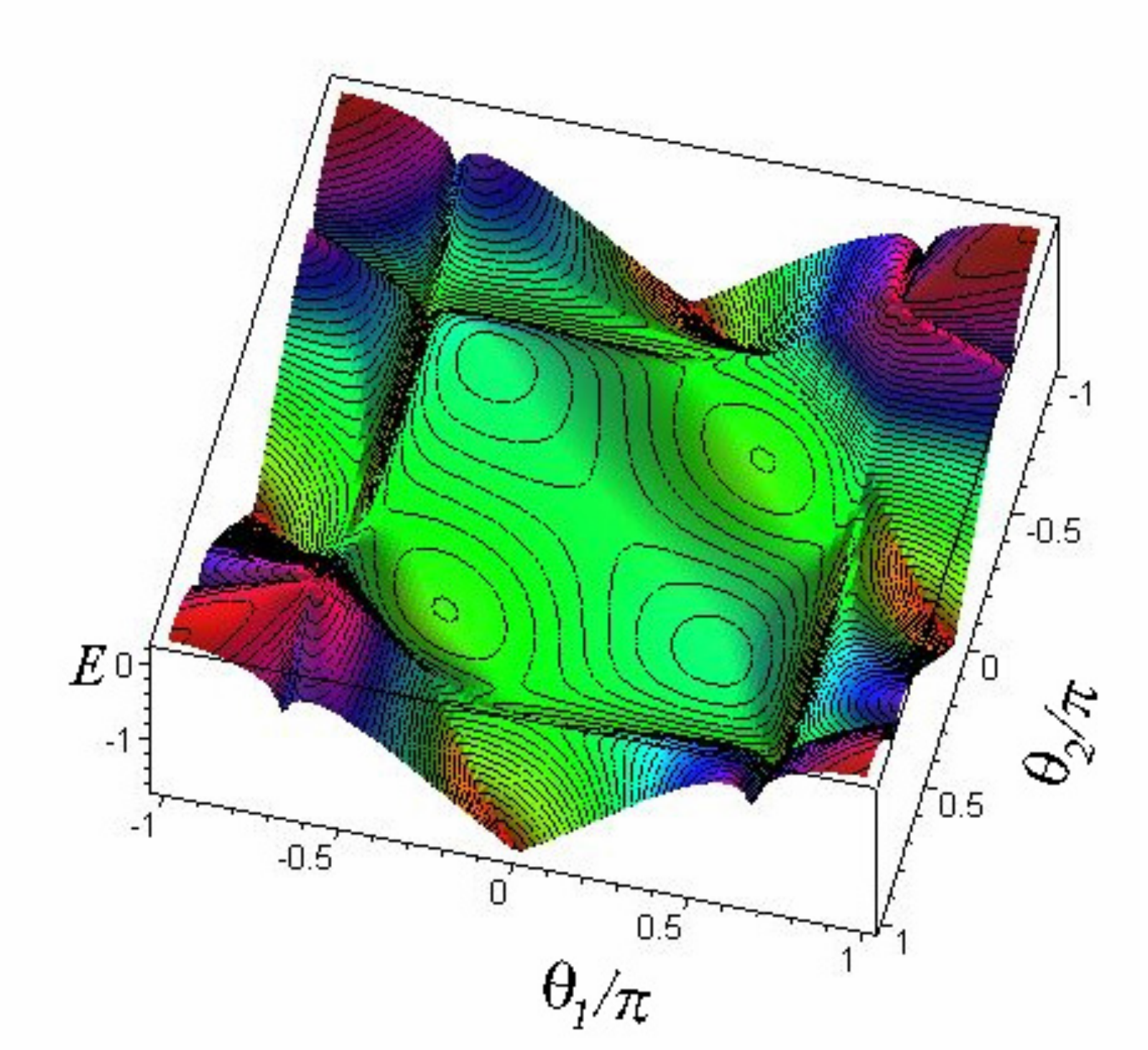}}
\caption{Energy Eq.~\eqref{Etot(theta)} of an orbital versus angles $\theta_1$ and $\theta_2$ at $J/t_0$ smaller (a) and larger
(b) than the critical value $\simeq 0.0075$ ~\cite{KugSboKhomPRB2008orbitals}.} \label{FigE3D}
\end{figure}

\begin{figure}[H]
\begin{center}
\includegraphics[width=0.45\columnwidth]{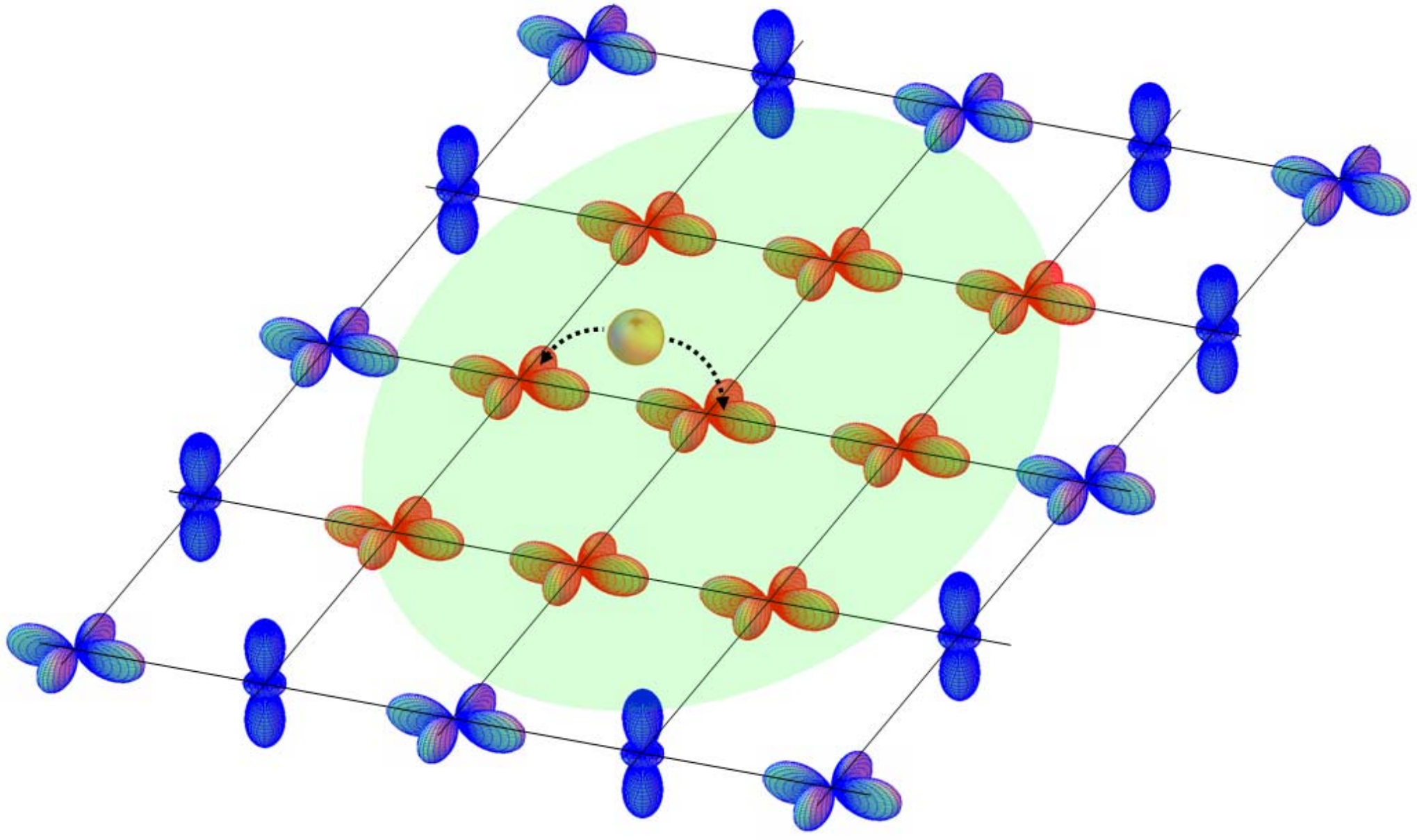}%FigPolaron_a.pdf
\includegraphics[width=0.45\columnwidth]{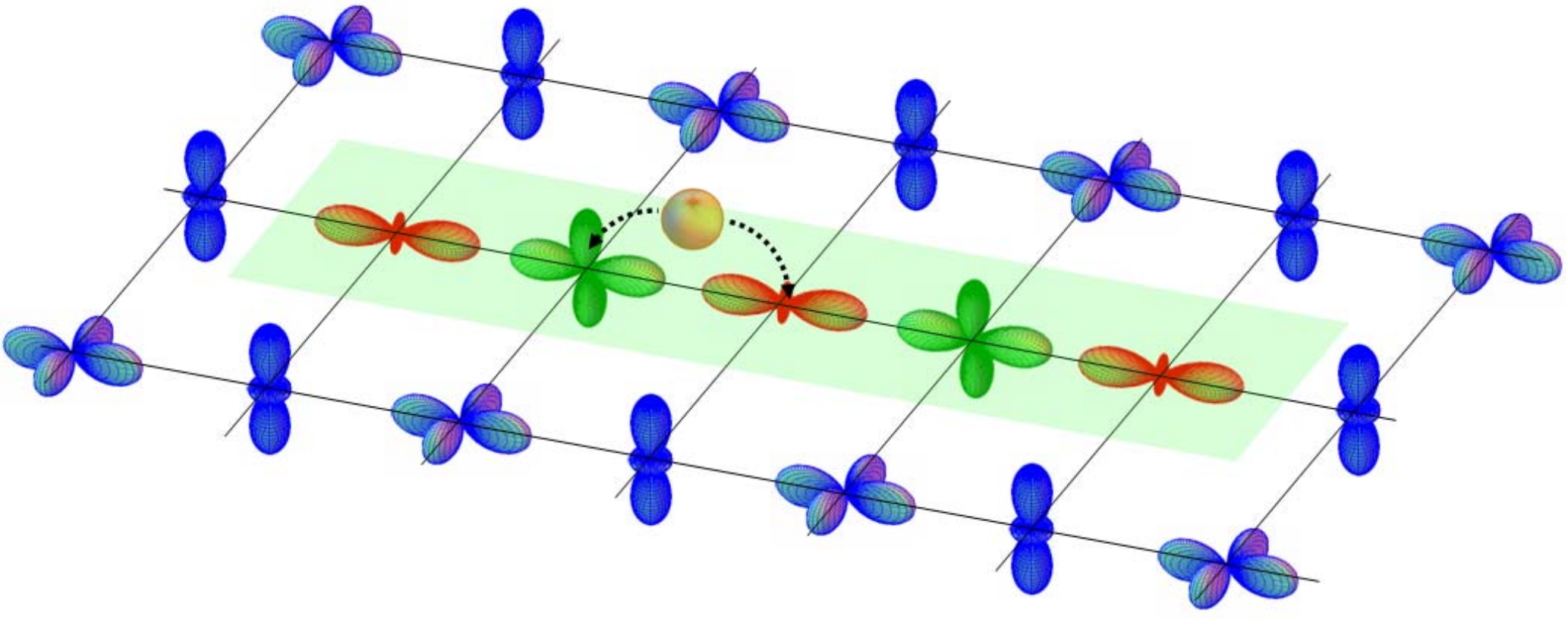}%FigPolaron_b.pdf
\end{center} \caption{\label{FigPolaron} (Color online) Possible types of orbital inhomogeneities arising due to doping:  circular (left panel) and needle-like (right panel) droplets. The motion of an itinerant charge carrier results in the formation of a ferro- or canted-OO structure within a limited region inside the host material with the antiferro-OO.~\cite{KugSboKhomPRB2008orbitals}.} \end{figure}

The physical mechanism underlying the formation of two types of droplets with different shapes is quite clear. Indeed, the ferro-OO state with $|x^2-y^2>$ orbitals provides the maximum gain in the kinetic energy. This gain prevails
at small $J/t_0$ favoring circular droplets with such the ferro-OO. In the opposite case of dominant potential energy (large $J/t_0$), nearly one-dimensional structures turn out to be favorable. Therefore, the optimum droplets have now cigar- or needle-like shape. The orbital configuration in a needle-like droplet provides the  maximum gain in the kinetic energy in the absence of hopping between neighboring chains.

It can be shown that in a certain doping range, $0<n<n^*_c$, the inhomogeneous state including circular or needle-like OO droplets within the antiferro-OO host material corresponds to the energy lower than that of the homogeneous state.  With the increase in the charge carrier density, the
droplets can  overlap; therefore, at some threshold value $n=n^*_c$, the the considered droplet type disappears. However, the phase separation can exist in a broader doping range  (see the previous subsection). For the threshold charge carrier density in the case of circular droplets, we obtain an estimate $n^*_c\sim1/\pi\rho_0^2$. Thus, at  $J/t_0=0.005$, we have $\rho_0\approx2$ (in units of lattice constant) and $n^*_c\sim0.08$.

For the needlelike droplets, $A_x=0$ (or $A_y=0$), and Eq.~\eqref{rho_0} gives us the chains of infinite length (but with zero volume $v$), $\rho_0=\infty$. This rather unphysical result stems from the assumption that the potential is proportional only to the droplet volume only. To estimate the characteristic length $L$ of the needle, we should consider also the surface contribution to the energy (proportional to the length of droplet) to the potential energy $E_{pot}$.  For such an estimate, we can consider, for example, an array of $|y^2-z^2>$ and $|2x^2 -y^2-z^2>$
orbitals stretched along the $x$ axis. For this array, we have
$A_x=\sqrt{3}$, $A_y=0$. The corresponding effective Hamiltonian \eqref{Heff} takes the form $$ \hat{H}_{eff}=-t_0\sqrt{3}+\frac{t_0\sqrt{3}}{2}
\frac{\partial^2}{\partial x^2}\,, $$ and the kinetic energy of
a charge carrier in the chain of length $L$  can be written as
$E_{kin}=-t_0\sqrt{3}(1-\pi^2/2L^2)$. The surface energy results from the interaction between the orbitals in the chain and in the host material. It The minimum surface energy corresponds to the configuration illustrated in Fig.~\ref{FigPolaron}, for which each $|y^2-z^2>$ ($|2x^2 -y^2-z^2>$)
orbital in the chain has its nearest neighbor $|x^2-y^2>$ ($|2z^2
-x^2-y^2>$) orbital in the host. Using the continuum approximation, we can write the potential energy as $E_{pot}=9JL/4$. Then, we minimize the total energy $E_{kin}+E_{pot}$ with respect to $L$, and obtain the expression for the characteristic length of a needle-like droplet
\begin{equation}\label{L0} L_0
=\left(\frac{4\pi^2t_0\sqrt{3}}{9J}\right)^{1/3}\!\!.
\end{equation}
 For$J/t_0=0.05>J_{cr}/t_0$, we obtain $L_0\approx5.5$. If we assume a  random distribution of the a needle-like droplet in host (with the needles directed both along $x$ and $y$ axes), the critical density is about $n^*_c\sim1/L_0^2$. Note, however, that this density can be higher in a more regular array such as that corresponding to stripes.

Let us recall that for spin polarons discussed in the previous sections, the existence of the gapless (Goldstone) excitations leads to slowly decaying spin structure distortions around magnetic defect (magnetic polaron)~\cite{deGennesPR1960}. Therefore, it is natural to expect that orbital polarons should give rise to a long-range strain fields will be created around them. In analogy to the case of spin polarons, these long range fields should not affect the existence of orbital polarons themselves. They can only modify the parameters of inhomogeneous state and lead to an additional long-range interaction between inhomogeneities.

In this and in the previous subsections, we were dealing with a two-dimensional case of the anisotropic model. However, in contrast to the isotropic case treated in subsection~\ref{SymModel}, the space dimensionality can drastically affect the obtained results. Indeed, the hopping integrals for $e_g$ electrons strongly anisotropic. This can produce a pronounced effect on the OO type in the homogeneous state as well as on  the shape and orbital order of the inhomogeneities. That is why, the detailed study of the three-dimensional case is a separate and quite complicated task.

\subsection{Simplified two-band model with the orbital ordering}\label{Model}

\subsubsection{Formulation of the problem}\label{problem}

In the previous subsections, we studied the effect of doping on already existing orbital ordering. Now, we shall treat a more natural situation when the same charge carriers are involved both in the orbital ordering and in charge transfer. We start with the analysis of the system with JT ions having double-degenerate ground state. As usual, such degeneracy is lifted by local lattice distortions. Then, each lattice site, can correspond to to two different ground states, $a$ or $b$ (in perovskites, for example
$a$ ($b$) state can be related to stretching (compression) of the surrounding anion octahedron). The $a$ and $b$ states at site $\mathbf{n}$ are, in fact, possible orbitals filled by a charge carrier. Of course, the local distortions are not independent, and their interaction can lead to certain  regular structure. The simplest Hamiltonian describing the orbital interaction can be described in an Ising-like form~\cite{KugSboKhoJSNM2009orbitals}
\begin{equation}
H_{\text{OO}}=J\sum_{\langle{\bf n}{\bf m}\rangle}
\tau^z_{\bf n}\tau^z_{\bf m}\,.
\end{equation}
Here, the $\tau^z_{\bf n}$ operators have the eigenvalues
$1$ or $-1$ depending on the state of site ${\bf n}$ ($a$ or $b$, respectively). In such system, we can consider two simplest types of the orbital ordering: either ferro-OO (with the same orbitals occupying each site) and antiferro-OO (corresponding to the alternation of orbitals).

The doping gives rise to nonzero density, $n\neq 0$, of itinerant charge carriers appear in the system. As a result, the system includes localized  charge carriers responsible for OO and itinerant ones. The electron motion is determined by hopping integrals depending on the states of the neighboring lattice sites. In the model under study, there are three possible hopping integrals, namely, $t^{aa}$, $t^{bb}$, and $t^{ab}=t^{ba}$. For simplicity, let us assume that $t^{aa}=t^{bb}>t^{ab}$. At both types of orbital ordering (ferro- and antiferro-OO) the charge carriers are described by the Hamiltonian
\begin{equation}\label{Hel}
H_{\text{el}}=-t\sum_{\langle{\bf n}{\bf
m}\rangle,\sigma}\left(a^{\dag}_{{\bf n}\sigma}a_{{\bf
m}\sigma}+h.c.\right)+\frac{U}{2}\sum_{{\bf n}\sigma}n_{{\bf
n}\sigma}n_{{\bf n}\bar{\sigma}}\,. \end{equation}
The first term in $H_{\text{el}}$ describes the kinetic energy of itinerant charge carriers. Symbol $t$ denotes the hopping integral $t^{aa}$ in for the ferro-OO and $t^{ab}$ for antiferro-OO. The second (Hubbard) term in
Eq.~\eqref{Hel} takes into account  the onsite Coulomb repulsion between electrons and the bar above $\sigma$ means the spin component of the opposite sign. Angular brackets $\langle ... \rangle$ denote the summation over nearest neighbor sites. We also assume that the parameters of Hamiltonian~\eqref{Hel} are independent of the doping level $n$.

\subsubsection{Homogeneous states and phase separation}\label{Hom}

Hamiltonian~\eqref{Hel}, is the usual Hubbard Hamiltonian for charge carriers moving on the ferro- or antiferro-OO background. We begin our analysis from the possible homogeneous states corresponding to such Hamiltonian. At $J>0$ and in the absence of doping($n=0$), we have the uniform antiferro-OO ground state, which is the ground state if $J>0$ (the case of ferro-OO state can be treated in a similar way). In the antiferro-OO state we can consider only the interorbital hopping $t=t^{ab}$ in Eq.~\eqref{Hel}. We determine the energy spectrum of charge carriers using the Hubbard I approximation~\cite{HubbardPrRoySocA1963} in equation of motion for one-electron Green's function $G_\sigma({\bf n,n}_0;\,t-t_0)=-i\langle Ta_{{\bf n}\sigma}(t)a^{\dag}_{{\bf n}_0\sigma}(t_0)\rangle$. In the frequency--momentum representation, this Green's function can be written as~\cite{HubbardPrRoySocA1963}
\begin{equation}\label{Gappr1}
G_{\sigma}({\bf k},\omega)=\frac{\omega+\mu-U(1-n/2)}{\left(\omega+\mu-E_1({\bf
k})\right)\left(\omega+\mu-E_2({\bf k})\right)}=
\!\frac{\omega\!+\!\mu\!-\!U(1\!-\!n/2)}{E_1({\bf k})-E_2({\bf
k})}\!\left\{\!\!\frac{1}{\omega\!+\!\mu\!-\!E_1({\bf
k})}\!-\!\frac{1}{\omega\!+\!\mu\!-\!E_2({\bf
k})}\!\!\right\}\!\!,
\end{equation}
where
\begin{equation}\label{E12a}
E_{1,2}({\bf k})=\frac{U+\varepsilon({\bf
k})}{2}\mp\sqrt{\left(\frac{U-\varepsilon({\bf
k})}{2}\right)^2+U\varepsilon({\bf k})\frac{n}{2}}\,,
\end{equation}
and $\varepsilon({\bf k})$ is the ${\bf k}$-dependence of  electron energy in the absence of correlations ($U=0$).  We do not consider any spin ordering and assume that the numbers of electrons with each spin  projections are equal, $n_\sigma=n_{-\sigma}=n/2$. In the case of  simple cubic lattice, we have
$\varepsilon({\bf k})=-2t^{ab}(\cos k_x+\cos k_y+\cos k_z)$, where for $k_i$, we use the  the units of inverse lattice constant $d$.

The charge carrier density can be determined using  Green's functions \eqref{Gappr1}
\begin{equation}C
n=-i\int\limits\frac{d\omega}{\pi}\int\frac{d^3{\bf
k}}{(2\pi)^3}G_\sigma({\bf
k},\omega+i0\;\textrm{sign}\omega)e^{i\omega 0}.
\end{equation}
 Changing in  \eqref{Gappr1} $\omega$ by
$\omega+\mu$ and using spectrum \eqref{E12a}, we find
\begin{equation}\label{delf}
n = 2\int_{\tilde{w}}^\mu d\omega\int\frac{d^3{\bf k}}{(2\pi)^3}\frac{U\left(1-n/2\right)-\omega}
{\sqrt{(U-\varepsilon({\bf k}))^2+2\varepsilon({\bf k}) nU}} \left\{\delta\left(\omega-E_1({\bf
k})\right)-\delta\left(\omega-E_2({\bf k})\right)\right\}\,,
\end{equation}
where $\tilde{w}=\tilde{w}(n)$ is the bottom of the lower band. When the lower band is not fully occupied filled, the second  $\delta$-function in \eqref{delf} does not work. contribute to the result. In fact, Eq.~\eqref{delf} allows us to find the the dependence of chemical potential on the doping level $n$.  The, the  kinetic energy per lattice sire has the form
\begin{equation}\label{Ekin1}
E_{\text{kin}} = 2\int_{\tilde{w}}^\mu \omega
d\omega\int\frac{d^3{\bf
k}}{(2\pi)^3}\frac{U\left(1-n/2\right)-\omega}
{\sqrt{(U-\varepsilon({\bf k}))^2+2\varepsilon({\bf k}) nU}}
\left\{\delta\left(\omega-E_1({\bf
k})\right)-\delta\left(\omega-E_2({\bf k})\right)\right\}\,,
\end{equation}

After that, we can calculate the total energy of the homogeneous antiferro-OO state
\begin{equation}\label{EtotAF}
E_{AFOO}=E_{kin}(t^{ab})-Jz/2.
\end{equation}
The plot of such energy as function of doping $n$ shown in Fig.~\ref{Fig1}, where it is compared to the corresponding plot for the ferro-OO state (see below)

\begin{figure}[H]
\begin{center}
\includegraphics*[width=0.45\textwidth]{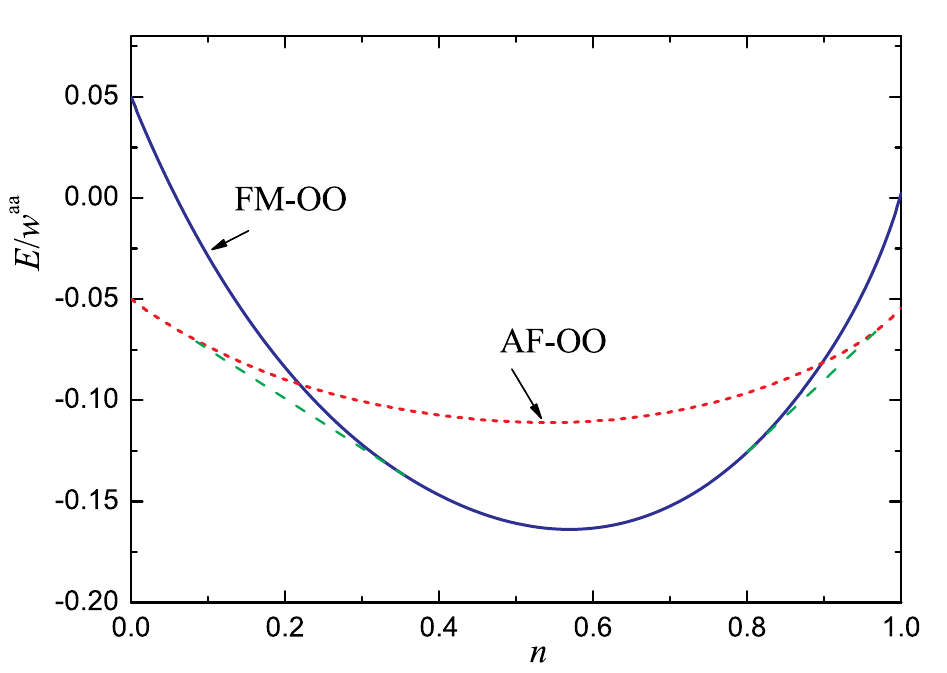} \end{center}
\caption{\label{Fig1} (Color online) Dependence of the  energies of homogeneous ferro-OO (blue solid line) and antiferro-OO (red dashed line) states onthe doping level$n$. Green dashed lines describe the energy of  an inhomogeneous state. The plots correspond to the following choice of parameters: $t^{ab}/t^{aa}=0.3$, $U/t^{aa}=3$, and $J/t^{aa}=0.1$~\cite{KugSboKhoJSNM2009orbitals}.}
\end{figure}

In the case of a homogeneous  ferro-OO state (the charge carriers occupy similar orbitals), the energy loss per site  related to the  unfavorable orbital ordering is $zJ/2$. However, the gain in kinetic energy is larger because the hopping integral between similar orbitals  exceeds that between different orbitals
($t^{aa}>t^{ab}$). Here, the kinetic energy is described by the same expression as \eqref{Ekin1} but with the different hopping integral, $t^{aa}$. Thus, instead of Eq.~\eqref{EtotAF}, we have
\begin{equation}\label{EtotFM}
E_{FMOO}=E_{kin}(t^{aa})+Jz/2.
\end{equation}

In Fig.~\ref{Fig1}, we  illustrate the behavior of energies for the  system in ferro-OO and  antiferro-OO states within the $0<n<1$  doping range. We can clearly see that at some dopings,  antiferro-OO is unfavorable and  the ferro-OO becomes the ground state owing to a larger  kinetic energy gain. In the vicinity of  the crossover range,  we can see that  the homogeneous state turns out to be unstable toward the phase separation into a mixture of ferro-OO and antiferro-OO domains characterized by different values of charge carrier density (see green dashed lines in Fig.~\ref{Fig1}).

\subsection{A model with charge carriers occupying $e_g$ orbitals: 2D case} \label{e_g}

\subsubsection{A model}

Let us now generalize the previous analysis  to the case of some specific orbitals such as  $e_g$ orbitals on the square 2D lattice~\cite{KugSboKhoJSNM2009orbitals}. This situation is relevant, e.g. to such compounds as  layered cuprates,( K$_2$CuF$_4$,) or manganites (La$_2$MnO$_4$ and La$_2$Mn$_2$O$_7$).  As it was discussed above, any $e_g$ orbital can be represented as a linear combination of two basis states $|a>=|x^2 - y^2>$ and $|b>=|2z^2 - x^2 - y^2>$
\begin{equation}\label{theta1}
|\theta>=\cos\frac{\theta}{2}\,|x^2-y^2>+\sin\frac{\theta}
{2}\,|2z^2-x^2-y^2>. \end{equation}

In the case under study,  the same charge carries  are responsible  both for the formation of orbital order and for the charge transfer.  This means that the Hamiltonian of the model should involve a, all possible types of of the intersite hopping integrals.  Of course, such hopping integrals  $t^{\alpha\beta}$ ($\alpha,\beta=a,b$) should depend  on the direction of of hopping. We can write the corresponding Hamiltonian in the form
\begin{equation}\label{HelEg}
H_{el}=-\sum_{\langle{\bf nm}\rangle\alpha\beta\sigma}%
t_{\mathbf{n}-\mathbf{m}}^{\alpha,\beta}d^{\dag}_{{\bf
n}\alpha\sigma}d_{{\bf m}\beta\sigma}+
\frac{U}{2}\sum_{{\bf n}\alpha\sigma}n_{{\bf n}
\alpha\sigma}n_{{\bf n}\alpha\bar{\sigma}}+\frac{U}{2}\sum_{{\bf
n}\alpha\sigma\sigma'}n_{{\bf n}\alpha\sigma}n_{{\bf n}
\bar{\alpha}\sigma'}\,,
\end{equation}
where for the the hopping integrals, we have $t_{\mathbf{n}-\mathbf{m}}^{\alpha\beta}=t_x^{\alpha\beta}$
($t_{\mathbf{n}-\mathbf{m}}^{\alpha\beta}=t_y^{\alpha\beta}$) for
$\mathbf{n}-\mathbf{m}$ directed along the $x$ ($y$)
axis. As usual, in Eq.~\eqref{HelEg}, $\bar{\sigma}$ and $\bar{\alpha}$ denote not $\sigma$ and not $\alpha$, respectively. The hopping integrals
$t_{x,y}^{\alpha\beta}$ can be represented in the matrix form
\begin{eqnarray}\label{t_xy1}
(t_{x,y})^{\alpha\beta} =\frac{t_0}{4}\left(%
\begin{array}{cc}
3 & \mp\sqrt{3} \\
\mp\sqrt{3} & 1 \\
\end{array}%
\right)\, .
 \end{eqnarray}
Here, signs minus and plus correspond to $x$ and $y$ hopping directions, respectively. Further on, we limit ourselves to the case of large $U/t_0$ in Hamiltonian~\eqref{HelEg}.

\subsubsection{Antiferro-orbital ordering near half filling}\label{AFOO}

The case  of one electron per site (half filling) is of special interest. Here, we study the situation near half filling, $x=1-n\ll1$. It can be demonstrated that at $n=1$ and large $U$, Hamiltonian~\eqref{HelEg} has the insulating the ground state with the antiferro-orbital ordering. Indeed, the second order of perturbation theory in terms of $t_0/U$ gives rise to the effective Hamiltonian in the form~\cite{KugKhomUFN1982}
\begin{eqnarray}\label{Hoo}
H_{OO}&=&J\sum_{\langle\mathbf{nm}\rangle}\left\{\mathbf{s_ns_m}
\left[\tau^z_{\mathbf{n}}\tau^z_{\mathbf{m}}
-2\tau^z_{\mathbf{n}}+1\pm2\sqrt{3}\tau^z_{\mathbf{n}}
\tau^x_{\mathbf{m}}\mp2\sqrt{3}\tau^x_{\mathbf{m}}+3\tau^x_{\mathbf{n}}
\tau^x_{\mathbf{m}}\right] \right.\\
&&\left.+\frac12%
\left[\tau^z_{\mathbf{n}}\tau^z_{\mathbf{m}}
+2\tau^z_{\mathbf{n}}\pm{\sqrt{3}\tau^x_{\mathbf{m}}}
2\sqrt{3}\tau^z_{\mathbf{n}}\tau^x_{\mathbf{m}}
\pm 2\sqrt{3}\tau^x_{\mathbf{m}}
+3\tau^x_{\mathbf{n}}\tau^x_{\mathbf{m}}\right]\right\} \,,
\end{eqnarray}
where $J=t_0^2/U$,
$\mathbf{s_n}=\sum_{\alpha\sigma\sigma'}d^{\dag}_{{\bf
n}\alpha\sigma}(\bm{\sigma})_{\sigma\sigma'}d_{{\bf
n}\alpha\sigma'}/2$ is the operator determining electron spin at site
$\mathbf{n}$, and $\tau^{z,x}_{\mathbf{n}}=\sum_{\alpha\beta\sigma}d^{\dag}_{{\bf
n}\alpha\sigma}(\sigma^{z,x})_{\alpha\beta}d_{{\bf
n}\beta\sigma}/2$ describe the occupation of orbitals, and
$\bm{\sigma}$ are the Pauli matrices. Similarly to Eq.~\eqref{t_xy1}, signs minus and plus correspond to $x$ and $y$ directions of the $\mathbf{n}-\mathbf{m}$ bond, respectively. In the absence of spin ordering, $\mathbf{s_ns_m}=0$, and the above Hamiltonian takes a simpler form
\begin{equation}\label{HooPara}
H_{OO}= \frac{J}{2}\sum_{\langle\mathbf{nm}\rangle}%
\left[\tau^z_{\mathbf{n}}\tau^z_{\mathbf{m}}+
2\tau^z_{\mathbf{n}}\pm2\sqrt{3}\tau^z_{\mathbf{n}}\tau^x_{\mathbf{m}}
\pm2\sqrt{3}\tau^x_{\mathbf{m}}%
+3\tau^x_{\mathbf{n}}\tau^x_{\mathbf{m}}\right]\,.
\end{equation}
Hamiltonian Eq.~\eqref{HooPara} has the antiferro-OO ground state corresopnding to the alternation of  $|\theta_0>$ and
$|-\theta_0>$ orbitals, where
\begin{equation}
\theta_0=\arccos\left(-\frac14\right)\,. \end{equation}
The orbital ordering leads to the energy gain per lattice site $E_{OO}=-13J/4$.

At the hole doping, there appear some empty lattice sites.  It is natural to assume that the long-range antiferro-orbital order is not destroyed at low doping levels, $x=1-n\ll1$, but the energy gain related to the ordering becomes smaller. In the framework of the mean-field approximation, we can write $\langle\tau^{z,x}_{\mathbf{n}}\rangle\propto1-x$ and
\begin{equation} E_{OO}=-\frac{13+11x^2}{4}J+6J\,x\,.
\end{equation}

To determine the kinetic energy of mobile holes in antiferro-OO structure, we can define the creation $b^{\dag}_{\mathbf{n}\sigma}$ and annihilation
$b_{\mathbf{n}\sigma}$ operators for such holes
\begin{equation}
b_{\mathbf{n}\sigma}=\cos\frac{\theta_0}{2}\,d^{\dag}_{na\sigma}\pm%
\sin\frac{\theta_0}{2}\,d^{\dag}_{nb\sigma}\,,\;\;b^{\dag}_{\mathbf{n}
\sigma}=(b_{\mathbf{n}\sigma})^{\dag}\,,
\end{equation}
where signs plus and minus sign correspond to the sites with $|\theta_0>$ and $|-\theta_0>$ orbitals, respectively. Then, the kinetic energy described by the first term in Hamiltonian~\eqref{HelEg}) can be written as
\begin{equation}\label{Hhole}
H_{el}=\frac{t_0}{8}\sum_{\langle{\bf n}{\bf
m}\rangle\sigma}\left(P_{{\bf n}\sigma}^{\dag}b^{\dag}_{{\bf
n}\sigma}b_{{\bf m}\sigma}P_{{\bf m}\sigma}+h.c.\right),
\end{equation}
where we projection operators $P$ are introduced to exclude a double occupation of each sites. Let us analyse the situation with the 2D square lattice. Then, the energy spectrum of free holes is
\begin{equation}\label{ehole}
\varepsilon({\bf k})=-\frac{t_0}{4}\left(\cos k_x+\cos
k_y\right)\,.
\end{equation}
In this case, the energy of of the doped antiferro-OO array can be written as
\begin{equation}\label{E_AFOO}
E_{AFOO}(x)=-\frac{13+11x^2}{4}J+\int_{-t_0/2}^{\mu(x)}
\!\!\!\!\!\!\!\!\!\!dE\,E\rho_0(E)+6Jx\,,
\end{equation}
where the hole density of states $\rho_0(E)$ for holes in given by the expression
\begin{equation}\label{rho0a}
\rho_0(E)=\!\int\!\frac{d^2\mathbf{k}}{(2\pi)^2}
\delta(E-\varepsilon(\mathbf{k}))=%
\!\int\limits_0^{\infty}\!\frac{ds}{\pi}\cos(Es)J_0^2\!
\left(\frac{t_0s}{4}\right)\!.
\end{equation}
The chemical potential $\mu$ is determined by the condition
$x=\int_{-t_0/2}^{\mu(x)}dE\rho_0(E)$. The calculated energy \eqref{E_AFOO} versus $n=1-x$ is plotted in Fig.~\ref{Fig7}, see the blue dashed curve and compared to the energy of the orbitally disordered state analyzed below.
\begin{figure}[H]
\begin{center}
\includegraphics*[width=0.45\textwidth]{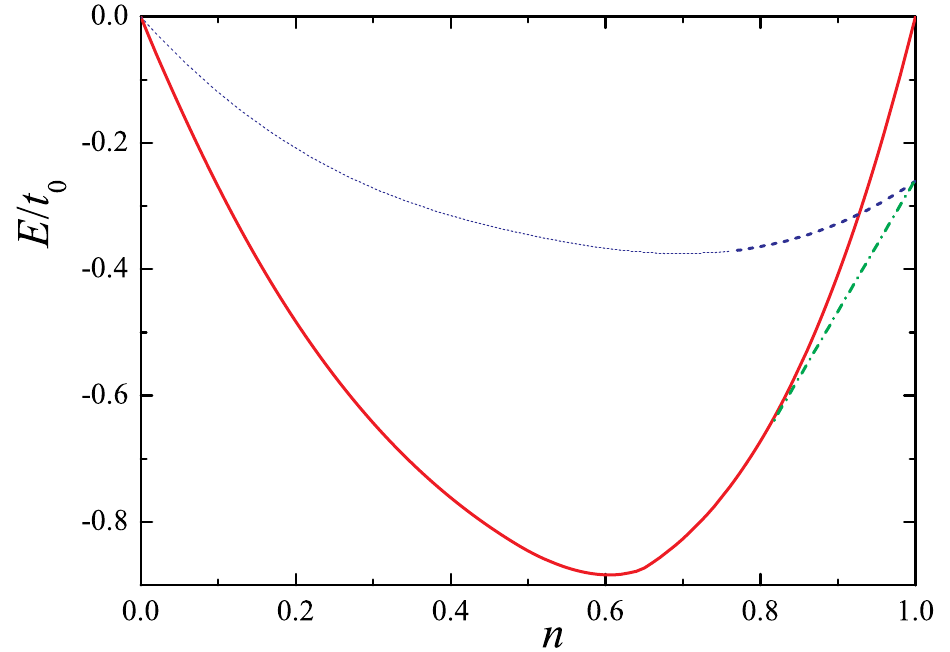} \end{center}
\caption{\label{Fig7}  (Color online) Dependence of energies on the total number $n$ of electrons for of orbitally disordered (red solid curve) and
antiferro-OO (dashed blue curve) states. $J/t_0=0.08$. The green
dot-ans-dash line shows  the energy of a possible
phase-separated state~\cite{KugSboKhoJSNM2009orbitals}.} \label{Fig7a}
\end{figure}

\subsubsection{Orbitally disordered state}

Now, let us discuss the situation characteristic of relatively high doping levels $x$, the the orbital ordering discussed above disappears. We shall consider the state corresponding to Hamiltonian~\eqref{HelEg} in the absence of the long-range OO at the electron densities $0<n<1$ and compare the corresponding energy \eqref{E_AFOO}. The one-particle Green's have the form similar to that considered above in subsubsection \ref{Hom} and analyze them using  the Hubbard I approximation~\cite{HubbardPrRoySocA1963}. The corresponding equations of motion can be written as
\begin{equation}\label{G(n)} (\omega + \mu)G_{\alpha\beta}({\bf
n}, \omega) = g_{\alpha} \left (1-\sum_{{\bf \Delta} \gamma}t_{\bf
\Delta}^{\alpha\gamma}G_{\gamma \beta}({\bf n +\Delta},
\omega)\right )\, , \end{equation} where
\begin{equation}\label{g_alpha}
g_{\alpha} = 1 - \frac{n_{\alpha}}{2} - n_{\bar{\alpha}},
\end{equation}
and the summation is taken over sites $\mathbf{m}$ , wihich are the neighbors of the chosen site $\mathbf{n}$, ($\bm{\Delta}=\mathbf{n}-\mathbf{m}$). We assume the absence of magnetic ordering, $n_{\alpha\uparrow} = n_{\alpha\downarrow}= n_{\alpha}/2$ and work at in the limit of strong correlations, $U \rightarrow\infty$.

Passing to the momentum representation, we can write Eq.\eqref{G(n)} as
\begin{equation}\label{G(n)1} (\omega + \mu)G_{\alpha\beta}({\bf
k}, \omega) = g_{\alpha} \left
(1-\sum_{\gamma}\varepsilon^{\alpha\gamma}({\bf k) }G_{\gamma
\beta}(\omega, {\bf k})\right )\,,
\end{equation}
where $\varepsilon^{\alpha, \beta}({\bf k)}$ is represented in the matrix form
\begin{equation}\label{t(k)}
\hat{\varepsilon}({\bf k})=\frac{-t_0}{2}\left(%
\begin{array}{cc}
  3(\cos k_x +\cos k_y) & \sqrt{3}(\cos k_y -\cos k_x) \\
 \sqrt{3}(\cos k_y -\cos k_x) & \cos k_x +\cos k_y \\
\end{array}%
\right).
 \end{equation}
Then, introducing the  matrix
\begin{equation}\label{Smatrix}
    \hat{S} =\sqrt{\hat{g}} = \left(%
\begin{array}{cc}
  \sqrt{g_a} & 0 \\
  0 & \sqrt{g_b} \\
\end{array}%
\right)\, , \end{equation}
it is possible to represent the Green's function in
the compact operator form
\begin{equation}\label{Gmatrix} \hat{G} = \hat{S}\frac{1}{\omega +
\mu -\hat{\varepsilon}_g}\hat{S} \, , \end{equation}
where
\begin{equation}\label{epsilon_g}
\hat{\varepsilon}_g = \left(%
\begin{array}{cc}
  g_a \varepsilon_{aa} & \sqrt{g_a g_b} \varepsilon_{ab} \\
  \sqrt{g_a g_b} \varepsilon_{ab} & g_b \varepsilon_{bb} \\
\end{array}%
\right) . \end{equation}

The energy spectrum is determined by the solution of equation
\begin{equation}\label{det} \det (\hat{\varepsilon}_g - E) = 0.
\end{equation}
Eventually, we find
\begin{equation}\label{Epm}
E_{\pm1}({\bf k}) =
-\frac{t_0(3g_a +
g_b)}{4} \left[\phantom{\sqrt{\frac12}}\!\!\!\!\!\!
(\cos k_x +\cos k_y) \pm\sqrt{(\cos k_x +\cos k_y)^2 -
  \frac{48 g_a g_b}{(3g_a + g_b)^2}\cos
k_x \cos k_y }\,\right]\,.
\end{equation}
Using the quasiparticle spectrum \eqref{Epm}, we can write the expression for the Green's function
\begin{equation}\label{Gab} G_{\alpha \beta} (\omega,\mathbf{k})=
\sum_{s=\pm 1}\frac{\sqrt{g_{\alpha} g_{\beta}}
v^{(s)}_{\alpha}({\bf k})v^{(s)}_{\beta}({\bf k})}{\omega + \mu -
E_s({\bf k})}\, . \end{equation}
In Eq.~\eqref{Gab}, $v^{(s)}_{\alpha}$ are the matrix elements of the transformation determined by the $\hat{v}^{(s)}$  obeying the relations
 \begin{equation}\label{v-defin}
\left[\hat{\varepsilon}_g({\bf k})\hat{v}^{(s)}\right]_{\alpha} =
E_s({\bf k})v^{(s)}_{\alpha} \, .\end{equation}
As a result of straightforward but cumbersome calculations, we find an explicit form of matrix $\hat{v}^{(s)}$ corresponding to the basis orbitals $|b> = |2z^2 - x^2 -y^2>$ and $|a>$ = $|x^2 - y^2>$ \begin{equation}\label{v-matrix}
 \hat{v}^{(s)}_{\alpha} = \frac{1}{\sqrt
 {(g_a\varepsilon_{aa}- E_s)^2 -
 g_ag_b\varepsilon^2_{ab}}}\left(%
\begin{array}{cc}
   -\sqrt{g_b g_a}\varepsilon_{ab} \\
   g_a\varepsilon_{aa}- E_s \\
 \end{array}%
 \right),
\end{equation}
where $\varepsilon_{aa}, \varepsilon_{ab}$, and
$\varepsilon_{ab}$ are matrix elements of matrix $-\hat{t}({\bf
k})$, see \eqref{t(k)}.

Using the above expressions, we can calculate the density of states corresponding to two new bands with indices $s=1$ and -1
\begin{equation}\label{rho_s}
    \rho^{(s)}(E) = \sum_{\alpha}\rho^{(s)}_{\alpha\alpha}(E),
\end{equation}
where \begin{equation}\label{rho_alpha}
\rho^{(s)}_{\alpha\alpha}(E) = \frac{1}{(2 \pi)^2}\int d{\bf
k}\left[v^{(s)}_{\alpha}(\mathbf{k})\right]^2 \delta(E-E_s({\bf
k})).
\end{equation}
Then, we can determine the partial filling of orbital states \begin{equation}\label{n_alpha}
    n_{\alpha} = 2g_{\alpha}\sum_{s=\pm
    1}\int_{-\infty}^{\mu}\rho^{(s)}_{\alpha\alpha}(E)dE.
    \end{equation}
The chemical potential $\mu$ as can be obtained using the condition
\begin{equation}\label{mu} n_a + n_b = 1 -x,
\end{equation}
where $x$ is the hole density (doping level). Substituting $\mu(n_a)$ to Eq.~\eqref{n_alpha}, we can determine the dependence of $n_a$ and
$n_b$ on the doping level $x$. Then, the total energy corresponding to Hamiltonian \eqref{Hel} is given by the expression
\begin{equation}\label{E_el}
    E_{el} = \sum_{\alpha, s=\pm
    1}(1+g_{\alpha})\int_{-\infty}^{\mu}\rho^{(s)}_
    {\alpha\alpha}(E)E dE.
    \end{equation}

In Fig.~\ref{Fig4} we illustrate the behavior of density of states for the new bands $s= \pm 1$, These densities of states exhibit the logarithmic van Hove singularity, which is related to the transition from the closed to open Fermi surface in each bands (the evolution of the Fermi surface is represented in Fig.~\ref{Fig6}).

\begin{figure}[H] \begin{center}
\includegraphics*[width=0.45\textwidth]{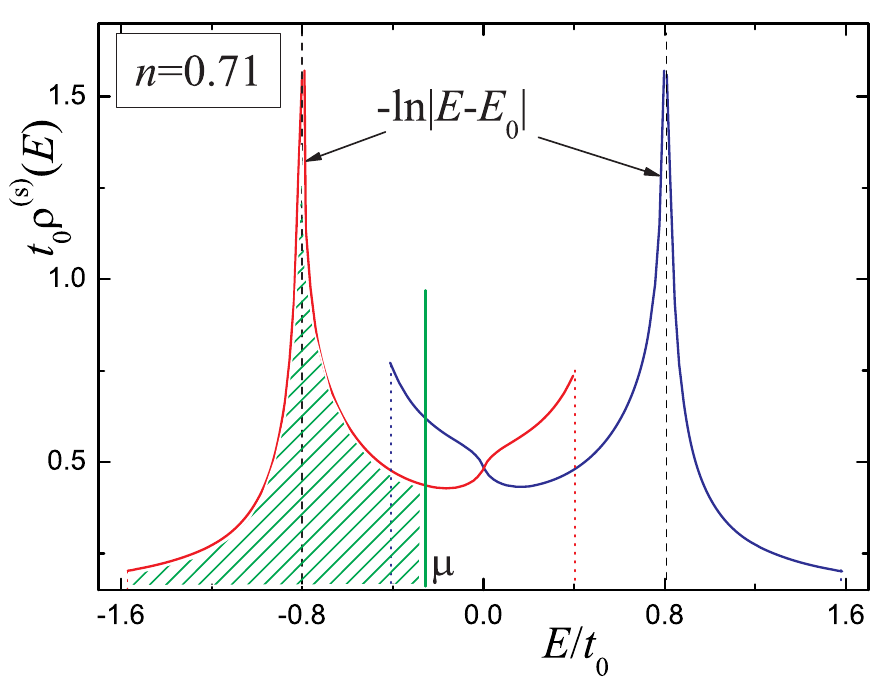} \end{center}
\caption{\label{Fig4} Densities of states for bands $s= -1$ (red solid line) and $s= 1$ (blue solid line). The occupied states at the total number of electrons per site $n= 0.71$ are shown by the shaded area~\cite{KugSboKhoJSNM2009orbitals}.}
\end{figure}

The behavior of occupation numbers $n_a$ and $n_b$ for $|a>$ and $|b>$ orbital states with the variation of the total number $n$ of electrons per site is illustrated in Fig.~\ref{Fig5}. With the increase in $n$, first only $|a> = |x^2-y^2>$ orbitals turn out to be occupied. This means that at high doping levels $1-n$, we have the occupation of only one sort of orbitals (a kind of ferro-OO) and these $|x^2-y^2>$ orbitals provide the largest
gain in the kinetic energy of electrons. The further increase in $n$ gives rise to electrons at $|b> = |2z^2 -x^2-y^2>$ orbitals. Eventually, at $n=1$, we get $n_a =1/4 = \sin^2(\pi/6)$ and $n_b =3/4 = \cos^2(\pi/6)$. This corresponds to the average occupation of orbitals at $n=1$ equal to $|\theta = \pm \pi/3>$. However, the 0btained results on the density of states and on the average occupation numbers $n_a$ and $n_b$ do not provide a conclusive evidence concerning the actual alteration of the $|\pm \theta>$ orbitals.

\begin{figure}[H] \begin{center}
\includegraphics*[width=0.45\textwidth]{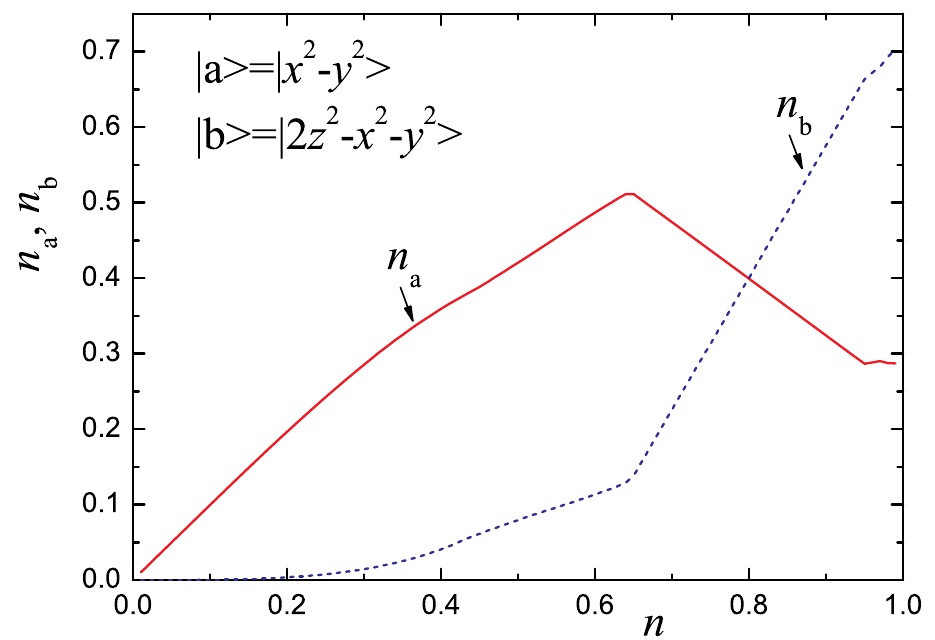} \end{center}
\caption{\label{Fig5} (Color online) Average occupation numbers of $|a>$ (red solid line) and $|b>$ (blue dashed line) orbitals versus total number $n$ of electrons per site. A kink near $n \simeq 0.6$ corresponds to
the doping level at which the chemical potential crosses the
bottom of the upper renormalized band with $s=1$~\cite{KugSboKhoJSNM2009orbitals}.} \end{figure}

Evolution of the Fermi surface with the increase in the total number $n$
of electrons is presented in Fig.~\ref{Fig6}.

\begin{figure} [H] \centering
   \subfigure[\quad $n = 0.15$]{
      \includegraphics[width=0.3\columnwidth]{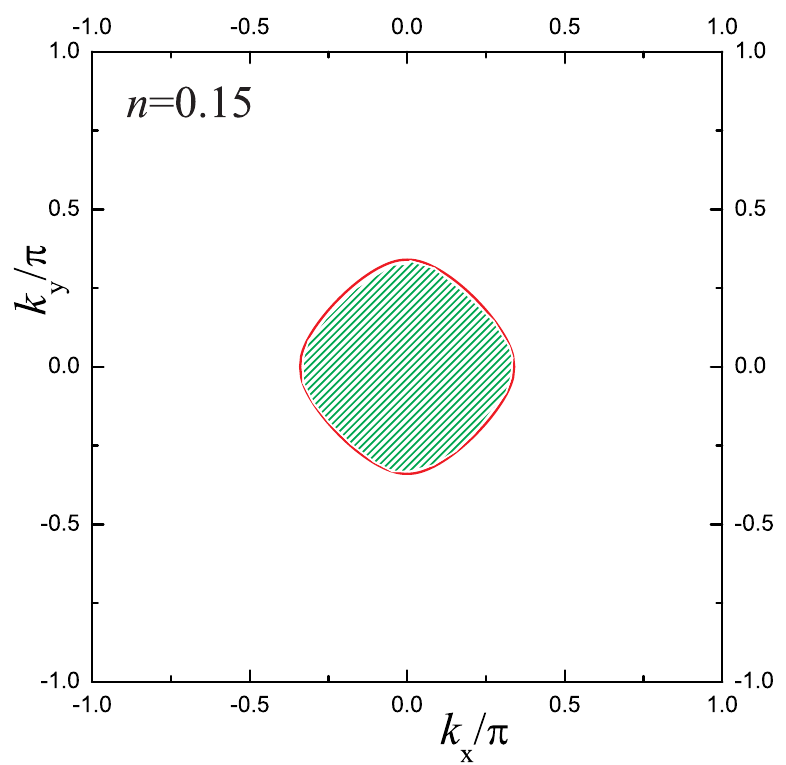}}
   \subfigure[\quad $n = 0.36$]{
      \includegraphics[width=0.3\columnwidth]{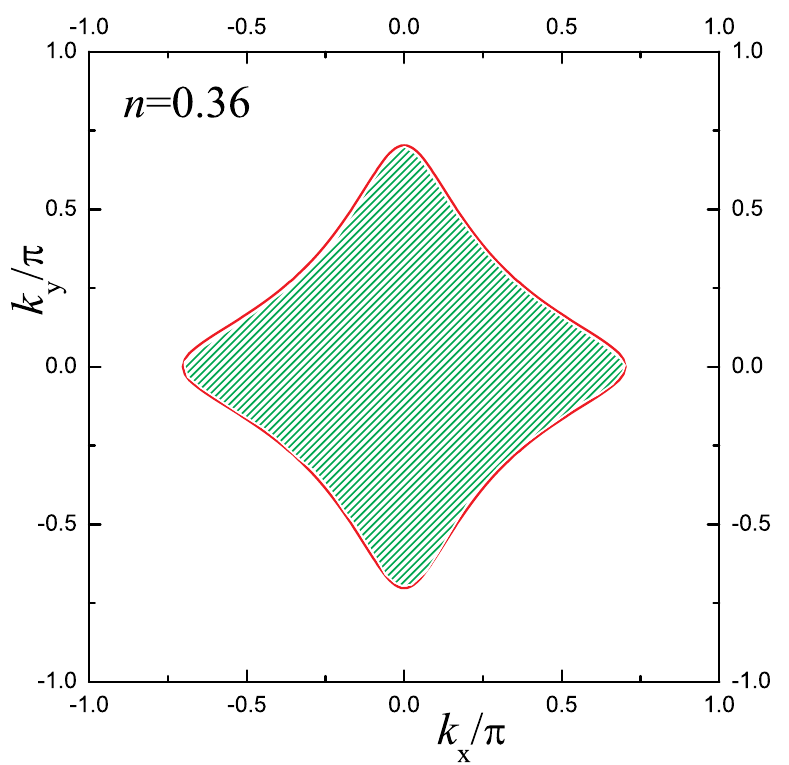}}
   \subfigure[\quad $n = 0.5$]{
      \includegraphics[width=0.3\columnwidth]{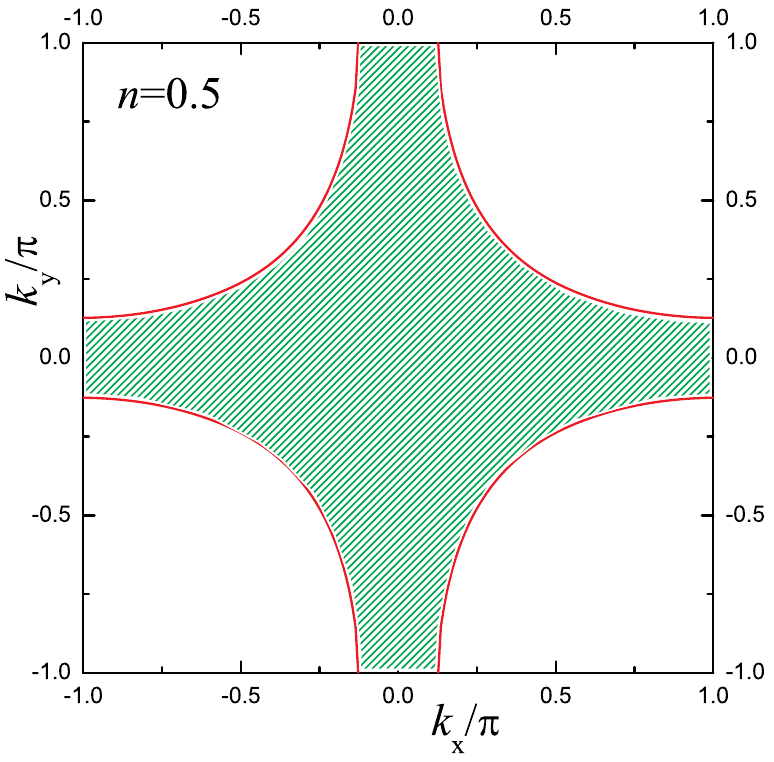}}
   \subfigure[\quad $n = 0.68$]{
      \includegraphics[width=0.3\columnwidth]{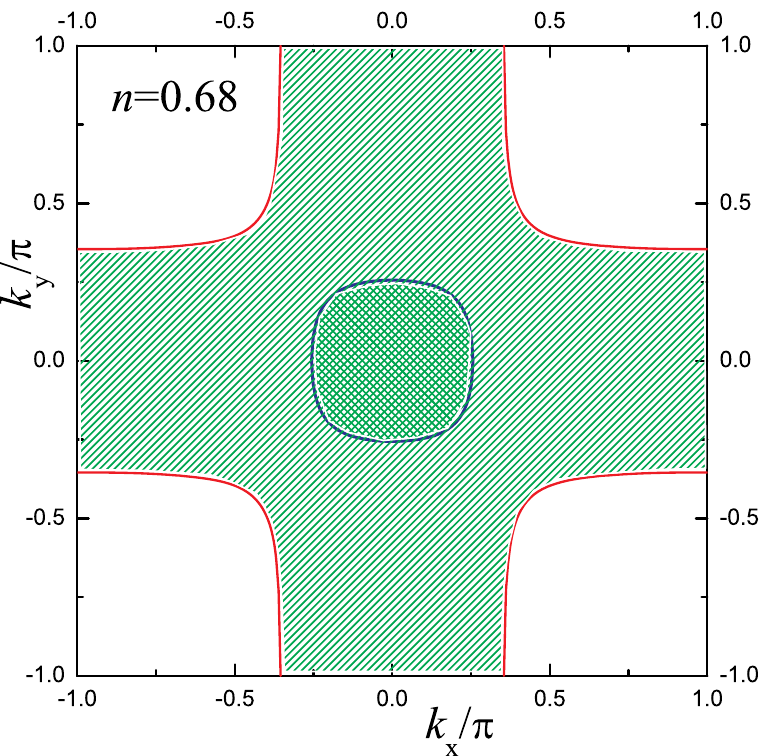}}
   \subfigure[\quad $n = 0.8$]{
      \includegraphics[width=0.3\columnwidth]{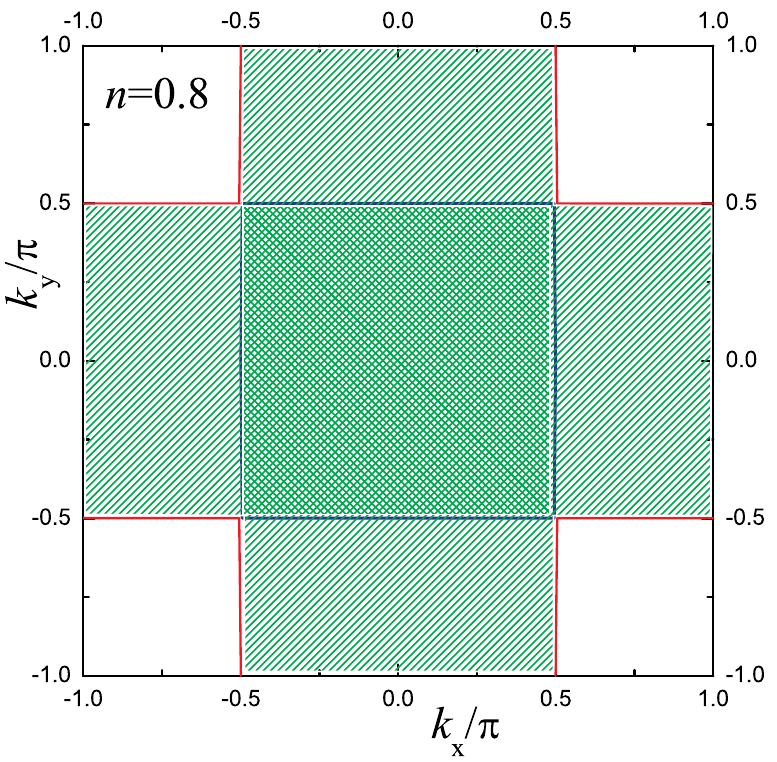}}
   \subfigure[\quad $n = 0.98$]{
      \includegraphics[width=0.3\columnwidth]{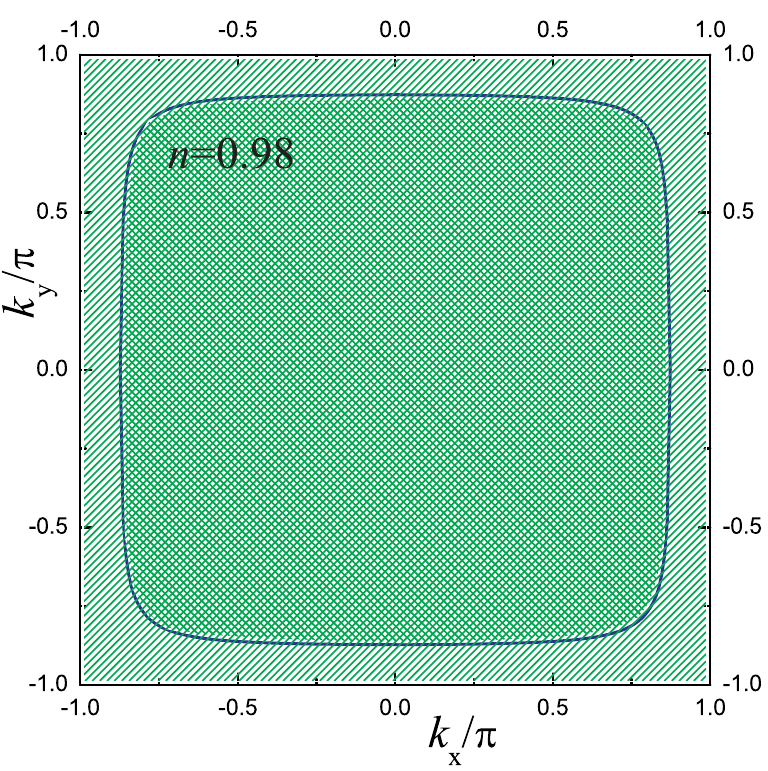}}
\caption{Evolution of the Fermi surface with the increase in the total number of electrons $n$. The red solid curves indicate the Fermi surface sheet corresponding to the $s = -1$ band and the blue solid curves correspond to the $s = 1$ band~\cite{KugSboKhoJSNM2009orbitals}.} \label{Fig6}
\end{figure}

At $n=0.8$ [see Fig.~\ref{Fig6}(e)] the Fermi surface involves several  flat portions. This quite interesting situation. In Fig~\ref{Fig5}, we see that  $n=0.8$ corresponds to $n_a=n_b=0.4$,  and hence, $g_a=g_b=0.4$. Substituting $g_a=g_b$ into Eqs.~\eqref{Epm} and~\eqref{rho_alpha}, we can find
out that in such particular case, the  quasiparticle spectrum and and the corresponding  density of states coincide with those for uncorrelated electrons up to a numerical factor. This means that for $n_a=n_b$, we are dealing with the compensation of  two main correlation effects, namely, the changes in bandwidths and in number of states within the bands.  Thus,  we have a clear similarity to the case of  half filling ($n=1$) for noninteracting electrons~\cite{EfrKhomPRB2005}.
As well as in Ref.~\cite{EfrKhomPRB2005}, at $n_a=n_b=0.4$, there appears the nesting of Fermi surface sheets, which corresponds to the translation by vector
$\mathbf{Q}=\{\pi,\,\pi\}$  implying a possibility of  arising some  orbital superstructure.  Owing to the suppression of correlation effects at this specific doping level, the appearing superstructure could be the same as that discussed in Ref.~\cite{EfrKhomPRB2005}, but, of course, this issue demands a special  analysis.

For the case of the disordered state under study, dependence of  energy on the total number $n$ of electrons per site is is shown in Fig.~\ref{Fig7a} ( red solid line). It is interesting that the shape of this curve is similar to that obtained in subsubsection~\ref{Hom} for the simplified model, see Fig.~\ref{Fig1}. In this case, the correlations themselves do not give rise to the phase separation. Nevertheless,  the comparison of such orbitally disordered model with the model that discussed in subsubsection~\ref{AFOO} reveals that within a certain doping range (at small hole doping), the phase separation could be provide a gain in energy. The energy of such possible phase-separated state is shown by the green dot-and-dash line in Fig.~\ref{Fig7a}.

\subsection{Conclusions} \label{concl-orbital}

In this section, we have analyzed a possibility of  electronic phase separation in the systems with itinerant charge carriers interacting with  an orbitally ordered array.  It was shown that such itinerant charge carriers favor the formation of an inhomogeneous state  with nanosize  inhomogeneities having  the orbital structure different from that in the host material. For the 2D lattice with $e_g$ orbitals, we have found the shapes and sizes of such inhomogeneities. It is shown that the inhomogeneities have he shape strongly dependent  on the ratio of the interorbital  interaction constant and  characteristic hopping integral for the charge carriers, $J/t_0$: there exists a threshold  value of $J/t_0$, corresponding to the transition from the circular inhomogeneities to needle-like ones.

The orbital model under discussion has certain similarities with that corresponding to the double exchange. Here, the operators describing the orbital states are similar to the spin operator. In the previous sections, we demonstrated that the double exchange model exhibits an instability toward phase separation into inhomogeneities with different types of magnetic structure. The inhomogeneous state with circular ferro-OO droplets is, in fact, similar to that involving magnetic polarons (ferromagnetic droplets within the antiferromagnetic host material. The latter state is inherent to  the double exchange~\cite{NagaevBook2002,KaganUFN2001,KaganEPJB1999}. However, this similarity is not trivial:  the orbital model has its own specific features owing to the contribution of  non-diagonal hopping  and the corresponding anisotropy. Such features give rise to the behavior  specific of  orbital systems, such as the kink in the energy of a homogeneous state and canted-OO needle-like droplets. Note also that the difference between spins and orbitals leads to many interesting effects, see, e.g., Refs.~\cite{DaghoferOlesPRL2008, NussinovOrtizEPL2008}.

Of course, in actual magnetic materials, we are often deal with  the usual  mechanism underlying the formation of magnetic polarons (in simple cases - ferromagnetic inhomogeneities in AF matrix), which ensures a gain in the kinetic energy of conduction electrons). However, even in this situation, the possible existence of  orbital degeneracy and orbital ordering present, it is necessary to consider the  orbital structure and its effect on the motion of charge carriers. For example, the  magnetic ordering (parallel spins) could not impede the electron motion, but if it occurs on the in the presence of antiferro-OO (such as. alternating $|2x^2-y^2-z^2\rangle$ and $|2y^2-z^2-x^2\rangle$ orbitals in the basal plane of manganites, which coexists with the ferromagnetic spin ordering in this plane), such antiferro-OO  itself should affect the motion of itinerant charge carriers. Therefore, this favor the formation of orbital polarons even at the ferromagnetic spin background.  This implies that  in the compounds with JT ions, the ferromagnetism does not exclude the formation of orbital inhomogeneities. Such reasoning is  in agreement with the Monte Carlo simulations for the two-orbital model in reported in Ref.~\cite{SenDagottoPRB2006}, which demonstrate the possibility of nanoscale phase separation even in the case of ferromagnetism. In this section, we have not considered any magnetic structure and spin ordering of  charge carriers
A due account taken for the spin degrees of freedom can give rise to a rich variety of  inhomogeneities with different orbital and spin structures.

In the first part of this section, we considered the localized electrons forming an orbital order, whereas conduction electrons or holes were supposed to be two different groups of electrons. However, we have demonstrated that that our main results are also valid for a model, where the same electrons take part both in the hopping and in the formation of orbitally ordered structure. For example, in magnetic oxides with Jahn--Teller ions, lattice distortions lift the orbital degeneracy, this leading to the formation of the ground state with an orbital ordering at $n=1$. In this case, it is natural to assume the long-range orbital ordering does not disappear at low doping levels, $x=1-n\ll 1$. Thus, we are dealing with holes at an orbitally ordered background. Using the mean-field approximation for the analysis of orbitally ordered arrays, we should introduce the factor $1-x$, thus  obtaining the effective interorbital exchange interaction $J(1-x)^2$. In the actual compounds with Jahn--Teller ions, the orbital Hamiltonians have more
complicated forms than the Heisenberg-like one. Calculations involving the superexchange mechanism of the orbital ordering~\cite{KugKhomUFN1982} show that the obtained results still remain valid.

In real substances, At high doping level, real materials can exhibit an
orbitally disordered state or an orbital liquid. For manganites, for example,  such state was discussed in Ref.~\cite{IshiharaNagaosaPRB1997}. This problem is rather controversial, and other possibilities have been also addressed
(e.g., complex orbitals~\cite{vdBrKhomPRB2001}). In the usual spin spin systems, a disordered state is as a rule more favorable from the viewpoint of of kinetic energy than the AF one (or in our case, the antiferro-OO state). However, the detailed analysis of spin systems undertaken in  Refs.~\cite{DagottoBook2003,NagaevBook2002,KaganUFN2001} clearly demonstrates that even in a disordered (i.e. paramagnetic) state, an additional energy gain can be achieved due to the formation  a ferromagnetic polaron (``ferron'') even at the paramagnetic background. It is natural to expect similar possibility for the systems with orbital degrees of freedom,
although this such situation demands a further analysis. To perform it accurately, it is necessary to take into account unusual specific features of the electron motion within disordered media. In particular, one should deal with the disorder, which adjusts itself in the course of electron motion. This is a quite interesting and complicated separate task.

\section{Systems with imperfect nesting and phase separation} \label{imperfect}

\subsection{Introduction} \label{intro-nesing}

Among the multiband materials with the tendency to electronic phase separation there exists an important family of the systems with the nesting of the Fermi surface. The Fermi surface nesting is a very popular and important concept in condensed matter physics~\cite{Khomskii_book2010}. The existence of two fragments of the Fermi surface, which can be matched upon translation by a certain reciprocal lattice vector (nesting vector $\mathbf{Q}_0$), entails an instability of a Fermi-liquid state. A superstructure or additional order parameter related to the nesting vector is generated due to the instability. The nesting is widely invoked for the analysis of charge density wave (CDW) states~\cite{Gruner_RMP1988_CDW,Monceau_AdvPh2012_CDW}, spin density waves (SDW) states~\cite{Overhauser_PR1962_SDW,Gruner_RMP1994_SDW}, mechanisms of high-$T_c$ superconductivity~\cite{RuvaldsPRB1995_nesting_supercond,
GabovichSST2001_SDW-CDW_supercond,TerashimaPNAS2009_nesting_Fe_based}, fluctuating charge/orbital modulation in magnetic oxides~\cite{ChuangScience_SO_fluct}, chromium and its alloys~\cite{ShibataniJPSJ1969_mag_field_chromium,ShibataniPR1969_first,
ShibataniJPSJ1970,RicePRB1970}, organic metals~\cite{NarayananPRL2014,LeePRL2005,VuleticEPJB2002,KangPRB2010}, etc. It is important to emphasize that in a real material, the nesting may be imperfect, i.e. the Fermi surface fragments can match only approximately. One of the earliest studies of imperfect nesting was performed by Rice~\cite{RicePRB1970} in the context of chromium and its alloys (see also  review articles~\cite{Tugushev_UFN1984_SDW_Cr,Fawcet_RMP1988_SDW_Cr}). Quite recently, it was demonstrated that the imperfect-nesting mechanism can be responsible for the nanoscale phase separation in chromium alloys~\cite{WeImperf}, iron-based superconductors~\cite{Sboychakov_PRB2013_PS_pnict}, organic metals (Bechgaard salt)~\cite{RakhmanovJSNM2020}, and in doped bilayer graphene~\cite{Sboychakov_PRB2013_MIT_AAgraph,Sboychakov_PRB2013_PS_AAgraph}. In this context, the studies of spin and charge inhomogeneities related to the imperfect nesting are currently especially active in the physics of low-dimensional compounds~\cite{NarayananPRL2014,CampiNature2015,ChenPRB2014,RakhmanovJETPL2017}.

Other types of inhomogeneous states (``stripes", domain walls) were also discussed in the literature in the framework of similar models~\cite{tokatly1992}. Moreover, it was shown that the possibility of SDW ordering in the systems with itinerant charge carriers results in very rich and complicated phase diagrams involving phase-separation regions~\cite{MachidaPRB1984,IgoshevJPCM2015spiral,Igoshev_JMMM2015}.

Physical mechanisms underlying the nucleation of the inhomogeneous state in materials with nesting are, in general, common to other systems considered in this review. The electronic spectrum instability related to the nesting gives rise to an order parameter and, consequently, to lowering of the system free energy. The better is nesting, the larger is this energy gain. Thus, it may be favorable for the system with imperfect nesting to break up into two phases having different density of itinerant electrons with the better and worse (or even absent) nesting.

\subsection{Phase separation in the Rice model for chromium and its alloys}\label{imperfect_Rice}

We start with the model proposed by Rice~\cite{RicePRB1970} to describe the incommensurate antiferromagnetism in chromium alloys (see also review article~\cite{Fawcet_RMP1988_SDW_Cr}). This model is rather simple and allows for a detailed examination of its phase diagram. Desite its simplicity, it can be applied to describe real important systems. In our consideration, we mostly follow Ref.~\cite{WeImperf}.

The model band structure corresponds to one spherical electron pocket and one spherical hole pocket with different radii (imperfect nesting). It includes as well, another band or bands, which do not participate in the magnetic ordering. All interactions are ignored except the repulsion between electrons and holes in the Fermi surface pockets giving rise to the ordering, since even a small coupling in this channel generates a band instability and opening a gap in the electron spectrum. The Hamiltonian of such model has the form
\begin{equation}\label{RiceHam}
\hat{H}=\sum_{\mathbf{k},\sigma,\alpha}\epsilon^\alpha(\mathbf{k})
n_{\mathbf{k}\sigma}^\alpha+\frac{V}{\cal{V}}\sum_{\mathbf{k,k',q},
\sigma,\sigma'} a^\dag_{\mathbf{k}+\mathbf{q}\sigma}a_{\mathbf{k}\sigma}
b^\dag_{\mathbf{k}'-\mathbf{q}\sigma'}b_{\mathbf{k}'\sigma'}
+\textrm{H.c.},
\end{equation}
where $\alpha$ is equal to either $a$ (electron pocket), $b$ (hole pocket), or $c$ (nonmagnetic bands), $a^\dag$ ($b^\dag$) are creation operators for an electron in the $a$ ($b$) pocket, $n$ is the number operator, $V$ is the Coulomb interaction, and ${\cal V}$ is the volume. The nonmagnetic $c$ band has a finite density of states $N_r$ at the Fermi energy. The energy spectra for the electron and hole pockets measured relative to the Fermi energy are taken as
\begin{eqnarray}\label{spektra}
  \epsilon^a(\mathbf{k}) &=& \hbar v_F(k-k_{Fa})=\hbar v_F(k-k_{F})-\mu, \\
\nonumber
  \epsilon^b(\mathbf{k}+\mathbf{Q}_0) &=& -\hbar v_F(k-k_{Fb})= -\hbar v_F(k-k_{F})-\mu,
\end{eqnarray}
where $k_F =(k_{Fa}+k_{Fb})/2$, $\mu = \hbar v_F(k_{Fa}-k_{Fb})/2$ is the chemical potential, and the wave vector $\mathbf{Q}_0$ connects the centers of the electron and hole pockets in the reciprocal space. We consider the weak-coupling regime: $VN_m \ll 1$, where $N_m = k^2_F/2\pi^2\hbar v_F$. We treat Hamiltonian (\ref{RiceHam}) using a mean-field approach, which gives an accurate result for weakly interacting electrons. In the case of perfect nesting, which corresponds to $\mu=0$, the radii of the electron and hole pockets are identical. If we translate the electron pocket by vector $\mathbf{Q}_0$, its Fermi sphere coincides perfectly with the Fermi sphere of the hole pocket. If we perform the transformations, $b_\mathbf{k} \rightarrow b^\dag_\mathbf{k},b^\dag_\mathbf{k} \rightarrow b_\mathbf{k}$, the interaction constant $V$ and the hole pocket dispersion Eq.~(\ref{spektra}) change sign. As a result, the Hamiltonian for $a$ and $b$ bands becomes identical to two copies of the BCS Hamiltonian. This mapping is useful since it allows us to use the familiar BCS mean-field approach to study the model.

Performing standard BCS-like calculations, it is easy to show that for $\mu=0$ the system is unstable with respect to the ordering with the AFM order parameter $\Delta_0=\frac{V}{{\cal V}}\sum_\mathbf{k}\langle a^\dag_{\mathbf{k} \sigma}b_{\mathbf{k+Q}_0 -\sigma}\rangle$. In the weak-coupling limit under study, we have
\begin{equation}\label{BCS}
\Delta_0=\epsilon_F\exp{\left(-1/N_mV\right)}\ll \epsilon_F,
\end{equation}
where $\epsilon_F=\hbar v_Fk_F$. The order parameter $\Delta_0$ couples electrons with unequal momenta. Consequently, in real space, the order parameter $\Delta_0$ corresponds to the rotation of the magnetization axis with wave vector $\mathbf{Q}_0$. Usually, the $a$ and $b$ pockets are located at the high-symmetry points of the Brillouin zone and the vector $\mathbf{Q}_0$ is related to the underlying lattice structure. Thus, this order is usually called commensurate.

When $\mu\neq 0$, the electron and hole Fermi spheres have different radii, and do not coincide upon translation. However, the difference between the translated spheres remains small, if $\mu$ is small and the system energy is optimized by treating the translation vector $\mathbf{Q}_1 =\mathbf{Q}_0 +\mathbf{Q}$ as a variation parameter. The SDW order parameter has the form
\begin{equation}\label{FFLO}
\Delta=\frac{V}{{\cal V}}\sum_\mathbf{k}\langle a^\dag_{\mathbf{k},\sigma}b_{\mathbf{k+Q}_1,-\sigma}\rangle.
\end{equation}
The vector $\mathbf{Q}$ is small: $|\mathbf{Q}|\ll|\Delta|/\hbar v_F\sim |\mathbf{Q}_0|$. Hence, the order parameter $\Delta$ describes an ordering with a slowly rotating AFM magnetization axis. This rotation is unrelated to the underlying lattice, and such order is referred to as incommensurate. If the transformation $b_\mathbf{k} \rightarrow b^\dag_\mathbf{k},b^\dag_\mathbf{k} \rightarrow b_\mathbf{k}$ is applied to  a system with nonzero $\mu$, the Hamiltonian of our magnetic system becomes similar to the Fulde--Ferrel--Larkin--Ovchinnikov (FFLO) Hamiltonian~\cite{FuldeFerPR1964,LarOvchJETP1965} of a superconductor in the Zeeman magnetic field. Note that we can introduce the order parameters of slightly different structure than in Eq.~(\ref{FFLO}), however, the general result does not change significantly (for details see, e.g., Refs.~\cite{WeImperf,LarOvchJETP1965,gor2010spatial}). In particular, the difference in the free energy and in the chemical potential calculated for different forms of the order parameter is small, and in actual systems, the balance may be shifted by factors unaccounted for by the present model (anisotropy, disorder, etc.)~\cite{WeImperf}.

In the model under discussion, it is convenient to calculate equilibrium parameters of the system by minimization of the grand thermodynamic potential
\begin{equation}\label{Omega}
\Omega=-T\ln{\left[\textrm{Tr}\,e^{-(\hat{H}-\mu \hat{N})}\right]},
\end{equation}
where $\hat{N}$ is the operator of the total number of particles. To evaluate $\Omega$ in the mean-field approximation, we need the eigenenergies of Hamiltonian (\ref{RiceHam}) in the mean-field approach. These are
\begin{equation}\label{eigenenergy}
E_{1,2}(\mathbf{k})=\frac{\epsilon^a(\mathbf{k})+
\epsilon^b(\mathbf{k+Q_1})}{2}\pm\sqrt{\Delta^2+
\left[\frac{\epsilon^a(\mathbf{k})-
\epsilon^b(\mathbf{k+Q_1})}{2}\right]^2}.
\end{equation}
Then, the grand potential equals
\begin{equation}\label{GrPot}
\frac{\Omega}{\cal{V}}=\frac{2\Delta^2}{V}\!-\!2T\!\sum_{s}
\!\!\int{\!\!\frac{d^3\mathbf{k}}{(2\pi)^3}
\ln{\!\left(1\!+\!e^{-E_s/T}\right)}}
-2TN_r\!\!\int{\!\!\ln{\!\left(1\!+
\!e^{-(\epsilon-\mu)/T}\right)}d\epsilon}.
\end{equation}
Here, the first and the second terms are the contributions of the bands responsible for the SDW ordering, while the third term corresponds to the nonmagnetic bands. To carry out the integration over $\mathbf{k}$, we expand the band energies in powers of $|\mathbf{Q}|$ and $\delta k = |\mathbf{k}|-k_F$
\begin{equation}\label{decayQ}
\epsilon^b(\mathbf{k+Q_1})+\epsilon^a(\mathbf{k})\approx 2\mu+2Q\eta,\quad\epsilon^b(\mathbf{k+Q_1})-\epsilon^a(\mathbf{k})\approx 2v_F\delta k+2Q\eta,
\end{equation}
where $Q=2v_F|\mathbf{Q}|/2$ and $\eta$ is the cosine of the angle between $\mathbf{k}$ and $\mathbf{Q}$. Performing the integration, one finds the following expression~\cite{WeImperf}
\begin{eqnarray}\label{deltaOm}
&&\!\!\!\!\!\!\delta\Omega=\!\! \Omega(\Delta,Q,\mu,T)-\Omega(0,Q,\mu,T) \\
\nonumber
&&\!\!\!\!\!\! =\!\! \frac{k_F^3{\cal V}}{\pi^2\epsilon_F}\!\left\{\Delta^2\!\left(\ln{\frac{\Delta}
{\Delta_0}}\!-\!\frac{1}{2}\right)\!+\!\frac{Q^2}{3}\!+\!\mu^2\!+\! \frac{\pi^2T^2}{3}\!+\!T\!\int_0^\infty \!\!\!\!d\xi\int_{-1}^1 \!\!\!\!d\eta\ln{\left[f(Q\eta\!-\!\mu\!-\!\epsilon)f(\mu\!
-\!Q\eta\!-\!\epsilon)\right]}\right\},
\end{eqnarray}
where $f(\epsilon)=1/(1+\exp{\epsilon/T})$ is the Fermi function and $\epsilon=\sqrt{\Delta^2+\xi^2}$.

The equations for the equilibrium values of the gap $\Delta$ and the magnitude of the structural vector $Q$ are determined by minimizing $\delta\Omega$~\cite{WeImperf}
\begin{eqnarray}\label{DeltaQ}
\nonumber
\ln{\frac{\Delta}{\Delta_0}} &=&\int_0^\infty\frac{d\xi}{2\epsilon}\int_{-1}^1 d\eta\left[f(\epsilon+\mu-Q\eta)+f(\epsilon-\mu+Q\eta)\right]  ,\\
\frac{2Q}{3} &=& -\int_0^\infty d\xi\int_{-1}^1\eta d\eta\left[f(\epsilon+\mu-Q\eta)+f(\epsilon-\mu+Q\eta)\right].
\end{eqnarray}
The total number of electrons per unit volume $n(\mu)$ is the sum of the magnetic $n_m(\mu)$ and nonmagnetic electrons $n_r(\mu)$ . The doping $x$ is defined as the difference $x= n(\mu)-n(0)$. Since $T,\mu\ll \epsilon_F$, we can write $n_r(\mu)-n_r(0)=N_r\mu$. For magnetic electrons, we have $n_m(\mu)=(2/{\cal V})\sum_{\mathbf{k},s}{f(E_s(\mathbf{k}))}$. After straightforward calculations, we derive
\begin{equation}\label{doping}
\frac{x}{x_0}=\frac{n\mu}{\Delta_0}+\int_0^\infty\frac{d\xi}
{\Delta_0}\int_{-1}^1d\eta\left[f(\epsilon-\mu+Q\eta)
-f(\epsilon+\mu-Q\eta)\right],
\end{equation}
where $x_0=4\Delta_0N_m$ and $n=N_r/2N_m$. Equations~(\ref{deltaOm}), (\ref{DeltaQ}), and (\ref{doping}) allow constructing a phase diagram of the system~\cite{WeImperf}.

The phase diagram of the model can be constructed by numerical solution of Eqs.~(\ref{deltaOm}), (\ref{DeltaQ}), and (\ref{doping}). First, one can calculate the N\'{e}el temperature $T_N(x)$ for homogeneous state of the system and the transition temperature $T_Q$ between the commensurate and incommensurate AFM phases, which corresponds to the highest doping, at which $Q =0$. Then, we can determine the region of the phase-separated state and the phases in this state analyzing the dependence of the grand potential $\Omega$ or chemical potential $\mu$ versus doping $x$ following the standard procedure described in the previous sections (see also for details Ref.~\cite{WeImperf}).

As an illustration, a typical dependence of the chemical potential $\mu$ on doping $x$ for the considered model is shown in Fig.~\ref{RiceChemPot}. It is seen from this figure that the function $\mu(x)$ has a descending part, which means a negative compressibility and, hence, instability of the homogeneous AFM phase. In the phase-separated state, the system segregates into two phases with different doping values $x_1$ and $x_2$. The values $x_1$ and $x_2$ can be found using the Maxwell construction (see, e.g., Ref.~\cite{LeBellacBook2004}). Figure~\ref{RiceChemPot} illustrates the latter concept: the horizontal dashed line is drawn in such a manner that the areas of the shaded regions, $S_1$ and $S_2$, are equal to each other.

\begin{figure}[H] \centering
  \includegraphics[width=0.5\columnwidth]{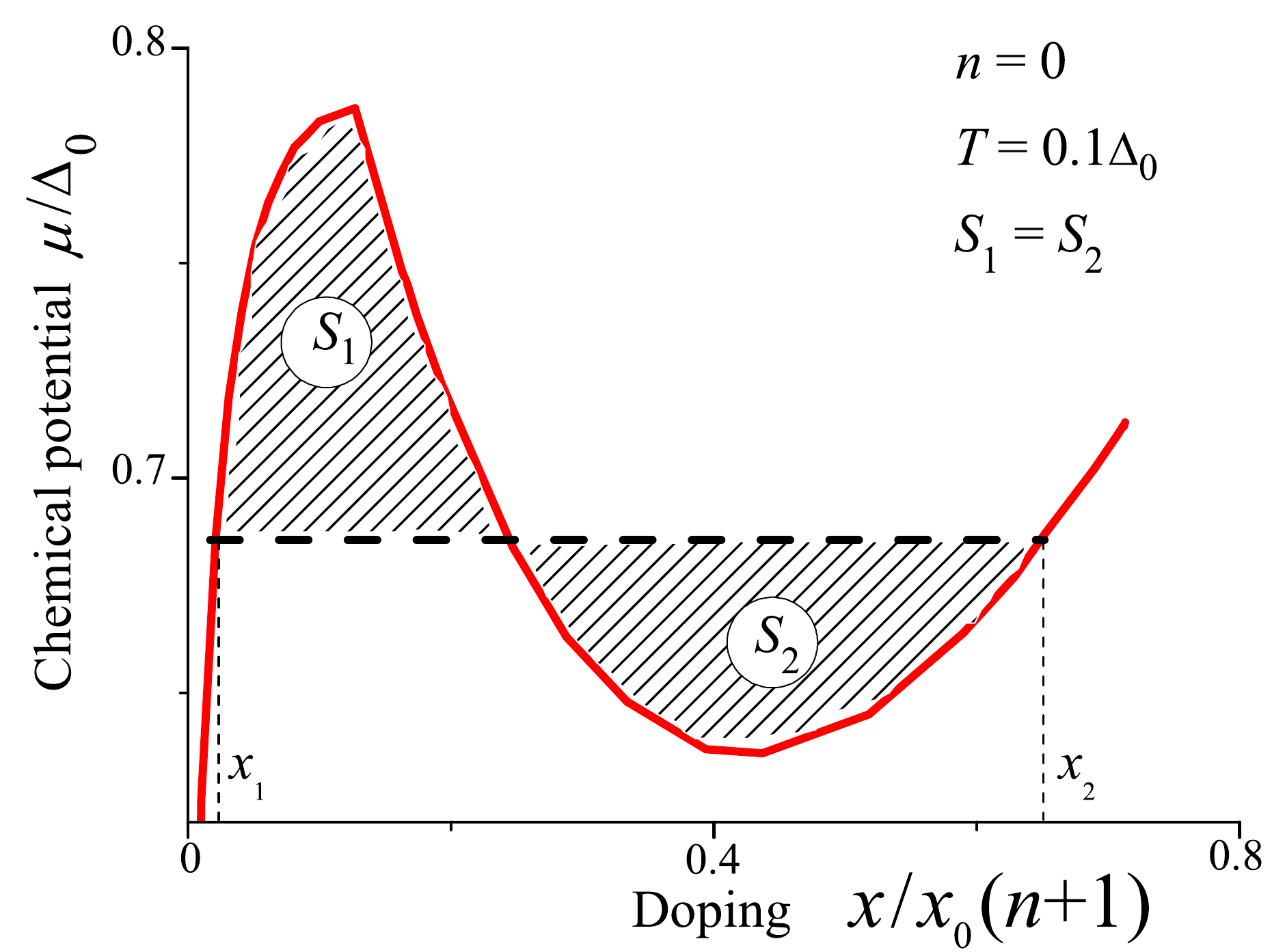}\\
  \caption{ (Color online)  Chemical potential $\mu$ versus doping $x$ for the homogeneous phase, $T/\Delta_0 =0.1$ and $n =0$ [solid (red) line]. The horizontal dashed (black) line shows the Maxwell construction, with shaded areas $S_1 = S_2$ \cite{WeImperf}.
  }\label{RiceChemPot}
\end{figure}

The calculated by the latter method boundary $T_{PS}(x)$ between the homogeneous and phase-separated states is shown by the dashed (red) curves in the $(x,T)$ phase diagrams drawn in Fig.~\ref{RicePsD}, for different values of $n$. The phase with lower doping, $x_1$, is the commensurate AFM ($Q =0$), whereas the phase with higher doping, $x_2$, is the incommensurate AFM ($Q \neq 0$). Thus, the phase separation occurs due to the competition between two AFM states with different structures.

\begin{figure}[H] \centering
    \includegraphics[width=0.5\columnwidth]{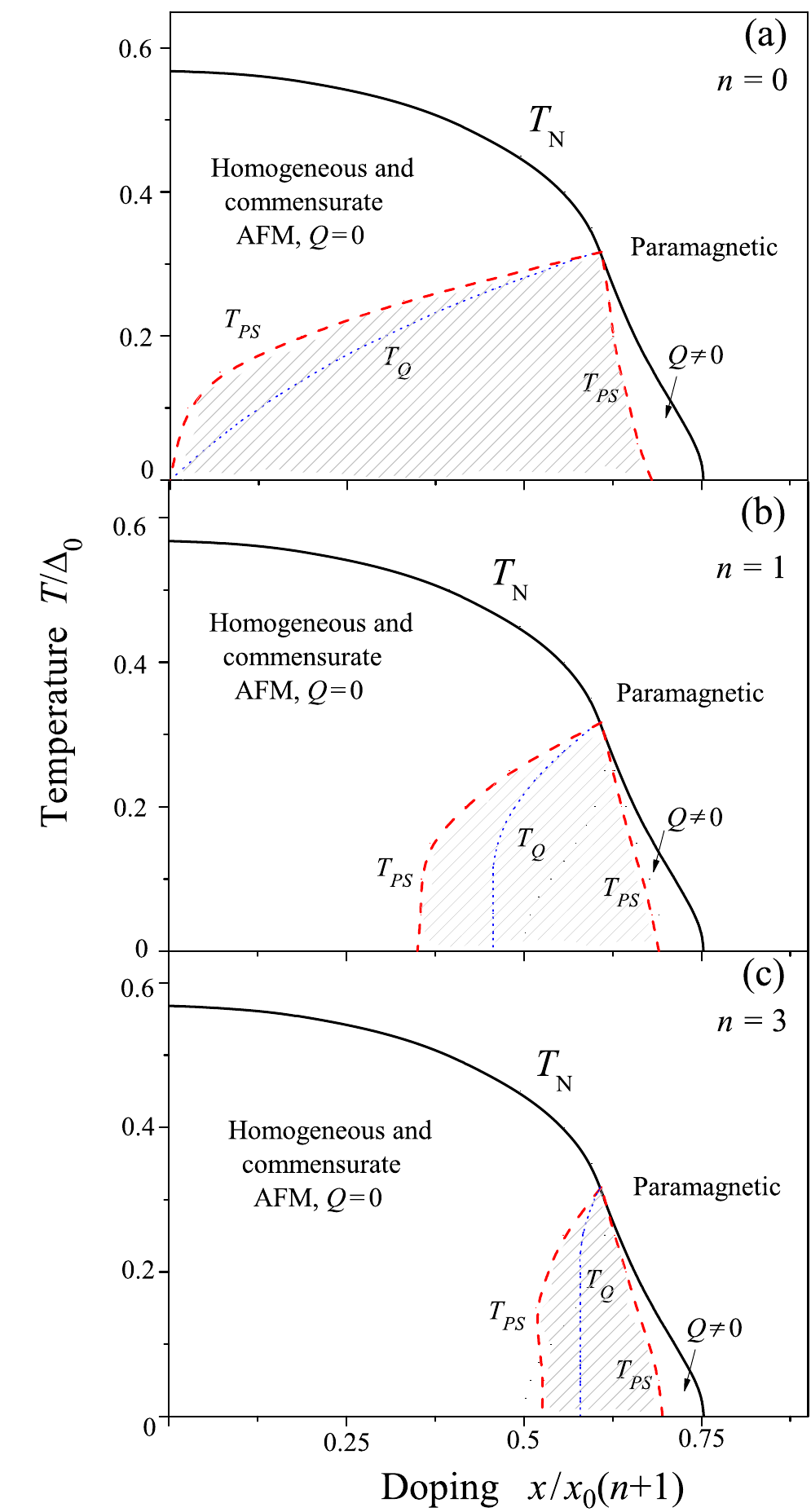}\\
  \caption{(Color online) Phase diagram of the Rice model for $n =0$ (no electrons in nonmagnetic pockets) (a), $n =1$ (b), and $n =3$ (c). The solid (black) curves represent the N\'{e}el temperature $T_N(x)$, which separates the paramagnetic ($\Delta=0$) and AFM ($\Delta \neq 0$) phases. The dotted (blue) curves denote $T_Q(x)$, i.e. the boundary between the commensurate ($Q =0$) and incommensurate ($Q \neq 0$) homogeneous AFM phases. The dashed (red) curve, $T_{PS}(x)$, is the boundary between the uniform and phase-separated (shaded areas) phases~\cite{WeImperf}.
}\label{RicePsD}
\end{figure}

Above, we have mentioned the mapping between the Rice model and the FFLO superconductor. However, the phase separation is absent from the phase diagram of the FFLO state in a superconductor. This has a very simple explanation: the Zeeman field in the FFLO superconductor corresponds to doping in the Rice model. The field, being an intensive thermodynamic quantity, does not allow for the phase separation. However, for cold atoms in an optical trap (where an analogue of the FFLO is predicted), it may be possible to control not the field, but the polarization, which is an extensive quantity. In this case, the phase separation occurs~\cite{sheehy2007bec}.

The FFLO state is very fragile with respect to the impurity scattering~\cite{takada1970superconductivity}. Thus, the incommensurate AFM, which is the mathematical analog of the FFLO phase, is expected to be susceptible to microscopic imperfections. Therefore, the disorder-induced modifications to the phase diagram is an open question.

The study of characteristic scales and geometry of the phase-separated state could be performed by the method similar to that described in the previous sections of the present review. It requires, as usual, to take into consideration an interplay between the long-range Coulomb interaction and the energy of the boundary between different phases. However, such a study is absent in literature up to date.

\subsection{The effect of magnetic field on the phase separation in systems with imperfect nesting}\label{imperfect_field}

Thus, the electron--electron coupling gives rise to the nucleation of the magnetic ordering and the phase separation in the system with nesting. Naturally, the magnetically ordered system is especially sensitive to the applied magnetic field. This problem was investigated in Ref.~\cite{SboychakovPRB2017} in the framework of the considered above Rice model. We analyze the effects of the applied magnetic field following the latter paper.

Considering  electrons or holes in the applied magnetic field $\mathbf{B}$, we should, first, replace the momentum operator $\mathbf{\hat{p}}$ by the gauge-invariant one $\mathbf{\hat{p}}+(e/c)\mathbf{A}$, where $\mathbf{A}$ is the vector potential of magnetic field. Second, we have to add the Zeeman term $g_\alpha\sigma\hbar\omega_\alpha$ in the single electron Hamiltonian, where $g_\alpha$ is the Land\'e $g$-factor and $\omega_\alpha=eB/ cm_\alpha$ is the cyclotron frequency of a charge carrier with mass $m_\alpha$. The energy spectrum of the Rice model is characterized by two single-particle energy scales. The first scale is the Fermi energy $\epsilon_F$, and the second one is $\hbar\omega_\alpha$, which is the distance between the Landau levels. The energy scale associated with the interactions will be characterized by the value of a spectral gap $\Delta_0$ and in the case of the weak electron--hole coupling $\Delta_0\ll \epsilon_F$. The Landau quantization is of importance in the range of high magnetic field, $\hbar\omega_\alpha>\Delta_0$, whereas at low fields, $\hbar\omega_\alpha<\Delta_0$, it can be neglected. Here, we limit ourselves by the regime of low magnetic fields, neglecting any corrections associated with the small ratio $\hbar\omega_\alpha/\epsilon_F\ll 1$. In the range of high magnetic fields, there exists a set of oscillation phenomena, which highly complicates the phase diagram even of this simple model.

Adding the Zeeman terms, we write instead of Eq.~(\ref{spektra}) the single-electron spectrum in the form
\begin{eqnarray}\label{spektra_B}
  \epsilon^a_\sigma(\mathbf{k}) &=& \hbar v_F(k-k_{F})-\mu+g_a\sigma\hbar\omega_a, \\
\nonumber
  \epsilon^b_\sigma(\mathbf{k}+\mathbf{Q}_0) &=&  -\hbar v_F(k-k_{F})-\mu +g_b\sigma\hbar\omega_b.
\end{eqnarray}

The applied magnetic field lifts the degeneracy with respect to the spin projection, and now we have two electron and two hole bands, see Fig.~\ref{NestMagBand}. In such case we have to introduce a two-component order parameter corresponding to the nesting vectors shown by the arrows in Fig.~\ref{NestMagBand}. Neglecting for simplicity the incommensurate SDW ordering, these components can be written as \cite{SboychakovPRB2017}

\begin{equation}\label{Delta_B}
\Delta_\uparrow=\frac{V}{\cal{V}}\sum_{\mathbf{k}}\langle a^\dag_{\mathbf{k} \uparrow}b_{\mathbf{k}\downarrow}\rangle,\quad \Delta_\downarrow=\frac{V}{\cal{V}}\sum_{\mathbf{k}}\langle a^\dag_{\mathbf{k} \downarrow}b_{\mathbf{k}\uparrow}\rangle.
\end{equation}

The case when the electron and hole bands are perfectly symmetric is the simplest. In such a situation, however, the effect of Zeeman field on the electron spectrum of the model is absent. Thus, some electron--hole asymmetry is necessary to obtain a non-trivial result. Following Ref.~\cite{SboychakovPRB2017}, we put $m_a = m_b = m$, hence, $\omega_a =\omega_b =\omega_H$, but $g_a \neq g_b$. Following the procedure described in the previous subsection, now we can find eigenenergies and calculate the grand potential $\Omega$. Minimization of $\Omega$ gives the expression for  components of the order parameter~\cite{SboychakovPRB2017}
\begin{equation}\label{Delta_sigma}
\ln{\frac{\Delta_0}{\Delta_\sigma}}=\int_0^\infty \!\!\!d\xi\,\frac{f(\sqrt{\Delta^2_\sigma+\xi^2}+\mu_\sigma)
+f(\sqrt{\Delta^2_\sigma+\xi^2}-\mu_\sigma)}{\sqrt{\Delta^2_\sigma+\xi^2}},
\end{equation}
where $\mu_\sigma=\mu-\sigma(g_a-g_b)\hbar\omega_H/2$. Neglecting the existence of the nonmagnetic part of the Fermi surface, one can write the expression for doping analogous to Eq.~(\ref{doping})
\begin{equation}\label{doping_B}
\frac{x}{x_0}=\sum_\sigma{\int_0^\infty\frac{d\xi}
{4\Delta_0}\left[f(\sqrt{\Delta^2_\sigma+\xi^2}
-\mu_\sigma)-f(\sqrt{\Delta^2_\sigma+\xi^2}+\mu_\sigma)\right]}.
\end{equation}

\begin{figure}[H] \centering
\includegraphics[width=0.5\columnwidth]{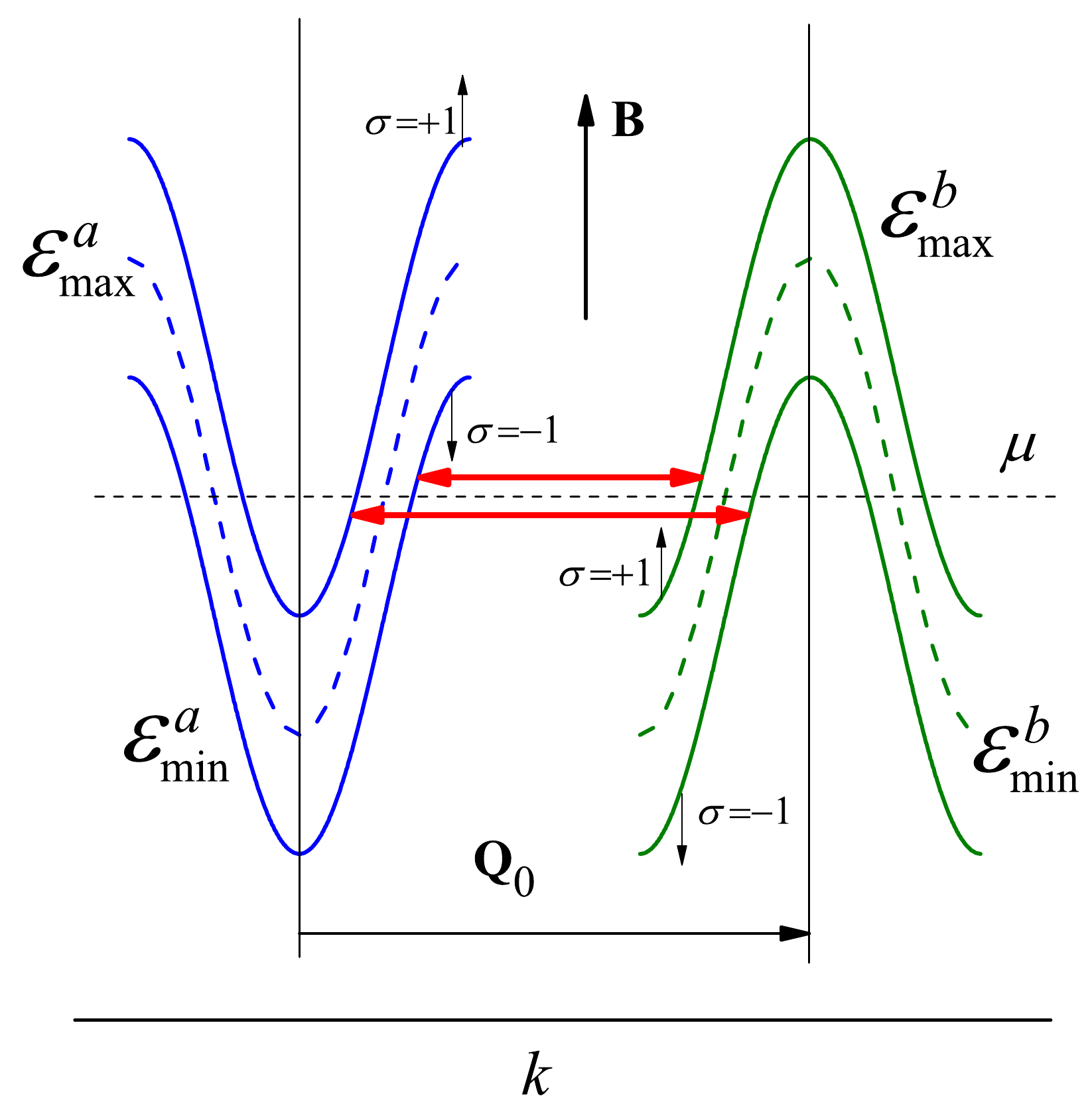}\\
\caption{ (Color online) Band structure of the model in an applied magnetic field. The applied magnetic field lifts the degeneracy of  electron-like ($a$) and hole-like ($b$) bands with respect to the electron spin. The red arrows indicate the interband coupling giving rise to the order parameters~\cite{SboychakovPRB2017}.
  }\label{NestMagBand}
\end{figure}

In Ref.~\cite{SboychakovPRB2017}, the phase diagram of the model in the plane (magnetic field, doping) was constructed in the limit $T=0$. The obtained results are illustrated in Fig.~\ref{PS_B}, where the dimensionless magnetic field is defined as $b=|g_a-g_b|\hbar\omega_H/2\Delta_0$. The homogeneous PM state exists in the system at high doping ($x/4x_0>2b +\sqrt{2}$) or at high magnetic field ($b>1/\sqrt{2}+x/8x_0$) (the value $x_0$ in Ref.~\cite{SboychakovPRB2017} corresponds to $4x_0$ in the present notation). In the range of the low doping and low magnetic field, the ground state of the system is the phase-separated one, which we refer to as PS1. The phase PS1 is a mixture of the two AFM phases. The first phase corresponds to SDW with zero doping, $x=0$, and $\Delta_\uparrow=\Delta_\downarrow=\Delta_0$. The second phase with nonzero doping, $\Delta_\uparrow=\Delta_0$, and $\Delta_\downarrow=0$ will be further referred to as AF1. The phase PS1 is the ground state of the model if $|b|<1/\sqrt{2}$ and $|x/4x_0|<1/\sqrt{2}$. The uniform SDW phase AF1 is stable within the intermediate magnetic fields range, $x/8x_0-\/2\sqrt{2}<b<x/8x_0+\/2\sqrt{2}$. The inhomogeneous state PS2 is stable in the regions between homogeneous PM and AF1 phases. The PS2 state consists of the mixture of the SDW phase AF1 and PM phase. It is interesting to note that the phase AF1 is some non-common type of the half-metal, since in this phase only electrons with a single spin projection exist at the Fermi surface~\cite{RozhkovPRL2017}.

Thus, we see that the magnetic field lifts the spin-projection degeneracy of the SDW order parameter in the system with imperfect nesting. As a result the number of possible homogeneous states increases, which, in turn, drastically affects the picture of electronic phase separation.

\begin{figure}[H] \centering
\includegraphics[width=0.5\columnwidth]{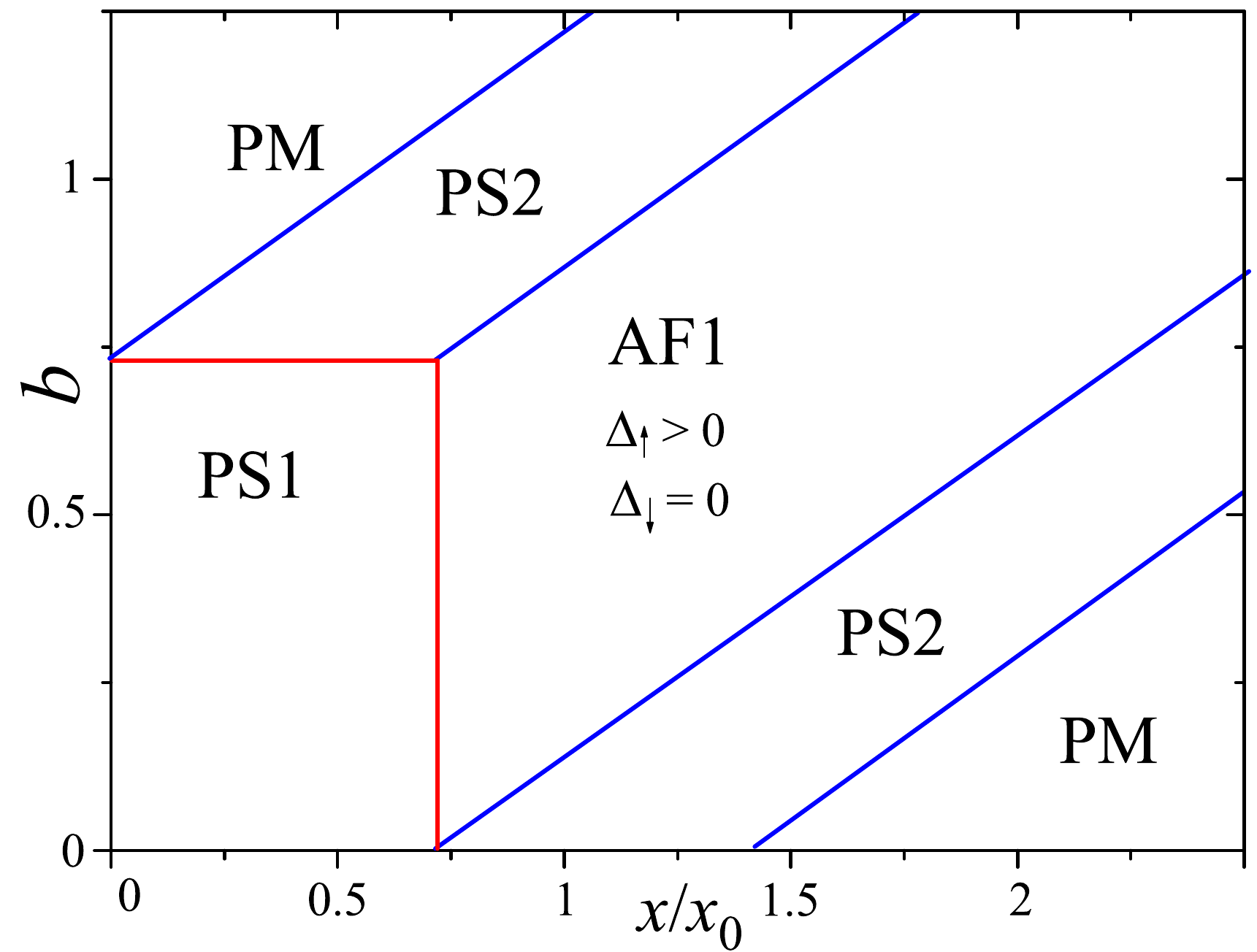}\\
\caption{ (Color online) Phase diagram of the model at zero temperature in the ($b,x/x_0$) plane \cite{SboychakovPRB2017}. The value $x_0$ in Ref.~\cite{SboychakovPRB2017} corresponds to $4x_0$ in the present notation. In the diagram, we see the homogeneous SDW phase AF1 (the definition is shown in the figure), two regions with the uniform PM phase, and the phase-separated states PS1 (mixture of two SDW phases, see the text) and PS2 (mixture of AF1 and the PM phases).\label{PS_B}}
\end{figure}

\subsection{Phase separation in the organic metals under pressure}\label{imperfect_Pressure}

In the above consideration we demonstrated that the system with nesting is very sensitive to the position of the Fermi level due to the doping and/or to the change of the electron spectrum due to application of the magnetic field. This conclusion can be formulated in a more general form: any impact that affects the value of the chemical potential or electron spectrum in a system with nesting can give rise to the crucial changes in its thermodynamic state. Here we discuss the effect of the pressure. This case is of a special interest due to the experiments with an archetypical Bechgaard salt (TMTSF)$_2$PF$_6$ under pressure~\cite{NarayananPRL2014,LeePRL2005,VuleticEPJB2002,KangPRB2010}. This system is commonly considered as a system with the nesting of the Fermi surface. In these experiments the coexistence of the SDW and PM metallic (superconducting at low temperatures) phases was clearly observed. Moreover, in Ref.~\cite{NarayananPRL2014} the SDW and PM metallic phases manifest itself in the form of phase separation with macroscopic domains of the high-pressure PM metallic phase embedded into the insulating SDW host, which are aligned along certain crystallographic axes. A natural explanation of this effect is the same as in the case of doping or magnetic field: the application of the pressure shifts the electron system from the position of a perfect nesting.

A qualitative explanation of the experiment~\cite{NarayananPRL2014} can be obtained by an incorporation of the pressure effect in Rice model~\cite{RakhmanovJSNM2020}. In the experiments~\cite{NarayananPRL2014}, an evolution with pressure of the geometry of the SDW-PM phase was also observed. To trace this evolution theoretically the electrostatic energy of the inhomogeneous state and the energy of interfaces between SDW and PM phases should be taken into account as it was described in section~\ref{CoulombSurface}.

In general, the pressure $P$ may affect all parameters of the system, e.g., electron coupling $V$, electron and hole effective masses, band positions $\epsilon^{\alpha}_{\text{min}}$, $\epsilon^{\alpha}_{\text{max}}$, etc. However, if the pressure is not very high, the main effect on the electron properties of the material with nesting occurs due to the shift of the Fermi energy closer or farther from the ideal nesting position. Following Ref.~\cite{RakhmanovJSNM2020}, we assume that the change in the Fermi energy arises due to the shift of the energy bands with respect to each other. For simplicity, in the cited work only the shift the reservoir band is taken into consideration while the shift of the magnetic bands is ignored. Below we discuss only the case of the commensurate SDW, that is, $\mathbf{Q}=0$. It occurs that this simple minimal model is sufficient to reproduce main features of the experimental observations reported in Ref.~\cite{NarayananPRL2014}.

If the position of the bottom of the reservoir band is shifted on $\delta\epsilon$ due to pressure, then, the corresponding variation of the the chemical potential is $\mu(P)-\mu(0)$. Naturally, the total number of electrons per unit cell (and, consequently, doping $x$) remains unchanged.
The variation in the number electrons per unit cell in the nonmagnetic band is
\begin{equation}\label{dnr}
\delta n_r=2n_r[\mu(P)-\mu(0)-\delta\epsilon]\,.
\end{equation}
while the variation in the number of magnetic electrons is
\begin{eqnarray}\label{dnm}
\delta n_m&=&4n_m\int\limits_0^\infty\!\!{d\xi\,\{f[\eta-\mu(P)]-f[\eta+\mu(P)]-f[\eta\!-\!\mu(0)]-f[\eta\!+\!\mu(0)]\}}.
\end{eqnarray}
We assume that the system is perfectly nested when $P=P_n$, that is, $\mu(P_n)=0$ and $x=0$. In the range of not too high pressures the value $\delta\epsilon$ depends linearly on the pressure
\begin{equation}\label{E_shift}
\delta\epsilon(P)=\frac{\partial\epsilon^{c}_{\text{min}}}{\partial P}\left(P-P_n\right)\,.
\end{equation}
We substitute Eqs.~\eqref{dnr}\,--\,\eqref{E_shift} into Eq.~\eqref{doping} with $Q=0$ and make the replacement $\mu+\delta\mu\to\mu$. Then, taking into account the condition $x(P)=const=0$, one derives~\cite{RakhmanovJSNM2020}
\begin{equation}\label{DopingP}
r\bar{P}=\frac{r\mu}{\Delta_0}+\frac{1}{\Delta_0}\!\int\limits_0^\infty
\!\!{d\xi\,[f(\eta-\mu)-f(\eta+\mu)]}\,,
\end{equation}
where the dimensionless variable $\bar{P}$ is
\begin{equation}\label{psmall}
\bar{P}=\frac{\partial\epsilon^{c}_{\text{min}}}{\partial P}\frac{P-P_n}{\Delta_0}\,.
\end{equation}
The sign of the derivative $\partial\epsilon^{c}_{\text{min}}/\partial P$ depends on the particular properties of the system. Below we assume that this value is positive. In the case of zero doping, $x=0$, Eqs.~\eqref{DeltaQ} at $Q=0$ and~\eqref{DopingP} form a closed system of equations determining the functions $\Delta(\bar{P},T)$ and $\mu(\bar{P},T)$. If $x\neq0$, one has to substitute $r\bar{P}\to r\bar{P}+x/x_0$ in the left-hand side of Eq.~\eqref{DopingP}, taking into account that under doping, pressure $P=P_n$ corresponds to a perfect nesting.

Comparing Eq.~\eqref{DopingP} with Eq.~\eqref{doping}, we see that the pressure acts as some effective doping. The only difference is that this `effective doping' is proportional to the density of electron states in the reservoir bands, $r$ and when the reservoir is absent, the pressure changes nothing. This is an artefact of the approximation used, where possible changes in the shapes of the Fermi surfaces of the magnetic bands were neglected. Under such assumption the only possible way to destroy the ideal nesting is to change the difference in the Fermi momenta $k_{Fa}-k_{Fb}$ of the magnetic bands, which is prohibited due to the charge conservation law. When the reservoir is present, the imbalance in the magnetic electrons appears due to the unequal  $k_{Fa}$ and $k_{Fb}$ can be compensated by the reservoir electrons.

The equation for the gap, Eq.~\eqref{DeltaQ}, at $Q=0$
and~\eqref{DopingP} can be solved numerically to obtain the functions $\Delta(p,T)$ and $\mu(\bar{P},T)$. Then, assuming that the ground state of the system is homogeneous, one can find the grand potential $\Omega$ and the free energy $F=\Omega+\mu n$, and construct the phase diagram of the model in the $p$--$T$ plane. Such calculations were performed in Ref.~\cite{RakhmanovJSNM2020} and the phase diagram of the model in the $(P,T)$ plane was obtained. The calculated dependence of the gap $\Delta$ on the pressure occurs similar to the dependence of the gap versus doping: the  gap decreases with $P-P_n$ and vanishes zero at some value of the applied pressure. However, the true ground state of the model is inhomogeneous similar to the cases considered in sections~\ref{imperfect_Rice} and \ref{imperfect_field}.

In the case of the finite doping $x$ Eq.~\eqref{DopingP} becomes~\cite{RakhmanovJSNM2020}
\begin{equation}\label{DopingPX}
\frac{x_{\text{eff}}(P)}{x_0}=\frac{r\mu}{\Delta_0}+\frac{1}
{\Delta_0}\!\int\limits_0^\infty\!\!{d\xi\,[f(\eta-\mu)-f(\eta+\mu)]}\,,
\end{equation}
where
\begin{equation}\label{xeff}
x_{\text{eff}}(P)=x+rx_0\bar{P}\,.
\end{equation}
First, we have to calculate the chemical potential $\mu$ as a function of $x_{\text{eff}}$. The obtained dependence $\mu(x_{\text{eff}})$ is similar to $\mu(x)$ shown in Fig.~\ref{RiceChemPot}. Thus, the value $\mu(x_{\text{eff}})$  behaves nonmonotonically indicating an instability of the homogeneous state toward the phase separation. The separated phases are the SDW phase with effective doping $x^{(1)}_{\text{eff}}$ and the PM phase with effective doping  $x^{(2)}_{\text{eff}}$. As above, the values $x^{(1,2)}_{\text{eff}}$ can be obtained using the Maxwell construction, see Fig.~~\ref{RiceChemPot}. These values depend on the model parameters and temperature, but do not depend on pressure, while ``bare'' doping in the in the phase-separated phases, $x_{1,2}$, depend on $P$. From Eq.~\eqref{xeff} it follows that $x_{1,2}=x^{(1,2)}_{\text{eff}}-rx_0p$. If the volume fraction of the PM phase is $p$, then, from the charge conservation law, we have
\begin{equation}\label{charge-cons}
(1-p)x_1+px_2=x\,.
\end{equation}
In the doping range, $x_1<x<x_2$, the ground state of the model is a mixture of SDW phase with the electron density $x_1=x^{(1)}_{\text{eff}}-rx_0\bar{P}$  and PM phase with the electron density $x_2=x^{(2)}_{\text{eff}}-rx_0\bar{P}$. This doping range shifts toward the smaller densities when the pressure increases. Suppose that $x<x^{(1)}_{\text{eff}}$, thus, at $P=P_n$ (when $\bar{P}=0$), the ground state is a homogeneous SDW phase. Increasing $\bar{P}$ above the value $\bar{P}_1=(x^{(1)}_{\text{eff}}-x)/(rx_0)$ makes the system inhomogeneous. The fraction of the PM phase, $p$, gradually increases with $\bar{P}$ and at $\bar{P}=\bar{P}_2=(x^{(2)}_{\text{eff}}-x)/(rx_0)$, the system becomes a homogeneous PM. As an illustration, the phase diagram of the model in the $(P,T)$ plane is shown in Fig.~\ref{fig3Pressure} for $r\ll 1$ (a) and for $r=2$ (b). The region of the possible inhomogeneous phase occupies a significant part of the phase diagram.

\begin{figure}[btp]
\centering
\leavevmode
\includegraphics[width=8.5 cm]{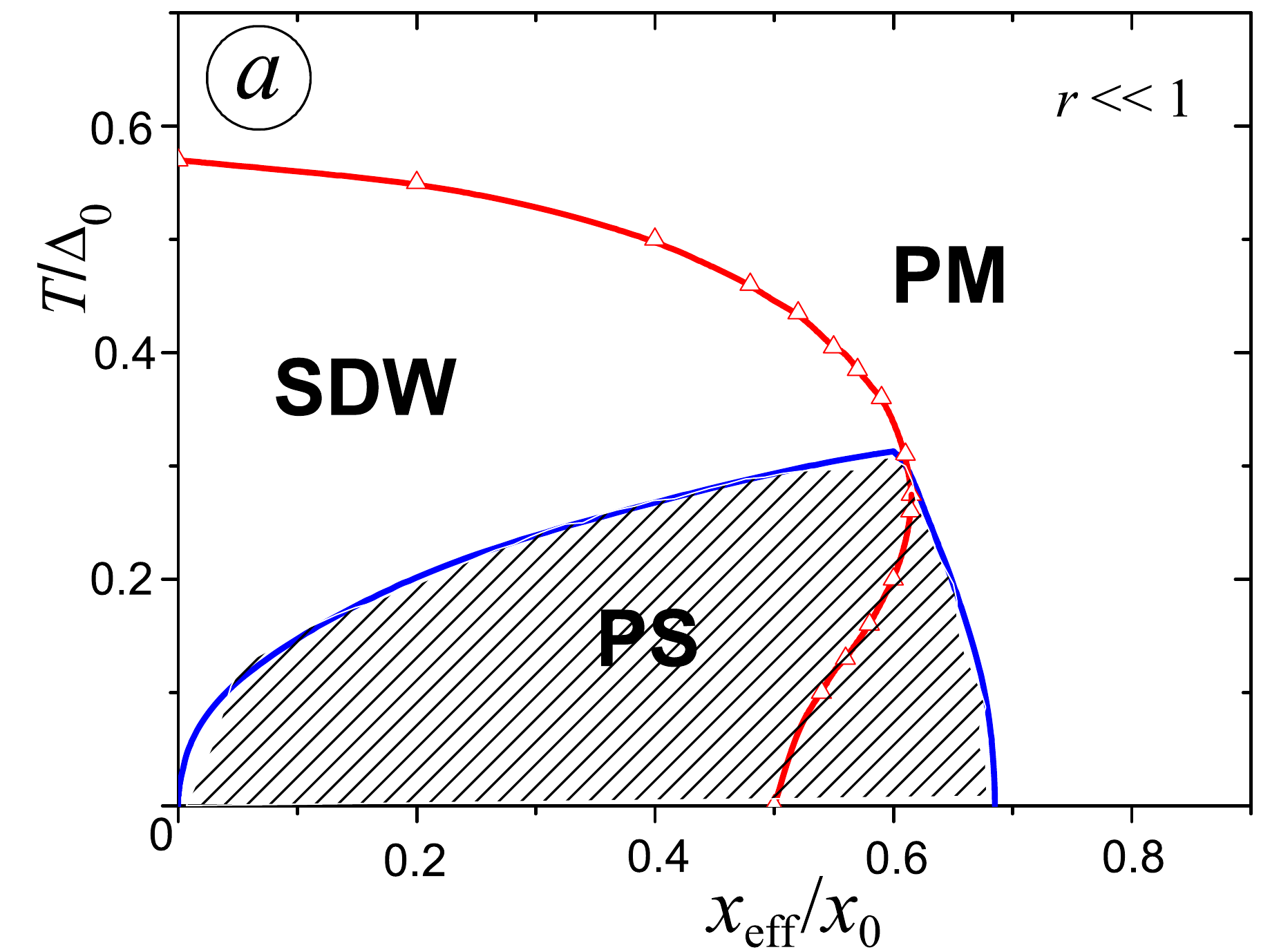}
\includegraphics[width=8.5 cm]{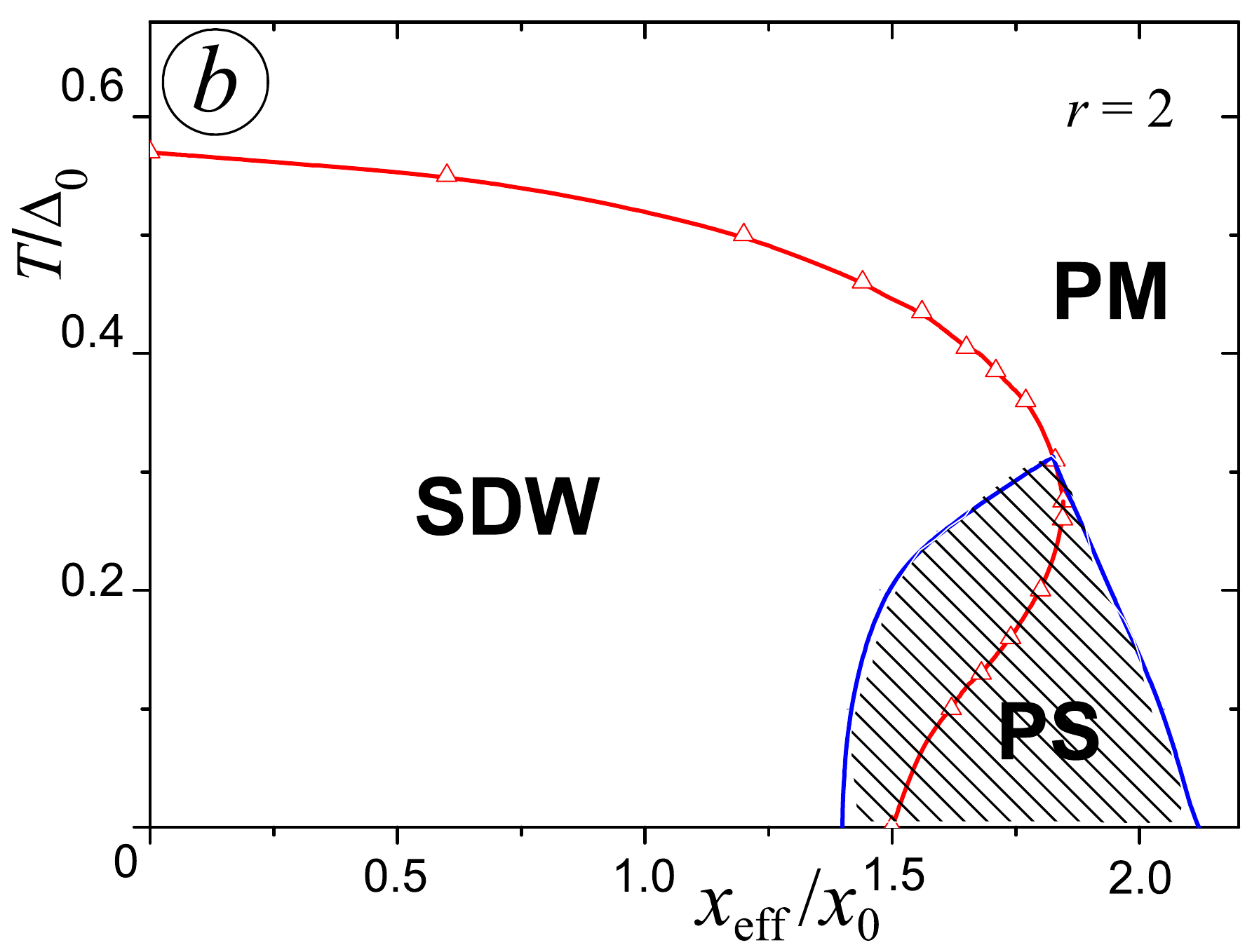}
\caption[]
{(Color online) Phase diagram of the model in the $(x_{\text{eff}},T)$ plane in the cases $r \ll 1$ (a) and $r=2$ (b). Solid (red) line with up triangles shows the boundary of the uniform stable SDW phase. The shaded area corresponds to the region with tendency to the phase separation (PS). (Fig.~3 from Ref.~\cite{RakhmanovJSNM2020})}\label{fig3Pressure}
\end{figure}

As it was discussed in section~\ref{CoulombSurface}, the change in the phases volume fractions is accomplished by the evolution of the inhomogeneous phase geometry, see Fig.~\ref{figGeom}. This evolution occurs due to change of the ratio of the long-range Coulomb interaction $E_c$ arising due to the violation of the charge neutrality and the energy $E_s$ of the interface between SDW and PM phases.~\cite{LorenzanaPRB_I_2001,KugelSuST2008}

\begin{figure}[btp]
\centering
\leavevmode
\includegraphics[width=8 cm]{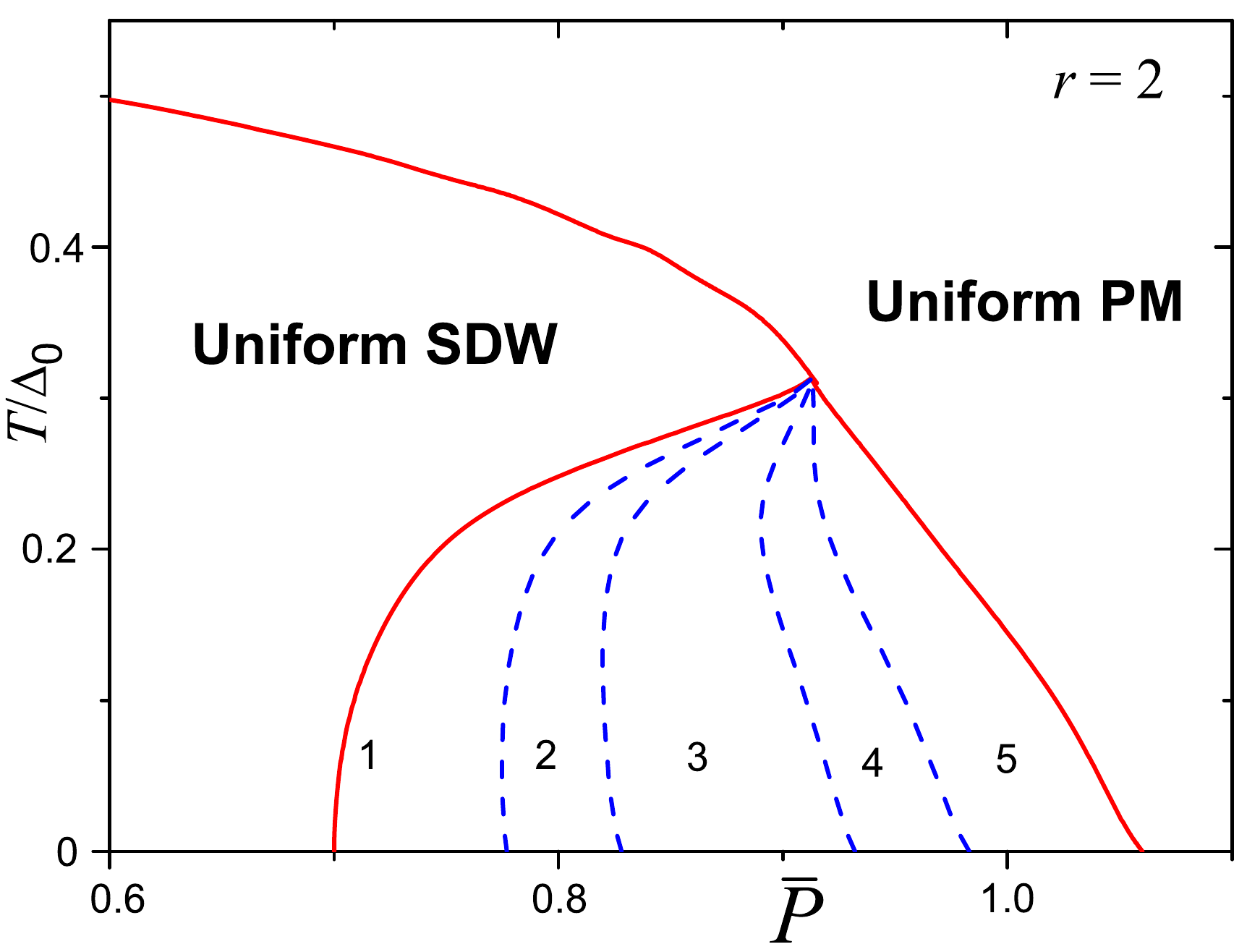}
\caption{(Color online) $(P,T)$ phase diagram in the case $x=0$ and $r=2$. Solid (red) lines show the boundary of the uniform stable phases. Blue dashed lines show the boundaries of the inhomogeneous phases with different shapes of the inhomogeneities: PM droplets in the SDW host exist in  region 1; PM pillars in the SDW host exist in  region 2; alternating PM and SDW slabs exist in region 3; SDW pillars in the PM host exist in region 4; and SDW droplets in the PM host exist in region 5. (Fig.~4 from Ref.~\cite{RakhmanovJSNM2020})}\label{fig5Pressure}
\end{figure}

The evolution of the inhomogeneous state geometry with pressure was analyzed in Ref.~\cite{RakhmanovJSNM2020} The content of the PM phase $p$ can be expressed as (for $x=0$)
\begin{equation}
p=\frac{rx_0\bar{P}-x^{(1)}_{\text{eff}}}{x^{(2)}_{\text{eff}}-x^{(1)}_{\text{eff}}}\,,
\end{equation}
where the effective dopings $x^{(1,2)}_{\text{eff}}$ are calculated according to the  Maxwell rule. Then, following the procedure described in section~\ref{CoulombSurface} the phase diagram of the system can be calculated. The results of such calculations are illustrated in Fig.~\ref{fig5Pressure}. The volume fraction of the PM phase, $p$, monotonically increases with pressure. As a result, the sequence of the geometries of inclusions changing with the growth of pressure is the following: uniform SDW phase at low pressure is changed by the phase separated state with PM islands in the SDW host (area 1 in Fig.~\ref{fig5Pressure}), then, by PM rods or pillars (region 2), PM slabs (region 3) and, finally, there arise inclusions of the SDW phase in the PM host (regions 4 and 5). The latter inhomogeneous states could be seen as a uniform PM metallic state in the transport measurements due to metal-insulator percolation transition.~\cite{RakhmanovJSNM2020}

\begin{figure}[btp]
\centering
\leavevmode
\includegraphics[width=8 cm]{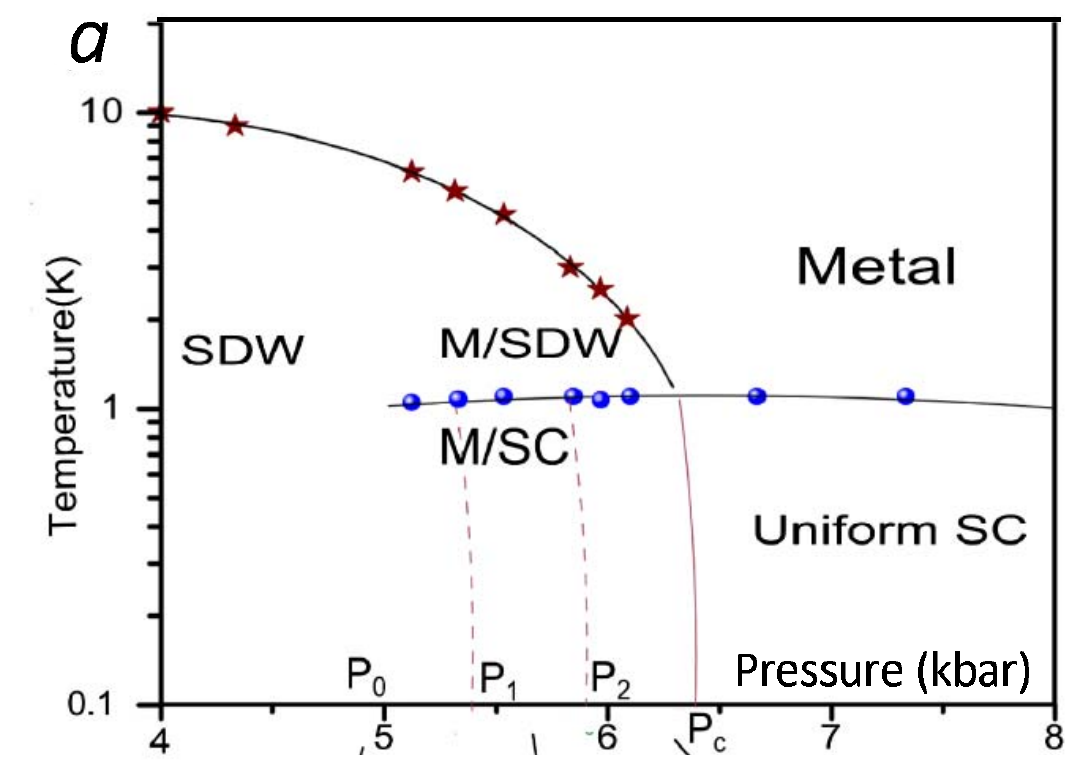}
\includegraphics[width=8 cm]{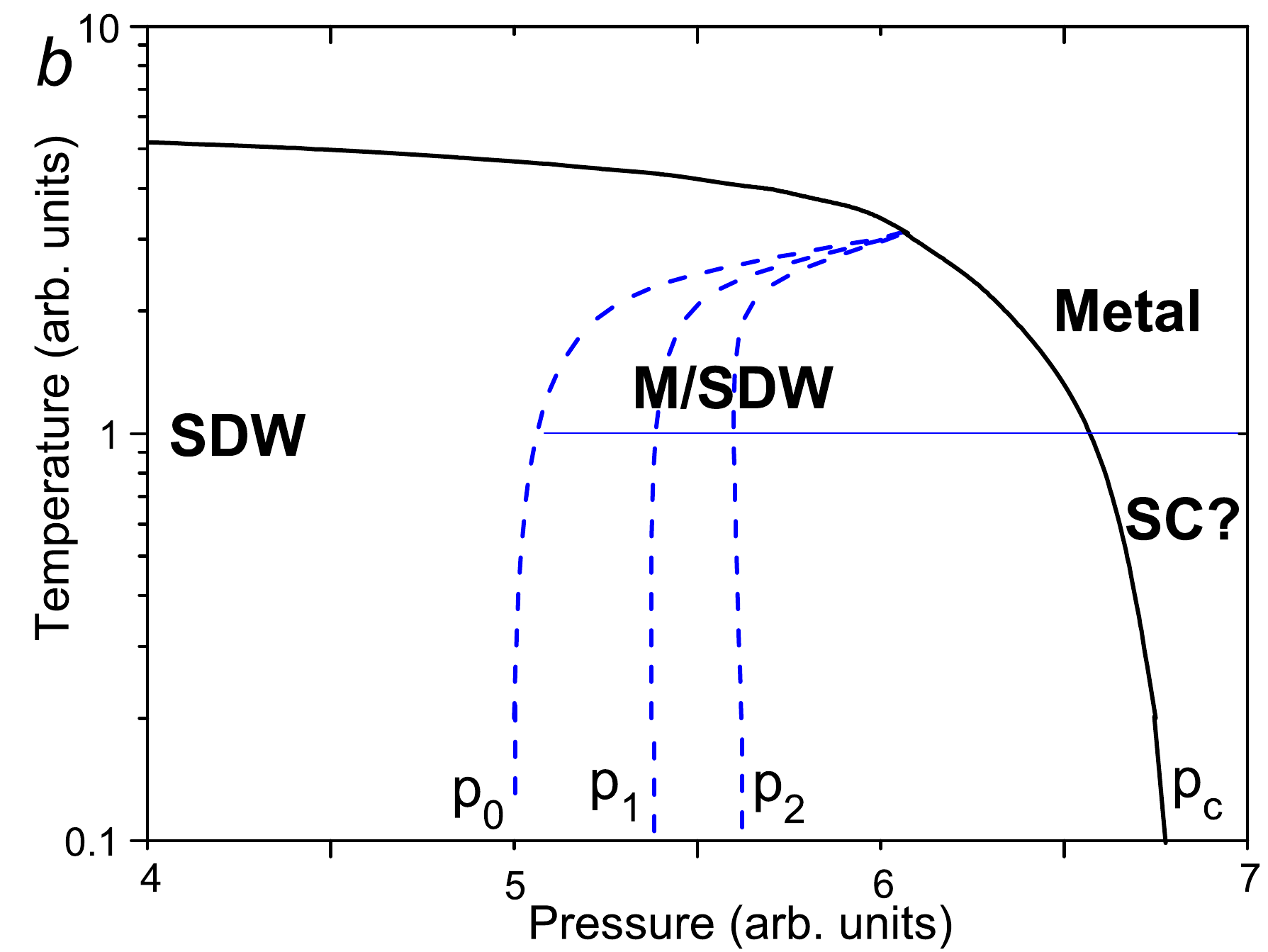}
\caption[]
{(Color online) (a) Experimental observation of the phase separation in the Bechgaard salt~\cite{NarayananPRL2014} (Fig.~1 from Ref.~\cite{NarayananPRL2014}). The uniform SDW phase is observed if pressure $P<P_0$. The inhomogeneous mixture of the SDW and PM metallic phases exists if $P_0<P<P_c$. The PS state includes metallic droplets in the SDW host if $P_0<P<P_1$, the metallic pillars in the SDW matrix if $P_1<P<P_2$, and the metallic slabs in the SDW host if $P_2<P<P_c$. When $P>P_c$ the sample is in the metallic state. The metal is superconducting if $T<T_c$, the dependence $T_c(P)$ is shown by (blue) solid line with circles. (b)  A part of the phase diagram shown in Fig.~\ref{fig5Pressure} redrawn in different coordinates, the notation corresponds to panel (a). (Fig.~5 from Ref.~\cite{RakhmanovJSNM2020}).}\label{fig6Pressure}
\end{figure}

In the above consideration, the pressure changes only the values of $k_{Fa}$ and $k_{Fb}$, but does not change the spherical geometry of the Fermi surface pockets. Naturally, it is a simplification. Below we discuss the phase separation observed in the Bechgaard salt~\cite{NarayananPRL2014} following Ref.~\cite{RakhmanovJSNM2020}. The Bechgaard salts are highly anisotropic and one may expect that its Fermi surface is also anisotropic, and its nested parts are probably not spherical. The pressure can affect not only the chemical potential, but also the geometry of the nested pockets. Thus, we need to include into consideration some additional geometrical factor, which governs the nesting. In this case, the de-nesting under pressure can occur even in the absence of the nonmagnetic reservoir. At the same time, if the electron and hole pockets are nested only partially, one can suggest that the SDW order parameter arises only in the nested parts of the Fermi surfaces, while other parts remain ungapped and play a role of the reservoir. Thus, one can expect that the model considered in this section could be relevant, at least, qualitatively, even in the case of nonspherical Fermi surfaces. (A more detailed analysis of the effects of the Fermi surface shape is given in the next section, where the phase separation in iron pnictide superconductors is discussed.)

The de-nesting of any nature gives rise to the decrease of the SDW order parameter and to the phase separation. With the increase of the pressure, the uniform SDW order is destroyed and the phase separation arises at some $P=P_0$. The volume concentration of the metallic phase increases with pressure and the described above evolution of the inhomogeneous phase geometry is observed, see Fig.~\ref{fig5Pressure}. The metallic phase may be superconducting at low temperature. In Fig.~\ref{fig6Pressure}(a), the experimental results on the phase separation in the Bechgaard salt under pressure taken from Ref.~\cite{NarayananPRL2014} are shown. In Fig.~\ref{fig6Pressure}(b), for comparison, a part of the phase diagram shown in Fig.~\ref{fig5Pressure} redrawn in the proper coordinates is presented. The inhomogeneous phases at high pressure where the sample is a metal from the experimental point of view are omitted. Qualitatively, the pictures illustrating the experimental and theoretical results are quite similar.

\subsection{Phase separation in iron pnictide superconductors} \label{imperfect_pnictides}

Superconducting iron-based pnictides form another system with the imperfect nesting, which we consider in this section. The phase diagram of iron pnictides is very rich and contains the regions of superconductivity and spin-density-wave order, both commensurate and incommensurate. Moreover, these materials often exhibit  spin and charge inhomogeneity with the characteristic features of the systems with electronic phase separation~\cite{DaiNatPh2012,ParkPRL2009electronic,InosovPRB2009suppression,
Lang2010nanoscale,ShenEPL2011intrinsic,LuoPRB2010neutron,Sboychakov_PRB2013_PS_pnict}. Superconducting iron chalcogenides also exhibit a phase separation~\cite{RicciPRB2011nanoscale,RicciSST2011intrinsic}. However, the physics related to the Fermi surface nesting may not be directly applicable to iron chalcogenides since some chalcogenides  have no hole pockets. %, but nevertheless have AFM states with rather high N\'{e}el temperatures.

\begin{figure}[H] \centering
\includegraphics[width=0.5\columnwidth]{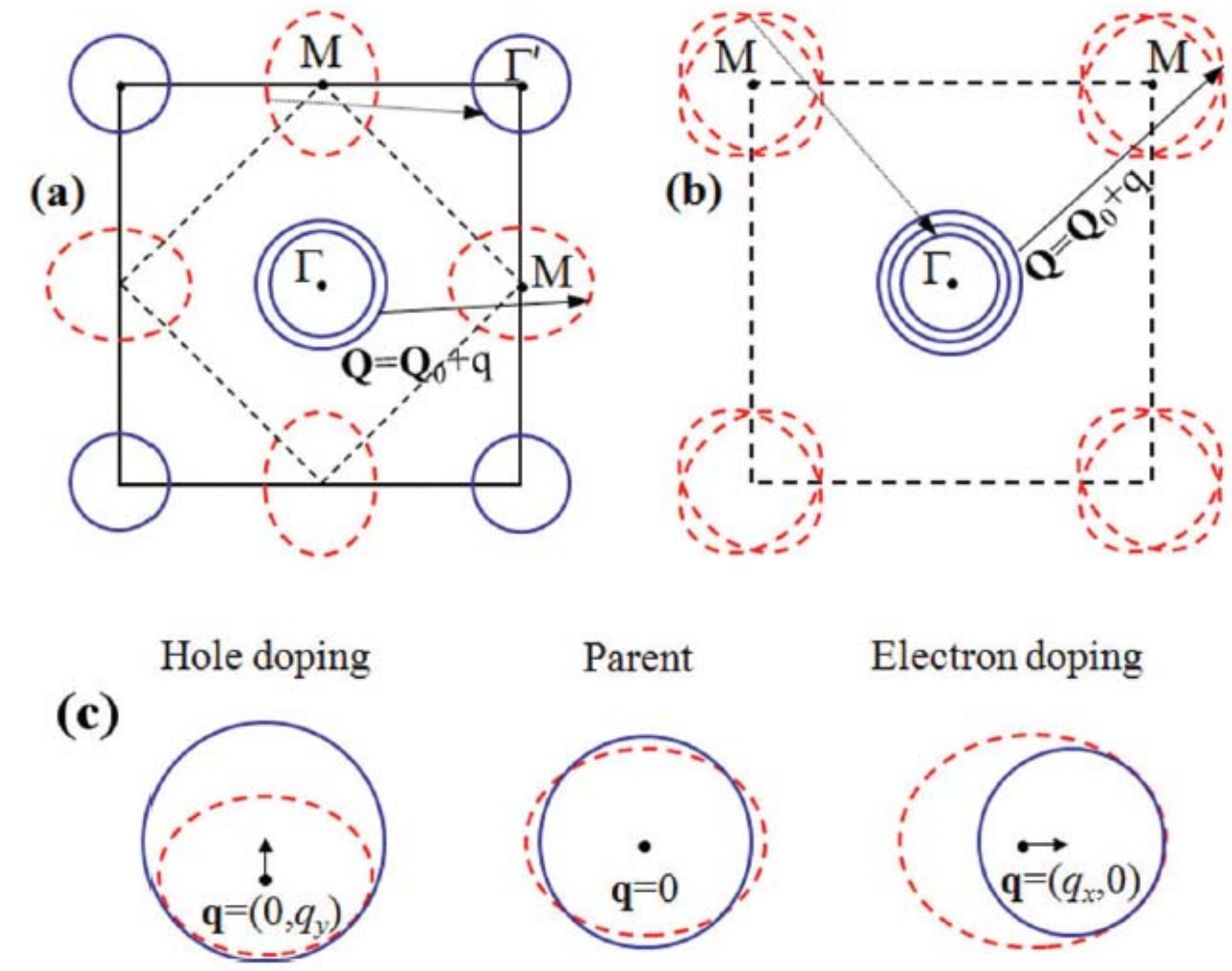}\\
\caption{ (Color online)  Schematic illustration of the Fermi surface of iron pnictides in the unfolded (a) and folded (b) Brillouin zone (BZ). We have three hole pockets located near the center of the folded BZ [shown by (blue) solid curves]. In the unfolded BZ, one of the hole pockets moves to the corner of the BZ. For the unfolded BZ (a), the electron pockets are elliptic [(red) dashed curves] and located near the ($0,\pi/a$) and ($\pi/a,0$) points, while in the folded BZ, they are represented by overlapping ellipses located at the corners. Arrows show the possible nesting vectors between hole and electron pockets, which give rise to the SDW order. (c) Illustration of the possible nesting at different doping levels. (Fig.~1 from Ref.~\cite{Sboychakov_PRB2013_PS_pnict}).
  }\label{PnFerSer}
\end{figure}

Unlike superconducting cuprates, which are believed to be in the strong electron--electron interaction regime, iron pnictides may be described by a weak-interaction model (see, e.g., Sec.IIIA of Ref.~\cite{stewart2011superconductivity}).

Currently, the Fermi surface of iron pnictides is typically described using two related representations briefly outlined below. In Fig.~\ref{PnFerSer}(a), we plot the Fermi surface within the so-called unfolded Brillouin zone~\cite{stewart2011superconductivity,HuPRX2012,richard2011fe}, which corresponds to the square lattice of iron atoms, with one Fe atom per unit cell and lattice constant $a$. In this representation, two quasi-two-dimensional nearly circular hole pockets are centered at the $\Gamma(0,0)$ point, one more circular hole pocket is located near the $\Gamma'(\pi/a,\pi/a)$ point, and two elliptically-shaped electron pockets are centered at the $M(0,\pi/a)$ and $M(\pi/a,0)$ points [see Fig.~\ref{PnFerSer}(a)]. At the same time, the actual unit cell of pnictides contains two Fe atoms, since in the crystal lattice, the pnictogen atoms are located at nonequivalent positions. The (folded) Brillouin zone corresponding to this unit cell is obtained by folding the Brillouin zone shown in Fig.~\ref{PnFerSer}(a) by dashed lines and consequent rotation by $\pi/4$. The Fermi surface in the folded Brillouin zone is shown in Fig.~\ref{PnFerSer}(b). In this figure, all three hole pockets are located near the $\Gamma(0,0)$ point, whereas the electron pockets represented by overlapping ellipses are located near the $M(\pm \pi/\bar{a},\pi/\bar{a})$ points, with $\bar{a} = a/\sqrt{2}$. Formulating this model, we will make several simplifications. First, we neglect the effects associated with nonequivalent positions of the Fe atoms. Consequently, the use of unfolded Brillouin zone is sufficient. Second, we will neglect the 3D structure of the material and study only the 2D model.

Following Ref.~\cite{Sboychakov_PRB2013_PS_pnict}, the Hamiltonian of the model can be written as $\hat{H}=\hat{H}_0+\hat{H}_{\textrm{int}}$, where the kinetic energy term has the form
\begin{equation}\label{HamKin}
\hat{H}_0=\sum_{\mathbf{k}\lambda\sigma}{\epsilon_{\lambda\mathbf{k}}^h a^\dag_{\mathbf{k}\lambda\sigma} a_{\mathbf{k}\lambda\sigma}}+\sum_{\mathbf{k}s\sigma}{\epsilon_{s\mathbf{k}}^e b^\dag_{\mathbf{k}s\sigma} b_{\mathbf{k}s\sigma}}.
\end{equation}
Here $a^\dag_{\mathbf{k}\lambda\sigma},a_{\mathbf{k}\lambda\sigma}$ ($b^\dag_{\mathbf{k}s\sigma}, b_{\mathbf{k}s\sigma}$) are the creation and annihilation operators for charge carriers in the holelike (electronlike) bands $\lambda = 1,2,3$ ($s = 1,2$) with spectra $\epsilon_{\lambda\mathbf{k}}^h$ ($\epsilon_{s\mathbf{k}}^e$). Assuming that the bands have quadratic dispersion laws near the Fermi level, for circular holelike bands we have
\begin{equation}\label{holeBand}
\epsilon_{1\mathbf{k}}^h=-\frac{\hbar v_F^h(k^2-k_F^2)}{2k_F}-\mu,\,\,
\epsilon_{2\mathbf{k}}^h=-\frac{\hbar v_{2F}^h(k^2-k_F^2)}{2k_F}-\Delta\epsilon_2-\mu,\,\,
\epsilon_{3\mathbf{k+\bar{Q}}}^h=-\frac{\hbar v_{3F}^h(k^2-k_F^2)}{2k_F}-\Delta\epsilon_3-\mu.
\end{equation}
Here, $v_F$ and $v_{2,3F}$ are the Fermi velocities for the hole bands, $\mathbf{\bar{Q}}= (\pi/a,\pi/a)$, and the energy shifts $\Delta\epsilon_{2,3}$ determine the difference in radii of the hole pockets. The electronlike components of the Fermi surface are elliptic and can be presented as follows~\cite{Sboychakov_PRB2013_PS_pnict}
\begin{equation}\label{electronBand}
\epsilon_{1\mathbf{k+Q_0}}^e=\epsilon_{1\mathbf{k+Q_0'}}^e=\frac{\hbar v_F^e(k^2-k_F^2)}{2k_F}+\frac{\alpha \hbar v_F^e}{k_F}(k_x^2-k_y^2)-\mu,
\end{equation}
where the centers of the elliptic bands are $\mathbf{Q}_0 = (\pi/a,0)$ and $\mathbf{Q_0'}= (0,\pi p/a)$, and $v_{F}^e$ is the Fermi velocity for the electron bands averaged over the Fermi surface. The parameter $\alpha$ defines the ellipticity of the electron pockets. The major axes of the ellipses are directed toward the $\Gamma$ point when $\alpha>0$ and perpendicular to this direction when $\alpha<0$.

Due to the multisheet structure of the Fermi surface, the interaction Hamiltonian $\hat{H}_{\textrm{int}}$, in general, includes a number of contributions. However, since we are interested in the SDW order and the phase separation, one can ignore the electron--electron and hole--hole interactions as in the previous analysis of the Rice model. It is known~\cite{DaiNatPh2012} that the value of local magnetic moment in pnictides oscillates along one of the crystal axes. To reproduce this stripy magnetic structure, it is sufficient to couple one electron and one hole band~\cite{EreminChubPRB2010}. For the definiteness sake, one can assume that these are the hole band $\epsilon^h_1$ and the electron band $\epsilon^e_1$. Thus, we can write
\begin{equation}\label{intHam}
\hat{H}_{\textrm{int}}=\frac{V}{{\cal V}}\sum_{\mathbf{kk'q}\sigma\sigma'}{a^\dag_{\mathbf{k+q}1\sigma} a_{\mathbf{k}1\sigma} b^\dag_{\mathbf{k'-q}1\sigma'} b_{\mathbf{k'}1\sigma'}}.
\end{equation}
The described model for pnictides is a generalization of the Rice model. As in the latter case, two bands participate in the magnetic transition, and there is a “reservoir” (nonmagnetic bands). However, the perfect nesting is impossible for pnictides since the forms of the hole and electron pockets are different for any doping level.

As we pointed out above, the Coulomb interaction in iron pnictides is weak and one can assume that $V/\epsilon_F\ll 1$, where the Fermi energy $\epsilon_F=(v_F^e+v_F^h)k_F/2$. Then, Hamiltonian (\ref{intHam}) can be considered in the mean-field approximation~\cite{Sboychakov_PRB2013_PS_pnict} and, as in the previous section, we will search the SDW order parameter in the form (\ref{FFLO}). In general, different parts of each pocket of the Fermi surface have different orbital composition. Such feature can be accounted by introducing a momentum-dependent coupling $V$ and would lead to a pronounced wave-vector dependence of the SDW order parameter. However, unless the variations of $V$ are extremely strong, they do not affect the phase separation qualitatively.

The magnetization corresponding to the chosen SDW order lies in the $xy$ plane. For the commensurate SDW case, the magnetization direction remains constant when one moves along the direction normal to $\mathbf{Q}_0$. When one moves parallel to $\mathbf{Q}_0$, the magnetization reverses its direction from one iron atom to the next iron atom. For incommensurate SDW, this “stripy” pattern slowly rotates in the $xy$ plane~\cite{Sboychakov_PRB2013_PS_pnict}.

Performing mean-field transformations in the interaction Hamiltonian (\ref{intHam}), we obtain the energy spectrum of the model similar to Eq.~(\ref{eigenenergy}) and grand potential in the form (\ref{GrPot}), where energies $\epsilon^{a,b}$ should be replaced by $\epsilon^{e,h}$. As usual, the minimization of the grand potential gives the equations for the SDW order parameter $\Delta$ and nesting vector $\mathbf{Q}$. The expression for $\Delta$ was derived in Ref.~\cite{Sboychakov_PRB2013_PS_pnict} at zero temperature
\begin{equation}\label{DeltaPnic}
\ln{\frac{\Delta_0}{\Delta}}=\int{\frac{\epsilon_Fd^2\mathbf{k}}{2\pi k_F^2}\frac{\Theta[E_1(\mathbf{k})]+
\Theta[-E_2(\mathbf{k})]}{\sqrt{\Delta^2+\frac{1}{4}
\left(\epsilon^e_{1\mathbf{k}+\mathbf{Q}_0}-
\epsilon^h_{1\mathbf{k}-\mathbf{q}}\right)^2}}},
\end{equation}
where $\Theta(x)$ is the step function, $\epsilon^{e,h}_{1\mathbf{k}}$ are defined by Eqs.~(\ref{holeBand}) and (\ref{electronBand}), and the usual limits $\Delta/\epsilon_F, |\mathbf{q}|/k_F\ll 1$ are assumed.

In the weak-coupling limit under discussion, the SDW order can exist only if the deviation from the perfect nesting is small, that is, $\alpha\ll 1$. Taking this into account and using Eqs.~(\ref{holeBand}) and (\ref{electronBand}), we can rewrite Eq.~(\ref{DeltaPnic}) in the form~\cite{Sboychakov_PRB2013_PS_pnict}
\begin{equation}\label{DeltaPnicDim}
\ln{\frac{1}{\delta}}=\int_0^{2\pi}{\frac{d\varphi}{2\pi}
Re\left\{\cosh^{-1}\left[\frac{\nu_0(\mathbf{p},\varphi)-
\nu}{\delta}\right]\right\}},
\end{equation}
where
\begin{eqnarray}\label{PnicNota}
\nonumber
  \nu_0(\mathbf{p},\varphi) &=& p_x\cos{\varphi}+ p_y\sin{\varphi}-\frac{\bar{\alpha}}{2}\cos{2\varphi},\quad \bar{\alpha}=\frac{\alpha\kappa\epsilon_F}{\Delta_0},\quad \kappa=\frac{2\sqrt{v_F^ev_F^h}}{v_F^e+v_F^h}, \\
  \delta &=& \frac{\Delta}{\Delta_0},\quad \nu=\frac{\mu}{\kappa\Delta_0},\quad \mathbf{p}=\frac{\kappa v_F}{2\Delta_0}\mathbf{q}.
\end{eqnarray}

Similarly, we obtain the equation for the nesting vector $\mathbf{Q}=\mathbf{Q}_0+2\Delta_0\mathbf{p}/v_F\kappa$:
\begin{equation}\label{NestPnicDim}
\left(
  \begin{array}{c}
    p_x \\
    p_y \\
  \end{array}
\right)=\int_0^{2\pi}{\frac{d\varphi}{\pi}\left(
                                              \begin{array}{c}
                                                \sin{\varphi} \\
                                                \cos{\varphi} \\
                                           \end{array}
                                           \right)\textrm{sgn}
                                        {[\nu_0(\mathbf{p},\varphi)-\nu]}
Re\sqrt{[\nu_0(\mathbf{p},\varphi)-\nu]^2-\delta^2}
}.
\end{equation}
Equations~(\ref{DeltaPnicDim}) and (\ref{NestPnicDim}) determine the SDW band gap and the nesting vector as functions of $\mu$. Experiments are performed at fixed doping, and we have to relate the electron density and $\mu$. Following the previously described approach [see Eq.~(\ref{doping}) and the text above it], we find in the  approximation under discussion~\cite{Sboychakov_PRB2013_PS_pnict}
\begin{equation}\label{dopingPnic}
\frac{x}{x_0}=n\nu-\int_0^{2\pi}{\frac{d\varphi}{2\pi}\textrm{sgn}
{[\nu_0(\mathbf{p},\varphi)-\nu]}
Re\sqrt{[\nu_0(\mathbf{p},\varphi)-\nu]^2-\delta^2}},
\end{equation}
where $x_0$ is the normalized characteristic doping level, $n$ is the density of electrons in nonmagnetic bands
\begin{equation}\label{pnicNota2}
x_0=\frac{2v_0k_F^2\Delta_0}{\pi\kappa\epsilon_F}, \quad n=\frac{(v_{2F}^h)^{-1}+(v_{3F}^h)^{-1}+(v_{F}^e)^{-1}}{(v_{F}^h)^{-1}+(v_{2F}^e)^{-1}},
\end{equation}
and $v_0$ is the unit cell volume. Note, that $x\sim \Delta_0/\epsilon_F\ll 1$ and considered below phases, existing when $x\sim x_0$, correspond to the low-doping regimes.

\begin{figure}[H]\centering
    \includegraphics[width=0.5\columnwidth]{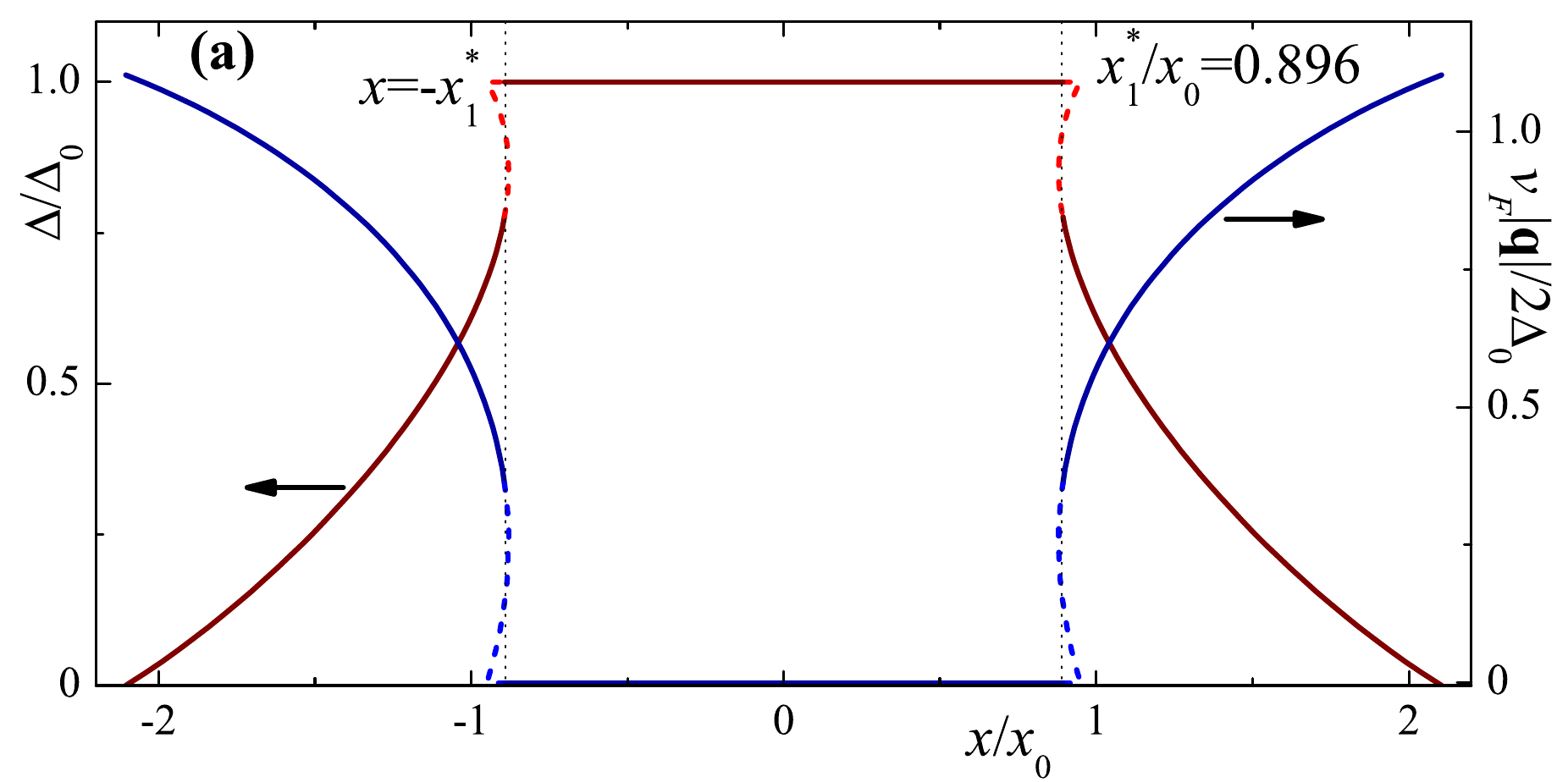}\\
    \includegraphics[width=0.5\columnwidth]{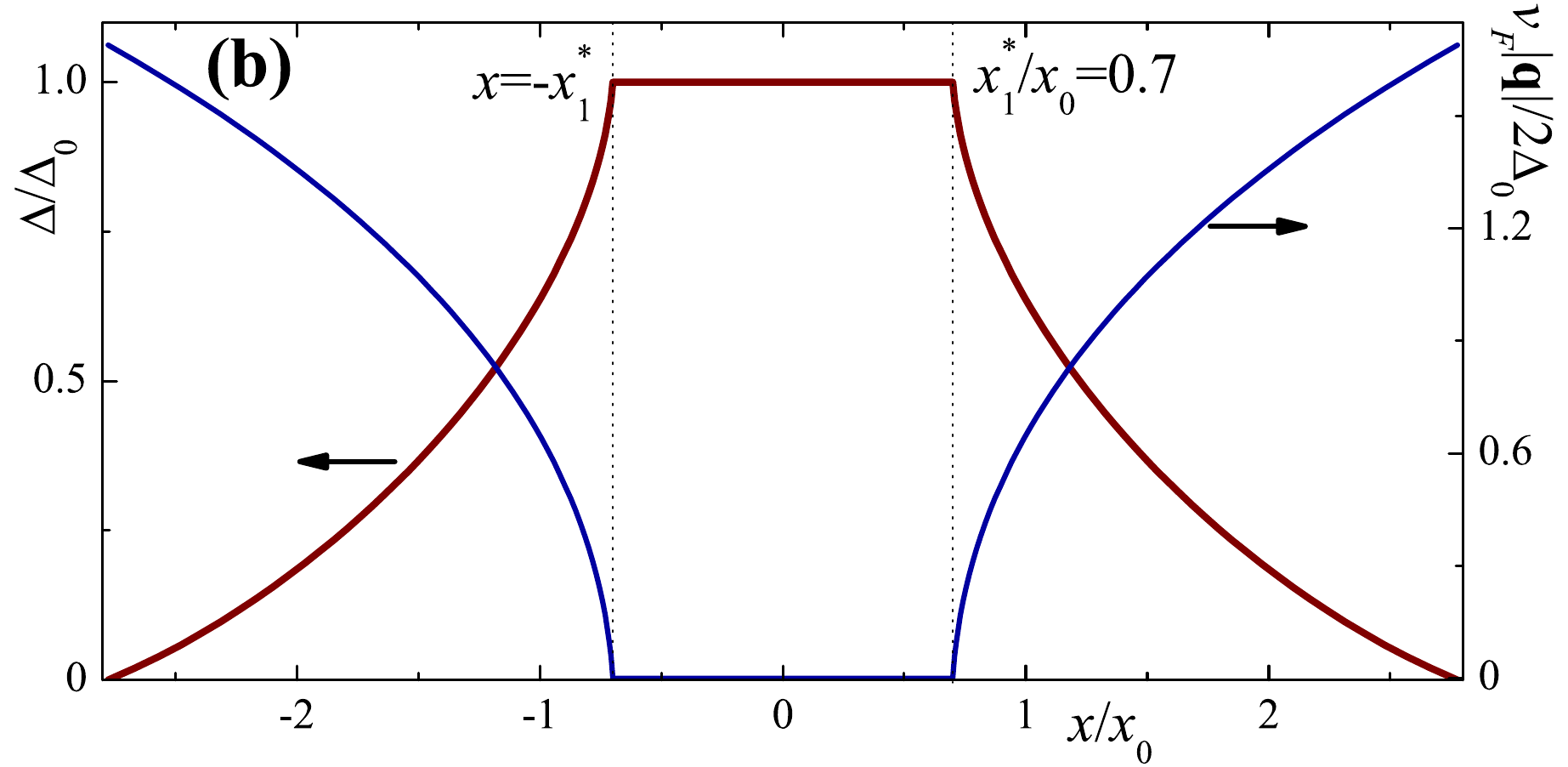}\vspace{5mm}\\
    \includegraphics[width=0.49\columnwidth]{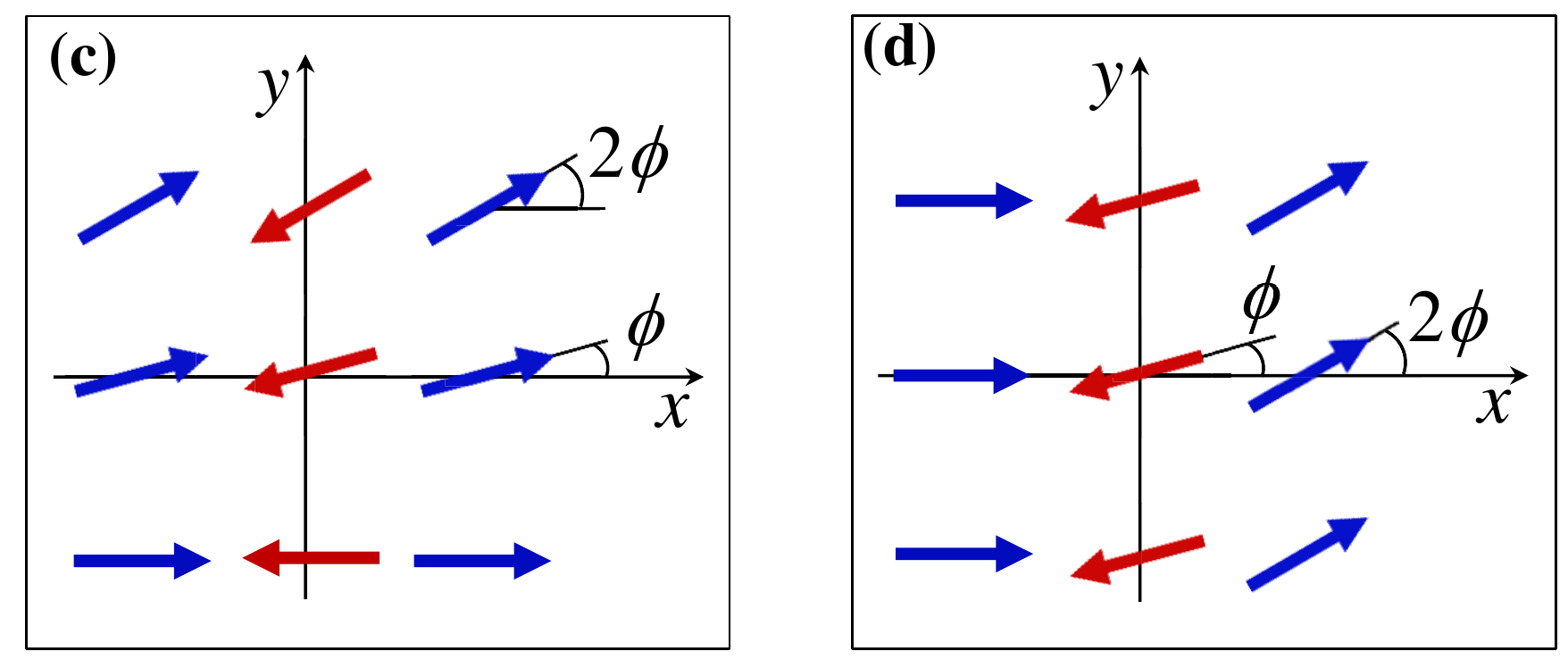}
 \caption{(Color online) Dependence of $\Delta$ and $|\mathbf{q}|$ on doping $x$,  $n=\kappa=1$. (a) $\bar{\alpha}=0.1$. The transition from commensurate to incommensurate SDW state is of the first order: both $\Delta$ and $|\mathbf{q}|$ rapidly change at $x=x_1^{*}$. Dashed lines show the functions $\Delta(x)$ and $|\mathbf{q}(x)|$ corresponding to the metastable  state. (b) $\bar{\alpha}=0.6$. The transition from commensurate to incommensurate SDW state is of the second order. Panels (c) and (d) show the schematics of the incommensurate SDW spin structure. If $\alpha>0$, then, (c) corresponds to the hole doping ($x<0$) and (d) to the electron doping ($x>0$).  If $\alpha<0$, then panel (c) corresponds to the electron doping, while panel (d) corresponds to the hole doping~\cite{Sboychakov_PRB2013_PS_pnict}.}
\label{FigDeltaQ}
\end{figure}

The computed functions $\Delta(x)$ and $q(x)$ are shown in Figs.~\ref{FigDeltaQ}(a) and \ref{FigDeltaQ}(b) for $n = \kappa = 1$ and two different values of elipticity $\alpha$. The numerical analysis reveals that the SDW order arises if $|\bar{\alpha}|<2$ or $\alpha<\alpha_c=2\Delta_0/\kappa\epsilon_F\ll 1$. At low doping, $|x|<x^*_1$, all the extra charge goes to the nonmagnetic bands. As a result, the order parameter is independent of $x$, $q(x)$ = 0 (a commensurate SDW order), and the chemical potential increases linearly with $x$. When $|x|>x^*_1$, electrons (holes) appear in the band $E_2$ ($E_1$). The SDW order becomes incommensurate, $q\neq 0$. If $n < n_1\approx 0.38$, the commensurate--incommensurate transition is of the second order for any $\bar{\alpha}$, while for larger $n$ it becomes the first order one if $\bar{\alpha}<\bar{\alpha}_1(n)$. The spin configurations for the incommensurate SDW are schematically shown in Figs.~\ref{FigDeltaQ}(c) and \ref{FigDeltaQ}(d) (for details, see Ref.~\cite{Sboychakov_PRB2013_PS_pnict}). For positive (negative) $\alpha$, Fig.~\ref{FigDeltaQ}(c) corresponds to $x > 0$ ($x < 0$), and Fig.~\ref{FigDeltaQ}(d) corresponds to $x < 0$ ($x > 0$).

\begin{figure}[H]\centering
    \includegraphics[width=0.5\columnwidth]{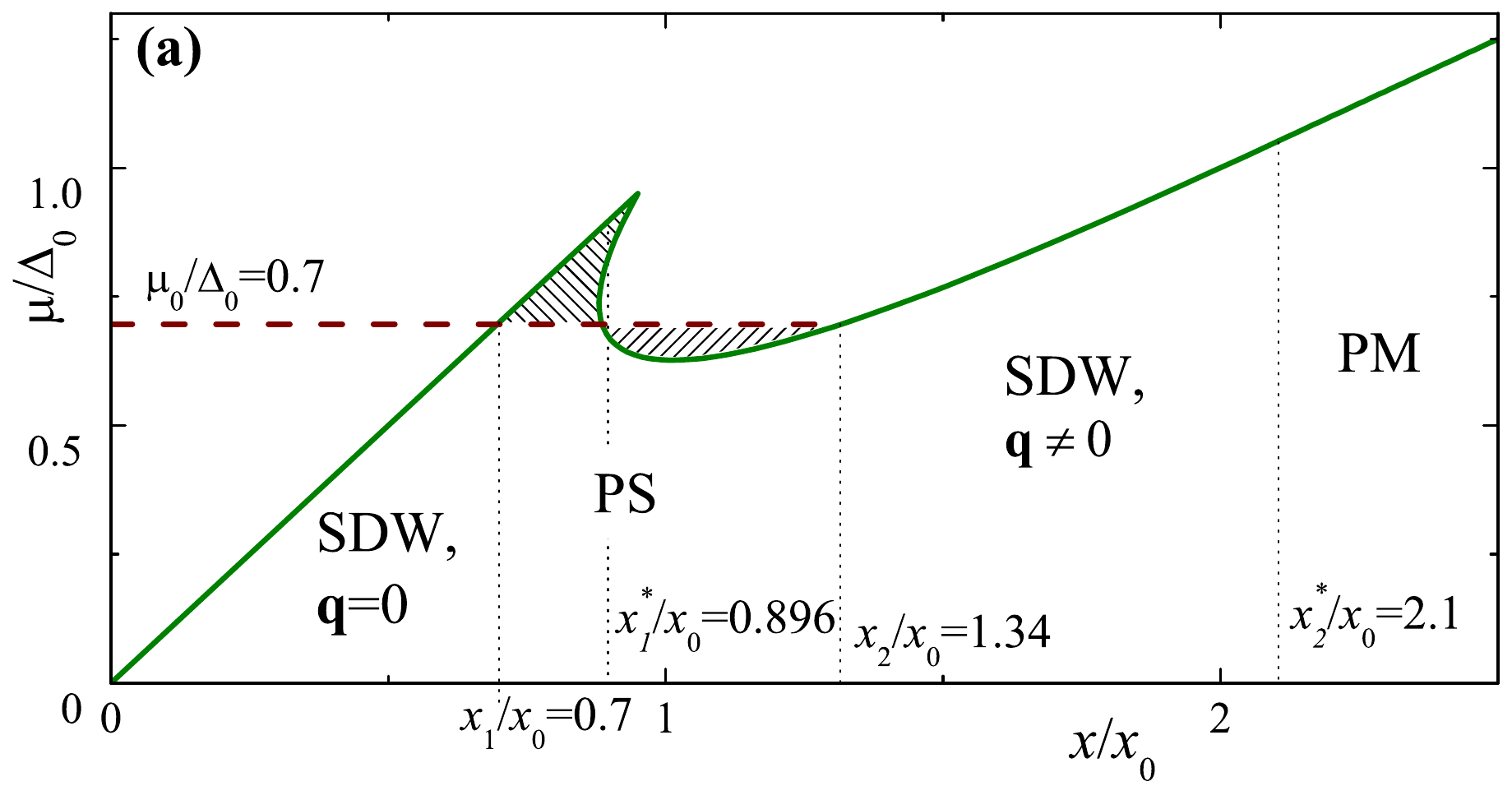}\\
    \includegraphics[width=0.5\columnwidth]{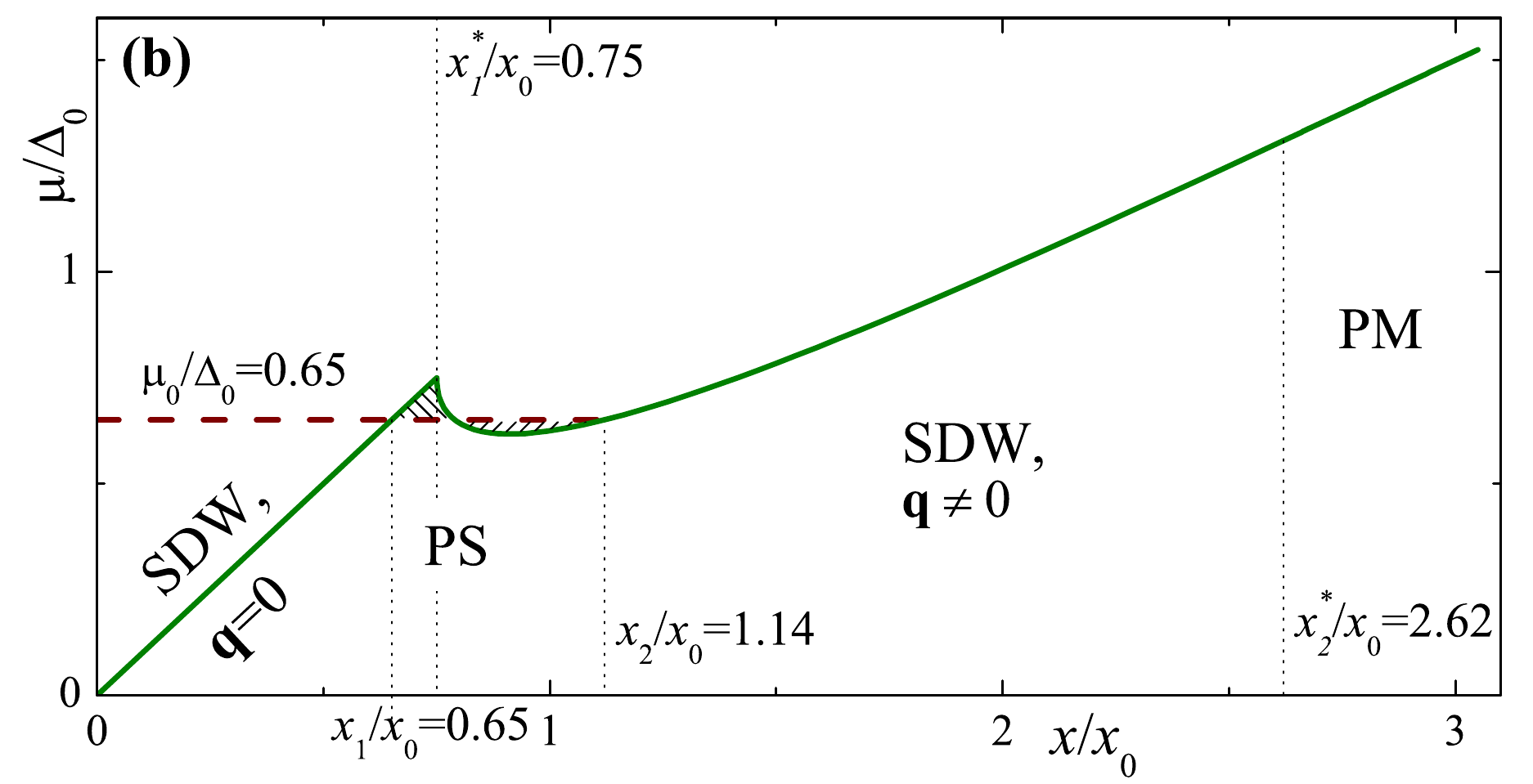}\vspace{5mm}\\
    \includegraphics[width=0.49\columnwidth]{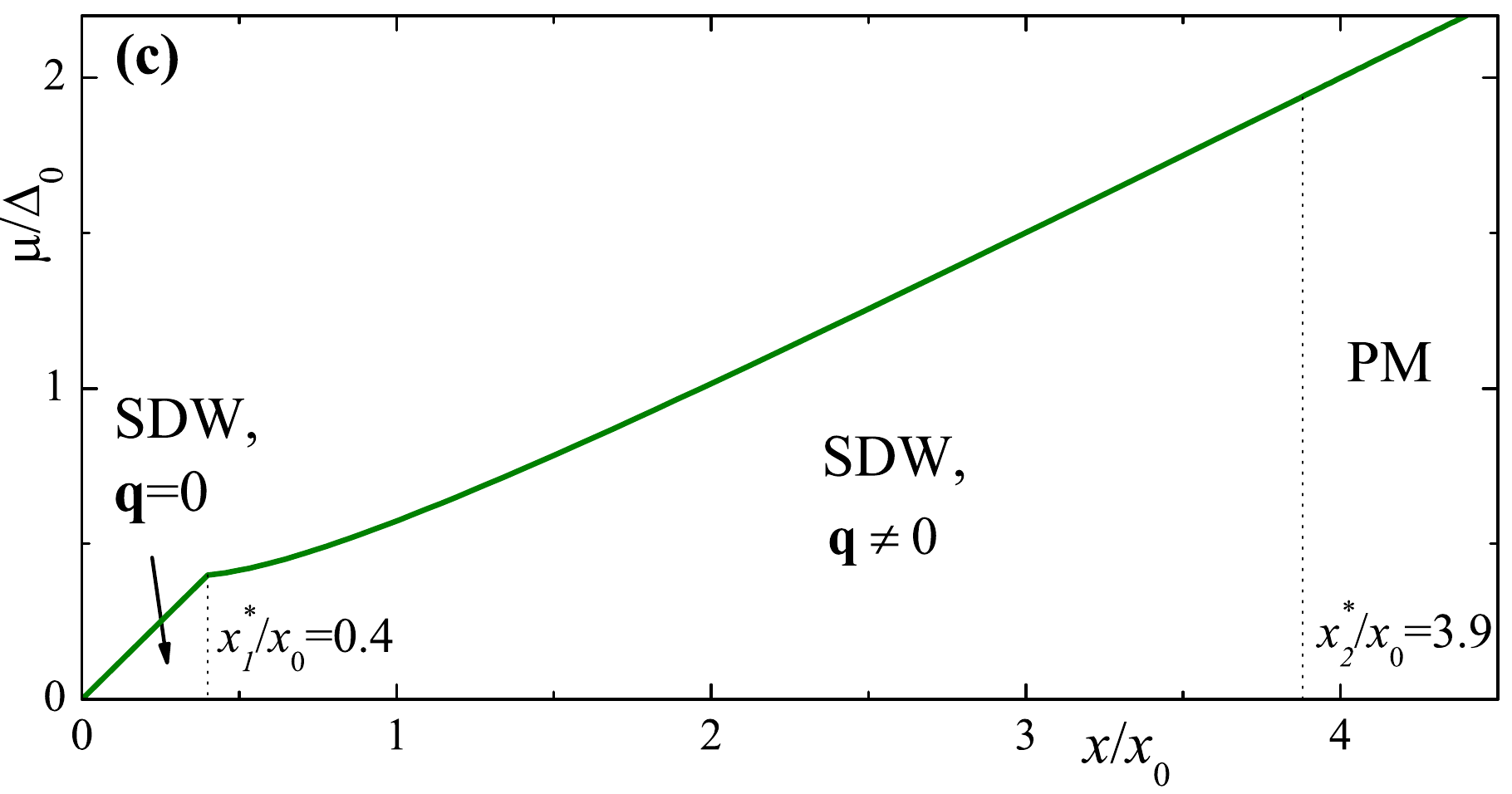}
 \caption{(Color online) Chemical potential $\mu(x)$ calculated at $\bar{\alpha}=0.1$ (a), $\bar{\alpha}=0.5$ (b), and $\bar{\alpha}=1.2$ (c). Here, $n=\kappa=1$. (a), (b) The homogeneous state is unstable toward the phase separation if $x_1 < x < x_2$. The dashed (red) curve corresponds to the $\mu_0$ found using the Maxwell construction. The shaded areas above and below $\mu_0$ are equal to each other. (c) $\mu(x)$ monotonically increases with $x$, no phase separation occurs. The homogeneous commensurate ($q = 0$) and incommensurate ($q \neq 0$) SDW, paramagnetic (PM), and inhomogeneous commensurate--incommensurate SDW (PS) states are separated by vertical dotted lines~\cite{Sboychakov_PRB2013_PS_pnict}.}
\label{FigMu}
\end{figure}

The computed in Ref.~\cite{Sboychakov_PRB2013_PS_pnict} plots of the chemical potential $\mu(x)$ are shown in Fig.~\ref{FigMu} for three different values of elipticity $\bar{\alpha}$ and $n=\kappa=1$. The $\mu(x)$ dependence for the hole doping can be determined using the particle--hole symmetry. For low $\bar{\alpha}$, the function $\mu(x)$ is nonmonotonic and multivalued near $x_1^*$, Fig.~\ref{FigMu}(a). At higher $\bar{\alpha}$, the multivaluedness disappears; however, the nonmonotonicity remains, Fig.~\ref{FigMu}(b). The latter vanishes at higher values of $\bar{\alpha}$, Fig.~\ref{FigMu}(c). Thus, if the elipticity is not too large, there are finite ranges of doping where the chemical potential decreases with doping and the homogeneous state is unstable with respect to the separation into two phases. However, the phase transitions between homogeneous commensurate and incommensurate SDW phases, which we described above, can be masked, at least partially, by the phase separation. The range of doping $x$, where the phase separation exists, is the largest when $\alpha=0$, that is, when the model under study is identical to the two-dimensional Rice model. The range of the phase separation $x_1 < x < x_2$ shrinks if the elipticity increases, and disappears at the critical value $\bar{\alpha}_c=1.15$.

\begin{figure}[H]\centering
    \includegraphics[width=0.5\columnwidth]{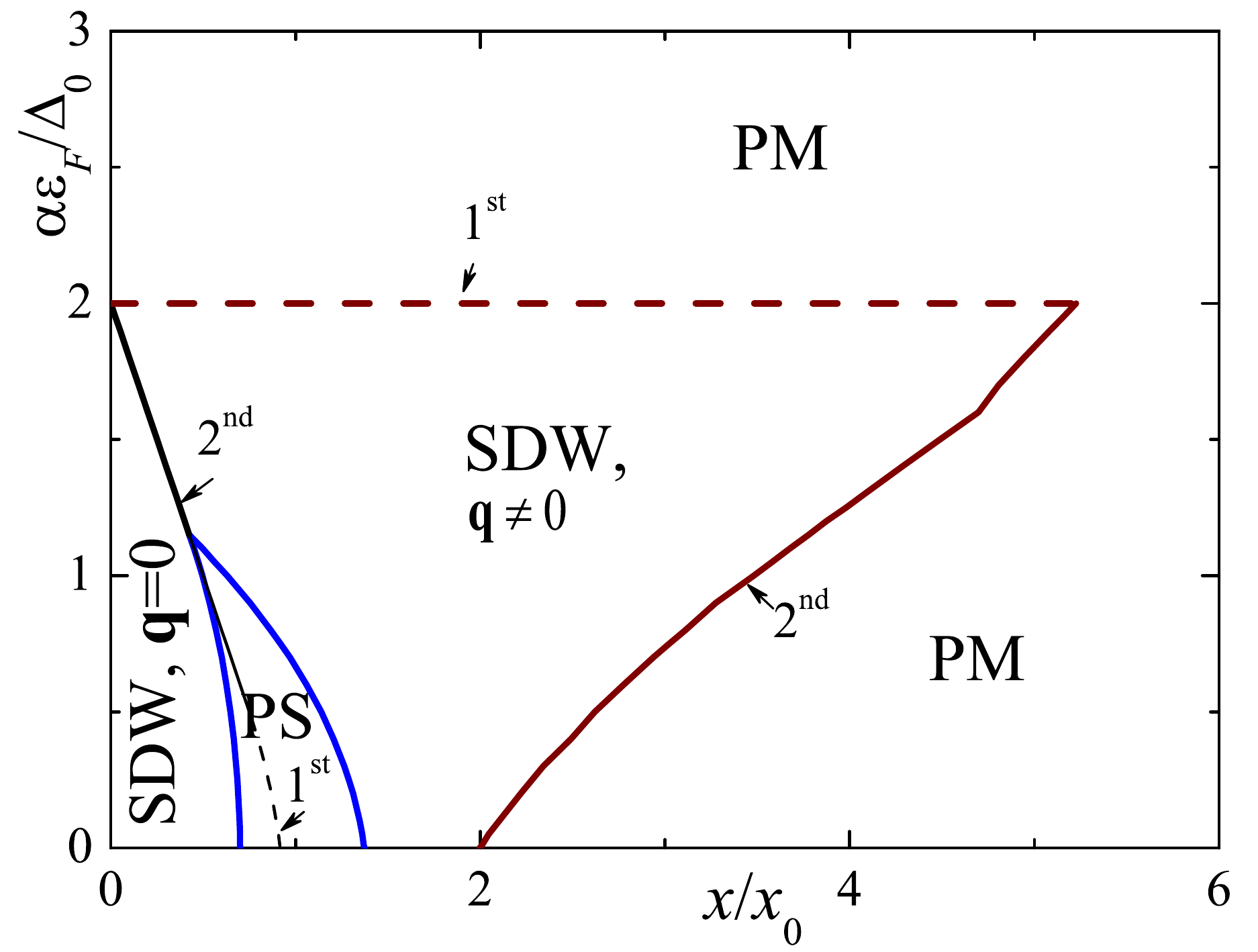}\\
\caption{(Color online) Phase diagram of the model for pnictides in the ($x,\alpha$) plane, for $n=\kappa = 1$ and $\alpha,x > 0$. It is symmetric with respect to the replacement $x\rightarrow-x$ and/or $\alpha\rightarrow-\alpha$. The boundary between the incommensurate SDW and paramagnetic (PM) states shown by the dashed (red) curve corresponds to the first order phase transition and and that shown by the solid (red) curve corresponds to the second order phase transition. Solid (blue) lines indicate the boundaries of the phase-separated (PS) state. The solid (black) curve (second order transition) and dashed (black) curve (first order transition) show the boundaries between commensurate and incommensurate homogeneous SDW phases. Note that the phase separation partially masks the transition line between the homogeneous SDW states~\cite{Sboychakov_PRB2013_PS_pnict}.}
\label{FigPhDiag}
\end{figure}

In the phase-separated state, there exist two phases, phase 1 with a lower charge carrier density is the commensurate one and phase 2 with a higher charge carrier density corresponds to the incommensurate SDW state. The phase diagram of the system can be constructed following the approach described in the previous subsections (see, e.g., subsection~\ref{imperfect_Rice}). The obtained results are summarized in the phase diagram in the $(x, \alpha)$ plane shown in Fig.~\ref{FigPhDiag}~\cite{Sboychakov_PRB2013_PS_pnict}. This phase diagram is calculated for $n = \kappa = 1$. It remains qualitatively the same if $n \neq 0$. If the nonmagnetic bands are absent, the homogeneous commensurate SDW phase exists only when doping is zero. Consequently, the electronic density $x_1$ vanishes in phase 1 of the phase-separated state for any $\alpha$.

The phase separation in iron-based superconductors was observed in several experiments~\cite{ParkPRL2009electronic,InosovPRB2009suppression,
Lang2010nanoscale,ShenEPL2011intrinsic}. For example, the inhomogeneous state with a commensurate AFM and nonmagnetic domains with characteristic sizes about 65~nm was observed in the hole-doped Ba$_{1-x}$K$_x$Fe$_2$As$_2$ compound~\cite{ParkPRL2009electronic}. The described above model predicts that the second phase is an incommensurate SDW rather than a nonmagnetic one. However, the proposed mechanism of phase separation can, in general, be consistent with the observations. The thermodynamic potentials of the incommensurate SDW and the metastable PM phases are very close to each other in the doping range $x > x_2$. The incommensurate SDW phase can be destroyed by an additional cause not taken into account in the model, e.g., by a disorder. In this case, the phase separation might occur between the commensurate SDW and the PM phases.

Here, we consider the SDW order parameter in the form~(\ref{FFLO}). As it was mentioned in the previous section, in the literature a different order parameter is also discussed (see, e.g., Ref.~\cite{gor2010spatial}). We must remember that the relative stability of different order parameters is likely a non-universal quantity, which depends on a variety of microscopic parameters (e.g., details of the band structure, interaction, disorder). Therefore, the type of suitable order parameter cannot be assuredly deduced without an input from experiments.

The calculations above show that there are only two possible equilibrium directions of the vector $\mathbf{q}$ characterizing the incommensurate SDW phase: it can be either parallel or perpendicular to the nesting vector $\mathbf{Q}_0$, depending on the type of doping and the sign of the ellipticity parameter $\alpha$. It is clear that both the magnitude and direction of $\mathbf{q}$ are sensitive to the shape of the Fermi surface, which is simplified in the present calculations. In actual pnictides, the shape of the hole pockets deviates from perfect circles and the spectrum of the bands depends on the transverse momentum $k_z$. However, the observation of the incommensurate SDW phase with $\mathbf{q}$ perpendicular to $\mathbf{Q}_0$ in the electron-doped Ba(Fe$_{1-x}$Co$_x$)$_2$As$_2$ was reported in Refs.~\cite{BonvilleEPL2010,LaplacePRB2009atomic,PrattPRL2011incommensurate} and double-peaked spin fluctuations corresponding to $\mathbf{q}$ parallel to $\mathbf{Q}_0$ have been observed in the hole-doped Ba$_{1-x}$K$_x$Fe$_2$As$_2$ compound in a wide doping range~\cite{LuoPRL2012coexistence,LeePRL2011incommensurate}. These observations are in agreement with the prediction of the model.

Thus, a simple model approximating the Fermi surface of pnictides implies the existence of electronic phase separation even in the weak-coupling regime. This is an important finding for the interpretation of the experimental data on phase inhomogeneity of iron pnictides: it proves that a purely electronic model with moderate interaction is sufficient to explain the observed inhomogeneities.

%%%%%%%%%%%%%%%%%%%%%%%%%%%%%%%%%%%%%%%%%%%%%%%%%%%%%%%%%%%%%%%

\section{Phase separation in graphene-based systems}
\label{graphene}

The formation of non-uniform electron states is a rather common feature for many complicated systems. In this section, we show that the reviewed above ideas of the phase separation can be implemented in the physics of graphene. We consider two examples. The first one is the so-called AA graphene bilayer and the second one deals with a graphene sheet with attached hydrogen atoms. If the hydrogenization of graphene is complete, the latter system is usually referred to as graphane. It turns out that AA graphene bilayer has the Fermi surface with evident nesting, whereas the electron system of hydrogenized graphene can be described in terms of the Falicov--Kimball model on the honeycomb lattice.

\subsection{AA graphene bilayer}\label{AAgraphene}

Graphene is a zero-gap semiconductor exhibiting a plethora of unusual electronic characteristics~\cite{NetoRMP2009electronic}. In addition to single-layer graphene, bilayer graphene is also actively studied~\cite{RozhkovPhRep2016}. This interest is driven by the desire to extend the family of graphene-like materials, which could be of interest for applications. Bilayer graphene exists in three stacking modifications. The most common is the so-called Bernal, or AB, stacking of bilayer graphene. In such a stacking, half of the carbon atoms in the top layer are located above the hexagon centers in the lower layer, and half of the atoms in the top layer lie above the atoms in the lower layer. A different layer arrangement, in which carbon atoms in the upper layer are located on top of the equivalent atoms of the bottom layer, is referred to as AA-stacked bilayer graphene (Fig.~\ref{AA_lattice}). The third modification is the twisted bilayer, in which the top graphene layer is rotated by some angle with respect to the bottom layer.

\begin{figure}[H]\centering
\includegraphics[width=0.24\columnwidth]{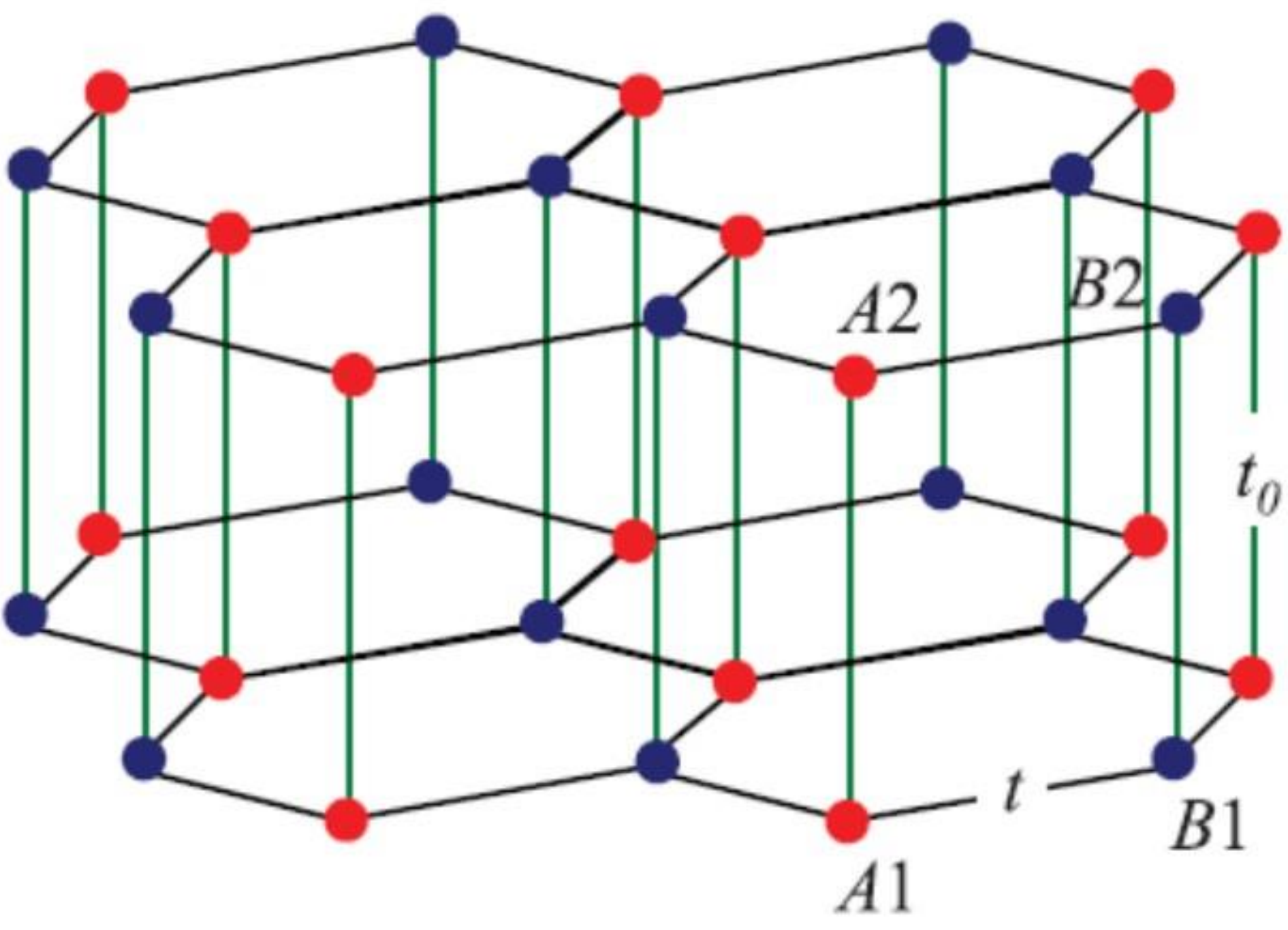}\\
\caption{(Color online) Crystal lattice of the AA graphene bilayer consists of two sublattices $A$ (red circles) and $B$ (blue circles). The unit cell of the crystal structure includes four atoms $A1$ and $B1$ in the top (1) layer and $A2$ and $B2$ in the bottom (2) layer. The values $t$ and $t_0$ are the in-plane and intrerplane nearest nearest-neighbor hopping integrals, respectively.}
\label{AA_lattice}
\end{figure}

\subsubsection{Single-electron spectrum}

The unit cell of the single-layer graphene crystal lattice consists of two carbon atoms. Consequently, two electron bands exist near the Fermi level~\cite{NetoRMP2009electronic}. The unit cell of the bilayer graphene crystal includes four carbon atoms, which forms sublattices $A1$ and $B1$ in the bottom (1) layer and $A2$ and $B2$ in the top (2) layers. The tight-binding analysis shows that both AA and AB bilayer have four bands (two hole bands and two electron bands). However, the structures of these bands are different~\cite{RozhkovPhRep2016}. In the undoped AB bilayer, if we neglect the so-called trigonal warping, two bands touch each other at two Fermi points, and the low-energy band dispersion is nearly parabolic. The AA-stacked bilayer graphene has two bands near the Fermi surface, one electron-like and another hole-like. The low-energy dispersion in the AA bilayer is linear, similar to the monolayer graphene. Unlike the latter, however, AA bilayer have Fermi surfaces instead of Fermi points (see Fig.~\ref{AA_band}). An important feature of the AA bilayer is that the hole and electron Fermi surfaces coincide in the undoped material, that is, there exist a perfect nesting, which becomes imperfect with doping. According the results reviewed in Section~\ref{imperfect}, in such a situation even small electron--electron coupling gives rise to the SDW instability resulting in  formation of the AFM commensurate and incommensurate ordering and to the phase separation (see also Refs.~\cite{Sboychakov_PRB2013_PS_AAgraph,RozhkovPhRep2016,
RakhmanovPRL2012instabilities}).

In the tight-binding approximation, the single-electron part of the Hamiltonian of AA-stacked graphene bilayer can be written as~\cite{RozhkovPhRep2016}
\begin{equation}\label{AA-Ham_0}
\hat{H}_0=-t\!\!\!\sum_{\langle\mathbf{nm}\rangle,i,\sigma}
{\!\!\left(a^\dag_{\mathbf{n}i\sigma}b_{\mathbf{m}i\sigma}\!
+\!\textrm{H.c.}\right)}
-t_0\sum_{\mathbf{n},\sigma}{\!\!\left(a^\dag_{\mathbf{n}1
\sigma}a_{\mathbf{n}2\sigma}+b^\dag_{\mathbf{n}1\sigma}
b_{\mathbf{n}2\sigma}\!+\!\textrm{H.c.}\right)}
-\mu\!\sum_{\mathbf{n},i,\sigma}
{\!\left(a^\dag_{\mathbf{n}i\sigma}a_{\mathbf{n}i\sigma}+
b^\dag_{\mathbf{n}i\sigma} b_{\mathbf{n}i\sigma}\right)},
\end{equation}
where $a^\dag_{\mathbf{n}i\sigma}$ ($b^\dag_{\mathbf{n}i\sigma}$) is the creation operator of an electron in sublattice $a$ ($b$) at site $\mathbf{n}$ in layer $i=1,2$ with spin projection $\sigma$, and the values $t$ and $t_0$ denote the in-plane and interplane nearest-neighbor hopping integrals, respectively. In graphene, next-nearest-neighbor hopping amplitudes are much smaller than $t$ and $t_0$ and hence, a longer-range hoping can be neglected~\cite{NetoRMP2009electronic,RozhkovPhRep2016}. The energy spectrum of Hamiltonian~(\ref{AA-Ham_0}) consists of four bands, which can be expressed in the form~\cite{RozhkovPhRep2016}
\begin{equation}\label{AA_spectrum}
\varepsilon^{1,2}_\mathbf{k}=-t_0\pm t|f_\mathbf{k}|,\quad \varepsilon^{3,4}_\mathbf{k}=t_0\pm t|f_\mathbf{k}|,
\end{equation}
where
\begin{equation}\label{fk}
f_\mathbf{k}=1+\exp{\!\!\left(\frac{3ik_xa_0}{2}\right)}
\cos{\!\!\left(\frac{\sqrt{3}k_ya_0}{2}\right)},
\end{equation}
and $a_0$ is the lattice constant for single-layer graphene.

The single-electron bands of the AA graphene bilayer (\ref{AA_spectrum}) are shown in Fig.~\ref{AA_band}. The bands $s = 2$ and 3 intersect the Fermi surface in the vicinity of the Dirac points $\mathbf{K}=2\pi(\sqrt{3},1)/3\sqrt{3}a_0$  and $\mathbf{K'}=2\pi(\sqrt{3},-1)/3\sqrt{3}a_0$ [see Fig.~\ref{AA_band}(b) and (c)]. When the doping is zero ($\mu = 0$, half filling), the Fermi energy is determined by the equation $|f_\mathbf{k}| = t_0/t$. Under condition $t_0/t\ll1$, the function $|f_\mathbf{k}|$ can be expanded near the Dirac points. Then, we obtain that the Fermi surface consists of two circles with radius $k_r = 2t_0/(3ta_0)$. In the case of a doped system, these Fermi surfaces transform into four circles. It is important that at half filling, the Fermi surfaces of the electron and hole bands are perfectly nested. This property is robust against changes in the Hamiltonian. For example, it survives if we take into account the longer-range hopping or consider a system with two nonequivalent layers~\cite{RakhmanovPRL2012instabilities}.

\begin{figure}[H]\centering
\includegraphics[width=0.4\columnwidth]{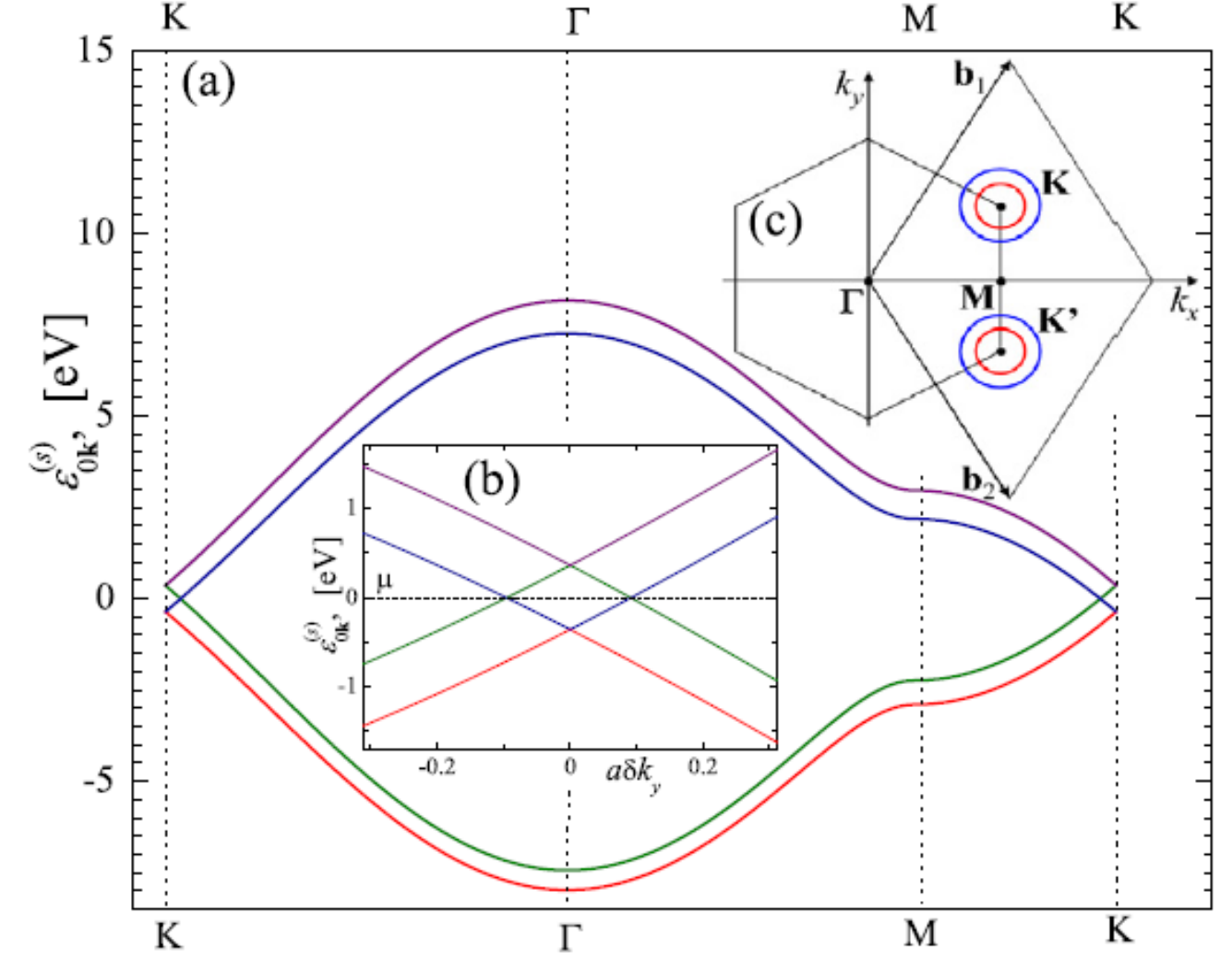}
\caption{(Color online) (a) Single-electron bands in the AA graphene bilayer. (b) The momentum dependence of the energy spectra $\varepsilon^s_\mathbf{k}$ near the Dirac point located at $\mathbf{K}$. Here, $\mathbf{k} = \mathbf{K} + \delta k_y\mathbf{e}_y$. The bands $s = 2$ and 3 intersect each other exactly at zero energy, which corresponds to the Fermi energy of the undoped graphene. (c) The first Brillouin zone (hexagon) and the reciprocal-lattice unit cell (rhombus) of the AA-stacked graphene bilayer. The circles near $\mathbf{K}$ and $\mathbf{K}'$ Dirac points show the Fermi surfaces of the doped sample~\cite{Sboychakov_PRB2013_PS_AAgraph}; $\varepsilon^{(s)}_{0\mathbf{k}}$ in this figure corresponds to $\varepsilon^s_\mathbf{k}$ in the present notation.}
\label{AA_band}
\end{figure}

\subsubsection{Symmetry-breaking phases} \label{sym-beak}

The spectrum with two bands having identical Fermi surfaces is unstable with respect to spontaneous symmetry breaking in the presence of a small electron--electron interaction. Such an instability opens a gap in the electronic spectrum, which gives rise to a decrease of the free energy of the system. We outline below only theoretical results since the experimental data on the electronic properties of the AA graphene bilayer now is nearly absent. The symmetries of the Hamiltonian can be used to make narrower the choices of the possible symmetry breakings and order parameters~\cite{RakhmanovPRL2012instabilities}. We can find the ground state of the system by minimization of the grand thermodynamic potential. We obtain that the ground state is the so-called G-type AFM order (all nearest neighbor spins in the lattice are antiparallel to each other) for parameter values characteristic of the AA graphene bilayer. The structure of the ground state is mainly controlled by the on-site Coulomb repulsion $U_0$.

The AFM order in AA-stacked graphene bilayer was studied in Refs.~\cite{RakhmanovPRL2012instabilities,Sboychakov_PRB2013_MIT_AAgraph,
Sboychakov_PRB2013_PS_AAgraph} using Hubbard  Hamiltonian with the interaction term in the form
\begin{equation}\label{aa_Hubb}
\hat{H}_{\textrm{int}}=\frac{U_0}{2}\sum_{\mathbf{m},i,\alpha,
\sigma}\left(n_{\mathbf{m}i\alpha\sigma}-\frac{1}{2}\right)\!
\left(n_{\mathbf{m}i\alpha\bar{\sigma}}-\frac{1}{2}\right).
\end{equation}
 Here $n_{\mathbf{m}i\alpha\sigma}$ is the operator of the particle number, $\alpha=A,B$ is the sublattice index, and $\overline{\sigma}$ means ``not $\sigma$``. Longer range Coulomb interaction was neglected. In this approach, it is commonly accepted that we should use the value for $U_0$ smaller than that predicted by the \textit{ab initio} calculations. A reasonable estimate is $U_0 = 5–-7$~eV~\cite{RozhkovPhRep2016}.

Thus, the commensurate order parameter can be written as
\begin{equation}\label{delta_com_AFM}
\Delta_{iA}=U_0\left\langle a^\dag_{\mathbf{m}i\uparrow}a_{\mathbf{m}i\downarrow}\right\rangle,\quad
\Delta_{iB}=U_0\left\langle b^\dag_{\mathbf{m}i\uparrow}b_{\mathbf{m}i\downarrow}\right\rangle,\quad \Delta_{1A}=\Delta_{2B}=-\Delta_{2A}=-\Delta_{1B}=\Delta,
\end{equation}
and $\Delta$ is real. The energy spectrum can be obtained by diagonalizing of $\hat{H}_0+\hat{H}_{\textrm{int}}$
\begin{equation}\label{mfAA_spectra}
\varepsilon_\mathbf{k}^{1,4}=\mp\sqrt{\Delta^2+(t|f_\mathbf{k}|+t_0)^2},\quad \varepsilon_\mathbf{k}^{2,3}=\mp\sqrt{\Delta^2+(t|f_\mathbf{k}|-t_0)^2}.
\end{equation}
The grand thermodynamic potential per unit cell can be written as
\begin{equation}\label{GrandAA}
\Omega=\frac{4\Delta^2}{U_0}-U_0(n^2-1)-2T\sum_{s=1}^4
{\int{\frac{d\mathbf{k}}{V_{BZ}}\ln{\left[1+e^{(\mu'-
\varepsilon^s_\mathbf{k})/T}\right]}}},
\end{equation}
where $n$ is the number of electrons per site, $x=n-1$ is the doping value, $V_{BZ}$ is the volume of the Brillouin zone, and $\mu'=\mu-U_0x/2$. To calculate the commensurate AFM gap we should minimize the grand thermodynamic potential~(\ref{mfAA_spectra}). Solving the obtained equation we can compute the gap as a function of doping, see (red) dot-and-dash line in Fig.~\ref{AAIncomGQ}. The gap decreases with the increase of the doping level $x$ and vanishes at some $x = x_c(T)$.

\begin{figure}[H]\centering
\includegraphics[width=0.44\columnwidth]{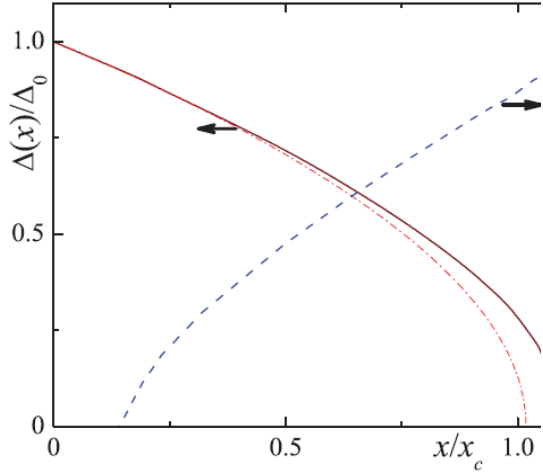}\\
\caption{(Color online) The AFM gap $\Delta$ (red solid curve) and $|q|$ (blue dashed curve) as functions of the doping $x$, calculated for $T/\Delta_0 = 0.06$ and $U_0 = 8$~eV. The gap calculated in the case of the commensurate AFM order, $q = 0$, is shown by the dot-and-dash curve. The value $x$ is normalized by the critical doping $x_c$ calculated for the commensurate AFM phase. The incommensurate AFM order occupies a slightly larger doping range than the commensurate one~\cite{Sboychakov_PRB2013_PS_AAgraph}.}
\label{AAIncomGQ}
\end{figure}

The state with G-type AFM order has the smallest value of the free energy among the commensurate magnetic phases. Further minimization of free energy can be achieved if we consider a incommensurate SDW ordering (see Section~\ref{imperfect}). We can write the complex order parameter for this state in the form~\cite{RozhkovPhRep2016}
\begin{equation}\label{delta_incom_AFM}
\Delta_{\mathbf{n}iA}=e^{i\mathbf{qn}}\Delta_{iA}, \qquad \Delta_{\mathbf{n}iB}=e^{i\mathbf{qn}}\Delta_{iB},
\end{equation}
where $\mathbf{q}$ describes the spatial variation of the direction of the AFM vector. The mean value of the electron spin $\mathbf{S}_{\mathbf{n}i\alpha}$ (in the lattice site $\mathbf{n}$, layer $i$, and sublattice $\alpha$) lies in the ($x,y$) plane. The equation $\mathbf{S}_{\mathbf{n}i\alpha}=\Delta_{i\alpha}[\cos{\mathbf{qn}},\sin{\mathbf{qn}}]/U_0$ relates the averaged spin and the order parameter. Now, the thermodynamic potential becomes a function of the vector $\mathbf{q}$. Thus, we should add the minimization condition $\partial\Omega/\partial\mathbf{q}=0$ to find the value of $\mathbf{q}$ in the ground state. The corresponding calculation was performed in Ref.~\cite{Sboychakov_PRB2013_PS_AAgraph}. The results for the gap $\Delta(x)$ and $|\mathbf{q}(x)|$ are presented in Fig.~\ref{AAIncomGQ}. We observe that the incommensurate AFM phase exists in a slightly wider range of doping than the commensurate one.

\subsubsection{Phase separation}

The phase separation is an inherent property of multiband electron systems and systems with nesting especially, as it was discussed in Section~\ref{imperfect}. According the results obtained in Refs.~\cite{Sboychakov_PRB2013_MIT_AAgraph,Sboychakov_PRB2013_PS_AAgraph}, the AA-stacked graphene bilayer can be separated into two phases with unequal electron densities $n_{1,2} = 1+x_{1,2}$ at non-zero doping. The chemical potential of the AA graphene bilayer depends nonmonotonically on doping and at nonzero temperature has the form similar to that shown in Fig.~\ref{RiceChemPot} for both commensurate and incommensurate AFM states provided the onsite Coulomb repulsion is not strong, $U_0<\pi\sqrt{3}t^2/t_0$. Therefore, the system is unstable and undergos the phase separation into commensurate ($\mathbf{q} = 0$, $x_1 < x$) and incommensurate ($\mathbf{q} \neq 0$, $x_2 > x$) AFM phases. The doping values $x_{1,2}$ can be calculated using the Maxwell construction.

In Fig.~\ref{AAPhaseDiag} we present the phase diagram of the proposed model of the AA-stacked graphene bilayer in the ($x,T$) plane for two values of the onsite Coulomb repulsion $U_0$. The boundary of the region of the phase separation is shown by (green) dot-and-dash curves. It is known that the incommensurate order is sensitive to disorder. In addition, the difference in the free energy between the incommensurate and commensurate AFM states is small. Therefore, if we ignore the possibility of an incommensurate AFM order, then, the phase separation occurs between the AFM insulator ($x_1=0$) and the metal ($x_2 > 0$), either PM ($U_0>6$~eV) or AFM ($U_0 < 6$~eV).

\begin{figure}[H]\centering
\includegraphics[width=0.6\columnwidth]{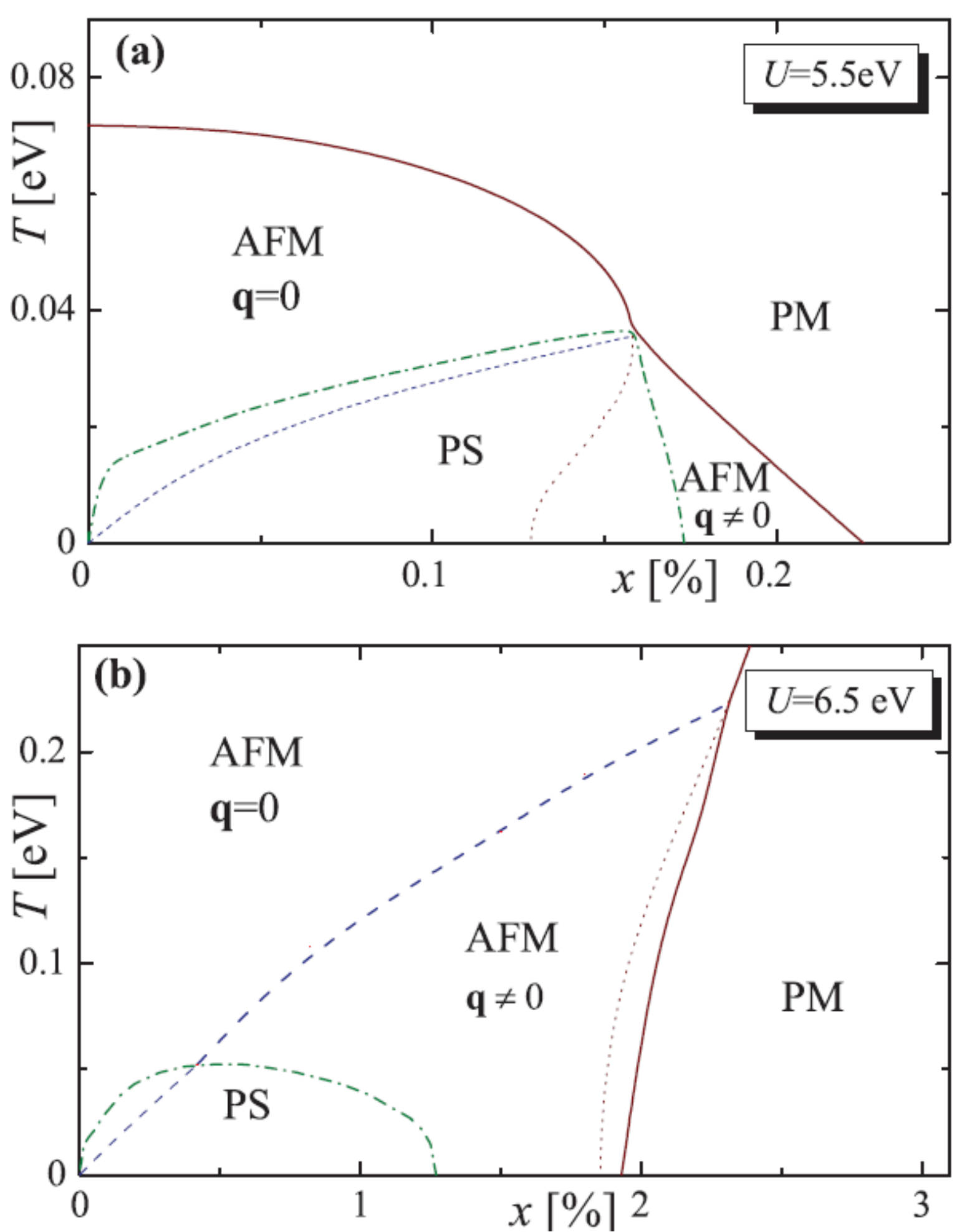}\\
\caption{(Color online) Phase diagram of the AA graphene bilayer; $U_0 = 5.5$~eV (a) and $U_0 = 6.5$~eV (b). Solid (red) curves show the transition temperature $T_{MF}(x)$ from AFM to PM phases calculated in the mean-field approach; (blue) dotted (a) and dashed (b) curves indicate the commensurate--incommensurate transition temperatures $T^q(x)$. The values of $T_{MF}(x)$ calculated without taking into account the incommensurate order are shown by the dotted (red) curves. The boundaries of the phase separated states are shown by the dot-and-dash (green) curves~\cite{RozhkovPhRep2016}.}
\label{AAPhaseDiag}
\end{figure}

If the electron-rich phase is a metal and the electron-poor phase is an insulator, then the insulator--metal percolation transition occurs when the doping level $x$ exceeds a percolation threshold. This threshold is about $0.5(x_1 + x_2)$ in the case of the considered 2D systems.

As it was discussed above, the phase separation will be destroyed by the long-range Coulomb repulsion since the separated phases have different electron densities. In the case of 2D system like considered here graphene, the long-range Coulomb interaction can be significantly affected by the environment, in particular, by the sample substrate. This allows for a tuning the insulator--metal transition in graphene systems.

Temperature range where the phase separation could be observed depends on the value $U_0$ and varies from 30--40~K to room or even higher temperatures (see Fig.~\ref{AAPhaseDiag}). The phase separation region shrinks along the doping axis with the growth of $U_0$, while the temperature range, where the phase separation (as well as AFM ordering) is observed, increases with $U_0$. However, one should keep in mind that the mean-field approach is not a good approximation in the case of large $U_0$.

\subsection{Twisted bilayer graphene} \label{TBLG}

Soon after the discovery of the graphene, it became clear that two graphene planes superimposed on one another (bilayer graphene) are no less interesting object than the single-layer graphene. The number of works dealing with it is already in the thousands. To understand, how vast this area is, one can just look at the review article~\cite{RozhkovPhRep2016}. For example, it is sufficient only to rotate the graphene layers relative to each other by a small angle, and a modulated structure will immediately appear, reminiscent of a moir{\'e} pattern in some of silken fabrics. At small angles of rotation, the unit cell of such a moir{\'e} superlattice can be very large (up to several thousands atoms).

An essential step in the study of such a twisted bilayer graphene was made in \cite{CaoNature2018,CaoNature2018_SC}. The authors of these works managed to develop a technology for making samples, in which the angle of rotation of the layers can be controlled with an accuracy of tenths of a degree. Such bilayer sample was placed in a structure that made it possible to control the density of charge carriers using the gate voltage. The researchers were interested in the ``magic angles" of rotation predicted earlier by theorists~\cite{BistritzerPNAS2011}. At the magic angle, the kinetic energy of electrons near the Dirac points is comparable to the energy of interlayer hybridization. As a result, the Fermi velocity drops to almost zero, and a flat (almost dispersionless) bands with a high electron density of states and a large effective mass of charge carriers appear near the Fermi level.

Band calculations show that the first such magic angle is approximately equal to  1.08$^{\circ}$. The authors of \cite{CaoNature2018,CaoNature2018_SC} made several samples with rotation angles close to this value and carried out their multifaceted study. First of all, they focused on the effects associated with the flat bands. Indeed, if the band is very narrow, then even with a relatively small Coulomb repulsion of electrons at one lattice site, we are dealing with the limit of strong electron correlations. Then, even when the conduction band is half-filled (one electron per site), a gap appears in the spectrum of elementary excitations, since the presence of two electrons at one site is extremely disadvantageous due to their strong Coulomb repulsion. This state with a gap is in fact a Mott insulator, and it was actually observed at magic angles. When the effective number of charge carriers is changed by the gate voltage, the gap in the spectrum manifests itself not only for completely filled minibands (arising due to the moir{\'e} superlattice), but also for half-filled ones, which is a clear manifestation of the Mott insulator state. Moreover, in twisted bilayer graphene, this state has no signs of any magnetic ordering, in contrast to cuprates, where it is antiferromagnetic. What is most curious is that when deviating from half-filling in one direction or another, graphene layers turned by a magic angle, become superconducting. The maximum superconducting transition temperature $T_c$ observed in \cite{CaoNature2018_SC} is 1.7 K. It seems to be low, but there are few charge carriers in the miniband. Therefore, the ratio of $T_c$ to the Fermi energy turns out to be higher than in cuprates, which indicates the relative strength of superconducting pairing. Thus, in the phase diagram, on both sides of the Mott insulator, there appear two superconducting ``domes". However, the prevalence of disorder-induced experimental features identified in the samples under study suggests that the correlated physics of twisted bilayer graphene is extremely sensitive to the structural details of the moir{\'e} pattern. For example, the applied pressure can be used for controlling the interparticle distance and thus the conditions for the formation of the flat bands. Indeed, more than 1 GPa of hydrostatic pressure induced superconductivity in a twisted bilayer device with a larger twist angle of 1.27$^{\circ}$ that did not otherwise exhibit any correlated behavior~\cite{YankowitzScience2019}. There also appeared the evidence that near three-quarters filling of the conduction miniband the electron correlations drive the twisted bilayer graphene into a ferromagnetic state. Near this 3/4 band filling, the system also exhibits a pronounced anomalous Hall effect and some indications of chiral edge states~\cite{SharpeScience2019}. The enhanced quality of twisted bilayer graphene samples allowed revealing various phases near the magic angle at different numbers of charge carriers per unit cell of the moir{\'e} superlattice: four superconducting domes and a number of correlated states between the domes, three of them were insulating, and three others seemed to be semimetallic; the noninsulating states were also topologically nontrivial ~\cite{LuNature2019}. Moreover, the screening of electron--electron interaction by introducing a metallic layer located close to twisted bilayer graphene provides a tool for controlling and tuning the form of the phase diagram~\cite{StepanovNature2020}.

The existence of various competing phases usually suggests a possibility of the phase separation. Indeed the capacitance measurements for the twisted bilayer graphene device reported in ~\cite{TomarkenPRL2019,ZondinerNature2020} demonstrate quite nonmonotonic behavior of the electronic compressibility $d\mu/dn$, where $\mu$ is the chemical potential and $n$ is the charge carrier density, as function of the number $\nu$ of charge carriers per the moir{\'e} unit cell. The corresponding plots exhibit the ranges of negative compressibility, whereas the chemical potential itself have the ranges with the negative curvature (see Fig.~\ref{mu_TBLG}). The latter is a clear evidence of the instability of homogenous phases and hence of the electronic phase separation, which could be not easy to reveal owing to large sizes of unit cells in moir{\'e} superlattices.

\begin{figure}[H] \centering
\includegraphics[width=0.3\columnwidth]{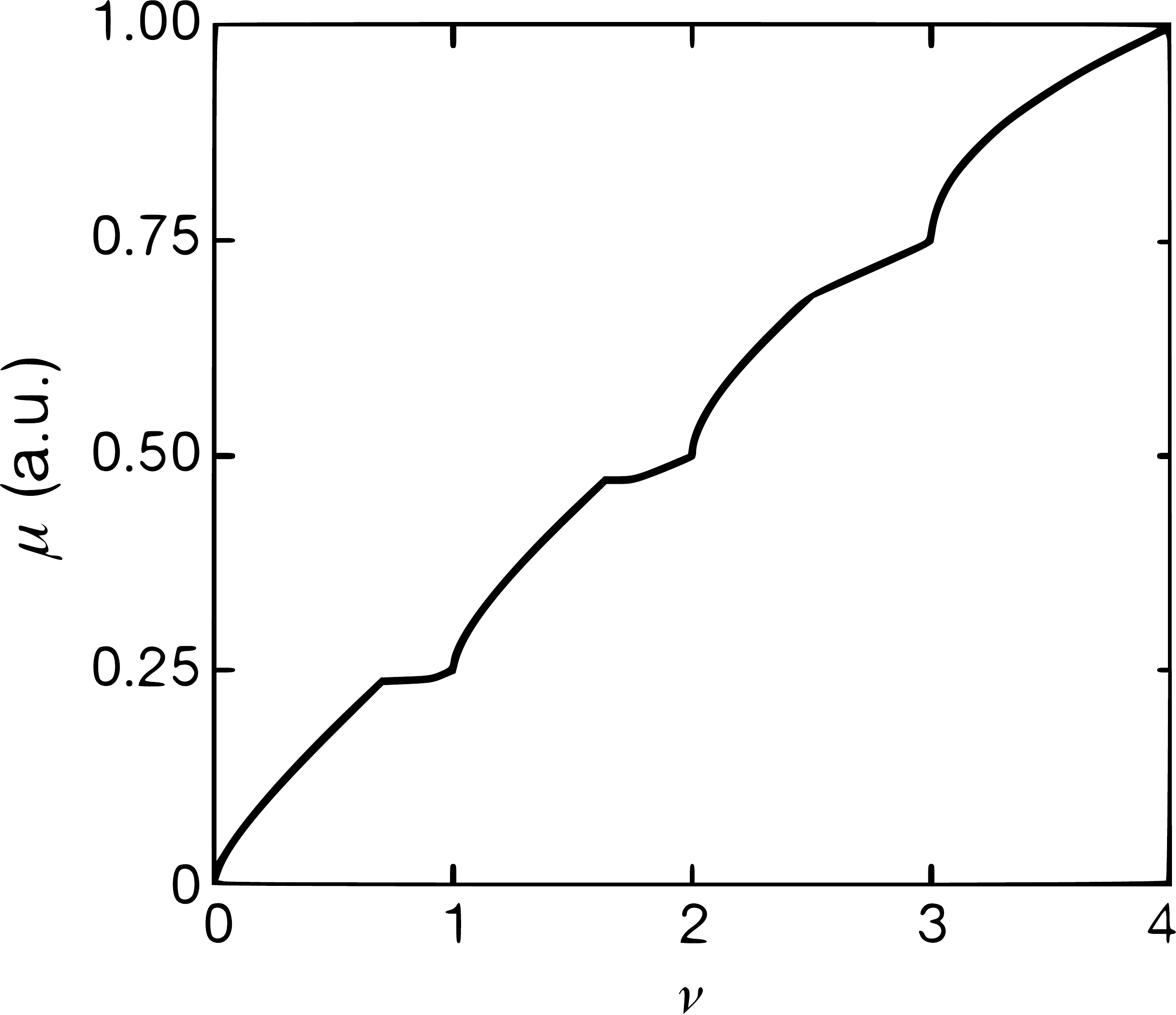}\\
\caption{ Chemical potential $\mu$ of twisted bilayer graphene as function of the number $\nu$ of charge carriers per the moir{\'e} unit cell.~\cite{ZondinerNature2020}}\label{mu_TBLG}
\end{figure}

The theoretical analysis of the possible inhomogeneous states in twisted bilayer graphene near the magic angle was undertaken in ~\cite{SboychakovJETPL2020}. It was assumed that the emerging nonsuperconducting order parameter is a spin density wave, and the evolution of such ordered state with doping was studied. It was shown that in the range of electron densities, where the order parameter is nonzero, the homogeneous state of the system can be unstable with respect to the phase separation. Phases in the inhomogeneous state are characterized by an even number  of electrons per a superlattice cell. This allows explaining some features in the behavior of the conductivity of the system with doping. Thus, it becomes possible to explain the fact that the conductivity minima, that could occur at doping levels corresponding to an odd number of electrons per supercell, are absent in some samples under study (phase separation occurs) and are present in other samples (phase separation is suppressed by the long-range Coulomb repulsion).

\subsection{Phase separation of hydrogen atoms adsorbed on graphene}\label{graphan}

Fully hydrogenated graphene is commonly referred to as graphane~\cite{sluiter2003cluster}. Graphane is an insulator. Its gap is about 5--6~eV.  In a sheet of graphane we can create a graphene patch of desired shape using a local dehydrogenation. In such a system low-energy charge carriers from graphene cannot penetrate the graphane host. Thus, the boundary between graphene and graphane can be considered as an effective edge of the graphene structure. Therefore, the stability of the graphene--graphane interface is an important issue. The numerical study indicates that such an interface is stable and that the adsorbed hydrogen atoms tend to join in clusters~\cite{OpenovJETPL2009spontaneous,AoAPL2010,roman2009high}. These features can be described in terms of the phase separation into hydrogen-rich and hydrogen-free regions. A specific type of coupling between electrons and adatoms is assumed in Ref.~\cite{shytov2009long} to demonstrate the phase separation. This assumptions are supported by some experimental and numerical arguments. In Ref.~\cite{rakhmanov2012phase}, the phase separation is discussed within the framework of a modified Falicov--Kimball model. The advantage of the latter approach is its generality since the phase separation is a well-known property of a ground state of the Falicov--Kimball model (Section~\ref{twobands}) and this property is robust against variation of the model details.

The model Hamiltonian for graphane can be written in the form~\cite{rakhmanov2012phase},
\begin{equation}\label{graphan_Ham}
\hat{H}_{A}=\hat{H}_E-\sum_{i,\sigma}{\left[t_{CH}
\left(P^\dag_{i\sigma}S_{i\sigma}+H.c.\right)+
\varepsilon_HS^\dag_{i\sigma}S_{i\sigma}+H.c.\right]},
\qquad \hat{H}_E=-t\sum_{i,j,\sigma}{\left(P^\dag_{i\sigma}
\hat{T}_{ij}P_{j\sigma}+H.c.\right)},
\end{equation}
where $P^\dag_{i\sigma}=(p^{A\dag}_{i\sigma},p^{B\dag}_{i\sigma})$, $S^\dag_{i\sigma}=(s^{A\dag}_{i\sigma},s^{B\dag}_{i\sigma})$, $\sigma$ is the spin projection. The Hamiltonian $\hat{H}_E$ ($\hat{H}_A$) corresponds to graphene (graphane). The Hamiltonian $\hat{H}_E$ is the usual Hamiltonian for the single-layer graphene rewritten in a convenient form for further consideration. It describes the hopping of $p_z$ electrons between nearest carbon atoms. Subscripts $i$ enumerate the unit cell of the crystall lattice. The spinor component $A$ ($B$) corresponds to a site on the $A$ ($B$) sublattice. In the spinor representation, the hopping matrix $\hat{T}_{ij}$ reads in the momentum space
\begin{equation}\label{Tij}
\hat{T}_{ij}=\left(
               \begin{array}{cc}
                 0 & f_\mathbf{k} \\
                 f^*_\mathbf{k} & 0 \\
               \end{array}
             \right).
\end{equation}
The Hamiltonian $\hat{H}_A$ describes the hybridization of the $p_z$ electrons of carbon and the $s$ electrons of hydrogen. Other bands are neglected.

The $p_z$--$s$ electron coupling constant $t_{CH} = 5.8$~eV is larger than the the carbon--carbon hopping amplitude $t = 2.8$~eV and the energy shift of the hydrogen $s$ orbital $\varepsilon_H = 0.4$~eV. The latter will be further neglected.

It is easy to diagonalize Hamiltonian $\hat{H}_A$ and to obtain four graphane bands~\cite{rakhmanov2012phase},
\begin{equation}\label{bands_graphane}
\varepsilon_m=\frac{1}{2}\left(\pm t|f_\mathbf{k}|
\pm\sqrt{4t_{CH}^2+t^2|f_\mathbf{k}|^2}|\right), \quad m = 1,2,3,4.
\end{equation}
In the model Hamiltonian, Eq.~(\ref{graphan_Ham}), only four bands are taken into account. However, this Hamiltonian captures the main features of graphane. In particular, at zero doping, Eq.~(\ref{graphan_Ham}) describes an insulator with a gap $E_g \simeq 6$~eV located at the $\Gamma$ point. This is consistent with the results of \textit{ab initio} calculations~\cite{LebeguePRB2009accurate}.

The numbers of hydrogen and carbon atoms are equal in the graphane and the Hamiltonian Eq.~(\ref{graphan_Ham}) should be modified if the hydrogen atoms are absent at some sites. The hydrogen $s$ orbitals are not available for electrons in these sites. This constraint may be mimic by an infinitely strong repulsion between hole and electron at the corresponding $s$ orbital
\begin{equation}\label{CH_graphane}
\hat{H}_{EA}=\hat{H}_{A}+U\sum_{i,\sigma}{S^\dag_{i\sigma}
\hat{N}_iS_{i\sigma}},\qquad \hat{N}_i=\left(
               \begin{array}{cc}
                 n_{Ai} & 0 \\
                 0 & n_{Bi} \\
               \end{array}
             \right),
\end{equation}
where $U\rightarrow\infty$ and $n_{A,Bi}$ are the numbers of hydrogen holes at site $i$. The  numbers $n_{A,Bi}$ randomly take the values 0 or 1. The mean value $\langle n_{A,Bi}\rangle=n^H$, where $n^H$ is the number of the hydrogen holes per one carbon atom. The obtained Hamiltonian $\hat{H}_{EA}$ is a version of the Falicov--Kimball model, in which mobile $p$ and $s$ electrons interact with immobile holes whose concentration $n_H$ is fixed externally.

Physically, the mechanism of the phase separation can be explained on the basis of the follows simple arguments~\cite{rakhmanov2012phase}. It is reasonable to consider a limit $t_{CH} \gg t$ since $t_{CH}$ is significantly larger than $t$. We introduce operators $a$ and $b$ diagonalizing the terms in $\hat{H}_{EA}$, which do not include the carbon--carbon electron hopping,
\begin{equation}\label{diag_graphane}
p_\alpha=\frac{b_\alpha-a_\alpha}{\sqrt{2}}\left(1-n_\alpha\right)+n_\alpha b_\alpha,\quad
s_\alpha=\frac{b_\alpha+a_\alpha}{\sqrt{2}}\left(1-n_\alpha\right)+n_\alpha a_\alpha,
\end{equation}
where index $\alpha$ labels the carbon atoms. Further expressions are the same for any $A$, $B$, and $\sigma$ and here we omit the sublattice and spin indices. Substituting Eq.~(\ref{diag_graphane}) in Hamiltonian~(\ref{CH_graphane}), we obtain that the on-site energy of the quasi-particles $a$ is much higher than that of $b$ if $t_{CH},U \gg t$. Thus, in the first approximation in $t/t_{CH}\ll 1$, these states are empty and can be neglected. In the considered limit we have
\begin{equation}\label{CH_graphane_approx}
\hat{H}_{EA}\approx -t_{CH}\sum_\alpha{b^\dag_\alpha b_\alpha(1-n_\alpha)}-\frac{t}{2}
\sum_{\langle\alpha\beta\rangle}{b^\dag_\alpha b_\beta\left[1+\gamma(n_\alpha+n_\beta)+\gamma^2n_\alpha n_\beta\right]},
\end{equation}
where $\gamma=\sqrt{2}-1$ and $\langle\alpha\beta\rangle$ denotes summation only over the nearest neighbors. It follows from this equation that the separation of two hydrogen holes located at neighboring sites costs an energy of the order of $t\gamma^2\langle b^\dag_\alpha b_\beta\rangle$. This evidently means an effective attraction between the hydrogen holes and, consequently, between the hydrogen atoms as in the model proposed in Ref.~\cite{shytov2009long}. The attraction between the hydrogen atoms induces the phase separation.

Hamiltonian~(\ref{CH_graphane}) was analyzed in Ref.~\cite{rakhmanov2012phase} using the exact diagonalization of finite clusters and the Hubbard-I approximation. Both approaches predict the existence of the phase separation.

The results are shown in Fig.~\ref{Graphane}. The system energy $E$ as a function of the concentration $n^H$ of the hydrogen adatoms has negative curvature in the whole range of $n^H$. This feature indicates on the possible instability of the homogeneous state and the phase separation. The free energy of the phase-separated state can be calculated using the Maxwell construction. In the considered cade, this construction is simply a single straight line which connects the energy of the graphene ($n^H = 0$) and the energy of the graphane ($n^H = 1$). Therefore, the separated states are graphene and graphane.

The phase separation occurs due to the redistribution of neutral hydrogen atoms. Thus, there is no violation of the charge neutrality and the geometrical structure of the inhomogeneous state depends only on the value and sign of the interface tension $\sigma_0$ between graphene and graphane. In the case of $\sigma_0<0$, it is favorable the existence of small clusters in the inhomogeneous state to maximize the boundary length. If $\sigma_0>0$, it is favorable to minimize the length of the graphene--graphane boundary.

 \begin{figure}[H]\centering
\includegraphics[width=0.6\columnwidth]{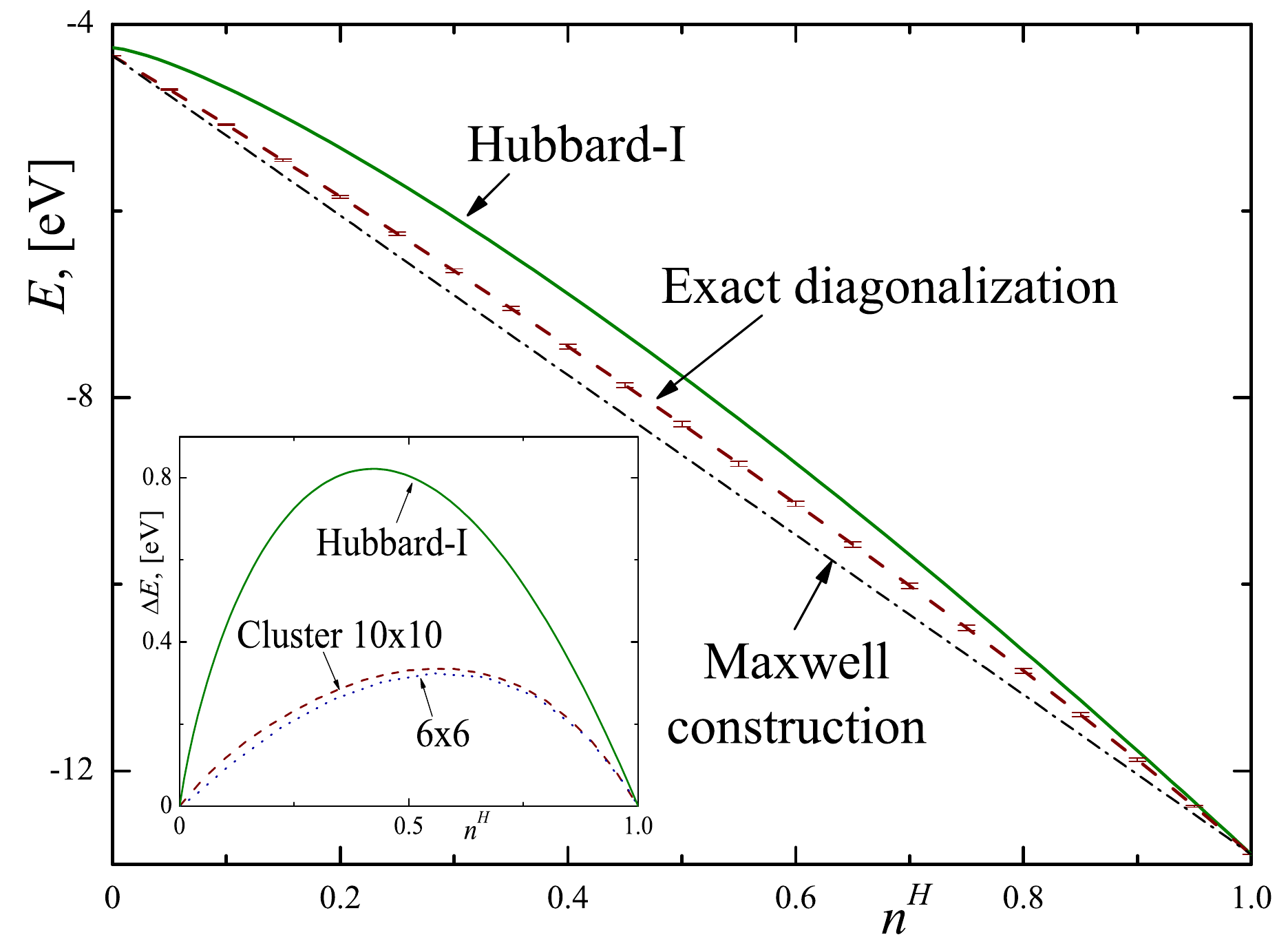}\\
\caption{(Color online) Electron energy $E$ as function of the concentration of hydrogen adatoms $n^H$; green solid curve -- calculation in the Hubbard-I approximation and red dashed curve -- exact diagonalization of cluster consists of $10\times10$ unit cells. The negative curvature of $E(n^H)$ indicates on the possible instability of the homogeneous system toward the phase separation into phases with $n^H = 0$ and $n^H = 1$. The blue dot-and-dash line shows the Maxwell construction. The difference in the energy between the homogeneous and phase-separated states is shown in the inset; green solid curve -- calculation in the Hubbard-I approximation, red dashed line -- exact diagonalizations of $10\times10$ cluster, and blue dotted curve -- exact diagonalizations of $6\times6$ cluster. The model parameters are: $t_{CH} = 5.8$~eV, $t = 2.8$~eV, and $\varepsilon_H = 0.4$~eV. For exact diagonalizations, we use $U = 400$~eV~\cite{rakhmanov2012phase}.}
\label{Graphane}
\end{figure}

Electrons in graphane are mainly localized at the C--H valence bonds since $t\ll t_{CH}$ and their contribution to the interface tension $\sigma_0$ is small. The electrons in graphene move from one carbon atom to its nearest neighbors but they cannot penetrate to graphane since electron have to overcome the energy gap, which is about $t_{CH}$. Therefore, each carbon--carbon bond connecting an atom in graphene with an atom in graphane does not contribute to the kinetic energy of the charge carriers in graphene. Effectively, this is is equivalent to an increase of the  electron kinetic energy in graphene. The number of broken bonds increases with the interface length. Thus, we have an estimate $\sigma_0\sim \varepsilon_b/a_0$, where $\varepsilon_b$ is the kinetic energy per single carbon--carbon bond
\begin{equation}\label{kinenGraphane}
\varepsilon_b=\frac{2t}{3}\int{d^2\mathbf{k}
\frac{V_{BZ}|f_\mathbf{k}|}{(2\pi)^2}}\sim t,
\end{equation}
where the integration is carried out over the first Brillouin zone, $V_{BZ}=3\sqrt{3}a_0^2/2$, the factor 2 arises due to two spin projections, and the factor of 1/3 appears here since there are three bonds in a graphene unit cell. Thus, $\sigma_0\sim t/a_0>0$. An accurate numerical computation gives $\sigma_0\approx 0.2t/a_0\approx 0.6$~eV~\cite{rakhmanov2012phase}. In case ubder study, we can neglect the effects of temperature on the phase separation since the characteristic energies of the problem are much higher than $k_BT$ for any realistic $T$.

Thus, the theoretical description of the partially hydrogenated graphene can be reduced to a Falicov--Kimball-like model. The thermodynamically stable state of this system is inhomogeneous. All hydrogen atoms are clustering together. As a result in the ground state we have two phases: pure graphane and pure (hydrogen-free) graphene. The graphane-graphene interface tension is large and positive. This means that the graphane-graphene boundary is stable and flat (privided the number of hydrogen adatoms in the sample is not small). Note, that relaxation time of the partially hydrogenated graphene to the ground state could be large.

\newpage
\section{Conclusions. Some additional problems}
\label{Concl}

We have demonstrated that during the passed half century, the field of electronic phase separation has evolved from an exotic effect manifesting itself in some magnetic semiconductors to a very broad range of phenomena inherent to all strongly correlated electron systems and not only to them. This research field is still rapidly progressing. This review article touches upon mostly the topics, to which we have contributed ourselves. Of course, some important issues related to the formation of spin and charge inhomogeneities turn out to be outside the scope of our review. Let us list some of them.

A very popular and well studied topic concerns the dynamic and static stripe structures in cuprates, nickelates, and other oxide compounds. There is a vast body of literature concerning this topic, see e.g., comprehensive reviews~\cite{BianconiStripes_Book2013,TranquadaNJPh2009,VojtaAdvPh2009}.
In addition to oxides, the existence of stripes has been also discussed recently for iron-based superconductors (where the stripe structures are closely related to the electron nematicity~\cite{FernandesNatPh2014}) and in borides~\cite{DemishevSciRep2017,SluchankoPRB2018}.

Another large and important issue is the formation of charge inhomogeneities owing to a random potential generated by impurities. This line of research has been especially active in connection to the two-dimensional electron gas in semiconductor heterostructures~\cite{TripathiPRB2011,PudalovPRL2012,PudalovPRB2016}. Such disorder-induced inhomogeneities are conceptually related to the long-standing problem of the insulator--superconductor transition accompanied by the formation of superconducting islands within a disordered insulator~\cite{GoldmanPhTod1998,DubiNature2007,DubiPhC2008,PoranNatCom2017}.
There are also vast research fields concerning heterophase fluctuations in different materials~\cite{YukalovPhRep1991,YukalovIJMPB2003} and the phase separation in ferroelectrics and multiferroics~\cite{YukalovJSCNM2016}.

A very interesting topic concerns the phase separation under nonequilibrium conditions (see e.g., \cite{ParmigianiPRB2016}), which has attracted a vivid interest of chemists and biologists~\cite{TrefzJChPh2016,DickmanNJPh2016}.

Note also that the recent discovery of the inconventional superconductivity in the twisted bilayer graphene at small twist angles~\cite{CaoNature2018_SC,CaoNature2018} has added a new impetus to the studies of inhomogeneous states in graphene-based systems.

Among more exotic possibilities, we could mention the phase separation in the spin-imbalanced  Fermi gas~\cite{FeiguinPRA2010,MitraPRL2016} and inhomogeneous states in the Fermi--Bose mixtures~\cite{MenushenkovJETP2001,MenushenkovJScNMag2016}. Quite recently, the formation of the superconducting droplets of the local bielectron pairs in the normal matrix of the unpaired electron states in the 2D attractive-$U$ Hubbard model in the quasibosonic limit of strong onsite
attraction $|U|\gg W$, low electron densities $n \ll 1$, and in the presence of strong diagonal disorder $V \gg W$ ($W$ and $V$ are the widths of the ``clean" and ``dirty" bands, respectively) was demonstrated in \cite{KaganMazurArxiv2020}.

The last but not the least issue is the application of the phase-separation concept to the nuclear matter including such scenarios as nuclear pasta-like structures~\cite{MaruyamaPRC2005,MaruyamaPRC2006} and  inhomogeneities in  rotating neutron stars~\cite{BastrukovAstroph1999} (similar to the situation with the vortices in superfluid helium~\cite{AndronRMP1966}).

We see that the problem of the phase separation is so vast and multifaceted that it cannot be covered within one, even large, treatise. Nevertheless, we believe that our paper provides a comprehensive overview of an important part of this problem.

\section*{Acknowledgements}
This work was supported by the Russian Foundation for Basic Research, project nos. 19-52-50015, 19-02-00421, 19-02-00509, and 20-02-00015. M.Yu.K. acknowledges the support from the Program of Basic Research of the National Research University Higher School of Economics.

\newpage

\bibliographystyle{apsrevlong_no_issn_url}

\section*{References}

%\bibliography{allrefs_new}

\end{document}